\documentclass[
twocolumn,
amsmath,amssymb,
aps,
longbibliography, superscriptaddress
]{revtex4-2}
\usepackage[utf8]{inputenc}
\usepackage{amsmath}
\usepackage{amsfonts}
\usepackage{amssymb}
\usepackage{braket}
\usepackage{graphicx}
\usepackage{float}
\usepackage{tikz}
\usepackage{bm}
\usetikzlibrary{arrows,decorations.pathmorphing,backgrounds,positioning,fit,petri,decorations.pathreplacing,decorations.markings}
\usepackage{xcolor}
\definecolor{mygreen1}{rgb}{0, 0.4, 0}
\usepackage[margin=0.7in]{geometry}
\newcommand{\mbeq}{\overset{!}{=}}
\usepackage{pifont}
\usepackage{hyperref}
\hypersetup{
	colorlinks=true,
	linkcolor=blue,
	filecolor=magenta,
	urlcolor=cyan,
	citecolor=blue
}

\begin{document}
	
	\title{Twisted Lattice Gauge Theory: Membrane Operators, Three-loop Braiding and Topological Charge}
	\author{Joe Huxford}
 \affiliation{Department of Physics, University of Toronto, Ontario M5S 1A7, Canada}
    \author{Dung Xuan Nguyen}
    \affiliation{Center for Theoretical Physics of Complex Systems, Institute for Basic Science (IBS), Daejeon 34126, Republic of Korea}
    \author{Yong Baek Kim}
 \affiliation{Department of Physics, University of Toronto, Ontario M5S 1A7, Canada}

	\begin{abstract}
		
		3+1 dimensional topological phases can support loop-like excitations in addition to point-like ones, allowing for non-trivial loop-loop and point-loop braiding statistics not permitted to point-like excitations alone. Furthermore, these loop-like excitations can be linked together, changing their properties. In particular, this can lead to distinct three-loop braiding, involving two loops undergoing an exchange process while linked to a third loop. In this work, we investigate the loop-like excitations in a 3+1d Hamiltonian realization of Dijkgraaf-Witten theory through direct construction of their membrane operators, for a general finite Abelian group and 4-cocycle twist. Using these membrane operators, we find the braiding relations and fusion rules for the loop-like excitations, including those linked to another loop-like excitation. Furthermore, we use these membrane operators to construct projection operators that measure the topological charge and show that the number of distinct topological charges measured by the 2-torus matches the ground state degeneracy of the model on the 3-torus, explicitly confirming a general expectation for topological phases. This direct construction of the membrane operators sheds significant light on the key properties of the loop-like excitations in 3+1 dimensional topological phases.

	\end{abstract}
	\maketitle
	\section{Introduction}

Among the rich tapestry of phases of matter that are known to us, topological phases stand out due to their ability to support quasiparticles with exchanges statistics that generalize those of fermions and bosons \cite{Leinaas1977, Wilczek1982, Arovas1984}. This property of topological phases is enabled by long-ranged entanglement between the local degrees of freedom, which cannot be disentangled by local unitary evolution over finite time \cite{Chen2010, Chen2013, Wen2013} and which allows the quasiparticles to transform non-trivially under exchange, even when the particles remain well separated. The form of these exchange statistics depends on the dimensions of space and the topology of the excitations themselves. In 2+1d, the exchange of point-like excitations is described by the (colored) braid group. On the other hand, in 3+1d, the exchange of such particles is described only by the permutation group, leading to the familiar bosonic and fermionic statistics \cite{Doplicher1971, Doplicher1974, Rao1992, Nayak2008}. If this were the full picture, then topological phases in 3+1d would not be nearly so interesting as their 2+1d counterparts. However, 3+1d phases can support loop-like excitations in addition to point-like ones, allowing for loop-loop and point-loop braiding that is described by the (colored) loop braid group \cite{Aneziris1991, Alford1992, Baez2007, McCool1986, Savushkina1996, Damiani2017}. Furthermore, loop-like excitations can be linked together and it has become apparent that the character of the loop-like excitations can be different in such a case \cite{Lin2015}. This can lead to distinct three-loop braiding (also called necklace braiding \cite{Bellingeri2016}), involving two loops undergoing an exchange process while linked to a third loop, called the base loop \cite{Lin2015, Wang2014, Jiang2014, Wang2015}. Indeed, even if the regular two-loop braiding is Abelian, this three-loop braiding can be non-Abelian. The non-Abelian character of this braiding can be characterized by a process called four-loop braiding \cite{Wang2015a, Wang2020, Tiwari2017, Putrov2017, Zhang2021}, involving multiple three-loop braids. In order to study such phenomena, it is useful to have tractable examples, even if they are only fixed points of a wider phase.

There have been several exactly solvable models that realize 3+1d topological phases, including the Walker-Wang model \cite{Walker2012, Keyserlingk2013, Chen2015, Wang2017} and the higher lattice gauge theory model \cite{Bullivant2017, Delcamp2018, Bullivant2020, Bullivant2020b, Huxford2023}. However, the model which perhaps best captures the properties of such phases is the Dijkgraaf-Witten model and its associated Hamiltonian constructions \cite{ Jiang2014, Wang2015, Dijkgraaf1990, Wan2015, Wen2017, Bullivant2019}, which are conjectured to include all 3+1d bosonic topological phases without emergent fermions \cite{Lan2018}. Already, much is known about this class of phases, both from field theory and from a Hamiltonian realization. Indeed, the Dijjkgraaf-Witten model and associated discrete gauge theory constructions have been used to study exotic statistics from the early days of the field \cite{Alford1992, Bais1980, Krauss1989, Bais1992, Propitius1999}. Each Dijkgraaf-Witten model is labeled by a group $G$ and a 4-cocycle (for the 3+1d case) in $H^4(G, U(1))$. When this cocycle is non-trivial, we say that the model is ``twisted" by the cocycle. The group determines the point-like charges, which are labeled by representations of the group, as well as the flux labels which are labeled by conjugacy classes in $G$. The point-loop braiding is also determined by the group and is given by a generalized version of the usual Aharanov-Bohm formula for a charge moving around a flux tube \cite{Ehrenberg1949, Aharonov1959, Wang2014}. However, it is known that the 4-cocycle is important for the properties of linked excitations. In particular, the three-loop braiding is captured by projective representations of the group $G$, with the factor system for these projective representations depending on the group cocycle \cite{Jiang2014, Wang2015}. In the Hamiltonian realization, the ground state degeneracy of the model on the 3-torus (which is equivalent to the partition function of the topological quantum field theory on the direct product of the 3-torus with a circle), is similarly described by such representations (along with two flux labels) \cite{Wan2015}. Despite this, there are some aspects of the model that are not yet clear. Studies of this phase have so far focussed on indirect arguments such as dimensional reduction \cite{Wang2014, Wang2015}, tube algebra \cite{Bullivant2019}, vacuum expectation values in field theory \cite{Tiwari2017, Putrov2017, Yoshida2017} and study of the ground states \cite{Jiang2014}. On the other hand, Ref. \cite{Lin2015} uses membrane operators to directly study the excitations in a model that appears to be equivalent to a Dijkgraaf-Witten type model for the group $\mathbb{Z}_2 \times \mathbb{Z}_2$. Ref. \cite{Lin2015} provides a clear exposition of some of the features of the excitations, especially those built from linked loops, but the specific group used restricts the type of excitations that are supported by the model. In particular, Ref. \cite{Lin2015} only finds Abelian three-loop braiding, but we expect more general (Abelian) groups to allow for non-Abelian braiding when twisted by an appropriate cocycle. In this work, we consider an explicit construction of the membrane operators that produce non-Abelian linked loop-like excitations. This allows us to study the properties of these excitations in more detail, including their conserved topological charge. It also allows us to see how the non-Abelian braiding can emerge from an Abelian group and how an internal space for the loop-like excitations may arise when they are linked to another loop.

In order to examine these excitations, we consider a Hamiltonian model for Dijkgraaf-Witten, which was constructed in \cite{Wan2015}. We provide a brief review of this model in Section \ref{Section_twisted_gauge_theory}, to familiarize readers with the key features which we will make use of throughout this work. In Section \ref{Section_summary_of_results}, we provide a summary of our results, before giving a more detailed description in the rest of the paper. The main focus of this work is on constructing the membrane operators that produce the emergent excitations in this model and using them to find various properties of these excitations. As we describe in Section \ref{Section_ribbon_and_membrane_operators}, these membrane operators differ from their counterparts in untwisted lattice gauge theory by additional phase factors or weights associated to the surface and the nearby tetrahedra. Of particular interest are the cylindrical membrane operators, which can be used to produce a pair of loop-like excitations that can be linked to an existing flux tube (the base loop). In Section \ref{Section_cylindrical_membranes}, we discuss these operators and construct a useful basis for these operators on a simple geometry. In Section \ref{Section_three_loop_braiding}, we will then use these operators to re-derive the braiding relations found in previous studies \cite{Wang2014, Jiang2014, Wang2015} and in Section \ref{Section_fusion_membranes} we discuss the fusion properties for these membranes to shed additional light on the different properties of the linked loops. In Section \ref{Section_topological_charge} we use these membrane operators to construct projection operators that measure the topological charge enclosed by a measurement surface and show that the number of distinct topological charges measured by the 2-torus matches the ground state degeneracy of the model on the 3-torus, matching a general expectation for topological phases. 

In the Supplemental Material, we provide various proofs that support the results in the main text. In Section \ref{Section_spherical_membranes}, we prove that spherical magnetic membrane operators can be expressed as a product of vertex transforms in the region enclosed by the membrane. This guarantees that these membrane operators can be deformed through an unexcited region by applying additional vertex transforms, which act trivially on the ground state and so do not affect the action of the membrane operator on the state. We call such membrane operators topological for this deformability property. In Section \ref{Section_more_general_membranes}, we consider more membrane operators of a more general topology and show that they are also topological. Finally, in Section \ref{Section_supplement_topological_charge}, we prove that the purported topological charge projectors described in Section \ref{Section_topological_charge} are indeed orthogonal projectors and span the space of measurement operators.

	\subsection{Twisted Lattice Gauge Theory Model}
	\label{Section_twisted_gauge_theory}
	The twisted lattice gauge theory model, introduced in Ref. \cite{Wan2015}, provides a Hamiltonian model for the Dijkgraaf-Witten topological quantum field theory \cite{Dijkgraaf1990}. Unlike Kitaev's quantum double model (or more properly, its 3+1d counterpart), the twisted lattice gauge theory allows for a ``twist" in the form of a 4-cocycle, which is crucial for the model to support novel three-loop braiding statistics.
	
	The input data for a particular twisted lattice gauge theory model is a finite group $G$ and a 4-cocycle $\omega \in H^4(G, U(1))$. The model is defined on a triangulation of a 3d manifold, with each (directed) edge carrying a local Hilbert space with basis states corresponding to elements of the group $G$. This means that one simple basis for the entire Hilbert space, which we call the configuration basis, has basis states where each edge is labeled by an element of $G$. The vertices of the lattice are indexed, with the edges pointing from the lower-indexed vertex to the higher-indexed one. The Hamiltonian is a sum of two types of terms, corresponding to the plaquettes (triangles) and vertices of the lattice. The Hamiltonian is given by \cite{Wan2015}
	\begin{equation}
		H= - \sum_{\text{plaquettes, }p} B_p - \sum_{\text{vertices, }v} A_v. \label{Equation_Hamiltonian}
	\end{equation}
	
	The plaquette term enforces flatness, analogous to the corresponding term from Kitaev's quantum double model \cite{Kitaev2003}:
	$$B_p = \delta( g(\text{boundary}(p)), 1_G),$$
	where $g(\text{boundary}(p))$ is the path element for the boundary of the plaquette, starting at any vertex on the plaquette and respecting the orientation of the edges, and $\delta$ is the Kronecker delta.

	The vertex term is also similar to the quantum double model equivalent. It can be written as a sum of gauge transforms, one for each element of $G$:
	$$A_v= \frac{1}{|G|} \sum_{g \in G} A_v^g.$$
	Each transform $A_v^g$ has two components, a multiplicative action on the surrounding edge elements and a phase factor associated to the surrounding tetrahedra. That is, we can write the transform as
	\begin{equation}
		A_v^g =A_v^g(0) \prod_{\text{tetrahedra }t \ni v} \theta_{v,t}^g,
	\end{equation}
where $A_v^g(0)$ is the vertex transform corresponding to a model with trivial cocycle, while $\theta_{v,t}^g$ is the phase factor for the tetrahedron $t$ and depends on the 4-cocycle. The untwisted transform affects the edges that are attached to that vertex, according to
	$$A_v^g(0) :g_i = \begin{cases} gg_i & \text{if $i$ points away from $v$} \\ g_i g^{-1} & \text{if $i$ points towards $v$.} \end{cases}$$
	
\begin{figure}[h]
		\centering
		\includegraphics[width=\linewidth]{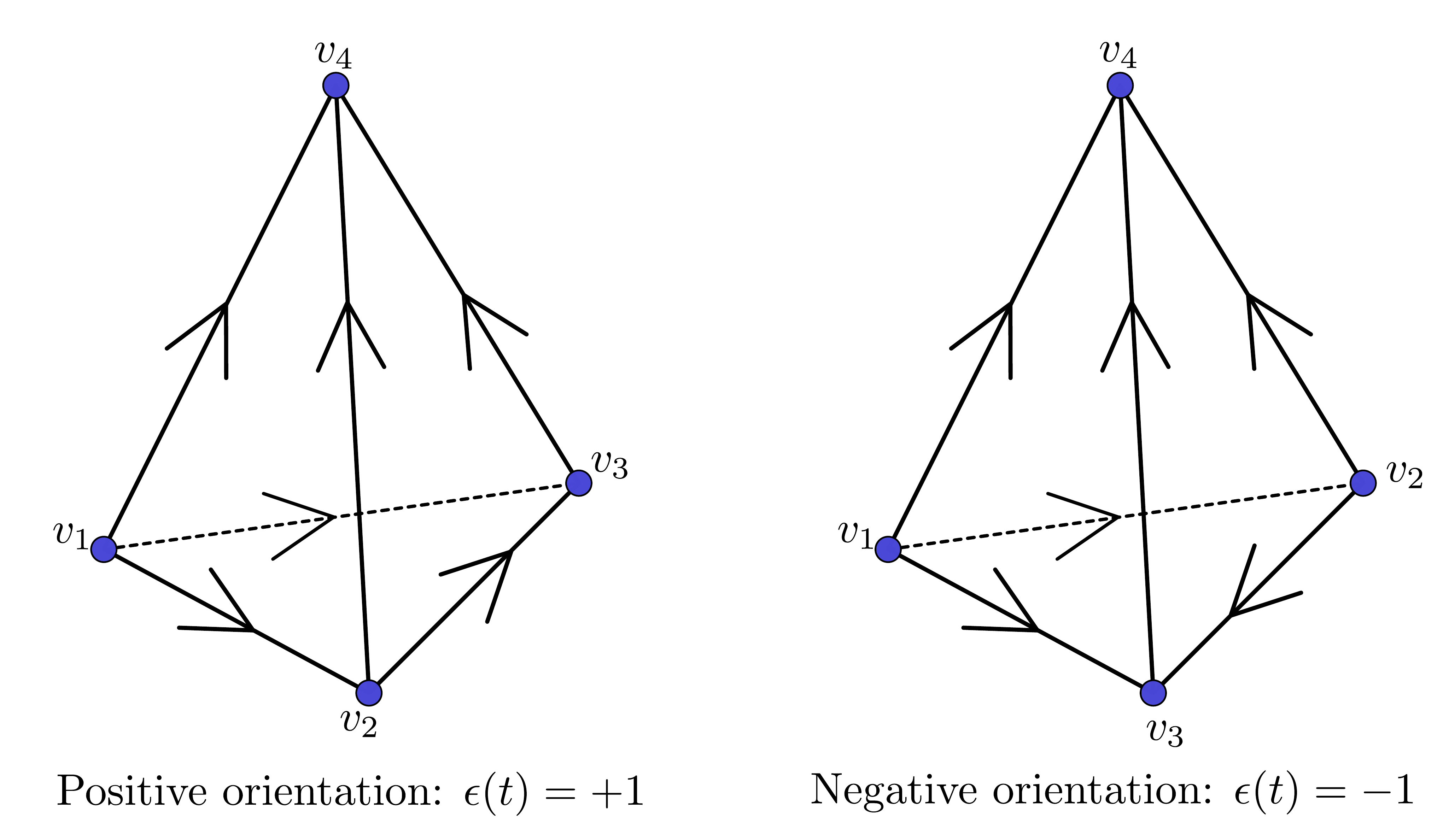}
		\caption{The two possible orientations of the indexed tetrahedron (up to rotation). Here $v_1$ is the lowest indexed vertex on the tetrahedron, $v_2$ is the next and so on. Note that the subscript $1$ is not itself its index (because we have to index the whole lattice, not just the tetrahedron).}
		\label{Figure_tetrahedra_orientation}
	\end{figure}

In order to define the phase $\theta_{v,t}^g$, we must first discuss the orientation of the tetrahedrons. Each tetrahedron has four vertices indexed by different integers. If we orient the tetrahedron so that the highest indexed vertex is uppermost, then we can determine the orientation of the tetrahedron using the right-hand rule. Pointing our thumb towards the highest indexed vertex, the orientation of the tetrahedron is positive if the index of the remaining vertices increases with the circulation of our hand and is negative if they decrease. The two possibilities, up to rotation, are illustrated in Figure \ref{Figure_tetrahedra_orientation}.

	The $\theta_{v,t}^g$ phase associated to a tetrahedron depends both on this orientation and on the index of the vertex in the tetrahedron, and is described by the cocycle $\omega$. Denoting the vertices on the tetrahedron, in ascending index order, by $v_1$, $v_2$, $v_3$ and $v_4$, the phase obtained from this tetrahedron by a transform on vertex $v_1$ is
	$$\theta_{v_1,t}^g=\omega(g, g_{v_1 v_2}, g_{v_2v_3}, g_{v_3v_4})^{\epsilon(t)},$$
	where $\epsilon(t)$ is $+1$ if the tetrahedron has positive orientation and $-1$ otherwise.
	For $v_2$ it is
	$$\theta_{v_2,t}^g=\omega( g_{v_1 v_2}g^{-1}, g, g_{v_2v_3}, g_{v_3v_4})^{-\epsilon(t) }.$$
	For $v_3$ it is
	$$\theta_{v_3,t}^g=\omega(g_{v_1 v_2}, g_{v_2v_3}g^{-1}, g, g_{v_3 v_4})^{\epsilon(t)}$$
	and for $v_4$ it is
	$$\theta_{v_4,t}^g=\omega(g_{v_1 v_2}, g_{v_2v_3}, g_{v_3 v_4}g^{-1}, g)^{-\epsilon(t)}.$$
One way of thinking about the phase is that the transform at $v$ introduces a new vertex $v'$, such that the index of $v'$ is very slightly less than that of $v$, and $g_{v'v}=g$. The evaluation of the 4-cocycle is then associated to a 4-simplex incorporating that vertex into the tetrahedron. Because the choice of 4-cocycle is fixed when the model is defined, we use the notation
	$$[g_1,g_2 ,g_3, g_4] := \omega(g_1, g_2, g_3, g_4).$$
	The 4-cocycle satisfies the cocycle condition:
	\begin{equation}
		\frac{[g_1, g_2, g_3, g_4] [g_0, g_1g_2, g_3,g_4][g_0, g_1,g_2, g_3g_4]}{[g_0g_1, g_2, g_3, g_4][g_0,g_1,g_2g_3,g_4][g_0,g_1,g_2,g_3]}=1 \label{Equation_4_cocycle_condition}
	\end{equation}
 for all $\set{g_i \in G} \ (i=0,1,2,3,4)$.
	In addition, it satisfies a normalization condition:
 \begin{align}
	1=[1_G,g_2,g_3,g_4]=[g_1,1_G,g_3,g_4]&=[g_1,g_2,1_G,g_4] \notag \\ &=[g_1,g_2,g_3,1_G],
 \end{align}
	for all $\set{g_i \in G}$ ($i=1,2,3,4$).

	With this definition, the vertex transforms satisfy the algebra $A_v^gA_v^h=A_v^{gh}$, which leads to the vertex terms $A_v= \frac{1}{|G|}A_v^g$ being projectors \cite{Wan2015}. The vertex terms commute with the plaquette terms because the plaquette terms depend on the label of a closed path, which is at most conjugated by the vertex terms (just as for Kitaev's quantum double model). The vertex terms also commute with each-other, but generally only do so in the subspace where the plaquette terms are satisfied. One solution to this problem is to define the vertex terms to be zero (and therefore the vertex to be excited) when an adjacent plaquette does not satisfy flatness, similar to how the plaquette terms in a string-net model are treated when adjacent to a vertex term that does not satisfy the fusion rules \cite{Levin2005}. Because we are usually only interested in which regions of the lattice do not satisfy the energy terms for a given state, rather than the actual energies of the states, and these additional vertex excitations are always adjacent to plaquette excitations, it is not particularly significant whether we take this approach or not.

	In addition to introducing the twisted lattice gauge theory models, Ref. \cite{Wan2015} describes many of their properties, including the ground state degeneracy, the equivalence between models defined with equivalent cocycles (up to a coboundary) and the consistency of the model under mutations of the lattice. However, Ref. \cite{Wan2015} does not directly describe the excitations or the membrane operators used to produce them. In the rest of this work, we will construct these membrane operators for a general Abelian group and use them to derive or re-derive the properties of the excitations.
	
	\subsection{Summary of Results}
	\label{Section_summary_of_results}
	
	Before we discuss our results in great detail, we will summarize them here. We consider the twisted lattice gauge theory model in 3+1d \cite{Wan2015} with an Abelian group $G$ (although many of our results can be extended to the non-Abelian case). As we explain in Section \ref{Section_ribbon_and_membrane_operators}, we construct the general form of the ribbon and membrane operators which produce the basic excitations in the twisted lattice gauge theory model. While the ribbon operators that produce the pure electric excitations are unchanged from the untwisted gauge theory case, the magnetic membrane operators are significantly different. Compared to the untwisted membrane operators, which simply multiply the edges cut by the membrane by an element $h^{\pm 1}$, where $h$ is the flux of the excitation produced by the membrane operator, the membrane operators in the twisted theory also must apply a weight depending on the labels of the edges near the membrane. This weight can be split into two parts: a dual phase, which depends on the edges of the tetrahedra cut by the membrane, and a surface weight, which depends on the edge labels on a surface called the direct membrane. This surface weight can be calculated from a set of reference diagrams by using graphical rules, which describe how the surface weight changes under the application of bistellar flips to the surface. The weights assigned to the reference diagrams are free variables, giving a space of membrane operators for a given flux. This is analogous to a charge for the membrane operators, with this charge distributed along the boundary of the membrane.
	
	A particularly interesting class of membrane operators are those applied on a cylindrical membrane, which we discuss in detail in Section \ref{Section_cylindrical_membranes}. These membrane operators produce a pair of loop-like flux tubes at the two ends of the cylinder. These two loops can be linked to an already existing third loop, called a base loop, which is threaded through the cylindrical membrane. For a particularly simple geometry, we find that the membrane operators obey a simple composition rule under concatenation of the cylinders, if the weights assigned to the reference diagrams are given by (matrix elements of) irreducible projective representations. This choice of the weights describes a basis for the space of such membrane operators, analogous to the use of linear irreps for electric ribbon operators. These projective representations obey the composition rule
	\begin{equation}
		\alpha^{k,h}(x_1) \alpha^{k,h}(x_2) \mbeq [x_1, x_2]_{k,h} \alpha^{k,h}(x_1x_2). \label{Equation_concatenation_relation_summary}
	\end{equation}
Here $k$ is the flux label of the base loop and $h$ is the flux label of the membrane operator that produces the other two loops, while $[x_1, x_2]_{k,h}$ is a 2-cocycle derived from the underlying 4-cocycle of the model by applying the slant product twice. That is, to obtain the 2-cocycle we first construct a 3-cocycle by \cite{Wan2015}
\begin{equation}
	[x_1, x_2, x_3]_k = \frac{[x_1,k,x_2, x_3] [x_1,x_2,x_3,k]}{[k,x_1,x_2,x_3] [x_1,x_2,k,x_3]}.
\end{equation}
Then we construct a 2-cocycle from that 3-cocycle in a similar way by
\begin{equation}
	[x_1,x_2]_{k,h} = \frac{[h,x_1,x_2]_k [x_1,x_2, h]_k}{[x_1,h,x_2]_k}.
\end{equation}
These 2-cocycles then form the factor system for the projective irreps, as indicated by Equation \ref{Equation_concatenation_relation_summary}. A 2-cocycle can be non-symmetric under exchange of $x_1$ and $x_2$, in which case the projective representations associated to that 2-cocycle are necessarily higher-dimensional, despite the Abelian nature of the group $G$. Note that the composition rule of the projective representations depends on the two flux labels $k$ and $h$, meaning that the allowed representations for the membrane operators (and so the charge labels for the excitations) are different depending on both the flux label of the membrane operator and the flux of the linked base loop. That is, the excitations linked to different loops are inherently different in general (as discussed in Ref. \cite{Lin2015}). In particular, in the case where the base loop is trivial ($k=1_G$), the 2-cocycle also becomes trivial and the projective representations become linear representations, reproducing the ordinary charge that we can attach to flux tubes. 

Using this simple class of membrane operators, we can find various properties of the flux excitations. In Section \ref{Section_three_loop_braiding}, we demonstrate how the braiding relations for these excitations can be calculated. Because the loop-like excitations may be linked to a base loop, we can find the three-loop braiding relations which describe loop braiding of two excitations while linked to a common base loop. We find that this braiding relation is similar to that obtained by braiding charges through each loop, except that instead of linear irreps, the charges are described by the projective irreps labeling each membrane operator. When the irreps are 1-dimensional, the braiding of a loop excitation with label $(a, \alpha_1^{k,a})$, where $a$ is the flux label and $\alpha_1^{k,a}$ the projective irrep for base loop $k$, with a loop excitation of label $(c, \alpha_2^{k,c})$ results in the accumulation of a phase
\begin{equation}
\theta_3 = \alpha_1^{k,a}( c)\alpha_2^{k,c}(a) .
\end{equation}
Note that when the base loop is trivial ($k=1_G$), $\alpha_1$ and $\alpha_2$ become linear irreps of $G$, indicating that two-loop braiding is independent of the 4-cocycle. For higher-dimensional irreps, which require a non-trivial base loop if $G$ is Abelian, the membrane operators carry additional labels corresponding to the matrix indices for the projective representations. Then the braiding relation results in a transformation described by the action of the representative matrices $\alpha_1^{k,a}( c)$ and $\alpha_2^{k,c}(a)$. This matrix action results in mixing membrane operators labeled by different matrix indices. This reflects the fact that the indices are not conserved quantities, although the irreps are. As we describe in Section \ref{Section_three_loop_braiding}, this agrees with results previously obtained from dimensional reduction and other arguments \cite{Wang2014, Jiang2014, Wang2015}.

In addition to the braiding, we demonstrate the fusion rules satisfied by these simple membrane operators, which describe what happens when we apply two membrane operators on the same membrane. This corresponds to the fusion of the excitations at the ends of the membrane operators. Under fusion of membrane operators labeled by $(a, \alpha_1^{k,a})$ and $(b, \alpha_2^{k,b})$, the result is a single membrane operator with label $(ab, \alpha_T^{k,ab})$, where
\begin{equation}
\alpha_T^{k,ab}(x)=[b,a]_{k,x} \alpha_1^{k,a}(x) \otimes \alpha_2^{k,b}(x).
\end{equation}
For higher-dimensional irreps $\alpha_1^{k,a}$ and $\alpha_2^{k,b}$, $\alpha_T^{k,ab}$ is generally a reducible projective representation, and its constituent irreps are the different possible fusion products. Note that when the base loop is trivial ($k=1_G$), the factor $[b,a]_{k,x}$ reduces to the identity and the fusion becomes the normal composition of irreps.

Finally, in Section \ref{Section_topological_charge}, we use the membrane operators to construct projectors to definite topological charges. These projection operators are applied on spherical or toroidal surfaces and measure the charge of excitations enclosed by the surfaces. We find that the spherical measurement operators are labeled by linear irreps of $G$, reflecting the nature of the point-like excitations. On the other hand, the toroidal measurement operators are labeled by the flux around the two handles of the torus and a corresponding projective representation (similar to the basis used for the cylindrical measurement operators). These toroidal projectors are in one-to-one correspondence with the ground states for the 3-torus found in Ref. \cite{Wan2015}, demonstrating that the connection between ground states on the 3-torus and charges measured by a 2-torus holds for this model. This is analogous to the connection between topological charge and the ground state degeneracy for a 2-torus from 2+1d topological phases. In the 3+1d case, because the topological charge depends on two flux labels, the charges are not in one-to-one correspondence with the simple loop-like excitations, but instead we must allow the loop-like excitations to be linked to different base loops for the correspondence to hold. 
	
	\section{Ribbon and Membrane Operators}
	\label{Section_ribbon_and_membrane_operators}
	\subsection{Ribbon Operators}
	\label{Section_ribbon_operators}
	
	We first consider the ribbon operators that produce the point-like electric excitations. These take the same form regardless of the cocycle twist and are equivalent to the operators from Kitaev's quantum double model \cite{Kitaev2003}, except that they are in 3+1d rather than 2+1d. An electric ribbon operator applied on a path $t$ has the form
	\begin{equation}
		S^{\vec{\alpha}}(t)= \sum_{g \in G} \alpha_g \delta(g,\hat{g}(t)),
	\end{equation}
where $\hat{g}(t)$ is the path element operator for path $t$ and $\alpha_g$ is a coefficient. To calculate the path element, we take the product (from left to right) of the labels of the edges along the path, with inverses for edges that are anti-aligned with the path. A useful basis for this space of operators is described by the irreps of $G$, with the trivial irrep labeling the identity operator and the operators labeled by non-trivial irreps producing excitations at the two ends of $t$. Such basis operators are given by

\begin{equation}
	S^{R,a,b}(t)= \sum_{g \in G} [D^R(g)]_{ab}\delta(g,\hat{g}(t)),
\end{equation}

where $R$ is an irrep of $G$, and $D^R(g)$ is the matrix representation of $g$ with matrix indices $a$ and $b$. The irreps $R$ describe a conserved topological charge carried by the excitations. In the case of non-Abelian groups, the irreps must be supplemented by the matrix indices in order to give a full basis for the space of operators. These indices are not conserved and instead describe an internal space for the topological charge. We can think of the two indices as corresponding to the internal space for the excitations at the two ends of the ribbon. On the other hand, for the Abelian groups that we consider here, the irreps are all one-dimensional and there is no internal space. The form of the ribbon operators is the same regardless of the 4-cocycle $\omega$, because the ribbon operators are diagonal in the configuration basis (the basis where each edge is labeled by an element of $G$) and so commute with the cocycle twist on the vertex terms (because the twist factor is similarly diagonal in the configuration basis). This means that the commutation relation between the electric ribbon operators and the vertex transforms is the same regardless of the cocycle twist.

	\subsection{Membrane Operators}
	\label{Section_membrane_operators}

\begin{figure}
			\centering
			\includegraphics[width=\linewidth]{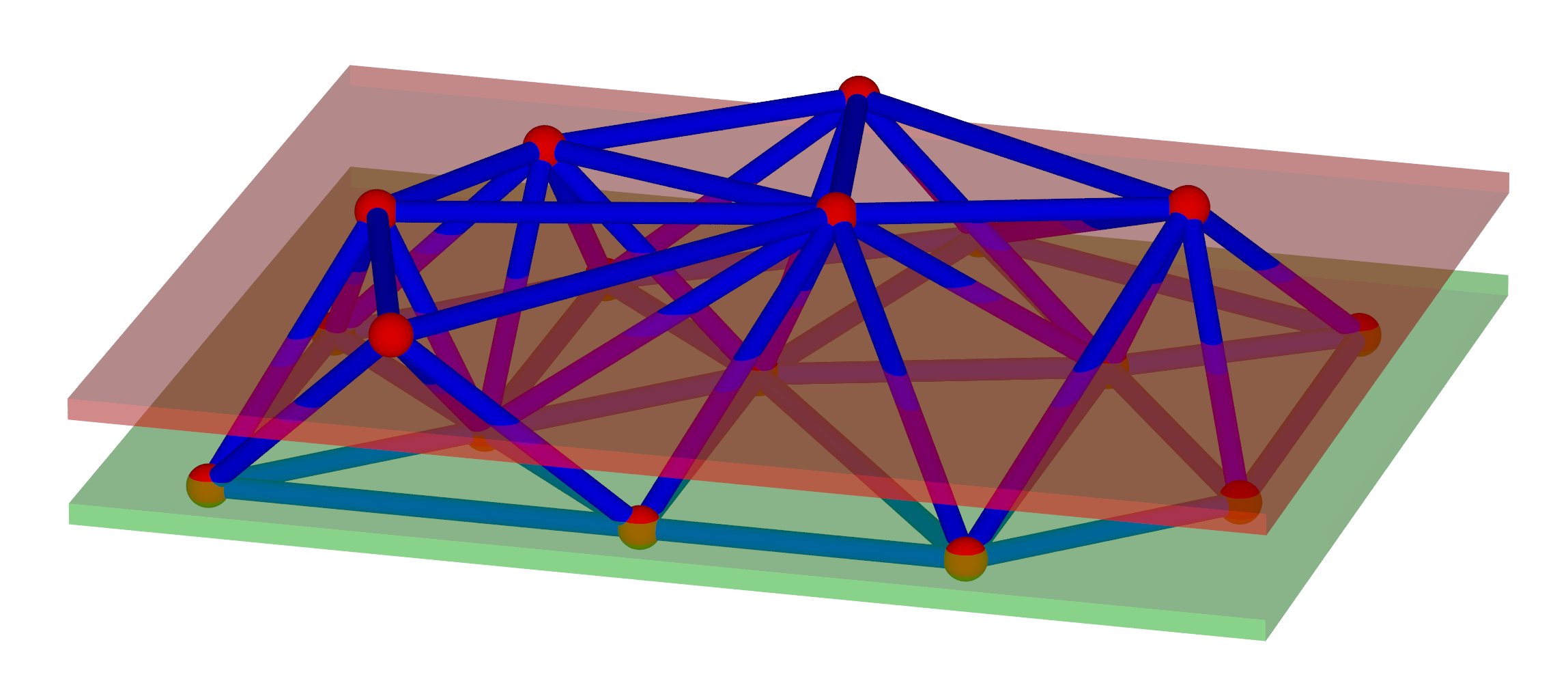}
			\caption{An example of a membrane operator acting across a fragment of lattice. The direct membrane (lower, green membrane) contains vertices, edges and plaquettes, while the dual membrane (upper, red membrane) cuts through edges, plaquettes and tetrahedra.}
			\label{fig:membraneanddualmembrane}
		\end{figure}

	Unlike the electric ribbon operators, the magnetic membrane operators strongly depend on the cocycle twist. In order to define these membrane operators, we first split them into three parts:
\begin{equation}
		F^{h, \vec{v}}(m)=C_0^h(m) \theta_D^h(m) \theta_S^{h, \vec{v}}(m), \label{Equation_membrane_operator_definition}
\end{equation}
		where $h$ is the flux label of the resulting magnetic excitations and $\vec{v}$ is a set of coefficients that determines the electric charge and which we will explain in more detail when we describe $\theta_S^{h, \vec{v}}(m)$. The membrane $m$ on which we define the operator has two parts: a direct membrane made up of plaquettes in the direct lattice and a dual membrane which bisects the edges of the lattice, as shown in Figure \ref{fig:membraneanddualmembrane}. These two membranes together form a thickened membrane and the support of the membrane operator lies in this region. This is analogous to how ribbon operators in the 2+1d quantum double model generally have both a direct path and dual path, with these paths together forming a ribbon of finite width.

		The first part of the membrane operator, $C_0^h(m)$, is the same regardless of the cocycle twist. This operator affects all edges cut by the dual membrane. When the group $G$ is Abelian, the membrane operator multiplies such an edge by $h$ if the edge points away from the direct membrane and by $h^{-1}$ if it points towards the direct membrane. This is the only part of the membrane operator that changes the edge labels, and so it is the only part that fails to commute with the plaquette energy terms. For a plaquette cut by the bulk of the dual membrane, two of the edges on the plaquette are affected by the multiplication, which leaves the total boundary label of the plaquette unaffected and so the plaquette term is left unexcited. On the other hand, for plaquettes cut by the boundary of the dual membrane, only one edge is affected and so there is no cancellation. This leaves the plaquettes cut by the boundary of the membrane excited by the action of the membrane operator. 
		
\begin{figure}
		\centering
		\includegraphics[width=0.9\linewidth]{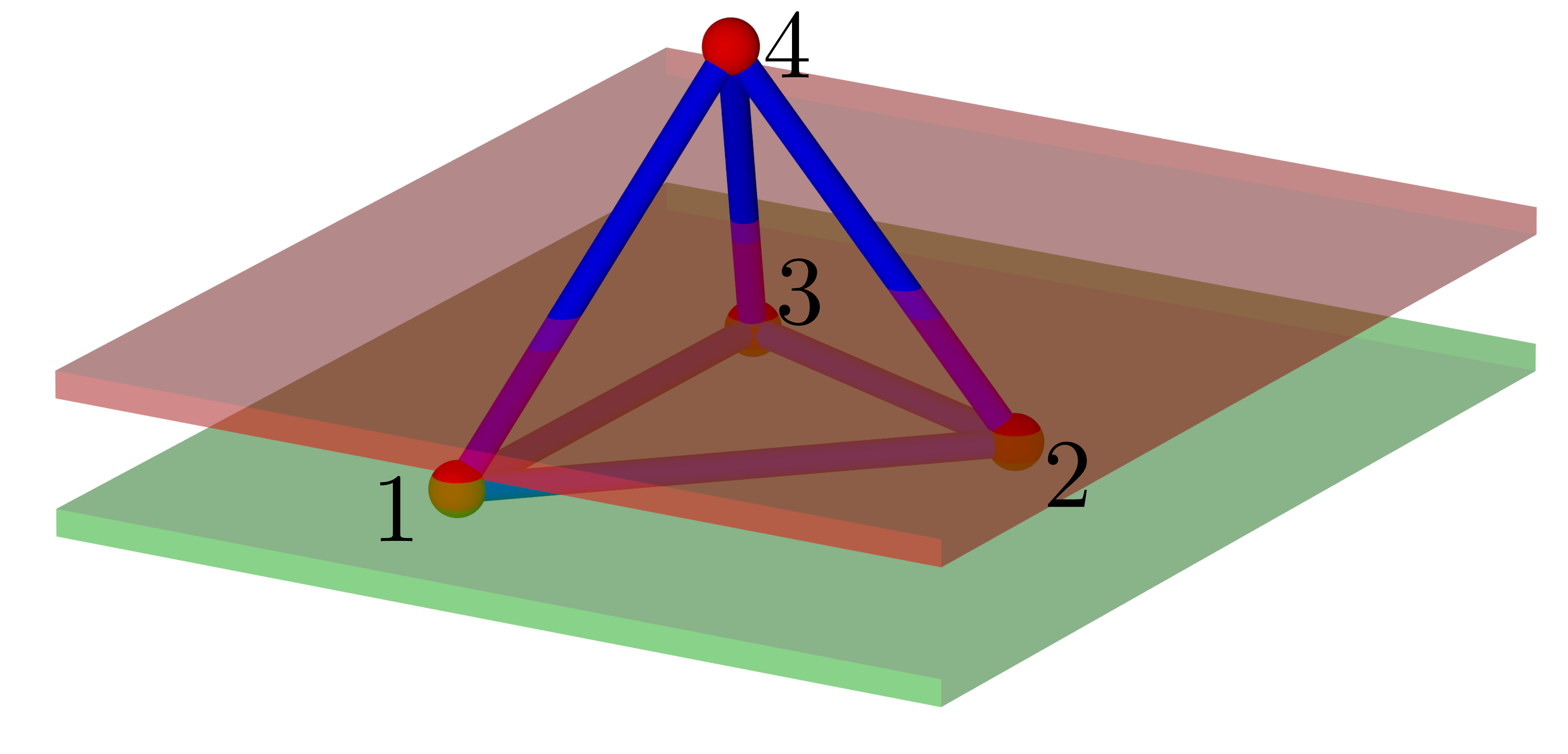}
		\caption{We consider the example of a tetrahedron $\set{1,2,3,4}$ that is cut by the dual membrane (the red upper surface) and that has a face $\set{1,2,3}$ on the direct membrane (the lower green surface). The contribution to the dual phase from this tetrahedron can be calculated by applying a vertex transform on each vertex on the direct membrane.}
		\label{Figure_example_tetrahedron_phase_factor}
	\end{figure}

		The second part of the membrane operator is $\theta_D^h(m)$, which we refer to as the dual phase. This applies a phase for each tetrahedron cut by the dual membrane, depending on the cocycle twist. As we explain in Sections \ref{Section_spherical_membranes} and \ref{Section_more_general_membranes} in the Supplemental Material, the magnetic membrane operators look locally like vertex transforms, and the phase associated to a tetrahedron cut by the dual membrane is equivalent to the phase obtained from that tetrahedron by applying a vertex transform on every vertex on the direct membrane. As an example, consider the tetrahedron shown in Figure \ref{Figure_example_tetrahedron_phase_factor}, which is cut by the dual membrane and has a face on the direct membrane. We consider a basis state $\ket{g_{12}, g_{13}, g_{23}, g_{14}, g_{24}, g_{34}}$ for the degrees of freedom on the tetrahedron, where $g_{ij}$ is the label of the edge from vertex $i$ to $j$. Applying the series of vertex transforms $A_3^h A_2^h A_1^h$, we get
		\begin{align*}
			&A_3^h A_2^hA_1^h \ket{g_{12}, g_{13}, g_{23}, g_{14}, g_{24}, g_{34}}\\
			 &=A_3^h A_2^h [h, g_{12}, g_{23}, g_{34}]\ket{hg_{12}, hg_{13}, g_{23}, hg_{14}, g_{24}, g_{34}}\\
			&= A_3^h [h, g_{12}, g_{23}, g_{34}] [(hg_{12})h^{-1}, h, g_{23}, g_{34}]^{-1} \\
   & \hspace{1cm}\ket{hg_{12}h^{-1}, hg_{13}, hg_{23}, hg_{14}, hg_{24}, g_{34}}\\
			&= [h, g_{12}, g_{23}, g_{34}] [g_{12}, h, g_{23}, g_{34}]^{-1} [g_{12}, hg_{23}h^{-1}, h, g_{34}]\\
   & \hspace{1cm} \ket{g_{12}, hg_{13}h^{-1}, hg_{23}h^{-1}, hg_{14}, hg_{24}, hg_{34}}\\
			&=[h, g_{12}, g_{23}, g_{34}] [g_{12}, h, g_{23}, g_{34}]^{-1} [g_{12}, g_{23}, h, g_{34}] \\
   & \hspace{1cm} \ket{g_{12}, g_{13}, g_{23}, hg_{14}, hg_{24}, hg_{34}}.
		\end{align*}

	We see that this recreates the action of $C_0^h(m)$ on the edges (multiplying the edges pointing away from the direct membrane by $h$, while leaving the edges lying on the direct membrane unaffected) and gives a phase factor of
	$$\theta_D^h(t)=[h, g_{12}, g_{23}, g_{34}] [g_{12}, h, g_{23}, g_{34}]^{-1} [g_{12}, g_{23}, h, g_{34}],$$
	which is the contribution to the dual phase from that tetrahedron. More generally, each tetrahedron cut by the dual membrane can be classified by which of its vertices $v_1(t)$, $v_2(t)$, $v_3(t)$ and $v_4(t)$ are on the direct membrane. The dual phases associated to the tetrahedron for each possible set of vertices on the direct membrane are given in Table \ref{Table_Dual_Phase}.

			\begin{table*}[t]
			\begin{center}
				\renewcommand{\arraystretch}{1.5}
				\begin{tabular}{ |c|c|} 
					\hline
					\textbf{Vertices on } & \\ \textbf{Direct Membrane} & \textbf{Dual Phase} \\
					\hline
						$\set{v_1(t)}$ & $[h, \hat{g}_{v_1(t) v_2(t)}, \hat{g}_{v_2(t)v_3(t)}, \hat{g}_{v_3(t)v_4(t)}]^{\epsilon(t)} $ \\
						
						\hline
						
				$\set{v_2(t)}$ & $[ \hat{g}_{v_1(t) v_2(t)}h^{-1}, h , \hat{g}_{v_2(t)v_3(t)}, \hat{g}_{v_3(t)v_4(t)}]^{-\epsilon(t)}$ \\
				
				\hline
					$\set{v_3(t)} $& $ [ \hat{g}_{v_1(t) v_2(t)}, \hat{g}_{v_2(t)v_3(t)}h^{-1}, h, \hat{g}_{v_3(t)v_4(t)}]^{\epsilon(t)} $ \\
					\hline
					
					$\set{v_4(t)}$ &	$[ \hat{g}_{v_1(t) v_2(t)}, \hat{g}_{v_2(t)v_3(t)}, \hat{g}_{v_3(t)v_4(t)}h^{-1}, h]^{-\epsilon(t)} $ \\
					
					\hline
					
					$\set{v_1(t),v_2(t)}$& $\big( [h, \hat{g}_{v_1(t) v_2(t)}, \hat{g}_{v_2(t)v_3(t)}, \hat{g}_{v_3(t)v_4(t)}] [ \hat{g}_{v_1(t) v_2(t)}, h , \hat{g}_{v_2(t)v_3(t)}, \hat{g}_{v_3(t)v_4(t)}]^{-1} \big)^{\epsilon(t)}$ \\
						\hline
					
					$	\set{v_1(t),v_3(t)}$& $\big( [h, \hat{g}_{v_1(t) v_2(t)}, \hat{g}_{v_2(t)v_3(t)}, \hat{g}_{v_3(t)v_4(t)}] [ h\hat{g}_{v_1(t) v_2(t)}, \hat{g}_{v_2(t)v_3(t)}h^{-1}, h, \hat{g}_{v_3(t)v_4(t)}]\big)^{\epsilon(t)}$ \\
						\hline

						$\set{v_1(t),v_4(t)}$& $ \big( [h, \hat{g}_{v_1(t) v_2(t)}, \hat{g}_{v_2(t)v_3(t)}, \hat{g}_{v_3(t)v_4(t)}] 	[ h\hat{g}_{v_1(t) v_2(t)}, \hat{g}_{v_2(t)v_3(t)}, \hat{g}_{v_3(t)v_4(t)}h^{-1}, h]^{-1} \big)^{\epsilon(t)} $\\
							\hline

		$\set{v_2(t),v_3(t)}$ & $ \big( [ \hat{g}_{v_1(t) v_2(t)}h^{-1}, h , \hat{g}_{v_2(t)v_3(t)}, \hat{g}_{v_3(t)v_4(t)}]^{-1} [ \hat{g}_{v_1(t) v_2(t)}h^{-1}, \hat{g}_{v_2(t)v_3(t)}, h, \hat{g}_{v_3(t)v_4(t)}] \big)^{\epsilon(t)} $\\
			\hline

	$\set{v_2(t),v_4(t)}$& \hspace{-2cm}$\big( [ \hat{g}_{v_1(t) v_2(t)}h^{-1}, h , \hat{g}_{v_2(t)v_3(t)}, \hat{g}_{v_3(t)v_4(t)}]^{-1}$ \\
	& \hspace{2cm}$[ \hat{g}_{v_1(t) v_2(t)}h^{-1}, h \hat{g}_{v_2(t)v_3(t)}, \hat{g}_{v_3(t)v_4(t)}h^{-1}, h]^{-1} \big)^{\epsilon(t)} $\\
		\hline

	$\set{v_3(t),v_4(t)}$& $\big( [ \hat{g}_{v_1(t) v_2(t)}, \hat{g}_{v_2(t)v_3(t)}h^{-1}, h, \hat{g}_{v_3(t)v_4(t)}] [ \hat{g}_{v_1(t) v_2(t)}, \hat{g}_{v_2(t)v_3(t)}h^{-1}, \hat{g}_{v_3(t)v_4(t)}, h]^{-1}\big )^{\epsilon(t)} $ \\
		\hline
	
	$\set{v_1(t),v_2(t), v_3(t)}$ 
		& $ \big( [h, \hat{g}_{v_1(t) v_2(t)}, \hat{g}_{v_2(t)v_3(t)}, \hat{g}_{v_3(t)v_4(t)}] [ \hat{g}_{v_1(t) v_2(t)}, h , \hat{g}_{v_2(t)v_3(t)}, \hat{g}_{v_3(t)v_4(t)}]^{-1} $ \\ & $ [ \hat{g}_{v_1(t) v_2(t)}, \hat{g}_{v_2(t)v_3(t)}, h, \hat{g}_{v_3(t)v_4(t)}] \big)^{\epsilon(t)} $ \\
			\hline

		$	\set{v_1(t),v_2(t), v_4(t)}$ & $\big( [h, \hat{g}_{v_1(t) v_2(t)}, \hat{g}_{v_2(t)v_3(t)}, \hat{g}_{v_3(t)v_4(t)}] [ \hat{g}_{v_1(t) v_2(t)}, h , \hat{g}_{v_2(t)v_3(t)}, \hat{g}_{v_3(t)v_4(t)}]^{-1} $ \\ & $ [ \hat{g}_{v_1(t) v_2(t)}, h \hat{g}_{v_2(t)v_3(t)}, \hat{g}_{v_3(t)v_4(t)}h^{-1}, h]^{-1} \big)^{\epsilon(t)} $\\
			\hline

		$\set{v_1(t),v_3(t), v_4(t)} $ &$\big( [h, \hat{g}_{v_1(t) v_2(t)}, \hat{g}_{v_2(t)v_3(t)}, \hat{g}_{v_3(t)v_4(t)}] [ h\hat{g}_{v_1(t) v_2(t)}, \hat{g}_{v_2(t)v_3(t)}h^{-1}, h, \hat{g}_{v_3(t)v_4(t)}] $ \\ & $[ h\hat{g}_{v_1(t) v_2(t)}, \hat{g}_{v_2(t)v_3(t)}h^{-1}, \hat{g}_{v_3(t)v_4(t)}, h]^{-1}\big)^{\epsilon(t)} $\\
			\hline

		$\set{v_2(t),v_3(t), v_4(t)} $& $\big( [ \hat{g}_{v_1(t) v_2(t)}h^{-1}, h , \hat{g}_{v_2(t)v_3(t)}, \hat{g}_{v_3(t)v_4(t)}]^{-1} [ \hat{g}_{v_1(t) v_2(t)}h^{-1}, \hat{g}_{v_2(t)v_3(t)}, h, \hat{g}_{v_3(t)v_4(t)}] $ \\ &$[ \hat{g}_{v_1(t) v_2(t)}h^{-1}, \hat{g}_{v_2(t)v_3(t)}, \hat{g}_{v_3(t)v_4(t)}, h]^{-1} \big)^{\epsilon(t)} $\\
			\hline

				$\set{v_1(t),v_2(t), v_3(t), v_4(t)}$ 
			& $ \big( [h, \hat{g}_{v_1(t) v_2(t)}, \hat{g}_{v_2(t)v_3(t)}, \hat{g}_{v_3(t)v_4(t)}] [ \hat{g}_{v_1(t) v_2(t)}, h , \hat{g}_{v_2(t)v_3(t)}, \hat{g}_{v_3(t)v_4(t)}]^{-1} $ \\ & $ [ \hat{g}_{v_1(t) v_2(t)}, \hat{g}_{v_2(t)v_3(t)}, h, \hat{g}_{v_3(t)v_4(t)}] [ \hat{g}_{v_1(t) v_2(t)}, \hat{g}_{v_2(t)v_3(t)}, \hat{g}_{v_3(t)v_4(t)}, h]^{-1} \big)^{\epsilon(t)} $ \\
			\hline
					
				\end{tabular}
				
				\caption{The dual phase associated to each configuration of vertices on the direct membrane. Here $\epsilon(t)$ is the orientation of the tetrahedron.}
				\label{Table_Dual_Phase}
			\end{center}
			
		\end{table*}

		While this describes the contribution to the dual phase from tetrahedra cut by the bulk of the dual membrane, there is some ambiguity for tetrahedra cut by the boundary of the dual membrane in the case of open membranes. There are two reasons for this. Firstly, as we described in Section \ref{Section_twisted_gauge_theory}, the vertex transforms themselves are not well-defined in regions that do not satisfy flatness, like the boundary of a magnetic membrane operator. Secondly, not all edges on the tetrahedron that would be affected by a vertex transform on the boundary are affected by the membrane operator. Because of this, we do not apply a dual phase for the tetrahedra on the boundary of the membrane. Another way of thinking about this is that we only define membrane operators up to operators on the boundary. Because these operators are local to the boundary (where we already have excitations in the form of plaquette excitations) they do not affect the topological properties of the membrane operators, such as braiding relations or topological charge, which can be measured far from the boundary for large membrane operators.
		
		The final part of the membrane operator is the surface weight, $\theta^{h, \vec{v}}_S(m)$, which depends on the degrees of freedom in the direct membrane. This quantity is analogous to the function $f_b$ used to define the membrane operators for the $\mathbb{Z}_2 \times \mathbb{Z}_2$ model considered in Ref. \cite{Lin2015} (although in that model, one copy of $\mathbb{Z}_2$ lies on the direct lattice and one lies on the dual lattice, so the presentation of the function is a little different). Similar to the dual phase, the properties of the surface weight are related to the connection between magnetic membrane operators and vertex transforms, as well as the topological property of the membrane operators. Deforming the membrane operator through a region $R$ is equivalent to applying vertex transforms in the region $R$. We can then break down the deformation through a region into a series of deformations over individual tetrahedra. As we explain in Section \ref{Section_more_general_membranes} of the Supplemental material, when we deform the membrane over a single tetrahedron, the surface weight changes by a factor equal to the phase acquired by applying vertex transforms on each vertex of that tetrahedron. This is the phase given in the last line of Table \ref{Table_Dual_Phase}, which we denote by $\theta^h_F(t)$. We can recognize this phase as being related to the slant product \cite{Wan2015},
		\begin{align}
	[g_1, g_2, g_3]_h &:= [h, g_1, g_2, g_3]^{-1} [g_1, h, g_2, g_3] \notag \\ & \hspace{1cm} [g_1, g_2, h, g_3]^{-1} [g_1, g_2, g_3,h],
		\end{align}
	by
	\begin{equation}
		\theta^h_F(t)= [\hat{g}_{v_1(t) v_2(t)}, \hat{g}_{v_2(t)v_3(t)}, \hat{g}_{v_3(t)v_4(t)}] _h^{- \epsilon(t)}. \label{Equation_full_tetrahedron_phase}
	\end{equation}
	For an Abelian group, this slant product obeys a 3-cocycle condition
\begin{equation}
	\frac{[x,y,z]_u [w, xy,z]_u [w,x,y]_u}{[wx,y,z]_u [w,x, yz]_u}=1, \label{Equation_3_cocycle_condition}
\end{equation}
as well as a normalization condition
\begin{equation}
	[1_G,y,z]_u =[x,1_G, z]_u = [x,y,1_G]_u = [x,y,z]_{1_G}=1,
\end{equation}
both of which can be derived from the 4-cocycle condition and normalization condition of the 4-cocycle \cite{Wan2015}.

\begin{figure}[h]
	\centering
	\includegraphics[width=0.8\linewidth]{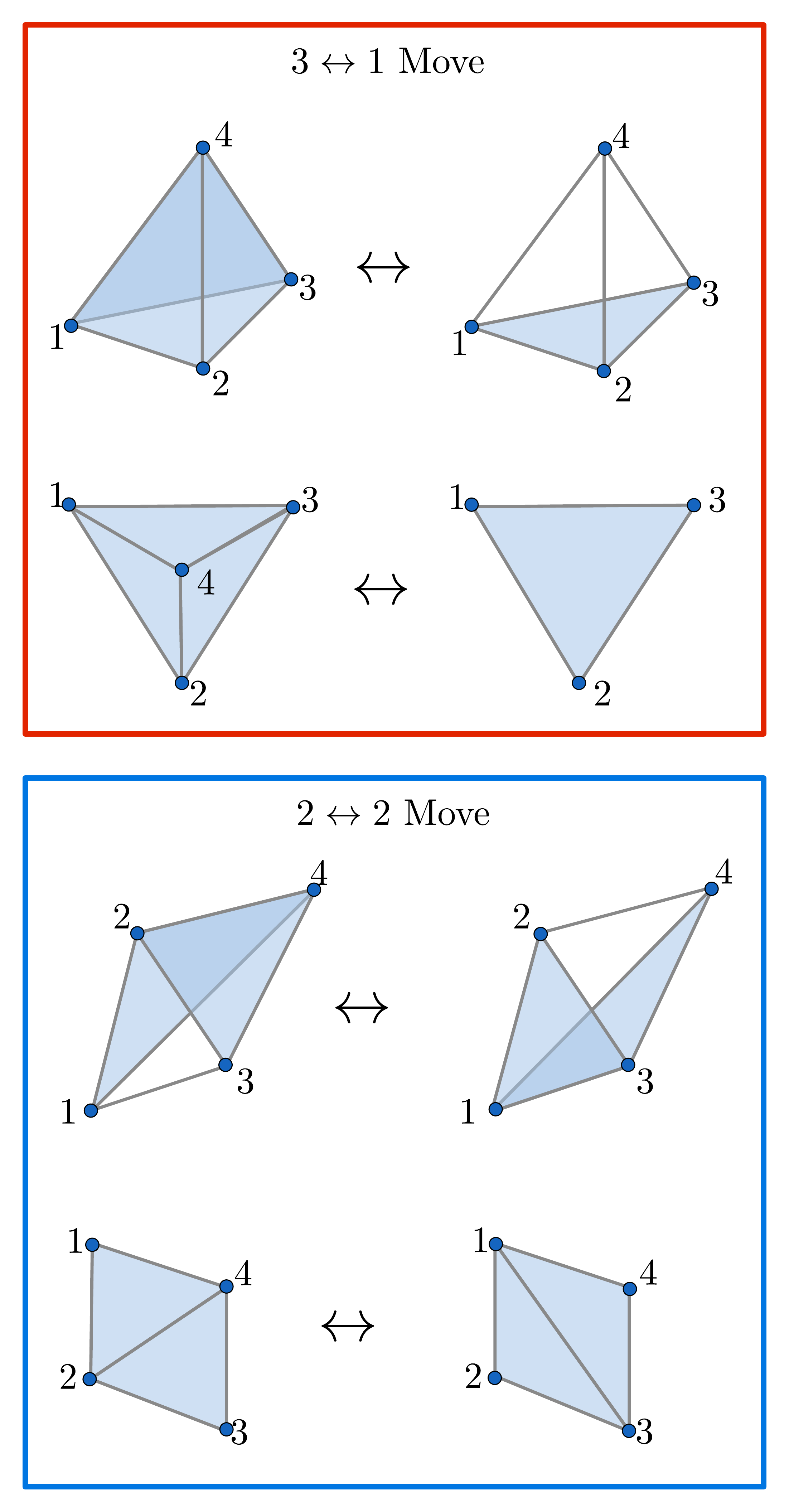}
	\caption{In the two boxes, we show the ways in which a membrane can be deformed over a tetrahedron. In the top row of each box, we show we show the move in 3d and in the lower row we show the corresponding change to a 2d diagram representing the surface. In the first case the membrane starts with three faces from the tetrahedron on the membrane and then ends with just the fourth face. The corresponding bistellar flip shown in the lower row is the $3 \leftrightarrow 1$ move. In the second case, the membrane starts with two faces from the tetrahedron and ends with the other two faces. The corresponding bistellar flip is therefore the $2 \leftrightarrow 2$ move.}
	\label{Figure_3d_deform_to_2d_move}
\end{figure}

\begin{figure}[h]
		\centering
		\includegraphics[width=0.9\linewidth]{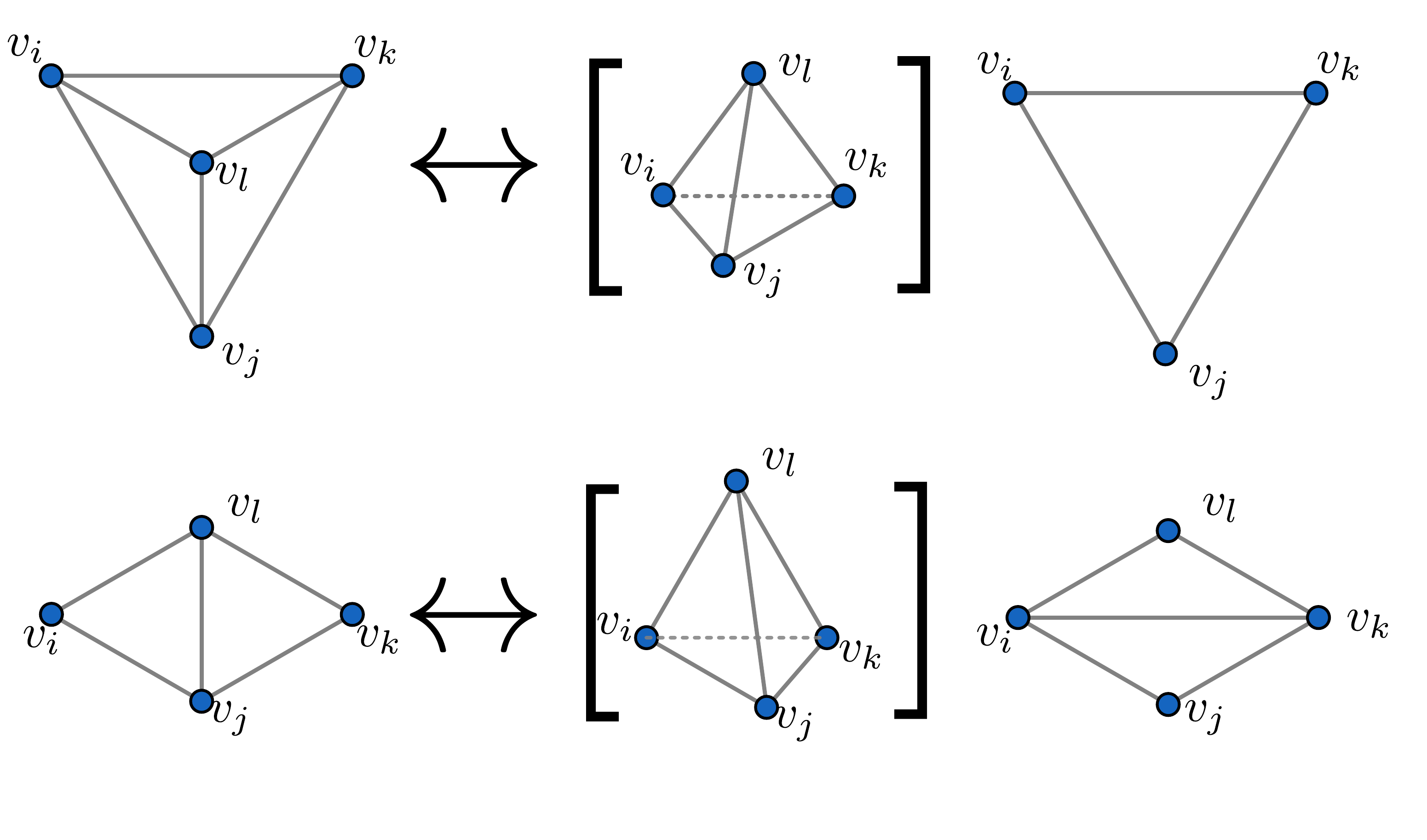}
		\caption{The graphical rules describe the phase gained upon performing 2d bistellar flips. Here the terms in squared brackets indicate the phase gained during the move, where the phase should be evaluated from the tetrahedron in the brackets according to Equation \ref{Equation_full_tetrahedron_phase}.}
		\label{Figure_2d_graphical_rules}
	\end{figure}

\begin{figure}[h]
			\centering
			\includegraphics[width=0.9\linewidth]{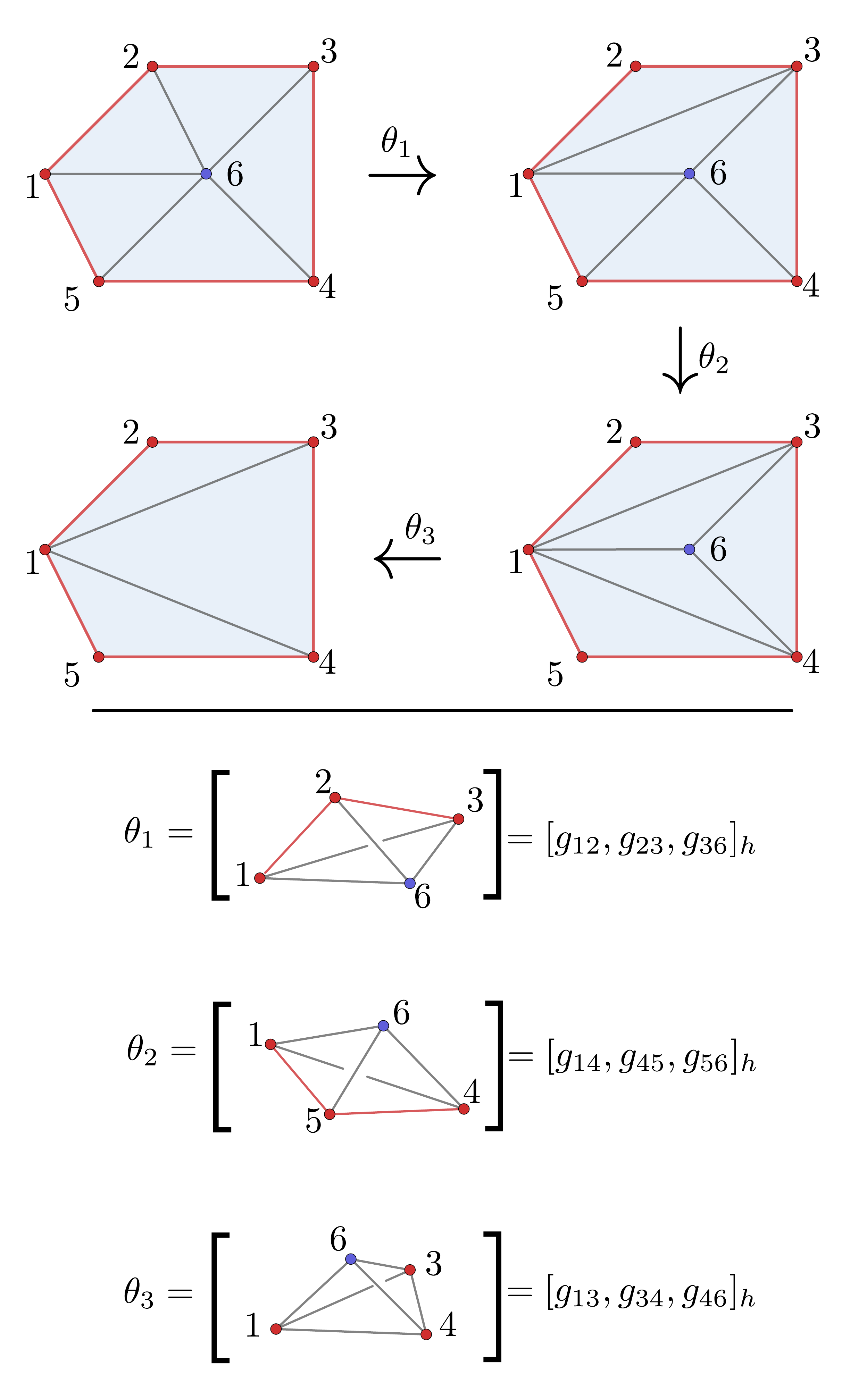}
			\caption{Here, we give an example of how a surface can be reduced to a simpler reference diagram. The edges and vertices on the boundary of the surface (shown in red online) cannot be removed by the diagrammatic moves. At each step, we gain a phase given by the slant product associated to a tetrahedron, as indicated in the lower part of the figure.}
			\label{Figure_diagram_reduction_example}
		\end{figure}

	When we deform the membrane over a tetrahedron, it induces a change to the 2d direct membrane, as shown in Figure \ref{Figure_3d_deform_to_2d_move}. These mutations can be recognized as 2d bistellar flips (also called Pachner moves) \cite{Pachner1990, Pachner1991, Casali1995}. Because the surface weight only depends on the degrees of freedom on the direct membrane, these bistellar flips and the phase that accompanies them become graphical rules that relate the surface weight that the membrane operator would associate to different diagrams, divorced from the physical lattice. These rules are shown in Figure \ref{Figure_2d_graphical_rules}. We can then use these bistellar flips to evaluate the surface weight, by systematically simplifying the diagram, similar to the process used in Ref. \cite{Lin2015}. An example of this reduction is shown in Figure \ref{Figure_diagram_reduction_example}. While we can greatly simplify the diagram this way, at some point, the diagram will become irreducible (for example, for open membranes, we cannot remove any vertices on the boundary of the membrane). This leaves us with a set of independent diagrams, which we call reference diagrams. The different values for these reference diagrams then describe a space of membrane operators for a given flux $h$, and a particular choice of these values defines the vector $\vec{v}$ which we used to label the membrane operator. That is, in order to calculate the surface weight of the membrane operator for a general diagram, we first use the diagrammatic rules to relate the weight to that of a reference diagram, then substitute in the weight associated to the reference diagram that is specified by the membrane operator. The allowed reference diagrams depend on the topology and boundary of the membrane, meaning that the surface weight carries this information. Furthermore, the value assigned to each reference diagram generally depends on the edge labels in that diagram, which means that the weight acts like a series of electric ribbon operators with end-points on the boundary of the membrane (as well as closed ribbon operators around any non-contractible curves on the membrane). The membrane operator should therefore be thought of as including both electric and magnetic components. 
	
		As an example of a membrane operator, consider a spherical membrane operator which does not enclose any excitations. Using the topological property, this sphere can be shrunk to nothing. Because deformation is equivalent to applying vertex transforms, this means that the membrane operator is equivalent to applying a series of vertex transforms in the region enclosed by the spherical membrane. The multiplicative action of the vertex transforms on the edges will cancel out in the interior of the region and will reproduce the action of the untwisted membrane on the boundary edges. We also obtain a phase for each tetrahedron attached to each vertex. We can group the tetrahedra into two types, those for which all vertices on that tetrahedron are in the region and those for which only a subset are in the region. The latter type are the boundary tetrahedra and their contribution to the overall phase is the dual phase we discussed, which comes from applying vertex transforms only on the vertices on the membrane, because these are the only vertices from those tetrahedra in the region enclosed by the sphere. For the former type of tetrahedron, the phase comes from applying transforms on each vertex of the tetrahedron. When we combine the phase from each such tetrahedron, it can be expressed in terms of variables on the surface by using the cocycle conditions. This combined phase is the surface weight. Because the membrane is spherical, it has no non-contractible loops and no boundary, which means it has only one possible reference diagram and the surface weight can be fully calculated in this way, without needing to define additional quantities in the membrane operator. We prove this relation between spherical membrane operators and vertex transforms in Section \ref{Section_spherical_membranes} of the Supplemental Material.

	\subsection{Cylindrical Membranes}
	\label{Section_cylindrical_membranes}
\begin{figure}[h]
		\centering
		\includegraphics[width=0.5\linewidth]{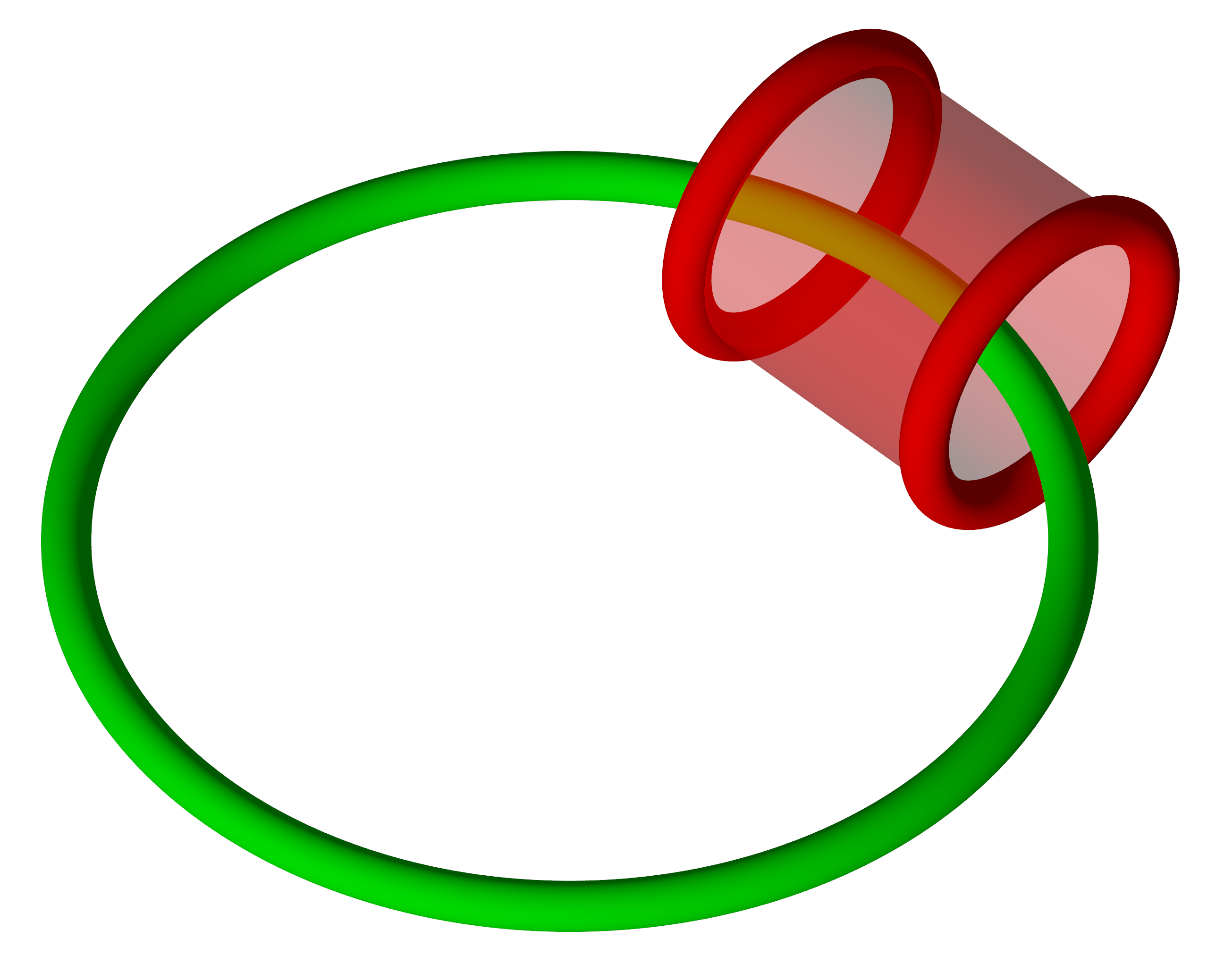}
		\caption{We can produce a pair of excitations linked with a given base loop (the large green loop) by applying a membrane operator that encloses that base loop.}
		\label{Figure_cylinder_membrane_linked_schematic}
	\end{figure}

	In the previous section, we considered membrane operators of a general topology. Now, we wish to consider specifically cylindrical (or annular) membrane operators, which produce flux loops at the two ends of the cylinder. These are of particular interest when considering linked loop-like excitations, because the cylindrical membrane operators can enclose an existing flux tube without crossing them, as shown in Figure \ref{Figure_cylinder_membrane_linked_schematic}. This means that cylindrical membrane operators can produce excitations that are linked to that existing flux tube, which we call the base loop. If we were to use a disk-shaped membrane, which produces a single loop-like excitation, it would have to intersect with the base loop in order to produce a loop-like excitation linked with the base loop. This means that the membrane operator would have a flatness-violation on its surface, which is incompatible with our approach for defining the membrane operators. This means that cylindrical membrane operators are the simplest way to obtain linked excitations. The cylindrical membrane operators that must be used depend explicitly on the base loop. This is because the path label of a cycle wrapping around the cylinder must match the flux of the base loop if no other excitations are present. As argued in Ref. \cite{Lin2015} for a $\mathbb{Z}_2 \times \mathbb{Z}_2$ model, this means that the loop-like excitations are fundamentally different depending on the base loop.
	
\begin{figure}[h]
		\centering
		\includegraphics[width=0.8\linewidth]{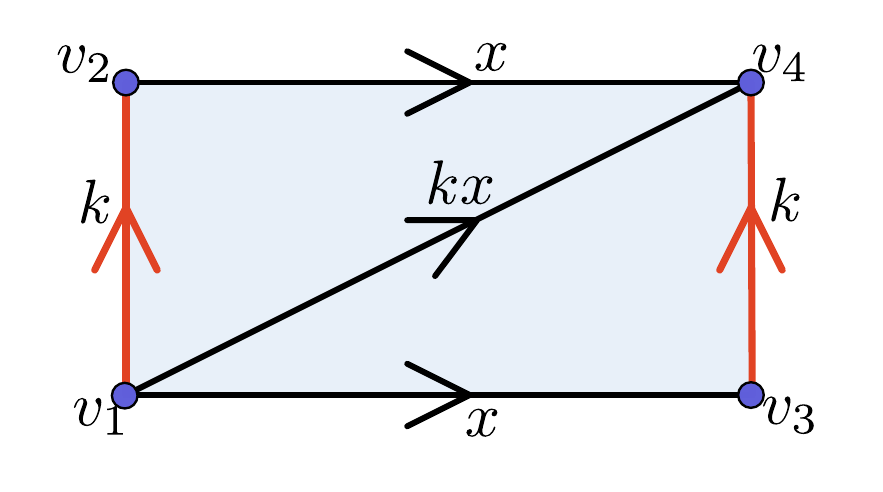}
		\caption{Here, we give an example of a simple reference diagram for a cylinder. There are periodic boundary conditions in the $k$ direction, so that the paths labeled by $k$, which form the boundaries of the membrane, are closed and the paths labeled by $x$ are identified.}
		\label{Figure_rectangle_reference_example}
	\end{figure}

	In order to explore this further, we will consider the reference diagrams for the cylindrical membrane operators in more detail. While a general reference diagram for the cylindrical membrane may have any number of vertices on the loop-shaped boundaries, it will be useful to consider the simplest possible situation, where there is only one vertex on each loop. As shown in Figure \ref{Figure_rectangle_reference_example}, these diagrams can be represented by rectangles with two of the edges identified. To specify a magnetic membrane operator of flux $h$ applied on this membrane, we must assign a coefficient for each possible label of the horizontal edge $x$, which we denote by $a^{k,h}(x)$. Here $h$ is a fixed quantity (the label of the operator we apply) while $k$ is the label of the closed path around the cylinder. While the label of this closed path is not fixed in a general state, it is $1_G$ in the ground state due to flatness. If we first create a flux tube with flux $k$ from the ground state and then apply the cylindrical membrane operator, then the cylindrical membrane operator will always measure the closed loop value to be $k$, so we can treat it as a fixed quantity. This means that $a^{k,h}(x)$ can be treated as an operator that depends only on $x$ and so is equivalent to an electric ribbon operator applied along the length of the cylinder.

	While any set of coefficients $a^{k,h}(x)$ gives a valid membrane operator, there is a useful basis that simplifies the properties of the membrane operators. This is analogous to the use of an irrep basis for the electric ribbon operators, which simplifies fusion rules and the algebra for laying the ribbon operators end-to-end. In the case of the electric ribbon operators, if we put two such operators labeled by the same irrep $R$ of $G$ end-to-end on paths $t_1$ and $t_2$, such that the paths concatenate to a path $t$, we have
	\begin{align*}
		\sum_{g \in G} R(g) \delta(\hat{g}(t_1), g) \sum_{k \in G} R(k)& \delta(\hat{g}(t_2),k) \\
  = R(\hat{g}(t_1)) R(\hat{g}(t_2))
		 &= R( \hat{g}(t_1)\hat{g}(t_2)) \\
   &= \sum_{g \in G} R(g) \delta(\hat{g}(t), g).
	\end{align*}
We see that the result is a single ribbon operator labeled by $R$ applied on the combined path $t$. Motivated by this, we wish to find a similar basis for the membrane operators.

\begin{figure}[h]
	\centering
	\includegraphics[width=\linewidth]{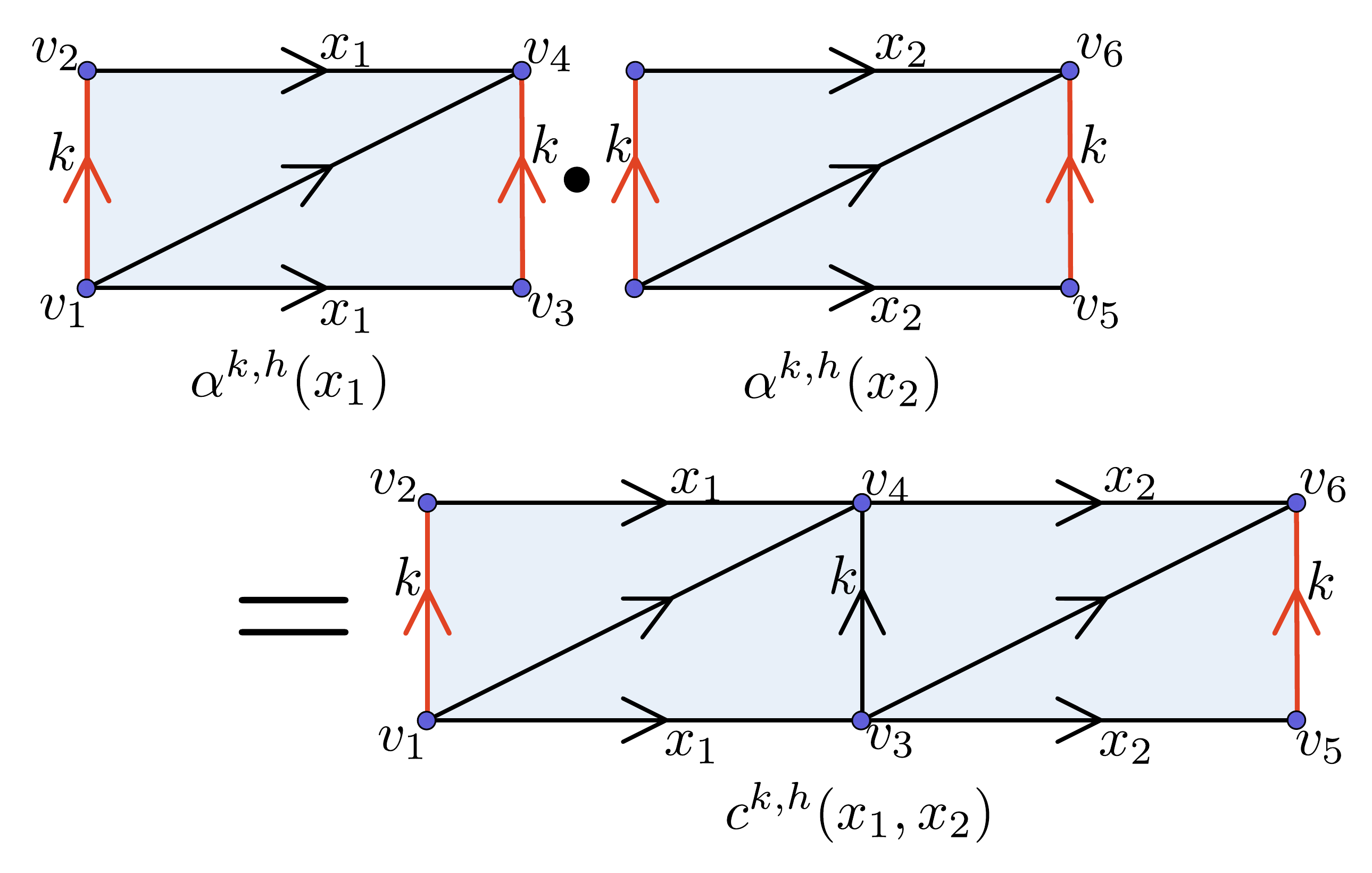}
	\caption{We look for a basis where the weights for two diagrams multiply to give the label of the concatenation of the two diagrams. In this case, that means $\alpha^{k,h}(x_1)\alpha^{k,h}(x_2)=c^{k,h}(x_1,x_2)$.}
	\label{Figure_rectangle_concatenation}
\end{figure}

\begin{figure*}[t]
	\centering
	\includegraphics[width=0.8\linewidth]{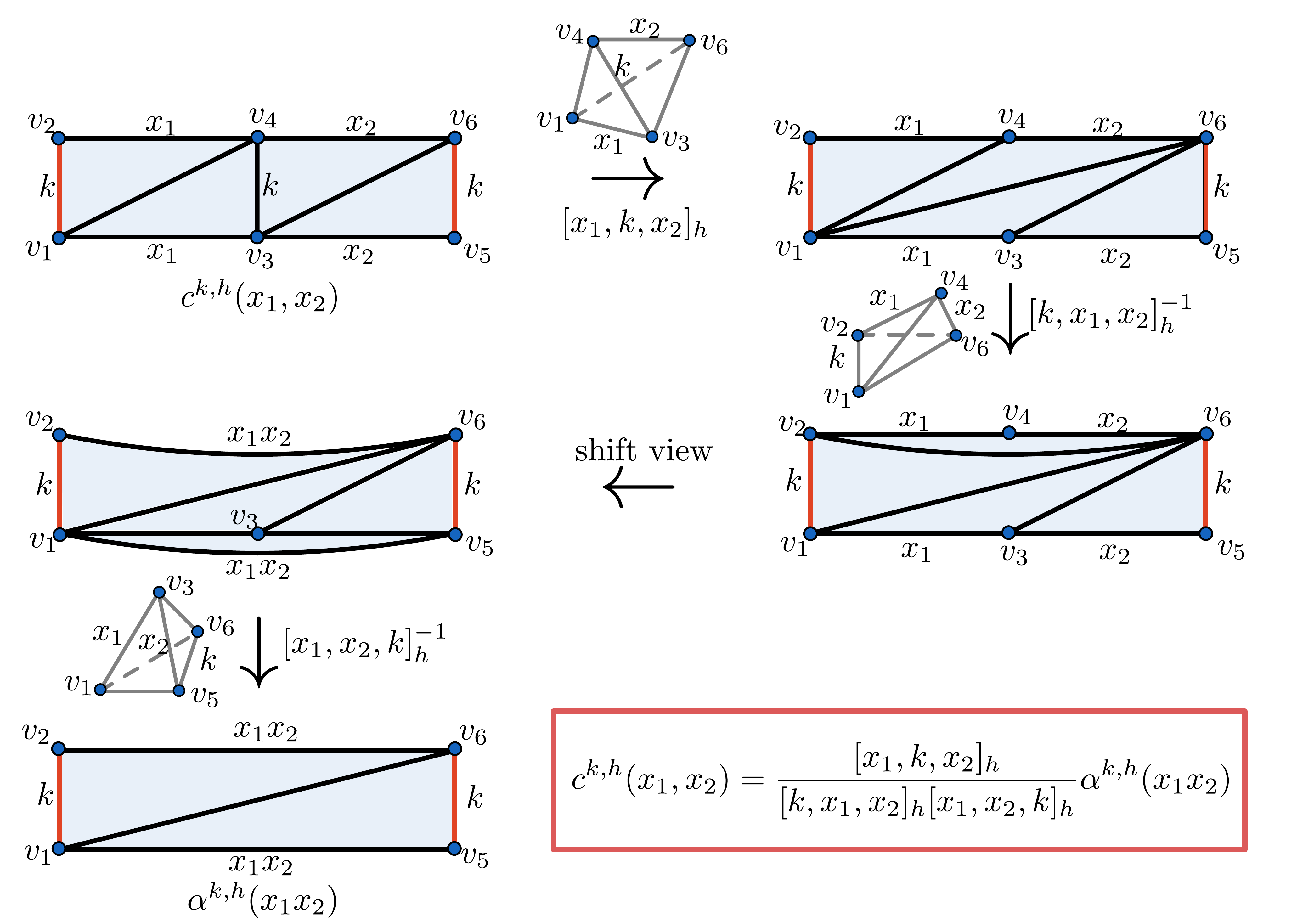}
	\caption{We can reduce the concatenated diagram to a single reference diagram by using the diagrammatic rules. This allows us to find the surface weight of the concatenated diagram in terms of the label of a reference diagram.}
	\label{Figure_concatenation_reduction}
\end{figure*}

That is, we want to find a basis where the surface weight given by the concatenation of two reference diagrams is equal to the product of the phase given by the two diagrams, as shown in Figure \ref{Figure_rectangle_concatenation}. We can then reduce the concatenated diagram to the same form as the reference diagram by using the diagrammatic rules, as shown in Figure \ref{Figure_concatenation_reduction}. Denoting the basis weight by $\alpha^{k,h}$ and the weight of the concatenated diagram by $c^{k,h}$, this gives us
\begin{align*}
	\alpha^{k,h}&(x_1) \alpha^{k,h}(x_2) \mbeq c^{k,h}(x_1, x_2) \\
 &= [x_1, k, x_2]_{h} [k, x_1, x_2]_h^{-1} [x_1, x_2, k]_h^{-1} \alpha^{k,h}(x_1x_2).
\end{align*}
Because $G$ is Abelian, we can write $[x_1, k, x_2]_{h} [k, x_1, x_2]_h^{-1} [x_1, x_2, k]_h^{-1}$ in terms of the slant product of the 3-cocycle
\begin{equation}
[x_1, x_2]_{h,k} = [k, x_1, x_2]_h [x_1, k, x_2]_{h}^{-1} [x_1, x_2, k]_h, \label{Equation_2_cocycle_definition}
\end{equation}
to obtain
\begin{equation}
	\alpha^{k,h}(x_1) \alpha^{k,h}(x_2) \mbeq [x_1, x_2]_{h,k}^{-1} \alpha^{k,h}(x_1x_2). \label{Equation_concatenation_relation_1}
\end{equation}
For Abelian $G$, this slant product satisfies the 2-cocycle conditions \cite{Wan2015} 
\begin{equation}
	\frac{[y,z]_{w,u} [x,yz]_{w,u}}{[xy,z]_{w,u} [x,y]_{w,u}}=1 \ \forall u,w,x,y,z \in G. \label{Equation_2_cocycle_condition}
\end{equation}
In addition, the slant product obeys the normalization condition,
\begin{equation}
	[1_G, y]_{w,u}=[x, 1_G]_{w,u}=[x,y]_{1_G, u} = [x,y]_{w, 1_G}, \label{Equation_2_cocycle_normalization}
\end{equation}
and the permutation condition,
\begin{equation}
	[x,y]_{w,u}= [x,y]_{u,w}^{-1}, \label{Equation_2_cocycle_permutation}
\end{equation}
for all $u,w,x,y \in G.$

Using the permutation property, we can write Equation \ref{Equation_concatenation_relation_1} as
\begin{equation}
	\alpha^{k,h}(x_1) \alpha^{k,h}(x_2) \mbeq [x_1, x_2]_{k,h} \alpha^{k,h}(x_1x_2). \label{Equation_projective_irrep_basis}
\end{equation}
This relation is similar to that satisfied by a representation of $G$, apart from the phase $ [x_1, x_2]_{k,h} $. Indeed, if we take the label $k$ of the base loop to be trivial, this phase becomes unity and so $\alpha$ is just a linear representation. In that case, the membrane operators can be labeled by a flux in $G$ and an irrep of $G$, just as for the untwisted case. More generally, we can recognize Equation \ref{Equation_projective_irrep_basis} as the defining relation for a projective representation of $G$, with factor system described by the 2-cocycle $\beta_{k,h}(x_1,x_2)=[x_1, x_2]_{k,h}$ \cite{Melnikov2022}. In order to specify the 2-cocycle, such a projective representation is called a $\beta_{k,h}$-projective representation. Several properties familiar from linear representations are also satisfied by the projective representations. Firstly, they can be chosen to be unitary \cite{Melnikov2022} and we will only consider such unitary representations from here on. Secondly, projective representations can be decomposed into irreducible projective representations (which we will call projective irreps). These projective irreps obey orthogonality and completeness conditions, which enables us to use their matrix elements as the basis for the space of weights for the reference diagrams. Schur's Lemma also applies to projective irreps, meaning that a matrix that commutes with all matrices in a projective irrep must be a scalar matrix.

On the other hand, projective representations do have some significant differences from linear representations. Firstly, the projective irreps may be higher-dimensional even for an Abelian group. This can be seen directly from Equation \ref{Equation_projective_irrep_basis}. Swapping the order of multiplication, we obtain
\begin{align*}
		\alpha^{k,h}(x_2) \alpha^{k,h}(x_1) &= [x_2, x_1]_{k,h} \alpha^{k,h}(x_2x_1)\\
		&=[x_2, x_1]_{k,h} \alpha^{k,h}(x_1x_2),
\end{align*}
where we used the fact that $G$ is Abelian to obtain the latter equality. We see that the two orders of multiplication give results that differ by a factor of
\begin{equation}
	\eta^{k,h}(x_1, x_2)= \frac{ [x_1, x_2]_{k, h}}{[x_2, x_1]_{k,h}}.
\end{equation}

If this factor is equal to unity (i.e., if the 2-cocycle defined in Equation \ref{Equation_2_cocycle_definition} is symmetric), then the two matrices always commute and by Schur's Lemma the irreps must be 1d. In this case, the (1d) irreps form a convenient basis for the membrane operators (using the completeness conditions) and are the phases assigned to the simple reference diagrams in that basis. On the other hand, if there are some values of $x_1$ and $x_2$ for which the factor is not equal to unity, then none of the irreps can be 1d. In this case, the matrix elements $[\alpha^{k,h}(x_1)]_{ij}$ of the irreps form a complete basis and are the numbers assigned to the reference diagrams. This results in the slightly more complicated concatenation rule
\begin{equation}
	\sum_{j=1}^{|\alpha^{k,h}|} [\alpha^{k,h}(x_1)]_{ij} [\alpha^{k,h}(x_2)]_{jk} = [x_1, x_2]_{k,h} [\alpha^{k,h}(x_1x_2)]_{ik}, \label{Equation_projective_irrep_basis_matrix}
\end{equation}
where $|\alpha^{k,h}|$ is the dimension of the irrep $\alpha^{k,h}$. This rule is analogous to the concatenation of electric ribbon operators for a non-Abelian group (although here the non-Abelian nature comes from using projective irreps of an Abelian group rather than linear irreps of a non-Abelian group).

Another property that differs from linear representations is that the representation of the inverse of a group element is not equal to the inverse of the representation of that group element. To see this, note that the projective representation $\alpha^{k,h}$ satisfies
\begin{equation}
	\alpha^{k,h}(g)\alpha^{k,h}(g^{-1}) = [g,g^{-1}]_{k,h} \alpha^{k,h}(1_G). \label{Equation_proj_inverse_intermediate}
\end{equation}
Then $\alpha^{k,h}(1_G)$ is the identity matrix, because
\begin{equation}
	\alpha^{k,h}(1_G) \alpha^{k,h}(g) = [1_G, g]_{k,h} \alpha^{k,h}(g) = \alpha^{k,h}(g), 
\end{equation}
from the normalization condition (Equation \ref{Equation_2_cocycle_normalization}). This means that Equation \ref{Equation_proj_inverse_intermediate} becomes
\begin{equation*}
	\alpha^{k,h}(g)\alpha^{k,h}(g^{-1}) = [g,g^{-1}]_{k,h} I,
\end{equation*}
so that
\begin{equation}
\alpha^{k,h}(g^{-1})= [g,g^{-1}]_{k,h} \alpha^{k,h}(g)^{-1}. \label{Equation_proj_inverse}
\end{equation}

\begin{figure}
	\centering
	\includegraphics[width=0.7\linewidth]{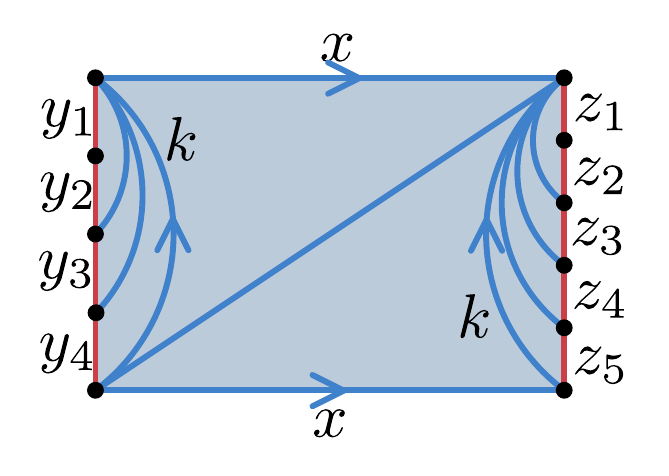}
	\caption{An example of a more complicated reference diagram. We can use a similar form for diagrams with any number of vertices on the boundaries of the cylinder.}
	\label{Figure_cylinder_reference_example}
\end{figure}

\begin{figure*}[t]
	\centering
	\includegraphics[width=0.7\linewidth]{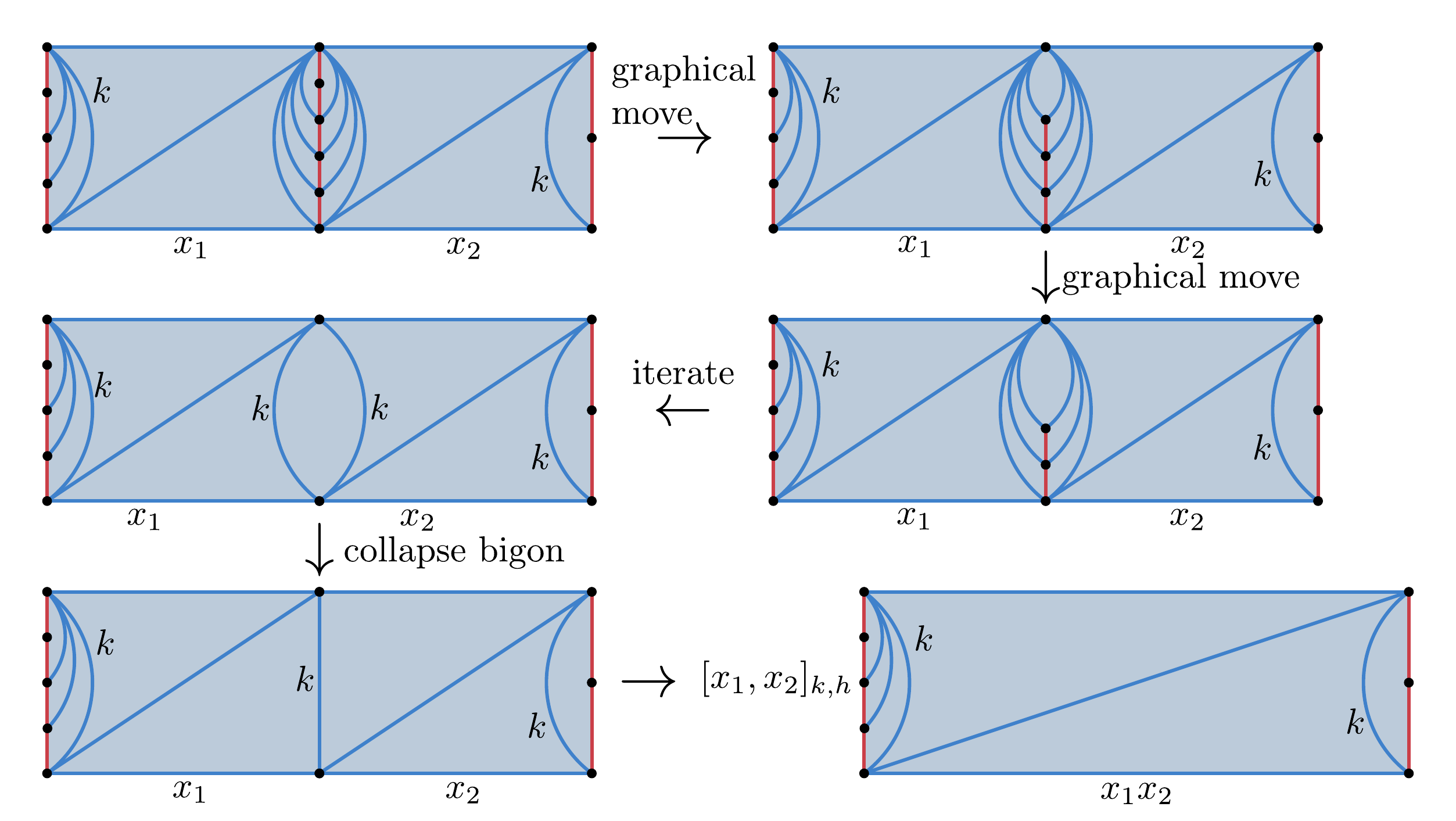}
	\caption{Using the form of the reference diagram from Figure \ref{Figure_cylinder_reference_example} allows two diagrams to be concatenated easily. We can remove the bubbles formed by the additional edges in the middle using the graphical moves, which do not result in a phase gain because the moves involve deformation over trivial tetrahedra, where or more of the faces is a bigon rather than a triangle. In the last step, we use the same reduction from the simple reference diagrams (shown in Figure \ref{Figure_concatenation_reduction}), which results in the phase $[x_1,x_2]_{k,h}$.}
	\label{Figure_cylinder_reference_concatenation}
\end{figure*}

So far, we have only considered the simplest reference diagrams, which we will use throughout for explicit calculations. However, we should briefly mention that more complicated reference diagrams are needed for general cylindrical membrane operators. This is because cylinders may have any number of vertices on the two ends of the cylinder, which cannot be removed by the graphical rules and so must be accounted for in the reference diagrams. In the case where there are multiple such vertices, it is convenient to choose reference diagrams that only have the minimal number of edges that cross the length of the cylinder, as shown in Figure \ref{Figure_cylinder_reference_example}. Then, the additional edges, when compared to the simple reference diagram from Figure \ref{Figure_rectangle_reference_example}, are local to the ends of the cylinder. The weight for the cylinder can then be decomposed into a part for the simple reference diagram, which is well labeled by the projective irreps as before, and an additional part which is equivalent to applying electric ribbon operators along parts of the ends of the cylinder. In order to compose two diagrams, these additional parts must be compatible, while the composition itself is described by the projective irreps as before. An example of this composition is shown in Figure \ref{Figure_cylinder_reference_concatenation}. In addition to the reference diagrams potentially having multiple vertices on each end of the cylinder, for a general reference diagram the edges along the boundary may have different orientations, which can change the composition rules (although they are still related to the projective irreps).

We have shown that a convenient basis for the membrane operators is described by projective irreps of the group $G$. As we show in Section \ref{Section_three_loop_braiding}, this basis also gives well-defined three-loop braiding statistics, suggesting that the basis operators carry a definite topological charge. Note that the projective irreps that label the membrane operators explicitly depend on the flux of the base loop. This implies that the loop-like excitations linked with different base loops are fundamentally different and are also different from the loop-like excitations that are not linked to any base loop, as found for another model in Ref. \cite{Lin2015}. This helps to explain why the three-loop braiding statistics are needed in addition to regular two-loop braiding in order to characterize a phase: the characteristics of unlinked loops alone are not sufficient to describe all of the excitations.

	\section{Braiding Relations}

\begin{figure}[h]
		\centering
		\includegraphics[width=0.7\linewidth]{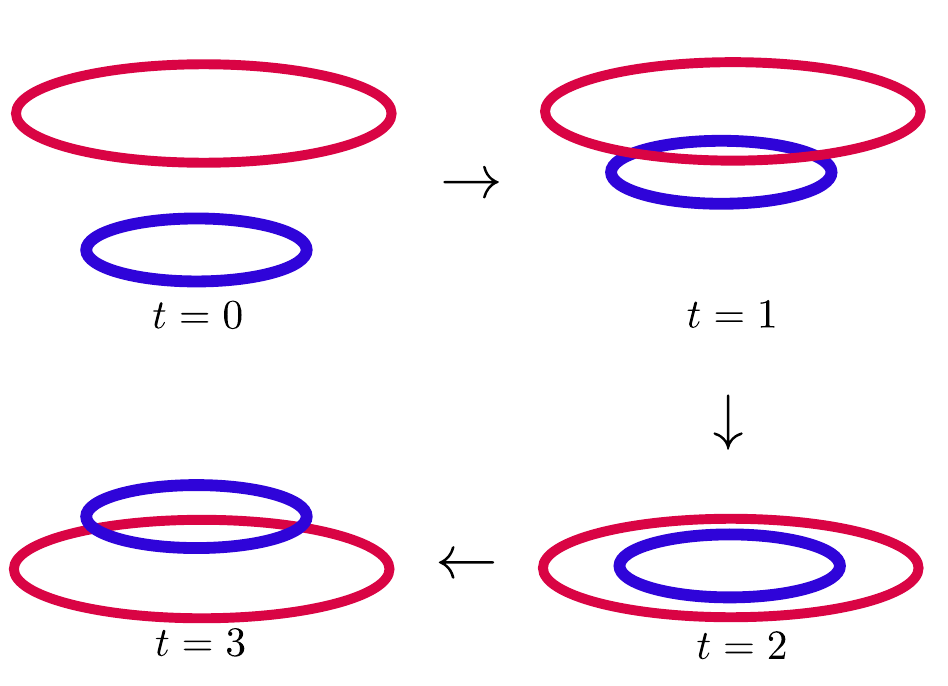}
		
		\caption{A time-lapse of an example braiding move, where the smaller (blue) loop is pulled up through another (red) loop.}
		\label{Figure_loop_braiding_timelapse}
	\end{figure}

	One of the key features of (long-range entangled) topological phases is that their excitations support exotic exchange statistics. However, in 3+1d, point-like particles only have fermionic or bosonic exchange statistics \cite{Doplicher1971, Doplicher1974}. This is because a process where one particle is moved around another can be continuously deformed into one where no such exchange is performed. On the other hand, 3+1d topological phases support loop-like excitations, some of which we have discussed in this work. Compared to the braiding that only involves point-like excitations, there are multiple types of braiding involving one or more loop-like excitations. The first type is the process mentioned earlier, where two excitations (either loop-like or point-like) are moved around each-other. This type of exchange, which we will refer to as permutation, is restricted to be bosonic or fermionic in 3+1d, regardless of whether it involves loops or points. On the other hand, there is also an exchange process (shown in Figure \ref{Figure_loop_braiding_timelapse}) where one excitation (either point-like or loop-like) is pulled \textit{through} a loop-like excitation, which we refer to as loop-braiding or simply as braiding \cite{Aneziris1991, Alford1992, Baez2007, McCool1986, Savushkina1996, Damiani2017}. If it is necessary to differentiate between the cases where a point or a loop is pulled through the other excitation, we use the terms point-loop and loop-loop braiding. If both of the excitations are loop-like, there is a generalization where a third loop, called the base loop, stays linked with the braiding loops during the process. This is referred to as three-loop braiding \cite{Wang2014} or necklace braiding \cite{Bellingeri2016}. Recently, it has become clear that this three-loop braiding is important for classifying a topological phase and two phases that have the same loop-loop and point-loop braiding may be distinguished through their three-loop braiding \cite{Lin2015, Wang2014, Jiang2014, Wang2015}. Furthermore, even if the loop-loop braiding is Abelian, the corresponding three-loop braiding may be non-Abelian. There has already been significant study of the three-loop braiding in lattice gauge theory \cite{Lin2015, Wang2014, Jiang2014, Wang2015}. However, most studies use indirect methods to extract the braiding statistics of the emergent quasiparticles and the few that explicitly construct the creation operators examine relatively simple cases \cite{Lin2015}. We aim to use the membrane operators constructed in the previous sections to reproduce these braiding relations.
	
	\subsection{Point-loop Braiding}
	\label{Section_point_loop_braiding}
	
\begin{figure}
		\centering
		\includegraphics[width=0.6\linewidth]{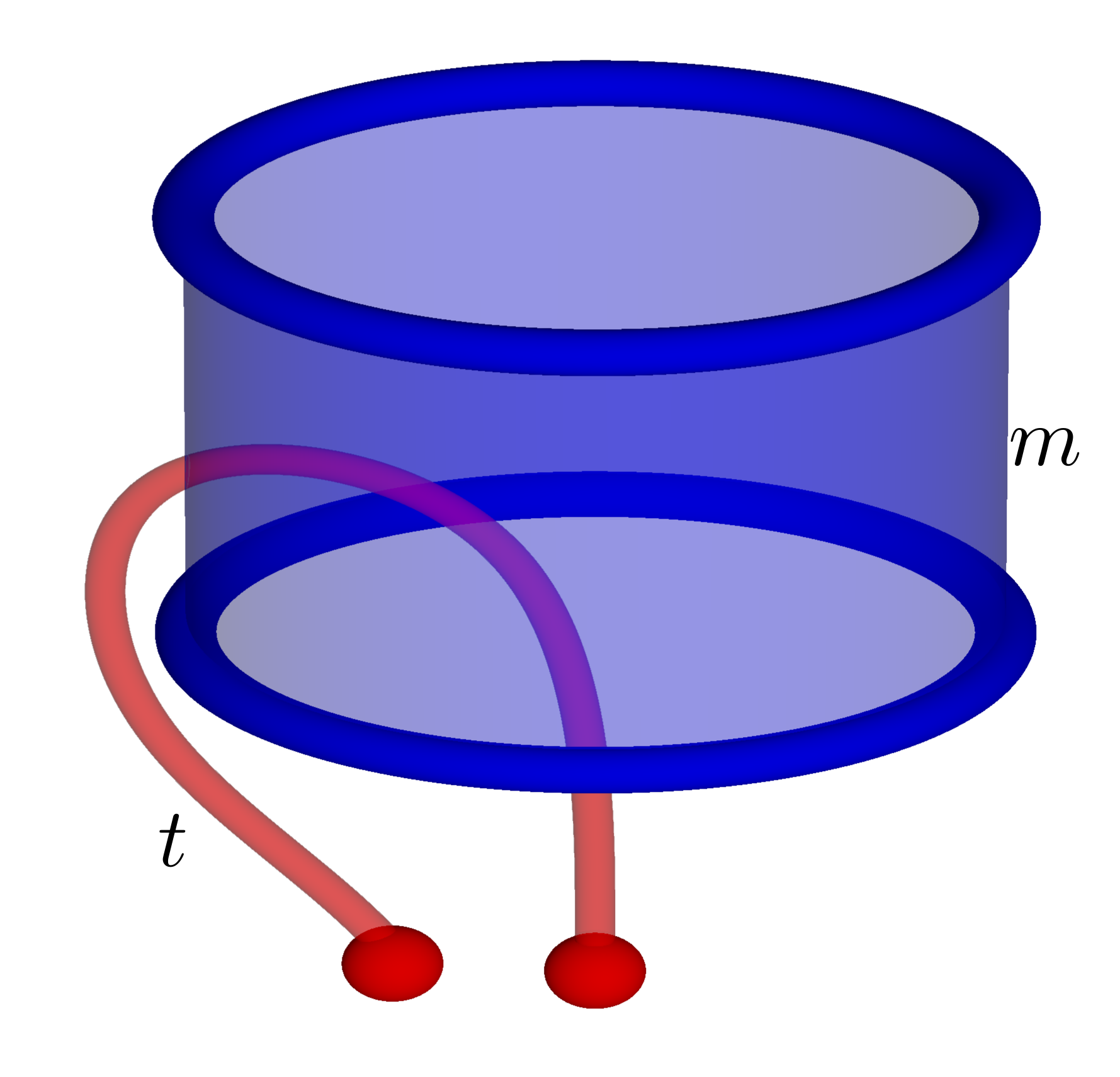}
		\caption{Point-loop braiding can be implemented using a ribbon operator applied on a path, $t$, which intersects with a membrane operator applied on a membrane, $m$. The loop-like excitation shown could be linked to another loop (not shown), but this would not affect the braiding relation unless the point-like particle also braids with this second loop}
		\label{Figure_point_loop_braiding}
	\end{figure}

\begin{figure}
		\centering
		\includegraphics[width=0.9\linewidth]{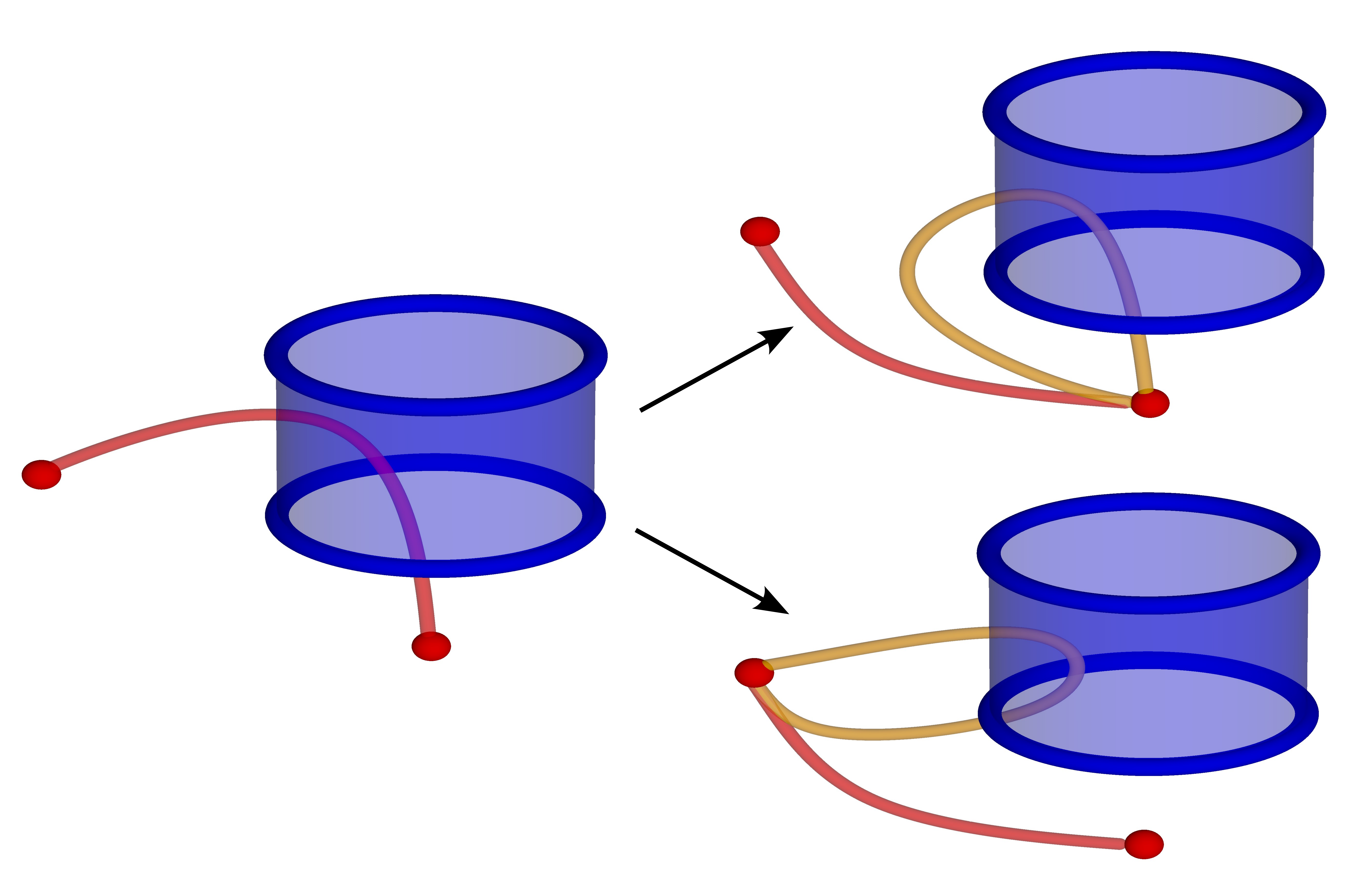}
		\caption{One way of conceptualizing the relationship between ribbon operators and the braiding relation is to deform the open ribbon into a composition of a closed ribbon (orange online) and an open part that does not intersect the membrane. The former encapsulates the braiding relation, while the latter moves the particles to their final positions. The closed ribbon operator can be placed at either end of the ribbon (as shown in the top and bottom images), which leads to different expressions for the closed ribbon part when the braiding is non-Abelian.}
		\label{Figure_particle_loop_deformation}
	\end{figure}

	The first type of braiding to discuss involves pulling a point-like excitation through a loop-like one. We only discussed the point-like excitations in this model briefly, because they have the same character as for untwisted gauge theory. This result extends to their braiding relations, as we will now see. In order to describe the braiding relations, we first relate the braiding relation to a commutation relation between ribbon and membrane operators. A ribbon operator can be interpreted as creating a pair of point-like excitations and bringing them to the ends of the ribbon. Similarly, a membrane operator creates some number of loop-like excitations and brings them to the boundary of the membrane. This means that the braiding can be described by a ribbon operator applied on a path that intersects with a membrane operator, as shown in Figure \ref{Figure_point_loop_braiding}. If we first apply a membrane operator to produce some loop-like excitation of interest, and then apply a ribbon operator that passes through the loop, we are considering a situation where we create the loop and then move a point-like excitation through it. On the other hand, if we first apply the ribbon operator and then the membrane operator, then we move the point-like excitation through empty space before creating the loop. Comparing these situations, by examining the commutation relation between the relevant operators, therefore gives us the transformation undergone during braiding. Another way of thinking about this is that we can deform the ribbon so that it intersects its start-point and then split it into a closed ribbon and an open one, as shown in Figure \ref{Figure_particle_loop_deformation}. Then the open part does not intersect the membrane operator and so commutes with the membrane operator. It describes the transport of the particle excitations along the new open path, regardless of the presence of the membrane operator. The closed part would be trivial if it acted directly on the ground state, but is non-trivial because it acts after the membrane operator. This closed part then describes how the particle excitation transforms under the braiding (as closed paths are usually required in order to obtain topological invariants). Instead of putting the closed part at the start of the combined ribbon operator, we could put it at the end (or indeed at any part), which gives a different expression when the braiding is non-Abelian.
	
	In this model, the ribbon operators simply measure the path element along the ribbon, as discussed in Section \ref{Section_ribbon_operators}. This means that they are not sensitive to the surface weight or dual phase of the membrane operator: the ribbon operators commute with these parts of the membrane operator. Instead, the commutation relation is entirely controlled by the flux of the membrane. This means that the braiding relation is independent of the 4-cocycle, which only enters through the dual phase and surface weight, and in particular is the same as in the untwisted gauge theory. Indeed, this braiding relation is just the same as the electric-magnetic braiding in the 2+1d quantum double model \cite{Kitaev2003}. Consider an electric ribbon operator applied on a path $t$ that passes through a membrane operator applied on membrane $m$, as shown in Figure \ref{Figure_point_loop_braiding}. In the case of an Abelian group $G$, if the membrane operator has flux $h$, then the membrane operator modifies the path label for the ribbon operator by multiplication by the flux label, $h$. Then, if the point-like excitation is labeled by an irrep $R$ of $G$ (which is one-dimensional), the commutation relation between the ribbon and membrane operator is given by

	\begin{align*}
		S^{R}(t)F^{h, \vec{v}}(m)&= \sum_{g \in G} R(g) \delta(g,\hat{g}(t)) F^{h, \vec{v}}(m)\\
		&= R(\hat{g}(t)) F^{h, \vec{v}}(m)\\
		&= F^{h, \vec{v}}(m) R(h\hat{g}(t)) \\
		&= F^{h, \vec{v}}(m) R(h) R(\hat{g}(t)). 
	\end{align*}
	
	We see that the result of the braiding is simply to pick up a phase $R(h)$ (or the inverse, depending on the orientation of the loop and direction of the braiding):
	
	\begin{equation}
			S^{R}(t)F^{h, \vec{v}}(m)= R(h) F^{h, \vec{v}}(m) 	S^{R}(t).
	\end{equation}
	
	In the case where $G$ is non-Abelian, the braiding depends on certain details of the ribbon and membrane operator, which control the total charge of the combined objects, but the transformation similarly only involves the irrep $R$ (and its matrix indices) and the flux $h$. A similar result holds for any more complicated process, where only exchange between a point-like excitation and one or more loop-like excitations (even if they are linked) is involved: the result only depends on the fluxes of the loops directly involved in the braiding, not the surface weight or dual phases of the membranes or the base loops that they may be linked to. One such process of significant interest is the so-called Borromean rings braiding \cite{Chan2018, Zhang2021, Zhang2023}, which measures the non-Abelian character of point-loop braiding and is therefore trivial in this model for Abelian $G$.

	\subsection{Three-loop Braiding}
	\label{Section_three_loop_braiding}
	
A more interesting process, which does depend on the additional quantities of the membrane operators in this model, is three-loop braiding. As discussed earlier, this is the process where one loop is passed through another while both loops are linked to a base loop. A special case of this process is ordinary loop-loop braiding, where the base loop is taken to be trivial, and we will later discuss how the braiding relation simplifies in this case.

	The first step in an explicit calculation of the braiding relations is to determine the geometry of the relevant membrane operators. For the three-loop braiding case there are three membrane operators: the one that produces the base loop that the others link to and then the two membrane operators that produce the loops directly involved in the braiding. We do not need to consider the former in detail, except to know its flux label, because it is not directly involved in the braiding process. For the other two membrane operators, we need to be careful because the membrane operators produce pairs of loops and we need the braiding process to only involve one loop from each pair.

\begin{figure}[h]
		\centering
		\includegraphics[width=\linewidth]{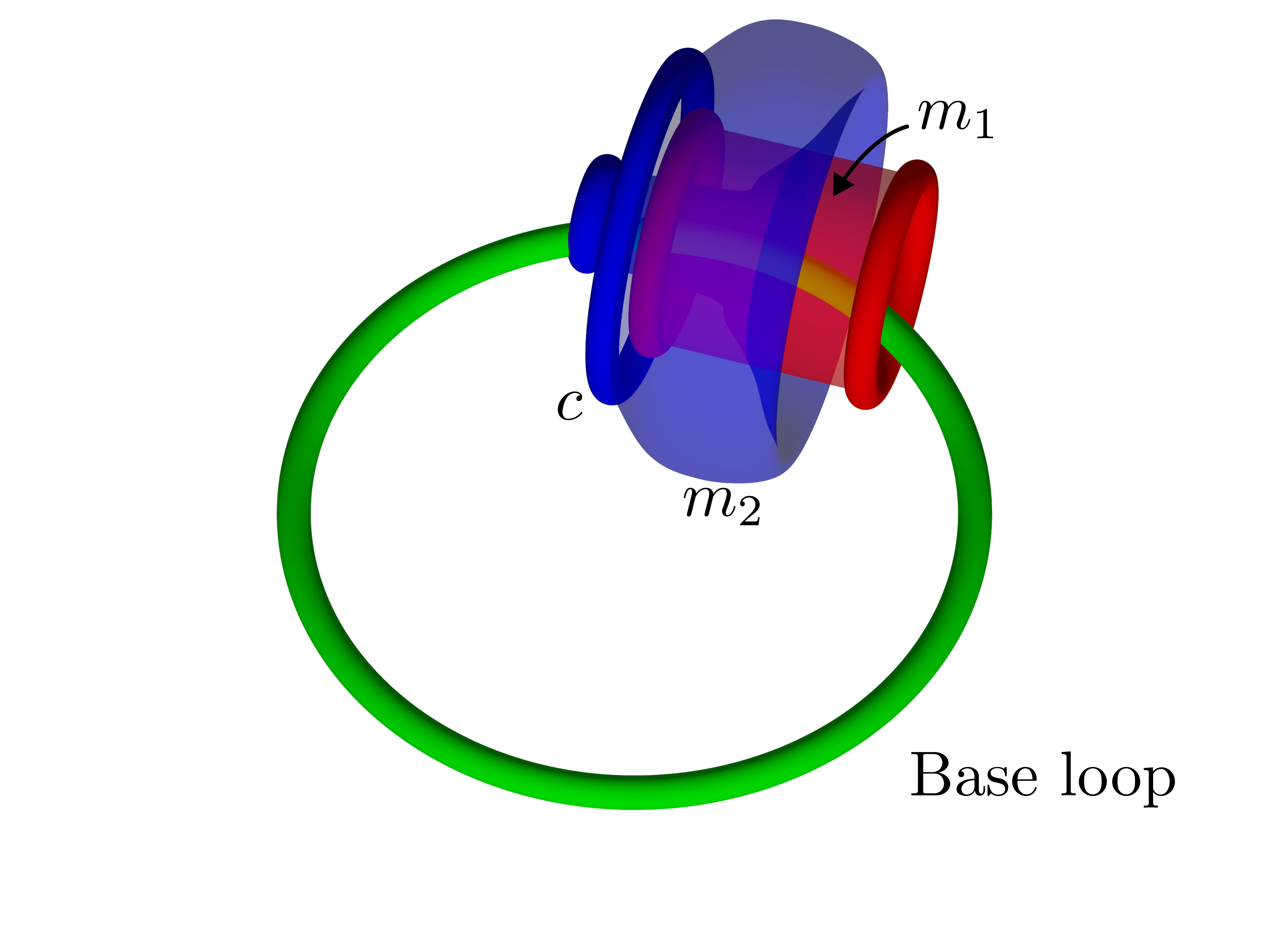}
		\caption{A possible geometry for the three-loop braiding relation, involving membrane operators on membranes $m_1$ (red) and $m_2$ (blue). The large (green) loop is the base loop to which the others are linked.}
		\label{Figure_three_loop_braiding}
	\end{figure}

	One possible geometry is shown in Figure \ref{Figure_three_loop_braiding}. In this case, we are considering a braiding move where the (blue) loop $c$ is pulled through the loop at the left side of $m_1$ and then back over it. To complete this braiding move, we could then close the (blue) membrane $m_2$, but the important part is the intersection between the two membranes. However, note that the loop-like excitations involved have an orientation, unlike point-like particles. In particular, because the loop-like excitation from $m_1$ is on the left of the membrane while the excitation from the $m_2$ is on the right, the two loop-like excitations have opposite orientations (reversing the orientation of a flux tube is the same as inverting its label). We will later consider the case where the excitations have the same orientation, but the calculation for this case is easier to follow and so we present this case first and then use it to obtain the result for the same-orientation case.

	With this geometry in mind, we can calculate the three-loop braiding relation using a commutation relation between the two membrane operators. Note that we will always consider situations where some membrane operator produces the base loop from the ground state before we apply any other operators. If we then apply the (red) cylindrical membrane operator on $m_1$ to produce a pair of (red) loops before applying the curved (blue) membrane operator on $m_2$, we consider a case where one (blue) loop $c$ is passed through the another (red) one and back over it. On the other hand, if we apply the membrane operator on $m_2$ first and then apply the membrane operator on $m_1$, the (blue) loop $c$ is passed through empty space before the other (red) loops are produced. Comparing these two orders of operations (i.e., calculating the commutation relation) therefore isolates the transformation from the loop $c$ passing through the other one and then back over it.
	
\begin{figure}
		\centering
		\includegraphics[width=\linewidth]{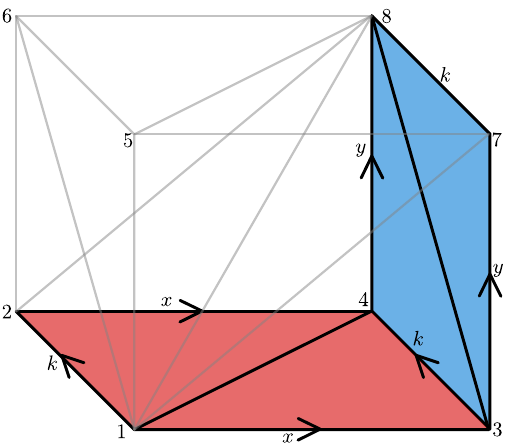}
		\caption{The direct membranes for the two operators in the intersection region are shown here, with the horizontal (red) one representing $m_1$ and the vertical (blue) one representing $m_2$. The grey lines indicate the rest of the intersection region, where both dual membranes act. There is a periodic boundary condition in the direction into the page, so that the edges labeled by $k$ are closed paths. This means that the vertices labeled 1 and 2 represent the same vertex, for example. The edge labels given describe a basis state.}
		\label{Figure_braiding_direct_membranes}
	\end{figure}

\begin{figure}
	\centering
	\includegraphics[width=\linewidth]{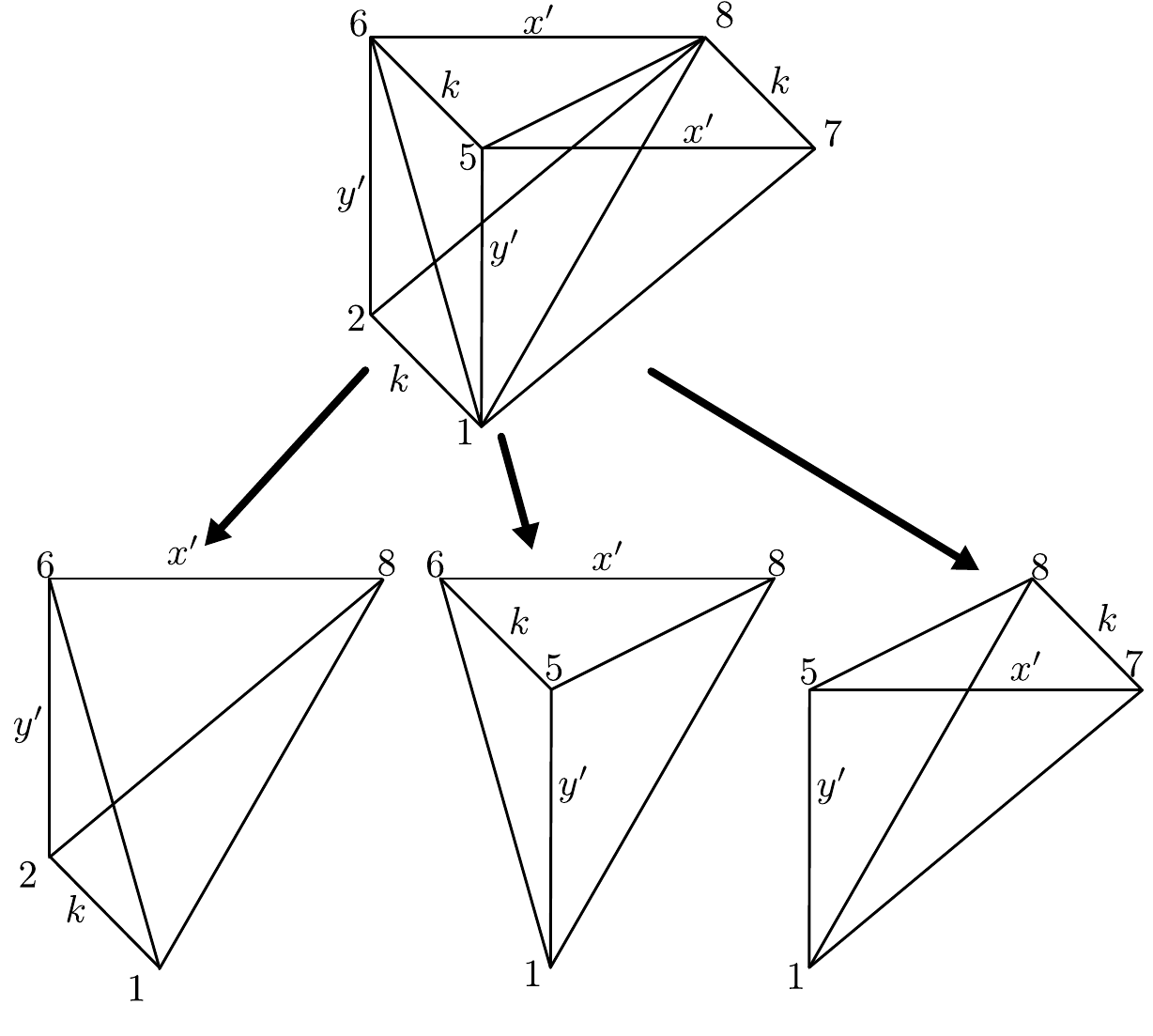}
	\caption{The cube-shaped intersection region can be split into two parts, of which this is the upper wedge. To find the dual phase associated to this region we split it into individual tetrahedra.}
	\label{Figure_braiding_upper_wedge}
\end{figure}

\begin{figure}
	\centering
	\includegraphics[width=\linewidth]{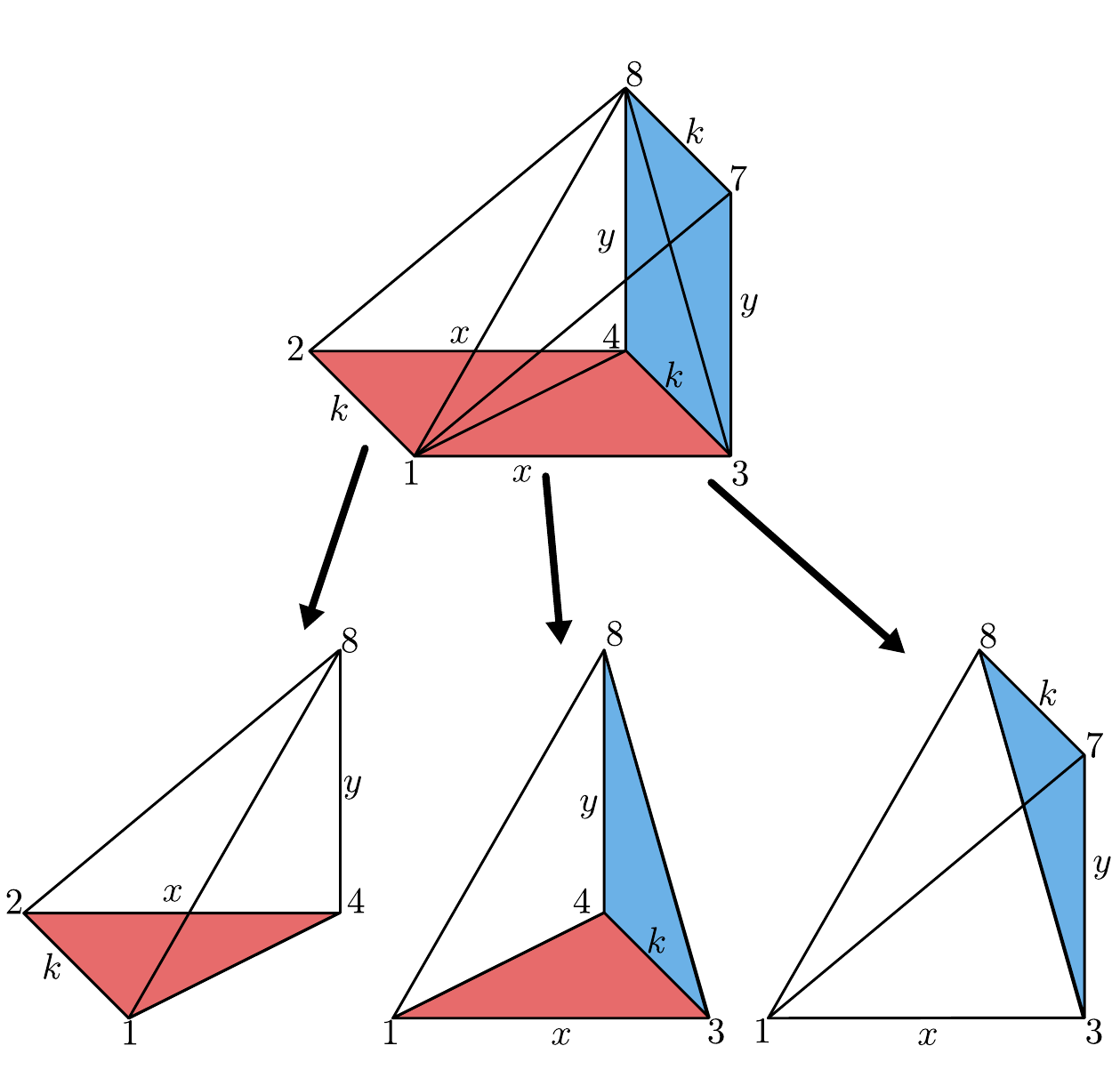}
	\caption{Here, we show the lower wedge of the intersection region and its constituent tetrahedra.}
	\label{Figure_braiding_lower_wedge}
\end{figure}

 For the intersection region, we consider the simple geometry shown in Figures \ref{Figure_braiding_direct_membranes}, \ref{Figure_braiding_upper_wedge} and \ref{Figure_braiding_lower_wedge}. Figure \ref{Figure_braiding_direct_membranes} shows the two direct membranes, while Figures \ref{Figure_braiding_upper_wedge} and \ref{Figure_braiding_lower_wedge} show the region affected by both dual membranes. In each case, the diagram is periodic in one direction, such that the path labeled by $k$ is closed, but not in the other directions. On the horizontal (red) membrane we apply the magnetic membrane operator (see Equation \ref{Equation_membrane_operator_definition})
 \begin{equation}
 F^{a, \alpha_1^{k,a}, i_1, j_1}(m_1)=C_0^a(m_1) \theta_D^a(m_1) \theta_S^{a, \alpha_1^{k,a}, i_1, j_1}(m_1),
 \end{equation}
 where $a$ is the flux label, while $\alpha_1^{k,a}$ is a $\beta_{k,a}$-projective irrep and $i_1$ and $j_1$ are its matrix indices. The surface weight for $m_1$ is then $[\alpha_1^{k,a}(x)]_{i_1j_1}$, because the diagram corresponding to the horizontal membrane shown in Figure \ref{Figure_braiding_direct_membranes} is already a reference diagram. On the vertical (blue) membrane we apply the operator
 \begin{equation}
 F^{c, \alpha_2^{k,c}, i_2, j_2}(m_2)=C_0^c(m_2) \theta_D^c(m_2) \theta_S^{c, \alpha_2^{k,c}, i_2, j_2}(m_2),
 \end{equation}
 for flux label $c$ and $\beta_{k,c}$-projective irrep $\alpha_2^{k,c}$ with matrix indices $i_2, j_2$.
	
Then we consider the state
\begin{align*}
F^{c, \alpha_2^{k,c}, i_2, j_2}(m_2)F^{a, \alpha_1^{k,a}, i_1, j_1}(m_1)\ket{\psi}
\end{align*}
obtained by applying the membrane operator on $m_1$ first, then on $m_2$, for a state $\ket{\psi}$ which has the base loop but no other excitations present. Decomposing the membrane operators into their constituent components (untwisted operator, dual phase and surface weight), we obtain
\begin{align*}
	F&^{c, \alpha_2^{k,c}, i_2, j_2}(m_2)F^{a, \alpha_1^{k,a}, i_1, j_1}(m_1)\ket{\psi}\\
 &\hspace{1.5cm}=C_0^c(m_2) \theta_D^c(m_2) \theta_S^{c, \alpha_2^{k,c}, i_2, j_2}(m_2) C_0^a(m_1) \\
 & \hspace{4cm} \theta_D^a(m_1) \theta_S^{a, \alpha_1^{k,a}, i_1, j_1}(m_1)\ket{\psi}.
	\end{align*}

Then we wish to commute the membrane operators past each-other, to see how they are affected by the braiding. The surface and dual phases are diagonal in the configuration basis (where each edge is labeled by a group element), so they commute with each-other. In addition, for an Abelian group the untwisted membrane operators also commute. Therefore,
	\begin{align}
		&F^{c, \alpha_2^{k,c}, i_2, j_2}(m_2)F^{a, \alpha_1^{k,a}, i_1, j_1}(m_1)\ket{\psi} \notag \\
		&\hspace{1cm}=C_0^c(m_2) C_0^a(m_1) \big[C_0^a(m_1) :\theta_D^c(m_2)\big] \notag \\
  &\hspace{2cm}\big[C_0^a(m_1) :\theta_S^{c, \alpha_2^{k,c}, i_2, j_2}(m_2)\big] \notag \\
  &\hspace{3cm} \theta_D^a(m_1) \theta_S^{a, \alpha_1^{k,a}, i_1, j_1}(m_1)\ket{\psi}, \label{Equation_braiding_setup_2}
	\end{align}
where $C_0^a(m_1) :\theta_D^c(m_2) =C_0^a(m_1)^{-1} \theta_D^c(m_2) C_0^a(m_1)$ is the evaluation of $\theta_D^c(m_2)$ after the action of $C_0^a(m_1)$ (and similar for the surface weight). We will now evaluate the dual phases and surface weights for the basis state shown in Figures \ref{Figure_braiding_direct_membranes}, \ref{Figure_braiding_upper_wedge} and \ref{Figure_braiding_lower_wedge}. Firstly, from the upper wedge shown in Figure \ref{Figure_braiding_upper_wedge}, the contribution to dual phase $\theta^a_D(m_1)$ is (using Table \ref{Table_Dual_Phase})
\begin{align}
\theta_D^U&(m_1) \notag\\
& =[a, k, y', x'] [k, a, y', x']^{-1} [a, y', k, x']^{-1}  [a, y', x', k] \notag\\
&= [a, y', x']_k. \label{Equation_braiding_dual_phase_upper_1}
\end{align}

On the other hand, the contribution to $\theta^c_D(m_2)$ is
\begin{align}
\theta_D^U(m_2)&= [k, y', x'c^{-1}, c]^{-1} [y', k, x'c^{-1},c] [y', x'c^{-1}, c, k] \notag \\
& \hspace{1cm}[y', x'c^{-1}, k, c]^{-1} \notag \\
&= [y', x'c^{-1}, c]_k. \label{Equation_braiding_dual_phase_upper_2}
\end{align}
If $C^a(m_1)$ acts first, it changes the labels of the vertical edges such that $C^a(m_1):y'=ay'$. Therefore,
$$C^a(m_1):\theta_D^U(m_2) = [ay', x'c^{-1}, c]_k.$$
If instead we acted with $C^c(m_2)$ first, then it would change the label of the horizontal edges such that $C^c(m_2):x'=x'c^{-1}$. Therefore,
$$C^c(m_2): \theta_D^U(m_1) = [a, y', x'c^{-1}]_k.$$

Denoting the total phase associated to the upper wedge by $$\theta_D^U\big(	F^{c, \alpha_2^{k,c}, i_2, j_2}(m_2)F^{a, \alpha_1^{k,a}, i_1, j_1}(m_1)\big)$$ and $$\theta_D^U\big(F^{a, \alpha_1^{k,a}, i_1, j_1}(m_1)F^{c, \alpha_2^{k,c}, i_2, j_2}(m_2)\big)$$ for the two orders of membrane operators, the ratio of these phases is given by
\begin{align*}
&\frac{\theta_D^U\big(	F^{c, \alpha_2^{k,c}, i_2, j_2}(m_2)F^{a, \alpha_1^{k,a}, i_1, j_1}(m_1)\big)}{\theta_D^U\big(F^{a, \alpha_1^{k,a}, i_1, j_1}(m_1)F^{c, \alpha_2^{k,c}, i_2, j_2}(m_2)\big)}\\
& \hspace{3cm} =\frac{\theta_D^U(m_1)C^a(m_1): \theta_D^U(m_2) }{\theta_D^U(m_2) C^c(m_2): \theta_D^U(m_1)} \\
& \hspace{3cm} = \frac{[a, y', x']_k [ay', x'c^{-1},c]_k}{ [a, y', x'c^{-1}]_k [y', x'c^{-1}, c]_k}.
\end{align*}
Applying the 3-cocycle condition Equation \ref{Equation_3_cocycle_condition}, this ratio becomes
\begin{equation*}
\frac{\theta_D^U\big(	F^{c, \alpha_2^{k,c}, i_2, j_2}(m_2)F^{a, \alpha_1^{k,a}, i_1, j_1}(m_1)\big)}{\theta_D^U\big(F^{a, \alpha_1^{k,a}, i_1, j_1}(m_1)F^{c, \alpha_2^{k,c}, i_2, j_2}(m_2)\big)} =[a, y'x'c^{-1}, c]
\end{equation*}
and noting that $y'x' =xy$ due to flatness, this can be written as
\begin{equation*}
\frac{\theta_D^U\big(	F^{c, \alpha_2^{k,c}, i_2, j_2}(m_2)F^{a, \alpha_1^{k,a}, i_1, j_1}(m_1)\big)}{\theta_D^U\big(F^{a, \alpha_1^{k,a}, i_1, j_1}(m_1)F^{c, \alpha_2^{k,c}, i_2, j_2}(m_2)\big)} =[a, xyc^{-1}, c].
\end{equation*}

Similarly, for the lower wedges, we have
\begin{align}
	\theta_D^L(m_1)&= [a, k, x, y]^{-1} [k,a,x,y] [k,x,a,y]^{-1} [a,x,k,y] \notag \\
  &\hspace{1cm} [x,a,k,y]^{-1} [x,k,a,y] [a,x,y,k]^{-1} [x,a,y,k] \notag\\
	&= [a,x,y]_k^{-1} [x,a,y]_k \label{Equation_braiding_dual_phase_lower_1}
\end{align}
and
\begin{align}
	\theta_D^L(m_2) &= [k,xc^{-1},c,y]^{-1} [k,xc^{-1},y,c] [xc^{-1}, c, k, y]^{-1} \notag \\ 
        & \hspace{1cm}  [xc^{-1}, k, c, y][xc^{-1}, k , y, c]^{-1} [xc^{-1}, c, y, k] \notag \\
        & \hspace{2cm}[xc^{-1}, y, c, k]^{-1} [xc^{-1}, y, k, c] \notag\\
	&= [xc^{-1}, c, y]_k [xc^{-1}, y, c]_k^{-1}. \label{Equation_braiding_dual_phase_lower_2}
\end{align}

Noting that $C^a(m_1):y =ay$ and $C^c(m_2):x=xc^{-1}$, this gives us the following ratio for the two orders:
\begin{align}
&\frac{\theta_D^L\big(	F^{c, \alpha_2^{k,c}, i_2, j_2}(m_2)F^{a, \alpha_1^{k,a}, i_1, j_1}(m_1)\big)}{\theta_D^L\big(F^{a, \alpha_1^{k,a}, i_1, j_1}(m_1)F^{c, \alpha_2^{k,c}, i_2, j_2}(m_2)\big)} \notag \\
& \ = \frac{ [a,x,y]_k^{-1} [x,a,y]_k [xc^{-1}, c, ay]_k [xc^{-1}, ay, c]_k^{-1}}{[a,xc^{-1},y]_k^{-1} [xc^{-1},a,y]_k [xc^{-1}, c, y]_k [xc^{-1}, y, c]_k^{-1}},
\end{align}
so the total ratio for the dual phases (including both upper and lower wedges) is
\begin{align}
&\theta_D^{\text{ratio}}= \notag \\
& \frac{ [a,x,y]_k^{-1} [x,a,y]_k [xc^{-1}, c, ay]_k [xc^{-1}, ay, c]_k^{-1} [a, xyc^{-1}, c]}{[a,xc^{-1},y]_k^{-1} [xc^{-1},a,y]_k [xc^{-1}, c, y]_k [xc^{-1}, y, c]_k^{-1}}.
\end{align}
Making use of Equation \ref{Equation_2_cocycle_definition}, we can simplify this by introducing 2-cocycles to obtain
\begin{align*}
	\theta_D^{\text{ratio}}&= \frac{ [x, y, a]_k [c,xc^{-1}, ay]_k [a, xyc^{-1}]_k}{[xc^{-1}, y, a]_k [c, xc^{-1},y]_k} \\
 & \hspace{2cm} \frac{[xc^{-1}, y]_{k,a}[xc^{-1},y]_{k,c}}{[x,y]_{k,a} [xc^{-1},ay]_{k,c}}\\
	&=[c, xyc^{-1}, a]_k [a, xyc^{-1}, c]_k\frac{[xc^{-1},y]_{k,a}[xc^{-1},y]_{k,c}}{[x,y]_{k,a} [xc^{-1},ay]_{k,c}},
\end{align*}
where we used the 3-cocycle condition Equation \ref{Equation_3_cocycle_condition} to obtain the second equality.

We can then write the remaining 3-cocycles in terms of 2-cocycles as
$$[c, xyc^{-1}, a]_k [a, xyc^{-1}, c]_k= [xyc^{-1}, a]_{k,c} [xyc^{-1}, c]_{k,a},$$
which can be verified by expanding out the 2-cocycles and observing the cancellation of terms. This means that
\begin{align}
	\theta_D^{\text{ratio}}&= \frac{[xc^{-1},y]_{k,a} [xyc^{-1}, c]_{k,a}}{[x,y]_{k,a}} \frac{[xc^{-1},y]_{k,c} [xyc^{-1},a]_{k,c}}{[xc^{-1},ay]_{k,c}}. \label{Equation_braiding_dual_phase_ratio}
\end{align}

Next, we consider the contribution from the surface weights. If we apply the membrane operator on $m_1$ first, the surface weight from the two membrane operators is
\begin{align*}
\theta_S\big(F^{c, \alpha_2^{k,c}, i_2, j_2}&(m_2)F^{a, \alpha_1^{k,a}, i_1, j_1}(m_1)\big)\\
&= [\alpha_1^{k,a}(x)]_{i_1j_1} [\alpha_2^{k,c}(ay)]_{i_2 j_2}.
\end{align*}
Using Equation \ref{Equation_projective_irrep_basis_matrix}, which defines the projective representations, we have
$$ [\alpha_2^{k,c}(ay)]_{i_2 j_2} = [a,y]_{k,c}^{-1} \sum_{m=1}^{|\alpha_2^{k,c}|} [\alpha_2^{k,c}(a)]_{i_2 m} [\alpha_2^{k,c}(y)]_{m j_2}$$
so
\begin{align*}
\theta_S&\big(F^{c, \alpha_2^{k,c}, i_2, j_2}(m_2)F^{a, \alpha_1^{k,a}, i_1, j_1}(m_1)\big)\\&= [a,y]_{k,c}^{-1} [\alpha_1^{k,a}(x)]_{i_1j_1} \sum_{m=1}^{|\alpha_2^{k,c}|} [\alpha_2^{k,c}(a)]_{i_2 m} [\alpha_2^{k,c}(y)]_{m j_2}.
\end{align*}

We wish to compare this to the total surface weight from the opposite order, which is
\begin{align*}
\theta_S\big(F^{a, \alpha_1^{k,a}, i_1, j_1}(m_1)&F^{c, \alpha_2^{k,c}, i_2, j_2}(m_2)\big)\\
&= [\alpha_1^{k,a}(xc^{-1})]_{i_1j_1} [\alpha_2^{k,c}(y)]_{i_2 j_2}.
\end{align*}
To do so, we write the surface weight from the original order as
\begin{align*}
	&\theta_S\big(F^{c, \alpha_2^{k,c}, i_2, j_2}(m_2)F^{a, \alpha_1^{k,a}, i_1, j_1}(m_1)\big)\\
 &= [a,y]_{k,c}^{-1} [\alpha_1^{k,a}(xc^{-1} c)]_{i_1j_1} \sum_{m=1}^{|\alpha_2^{k,c}|} [\alpha_2^{k,c}(a)]_{i_2 m} [\alpha_2^{k,c}(y)]_{m j_2}\\
	&=[a,y]_{k,c}^{-1} [xc^{-1},c]_{k,a}^{-1} \sum_{n=1}^{|\alpha_1^{k,a}|} [\alpha_1^{k,a}(xc^{-1})]_{i_1n} [\alpha_1^{k,a}( c)]_{nj_1} \\
 & \hspace{1cm} \sum_{m=1}^{|\alpha_2^{k,c}|} [\alpha_2^{k,c}(a)]_{i_2 m} [\alpha_2^{k,c}(y)]_{m j_2}\\
	&= [a,y]_{k,c}^{-1} [xc^{-1},c]_{k,a}^{-1} \sum_{n=1}^{|\alpha_1^{k,a}|} \sum_{m=1}^{|\alpha_2^{k,c}|} [\alpha_2^{k,c}(a)]_{i_2 m} [\alpha_1^{k,a}( c)]_{nj_1} \\
 & \hspace{1cm}\theta_S\big(F^{a, \alpha_1^{k,a}, i_1, n}(m_1)F^{c, \alpha_2^{k,c}, m, j_2}(m_2)\big).
\end{align*}
Including both the dual phase and surface weight, and noting that the dual phase only depends on the flux label, not the irrep or matrix indices, we see that	
\begin{widetext}
\begin{align}
		\theta_D&\big(F^{c, \alpha_2^{k,c}, i_2, j_2}(m_2)F^{a, \alpha_1^{k,a}, i_1, j_1}(m_1)\big)	\theta_S\big(F^{c, \alpha_2^{k,c}, i_2, j_2}(m_2)F^{a, \alpha_1^{k,a}, i_1, j_1}(m_1)\big) \notag \\
		 &= \sum_{n=1}^{|\alpha_1^{k,a}|} \sum_{m=1}^{|\alpha_2^{k,c}|} \frac{\theta_D\big(F^{c, \alpha_2^{k,c}, i_2, j_2}(m_2)F^{a, \alpha_1^{k,a}, i_1, j_1}(m_1)\big)}{\theta_D\big(F^{a, \alpha_1^{k,a}, i_1, n}(m_1)F^{c, \alpha_2^{k,c}, m, j_2}(m_2)\big)} \theta_D\big(F^{a, \alpha_1^{k,a}, i_1, n}(m_1)F^{c, \alpha_2^{k,c}, m, j_2}(m_2)\big) [a,y]_{k,c}^{-1} [xc^{-1},c]_{k,a}^{-1} \notag \\
		 & \hspace{0.5cm} [\alpha_2^{k,c}(a)]_{i_2 m} [\alpha_1^{k,a}( c)]_{nj_1} \theta_S\big(F^{a, \alpha_1^{k,a}, i_1, n}(m_1)F^{c, \alpha_2^{k,c}, m, j_2}(m_2)\big) \notag \\
		&= \frac{[xc^{-1},y]_{k,a} [xyc^{-1}, c]_{k,a}}{[x,y]_{k,a}} \frac{[xc^{-1},y]_{k,c} [xyc^{-1},a]_{k,c}}{[xc^{-1},ay]_{k,c}} [a,y]_{k,c}^{-1} [xc^{-1},c]_{k,a}^{-1}\sum_{n=1}^{|\alpha_1^{k,a}|} \sum_{m=1}^{|\alpha_2^{k,c}|} [\alpha_2^{k,c}(a)]_{i_2 m} [\alpha_1^{k,a}( c)]_{nj_1} \notag \\
		& \hspace{0.5cm} \theta_S\big(F^{a, \alpha_1^{k,a}, i_1, n}(m_1)F^{c, \alpha_2^{k,c}, m, j_2}(m_2)\big)\theta_D\big(F^{a, \alpha_1^{k,a}, i_1, n}(m_1)F^{c, \alpha_2^{k,c}, m, j_2}(m_2)\big) \notag \\
		&= \frac{[xc^{-1},y]_{k,a} [xyc^{-1}, c]_{k,a}}{[x,y]_{k,a} [xc^{-1},c]_{k,a}} \frac{[xc^{-1},y]_{k,c} [xyc^{-1},a]_{k,c}}{[xc^{-1},ay]_{k,c} [a,y]_{k,c}} \sum_{n=1}^{|\alpha_1^{k,a}|} \sum_{m=1}^{|\alpha_2^{k,c}|} [\alpha_2^{k,c}(a)]_{i_2 m} [\alpha_1^{k,a}( c)]_{nj_1} \notag \\
		& \hspace{0.5cm} \theta_D\big(F^{a, \alpha_1^{k,a}, i_1, n}(m_1)F^{c, \alpha_2^{k,c}, m, j_2}(m_2)\big)\theta_S\big(F^{a, \alpha_1^{k,a}, i_1, n}(m_1)F^{c, \alpha_2^{k,c}, m, j_2}(m_2)\big). \label{Equation_braiding_intermediate_1}
\end{align}
\end{widetext}
Applying the 2-cocycle condition Equation \ref{Equation_2_cocycle_condition}, we have
$$\frac{[xc^{-1},y]_{k,c} [xyc^{-1}, a]_{k,c}}{[xc^{-1},ay]_{k,c}[a,y]_{k,c}} = \frac{[y,a]_{k,c}}{[a,y]_{k,c}}.$$
Then we can expand the 2-cocycles into 3-cocycles (using Equation \ref{Equation_2_cocycle_definition}) to obtain
\begin{align}
	\frac{[y,a]_{k,c}}{[a,y]_{k,c}} &= \frac{ [c,y,a]_{k} [y,a,c]_k [a,c,y]_k}{ [y,c,a]_k [c,a,y]_k [a,y,c]_k} \notag \\
	&= \frac{[c,y]_{k,a}}{ [y,c] _{k,a}}. \label{Equation_1_cocycle_permutation}
\end{align}
Then, applying the 2-cocycle condition to both the numerator and denominator, we find
\begin{align*}
	\frac{[c,y]_{k,a}}{ [y,c] _{k,a}} &= \big( \frac{[c^{-1}, c]_{k,a} [c^{-1}c,y]_{k,a}}{[c^{-1}, cy]_{k,a}}\big) \big( \frac{[c^{-1},cy]_{k,a}}{[c^{-1},y]_{k,a} [c^{-1}y,c]_{k,a}} \big)\\
	&= \frac{[c^{-1},c]_{k,a}}{[c^{-1},y]_{k,a} [c^{-1}y,c]_{k,a}} = \frac{[c^{-1},c]_{k,a}}{[c^{-1},y]_{k,a} [yc^{-1},c]_{k,a}} .
\end{align*}
Applying the 2-cocycle condition again, we obtain
\begin{align}
	\frac{[c,y]_{k,a}}{ [y,c] _{k,a}} &= \frac{[y,c^{-1}]_{k,a}}{[c^{-1},y]_{k,a}}. \label{Equation_1_cocycle_permutation_2}
\end{align}
	Substituting this into Equation \ref{Equation_braiding_intermediate_1}, we find that
	\begin{align}
	\theta_D&\big(F^{c, \alpha_2^{k,c}, i_2, j_2}(m_2)F^{a, \alpha_1^{k,a}, i_1, j_1}(m_1)\big)	\notag \\
 &  \hspace{2cm} \theta_S\big(F^{c, \alpha_2^{k,c}, i_2, j_2}(m_2)F^{a, \alpha_1^{k,a}, i_1, j_1}(m_1)\big) \notag \\
		&= \frac{[xc^{-1},y]_{k,a} [xyc^{-1}, c]_{k,a}}{[x,y]_{k,a} [xc^{-1},c]_{k,a} } \frac{[y,c^{-1}]_{k,a}}{[c^{-1},y]_{k,a}} \notag \\& \hspace{0.7cm} \sum_{n=1}^{|\alpha_1^{k,a}|} \sum_{m=1}^{|\alpha_2^{k,c}|}  [\alpha_2^{k,c}(a)]_{i_2 m} [\alpha_1^{k,a}( c)]_{nj_1} \notag \\
  & \hspace{1.2cm} \theta_D\big(F^{a, \alpha_1^{k,a}, i_1, n}(m_1)F^{c, \alpha_2^{k,c}, m, j_2}(m_2)\big) \notag \\
		& \hspace{1.7cm} \theta_S\big(F^{a, \alpha_1^{k,a}, i_1, n}(m_1)F^{c, \alpha_2^{k,c}, m, j_2}(m_2)\big). \label{Equation_braiding_intermediate_2}
	\end{align}
Applying the 2-cocycle condition to the remaining 2-cocycles, we have
\begin{align*}
	&\frac{[xc^{-1},y]_{k,a} [xyc^{-1}, c]_{k,a}}{[x,y]_{k,a} [xc^{-1},c]_{k,a} } \frac{[y,c^{-1}]_{k,a}}{[c^{-1},y]_{k,a}}\\
 & \hspace{3cm}= \frac{[xc^{-1},y]_{k,a} [yc^{-1},c]_{k,a}}{ [xc^{-1},c]_{k,a} [x,yc^{-1}]_{k,a}} \frac{[y,c^{-1}]_{k,a}}{[c^{-1},y]_{k,a}} \\
	& \hspace{3cm}= \frac{[c^{-1},y]_{k,a} [yc^{-1}, c]_{k,a}}{[x,c^{-1}]_{k,a}[xc^{-1},c]_{k,a}}\frac{[y,c^{-1}]_{k,a}}{[c^{-1},y]_{k,a}} \\
	& \hspace{3cm} = \frac{[c^{-1},y]_{k,a} [yc^{-1},c]_{k,a}}{[c^{-1},c]_{k,a}[x, c^{-1}c]_{k,a}}\frac{[y,c^{-1}]_{k,a}}{[c^{-1},y]_{k,a}}.
\end{align*}
Using the normalization condition $[x, c^{-1}c]_{k,a}=[x,1_G]_{k,a}=1$, these 2-cocycles become
\begin{align*}
	\frac{[c^{-1},y]_{k,a} [yc^{-1},c]_{k,a}}{[c^{-1},c]_{k,a}[x, c^{-1}c]_{k,a}}\frac{[y,c^{-1}]_{k,a}}{[c^{-1},y]_{k,a}} &=	\frac{[y,c^{-1}]_{k,a} [yc^{-1},c]_{k,a}}{[c^{-1},c]_{k,a}}.
\end{align*}
Applying the 2-cocycle condition one last time, we find
\begin{align*}
\frac{[y,c^{-1}]_{k,a} [yc^{-1},c]_{k,a}}{[c^{-1},c]_{k,a}}=[y,c^{-1}c]_{k,a}=1.
\end{align*}
That is, all of the cocycles in Equation \ref{Equation_braiding_intermediate_2} cancel and the relationship between the two orders is just
\begin{align}
&\theta_D\big(F^{c, \alpha_2^{k,c}, i_2, j_2}(m_2)F^{a, \alpha_1^{k,a}, i_1, j_1}(m_1)\big) \notag \\
& \hspace{1cm} \theta_S\big(F^{c, \alpha_2^{k,c}, i_2, j_2}(m_2)F^{a, \alpha_1^{k,a}, i_1, j_1}(m_1)\big) \notag \\ &=\sum_{n=1}^{|\alpha_1^{k,a}|} \sum_{m=1}^{|\alpha_2^{k,c}|} [\alpha_2^{k,c}(a)]_{i_2 m} [\alpha_1^{k,a}( c)]_{nj_1} \notag \\& \hspace{2cm} \theta_D\big(F^{a, \alpha_1^{k,a}, i_1, n}(m_1)F^{c, \alpha_2^{k,c}, m, j_2}(m_2)\big) \notag \\
& \hspace{3cm}\theta_S\big(F^{a, \alpha_1^{k,a}, i_1, n}(m_1)F^{c, \alpha_2^{k,c}, m, j_2}(m_2)\big) .
\end{align}
Note that this relation is independent of the basis state used (i.e., it does not depend on $x$, $y$, $x'$ or $y'$) and so holds true for the whole state $\ket{\psi}$ which is a linear combination of basis states. Therefore the commutation relation Equation \ref{Equation_braiding_setup_2} becomes
\begin{align*}
	&F^{c, \alpha_2^{k,c}, i_2, j_2}(m_2)F^{a, \alpha_1^{k,a}, i_1, j_1}(m_1)\ket{\psi}\\
	&= \sum_{n=1}^{|\alpha_1^{k,a}|} \sum_{m=1}^{|\alpha_2^{k,c}|} [\alpha_2^{k,c}(a)]_{i_2 m} [\alpha_1^{k,a}( c)]_{nj_1} C_0^a(m_1) C_0^c(m_2) \\
 & \hspace{1cm}\theta_D\big(F^{a, \alpha_1^{k,a}, i_1, n}(m_1)F^{c, \alpha_2^{k,c}, m, j_2}(m_2)\big) \\
	& \hspace{2cm} \theta_S\big(F^{a, \alpha_1^{k,a}, i_1, n}(m_1)F^{c, \alpha_2^{k,c}, m, j_2}(m_2)\big) \ket{\psi}\\
	&= \sum_{n=1}^{|\alpha_1^{k,a}|} \sum_{m=1}^{|\alpha_2^{k,c}|} [\alpha_2^{k,c}(a)]_{i_2 m} [\alpha_1^{k,a}( c)]_{nj_1} C_0^a(m_1) \theta_D^a(m_1) \\
 & \hspace{1cm} \theta_S^{a, \alpha_1^{k,a}, i_1, n}(m_1) C_0^c(m_2) \theta_D^c(m_2) \theta_S^{c, \alpha_2^{k,c},m, j_2}(m_2) \ket{\psi}.
\end{align*}

Recombining the components into the full membrane operators, we see that the braiding relation is given by
\begin{align}
	&F^{c, \alpha_2^{k,c}, i_2, j_2}(m_2)F^{a, \alpha_1^{k,a}, i_1, j_1}(m_1)\ket{\psi} \notag \\
 & \hspace{0.5cm} = \sum_{n=1}^{|\alpha_1^{k,a}|} \sum_{m=1}^{|\alpha_2^{k,c}|} [\alpha_2^{k,c}(a)]_{i_2 m} [\alpha_1^{k,a}( c)]_{nj_1} \notag \\
 & \hspace{1.5cm}F^{a, \alpha_1^{k,a}, i_1, n}(m_1) F^{c, \alpha_2^{k,c},m, j_2}(m_2) \ket{\psi}. \label{Equation_three_loop_braiding_result_1}
\end{align}
This is the same as we would expect for the braiding of two dyonic excitations in the untwisted theory, except that we have projective irreps rather than linear irreps. Notice that the irreps $\alpha_1^{k,a}$ and $\alpha_2^{k,c}$, as well as the fluxes $a$ and $c$, are conserved under this braiding relation. This suggests that these are conserved quantities (i.e., label topological charges), as we will discuss further in Section \ref{Section_topological_charge}. If the irreps are one-dimensional, we obtain the simpler relation
\begin{align}
	&F^{c, \alpha_2^{k,c}}(m_2)F^{a, \alpha_1^{k,a}}(m_1)\ket{\psi} \notag \\
 & \hspace{0.5cm}= \alpha_2^{k,c}(a) \alpha_1^{k,a}( c)F^{a, \alpha_1^{k,a}}(m_1) F^{c, \alpha_2^{k,c}}(m_2) \ket{\psi}, \label{Equation_three_loop_braiding_result_1D}
\end{align}
which is Abelian braiding with only a phase gain. More generally, there is mixing within the spaces described by the irreps $\alpha_1^{k,a}$ and $\alpha_2^{k,c}$, indicating non-Abelian braiding. We note that these expressions agree with the results found from the modular $S$-matrix in Refs. \cite{Jiang2014, Wang2015}, up to conjugation of the irreps (likely due to a difference in convention for the braiding) and the fact that the expression in Ref. \cite{Wang2015} only involves the trace of the representation (due to being calculated from a ground state quantity). The braiding result also agrees with Ref. \cite{Wang2014}, for groups which are a product of multiple copies of the same cyclic group, as we show in Section \ref{Section_braiding_example_1}.

It is instructive to consider the case where the flux, $k$, of the base loop is taken to be trivial, giving ordinary loop-loop braiding. In this case, the projective irreps $\alpha_1^{k,a}$ and $\alpha_2^{k,c}$ are in fact linear irreps, which must be one-dimensional due to the Abelian nature of the group $G$. In this case, Equation \ref{Equation_three_loop_braiding_result_1D} becomes

\begin{align}
	&F^{c, \alpha_2}(m_2)F^{a, \alpha_1}(m_1)\ket{\psi} \notag \\
	& \hspace{0.5cm}= \alpha_2(a) \alpha_1( c)F^{a, \alpha_1}(m_1) F^{c, \alpha_2}(m_2) \ket{\psi}, \label{Equation_two_loop_braiding_result}
\end{align}
which is just the result of passing a charge $\alpha_1$ through a flux $c$ and charge $\alpha_2$ through a flux $a$. Notably, the 4-cocycle no longer enters anywhere in this expression, because it only enters the three-loop result through the projective irreps. We see that the two-loop braiding is independent of the cocycle twist, as documented in previous works \cite{Wang2014, Lin2015, Wang2015a}, illustrating the importance of three-loop braiding in distinguishing between different phases.

\begin{figure}
	\centering
	\includegraphics[width=\linewidth]{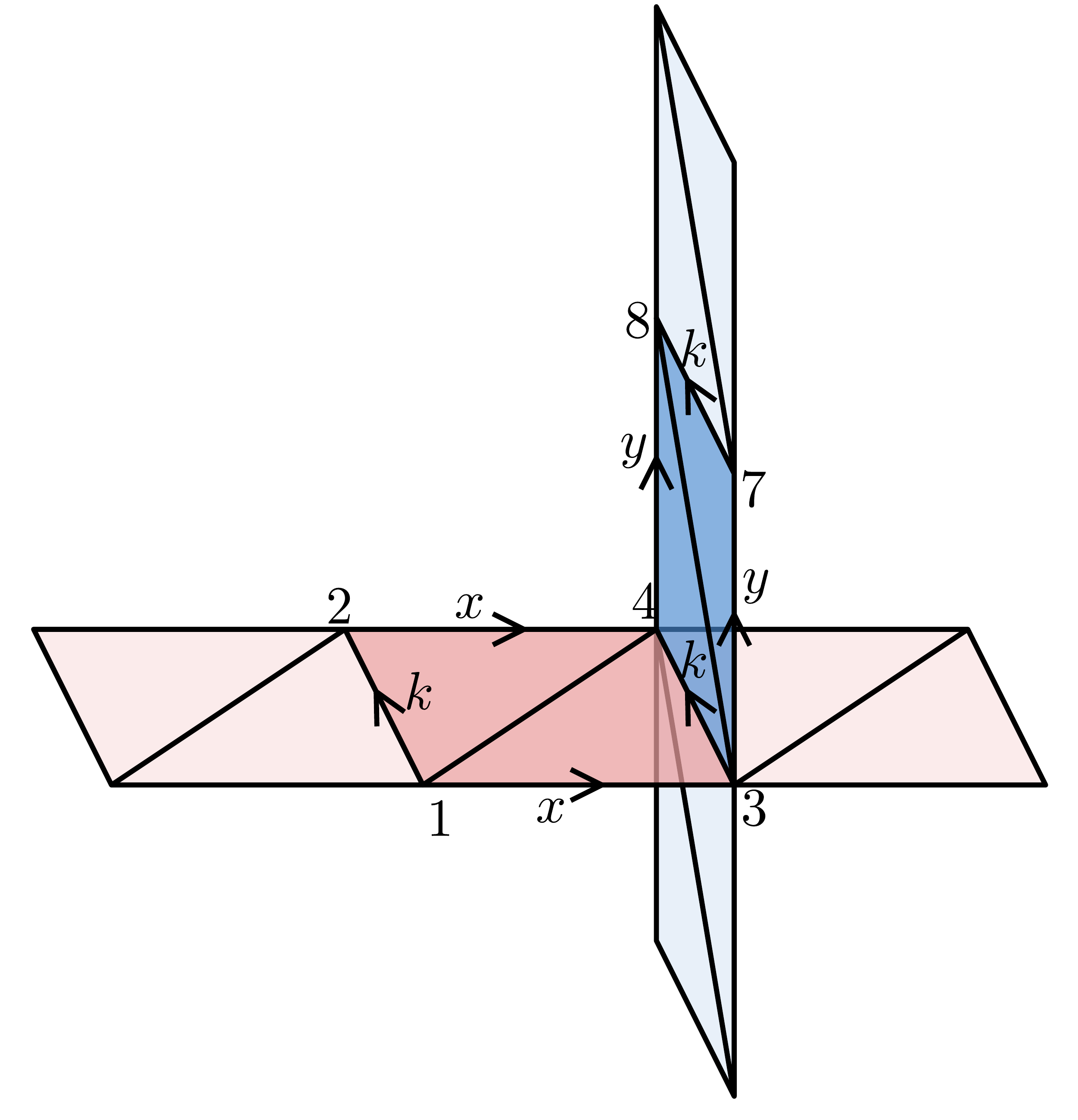}
	\caption{We give an example of two membranes (red and blue) which have the intersection region shown in Figure \ref{Figure_braiding_direct_membranes}.}
	\label{Figure_intersection_extended}
\end{figure}

 While the transformation given in Equation \ref{Equation_three_loop_braiding_result_1} is simple, there is an additional subtlety when the irreps involved are not one-dimensional. Because the braiding is non-Abelian, we only expect this type of simple relation when the excitations involved have a definite fusion channel. For example, in the non-Abelian quantum double model \cite{Kitaev2003}, the braiding between
 fluxes is simple when the ribbon operators for the two fluxes have the same start-point. In non-Abelian gauge theory, the flux measured by a charge traversing a closed cycle around a flux depends on the start of that path, with the measured flux changing by conjugation if the start of the closed path is changed. Generally, when a flux excitation is created, the flux is only well-defined when measured from a certain point. The ribbon operators that create the fluxes in the non-Abelian Quantum Double model therefore have special points, called the start-point of the ribbon, from which the flux is well-defined. Then braiding is relatively simple if fluxes share this start-point, which gives them a well-defined combined flux. Similarly, braiding between a flux and a charge is simpler when the start-point of the magnetic ribbon operator matches the start of the electric ribbon operator that produces the charge. To see that a similar phenomenon occurs for Abelian twisted lattice gauge theory in 3+1d, note that we have so far only considered the region in which the two membrane operators performing the braiding intersect, because the other parts of the membrane operators commute. However, when the irreps labeling the membrane operators are higher-dimensional, the matrix multiplication rule for composition of the parts of the surface weight means that we cannot simply pull the factor gained by the intersection region from the braiding to the front of the membrane operator. Suppose that there are additional parts to the membrane operators, such as the situation shown in Figure \ref{Figure_intersection_extended}. Then the surface weight of $F^{a, \alpha_1^{k,a}, i_1, j_1}(m_1)$ has the form
$$\sum_{p,q=1}^{|\alpha_1^{k,a}|} \hat{M}_{i_1p} [\alpha_1^{k,a}(\hat{x})]_{p q} \hat{N}_{q j_1},$$
where $\hat{M}$ and $\hat{N}$ are the contributions from the parts of the diagram before and after the intersection respectively and are operators (they depend on the edge labels).

Under the braiding, we know from Equation \ref{Equation_three_loop_braiding_result_1} that $[\alpha_1^{k,a}(\hat{x})]_{pq} $ becomes $ \sum_{n=1}^{|\alpha_1^{k,a}|} [\alpha_1^{k,a}(\hat{x})]_{p n} [\alpha_1^{k,a}( c)]_{nq} $ and so the full surface weight should be replaced with
$$\sum_{p,q=1}^{|\alpha_1^{k,a}|} \hat{M}_{i_1p} \sum_{n=1}^{|\alpha_1^{k,a}|} [\alpha_1^{k,a}(\hat{x})]_{p n} [\alpha_1^{k,a}( c)]_{nq} \hat{N}_{q j_1} ,$$
which we can write more clearly in terms of matrices as
\begin{align*}
[\hat{M} \alpha_1^{k,a}(\hat{x}) \alpha_1^{k,a}( c)& \hat{N}]_{i_1 j_1}\\
&=[\hat{M} \alpha_1^{k,a}(\hat{x}) \hat{N} (\hat{N}^{-1}\alpha_1^{k,a}( c)\hat{N} )]_{i_1 j_1}.
\end{align*}

From this, we see that the braiding transformation is by matrix multiplication by $\hat{N}^{-1}\alpha_1^{k,a}( c)\hat{N}$ rather than $\alpha_1^{k,a}( c)$. This matrix is in the same equivalence class of irreps, because it is related to the original matrix representation by conjugation, reflecting the fact that the class of irreps is a conserved quantity. We can think of the conjugating matrices as mixing different matrix indices, which are internal degrees of freedom for the topological charge rather than conserved quantities. A similar result will hold for the other membrane operator. The conjugating matrices depend on the edge labels, reflecting the non-coherence of the two excitations (again, this is familiar from non-Abelian untwisted lattice gauge theory). Note that when we take 1d irreps, the conjugation becomes trivial and we just obtain the same phase we reported before.

\begin{figure}
	\centering
	\includegraphics[width=0.8\linewidth]{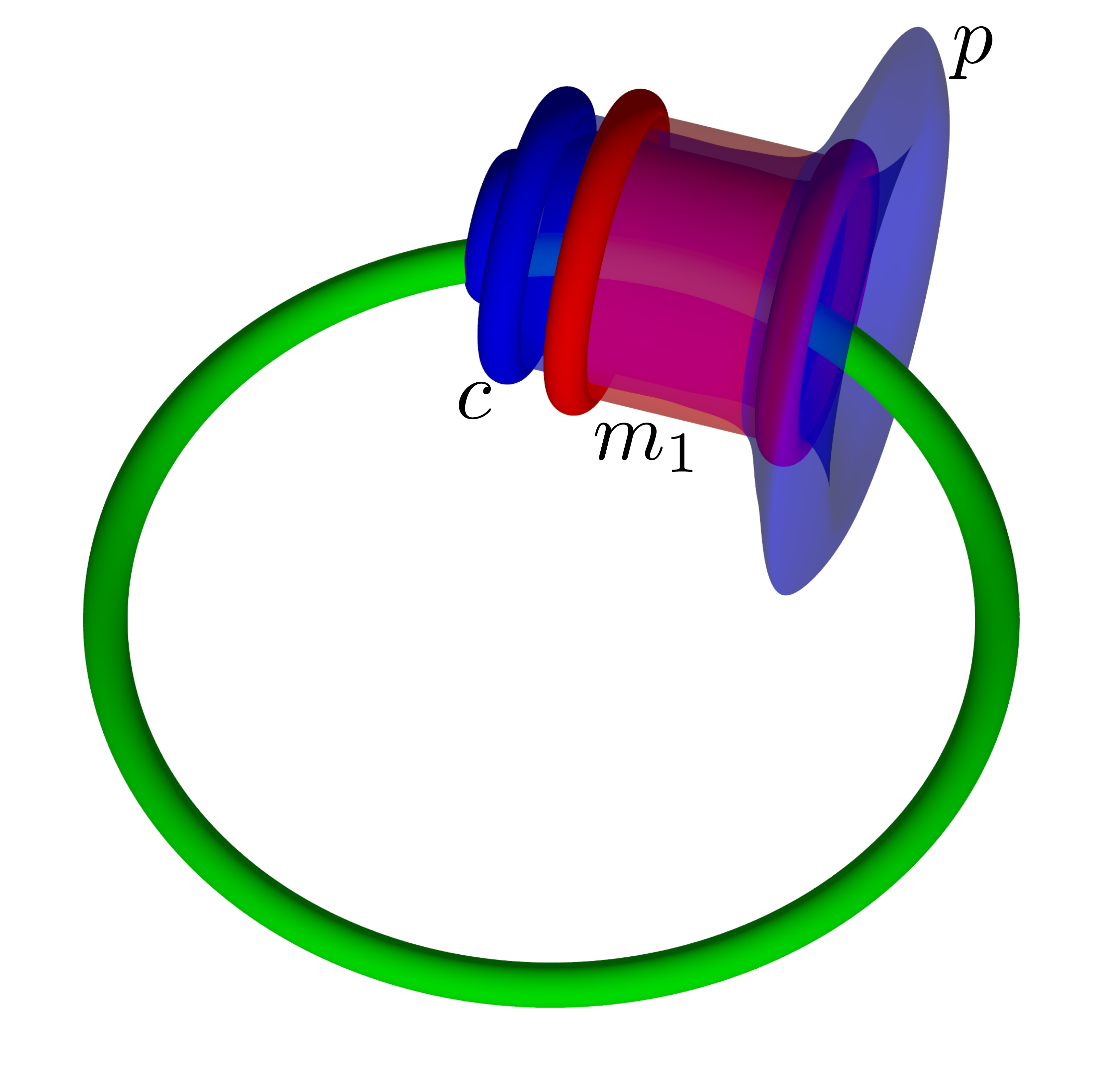}
	\caption{Instead of braiding the (blue) loop $c$ around the loop on the left-hand side of $m_1$, which has the opposite orientation to $c$, we can braid it around the loop at the right-hand side of $m_1$. We can do this by applying a membrane operator on membrane $p$.}
	\label{Figure_alternate_braiding}
\end{figure}

\begin{figure}
	\centering
	\includegraphics[ width=\linewidth]{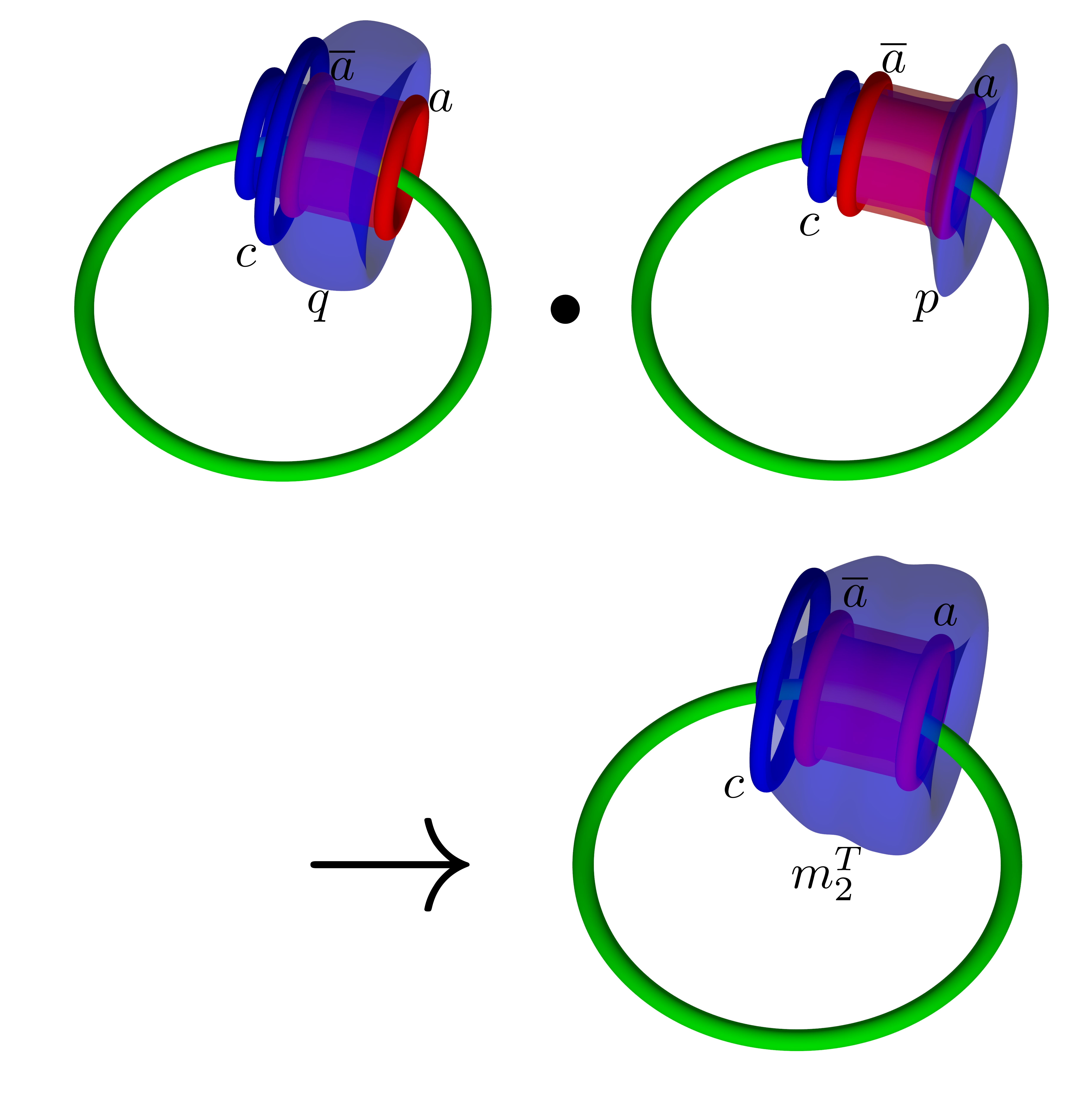}
	
	\caption{If we perform both braiding moves sequentially (the braiding around the loop at the right-hand of $m_1$, using the membrane $p$, followed by the braiding around the loop at the left-hand of $m_1$, using the membrane $q$), it is equivalent to braiding around both loops simultaneously, using the membrane $m_2^T$. Because the total flux of the combined loops is trivial, this combined braiding is trivial.}
	\label{Figure_consecutive_braiding}
\end{figure}

Another subtlety concerning Equation \ref{Equation_three_loop_braiding_result_1} relates to the geometry of the situation. As indicated in Figure \ref{Figure_three_loop_braiding}, we considered the situation where we braided the loop from the right of the membrane $m_2$ with the loop from the left of the membrane $m_1$. These loops have opposite orientations, so it is more natural to braid the loop from the right of membrane $m_2$ with the one from the right of $m_1$ (although the geometry is slightly more complicated). Such a process is shown in Figure \ref{Figure_alternate_braiding}. As shown in Figure \ref{Figure_consecutive_braiding}, if we apply both braiding moves sequentially it is equivalent to braiding the loop labeled by $c$ around both the loop and anti-loop labeled by $a$. Such a braiding move can be done without any intersection of the membrane operators, so the braiding move must be trivial. We can use this to deduce the braiding relation with the right-hand loop. If we take $m_2^T$ to be the total membrane on the right-hand side of Figure \ref{Figure_consecutive_braiding}, then we have
\begin{align}
	&F^{c, \alpha_2^{k,c}, i_2, l_2}(m_2^T) F^{a, \alpha_1^{k,a}, i_1, j_1}(m_1) \ket{\psi} \notag \\
 & \hspace{1cm}= F^{a, \alpha_1^{k,a}, i_1, j_1}(m_1) F^{c, \alpha_2^{k,c}, i_2, l_2}(m_2^T)\ket{\psi}, \label{Equation_trivial_total_braid_move}
\end{align}
because the total membrane $m_2^T$ does not intersect with $m_1$ and so the two operators commute. We can then split $m_2^T$ into the two parts $p$ and $q$, which are shown on the first line of Figure \ref{Figure_consecutive_braiding} ($q$ is the one corresponding to the braiding move we previously considered, while $p$ corresponds to the new braiding move):
\begin{equation}
		F^{c, \alpha_2^{k,c}, i_2, l_2}(m_2^T) = \sum_{j_2=1}^{|\alpha_2^{k,c}|} 		F^{c, \alpha_2^{k,c}, i_2, j_2}(p) F^{c, \alpha_2^{k,c}, j_2, l_2}(q).
\end{equation}
Then we already calculated the commutation relation between the membrane operators $$F^{c, \alpha_2^{k,c}, j_2, l_2}(q)$$ and 
$$F^{a, \alpha_1^{k,a}, i_1, j_1}(m_1)$$ in Equation \ref{Equation_three_loop_braiding_result_1}, so we have
\begin{align*}
	&F^{c, \alpha_2^{k,c}, i_2, l_2}(m_2^T) F^{a, \alpha_1^{k,a}, i_1, j_1}(m_1) \ket{\psi}\\
 &= \sum_{j_2=1}^{|\alpha_2^{k,c}|} 		F^{c, \alpha_2^{k,c}, i_2, j_2}(p) F^{c, \alpha_2^{k,c}, j_2, l_2}(q) F^{a, \alpha_1^{k,a}, i_1, j_1}(m_1) \ket{\psi}\\
	&= \sum_{j_2=1}^{|\alpha_2^{k,c}|} 		F^{c, \alpha_2^{k,c}, i_2, j_2}(p) \sum_{n=1}^{|\alpha_1^{k,a}|} \sum_{m=1}^{|\alpha_2^{k,c}|} [\alpha_2^{k,c}(a)]_{j_2m} [\alpha_1^{k,a}(c)]_{nj_1}  \\
 & \hspace{1cm}F^{a, \alpha_1^{k,a}, i_1, n}(m_1) F^{c, \alpha_2^{k,c}, m, l_2}(q) \ket{\psi}.
\end{align*}
Requiring this to agree with Equation \ref{Equation_trivial_total_braid_move}, we obtain
\begin{align}
&F^{c, \alpha_2^{k,c}, i_2, j_2}(p) F^{a, \alpha_1^{k,a}, i_1, n}(m_1) \ket{\psi}\notag \\
&\hspace{1cm}= \sum_{m'=1}^{|\alpha_2^{k,c}|} \sum_{n'=1}^{|\alpha_1^{k,a}|} [\alpha_1^{k,a}(c)]^{-1}_{n'n} [\alpha_2^{k,c}(a)]^{-1}_{m' j_2} \notag \\
&\hspace{2cm} F^{a, \alpha_1^{k,a}, i_1, n'}(m_1) F^{c, \alpha_2^{k,c}, i_2, m'}(p)\ket{\psi}. \label{Equation_three_loop_braiding_reverse}
\end{align}
Similar to the result for the other geometry, there may generally be matrices conjugating $\alpha_1^{k,a}(c)$ and $\alpha_2^{k,c}(a)$ to account for the possibility that the membranes are extended before the intersection. Note that this braiding transformation acts on the opposite index of $F^{c, \alpha_2^{k,c}, i_2, j_2}(p)$ (the $j_2$ index) compared to Equation \ref{Equation_three_loop_braiding_result_1}, corresponding to matrix post-multiplication by $[\alpha_2^{k,c}(a)]^{-1}$, rather than pre-multiplication by $[\alpha_2^{k,c}(a)]$. This is not a fundamental property, as post-multiplication can be converted to pre-multiplication by appropriate matrix conjugation. In addition, processes involving open membranes are often sensitive to the details of the process. If we take $p$ to be a closed membrane, corresponding to the situation where the loop-like excitations it produces are brought back to their initial position and fused to the vacuum (for $i_2=j_2$), the membrane operator acts trivially on $\ket{\psi}$, even in the presence of the base loop. Then the contribution from $F^{c, \alpha_2^{k,c}, i_2, j_2}(p)$ to the right-hand side of Equation \ref{Equation_three_loop_braiding_reverse} is just the matrix $[\alpha_2^{k,c}(a)]^{-1}$ and post-multiplication or pre-multiplication becomes irrelevant. More generally, we can think of $[\alpha_2^{k,c}(a)]^{-1}$ as the contribution from the closed braiding motion and the remaining operator $F^{c, \alpha_2^{k,c}, i_2, m'}(p)$ on the right-hand side of Equation \ref{Equation_three_loop_braiding_reverse} as a process where no braiding occurs but the loop-like excitations are moved into their final positions, as we discussed for point-like braiding in Section \ref{Section_point_loop_braiding} (see Figure \ref{Figure_particle_loop_deformation}). Then the order of multiplication appears to correspond to the order in which these operations are carried out.

\subsubsection{Three-loop Braiding Example: the Group $\mathbb{Z}_N^W$}
\label{Section_braiding_example_1}
To further explore the braiding relation described by Equation \ref{Equation_three_loop_braiding_result_1D}, we will consider the example of a group $\mathbb{Z}_N^W$ made of $W$ copies of a cyclic group $\mathbb{Z}_N$. In particular, we would like to compare our result to the braiding found in Ref. \cite{Wang2014} from dimensional reduction. Ref. \cite{Wang2014} considers 4-cocycles of the form
\begin{equation}
	[a,b,c,d] = \exp (\frac{2 \pi i}{N^2} \sum_{i,j,k=1}^W M_{ijk} a_i b_j (c_k+d_k - [c_k+d_k])). \label{Equation_cocycle_form}
\end{equation}
Here $M_{ijk}$ is a three-index tensor with integer entries and the group elements $a,b,c,d \in \mathbb{Z}_N^W$ are written in the form of vectors as $a=(a_1, a_2, ..., a_W)$. The identity element for the $i$th copy of the group is denoted by $a_i=0$ and group multiplication on each copy is performed by addition modulo $N$. $[c_k+d_k]$ is the group multiplication of $c_k$ and $d_k$ (i.e., addition modulo $N$), while $c_k+d_k$ is simple addition (without the modular arithmetic, meaning it does not correspond to group multiplication).

From this 4-cocycle is defined as a dimensionally reduced cocycle, which is equivalent to the twisted 3-cocycle we have discussed previously:
\begin{equation}
[h_1,h_2,h_3]_h=\frac{[h_1,h_2,h_3, h] [h_1, h, h_2, h_3]}{[h_1,h_2,h,h_3] [h, h_1, h_2, h_3]}.
\end{equation}
With the form of the 4-cocycle from Equation \ref{Equation_cocycle_form}, this 3-cocycle is given by 
\begin{equation}
[a,b,c]_h= \exp\big(\frac{2 \pi i}{N^2} \sum_{i,j} P_{ij}^h a_i (b_j + c_j - [b_j + c_j])\big), \label{Equation_dim_red_cocycle_form}
\end{equation}
where 
\begin{equation}
P_{ij}^h= \sum_k (M_{ikj} - M_{kij})h_k \label{Equation_P_definition},
\end{equation}
as described in the Supplemental Material \cite{Wang2014Sup} for Ref. \cite{Wang2014}. The doubly twisted 2-cocycle is therefore given by
\begin{align}
	[a,b]_{c,d}=& \frac{[d,a,b]_c [a,b,d]_c}{[a,d,b]_c} \notag \\
	= \exp\bigg(&\frac{2 \pi i}{N^2} \sum_{i,j} P_{ij}^c \big(d_i (a_j + b_j - [a_j + b_j]) \notag \\  &\hspace{1cm} + a_i(b_j +d_j - [b_j +d _j]) \notag \\
 & \hspace{1.5cm}-a_i(d_j +b_j - [d_j+b_j])\big)\bigg) \notag \\
	=\exp\bigg(&\frac{2 \pi i}{N^2} \sum_{i,j} P_{ij}^c d_i (a_j + b_j - [a_j + b_j])\bigg). \label{Equation_dim_red_2_cocycle}
\end{align}
This is symmetric under the exchange of the two elements $a$ and $b$ (indicating that the underlying 4-cocycle is Type III).

In terms of the data that define the cocycles, Ref. \cite{Wang2014} presents the 3-loop braiding statistics as
\begin{align*}
\theta_{\alpha \beta, c} =& \frac{2 \pi}{N^2} \sum_{i, j}( P_{ij}^c + P_{ji}^c - P_{ij}^0 - P_{ji}^0)a_i b_j \\ &+ \frac{2 \pi}{N} \sum_i ((m_i'+ m_i '' )b_i + (n_i' + n_i'')a_i),
\end{align*}
where $\alpha=(a,m)$ and $\beta=(b,n)$ describe the two loop-like excitations undergoing the braiding (the first label is the flux and the second label is the charge), $c$ is the flux label of the base loop and $m', m''$ and $n', n''$ are additional flux labels related to the dimensional reduction procedure. We can remove those additional labels by multiplying the 3-loop braiding phase by $N$ to obtain $N \theta_{\alpha \beta,c}$. This means that the second term in the braiding phase vanishes (modulo $2 \pi$) and we just have
$$N\theta_{\alpha \beta, c} = \frac{2 \pi}{N} \sum_{i, j}( P_{ij}^c + P_{ji}^c - P_{ij}^0 - P_{ji}^0)a_i b_j. $$

In addition, $P_{ij}^0$ is just zero because $0_k=0$ for all $k$, so we find
\begin{equation}
N\theta_{\alpha \beta, c}= \frac{2 \pi}{N} \sum_{i, j}( P_{ij}^c + P_{ji}^c)a_i b_j. \label{Equation_Wang2014_braiding_1}
\end{equation}
Multiplying by $N$ also removes the contribution from any regular (non-projective) charge, because any linear representation would give terms like $R(a)$ for representation $R$, which satisfy $R(a)^N=R(a^N)=R(1_G)=1$. For the case where the three flux tubes have unit flux in one of the copies of $\mathbb{Z}_N$ (copy $i$ for $\alpha$, $j$ for $\beta$ and $k$ for $c$), $N\theta_{\alpha \beta, c}$ takes the form
\begin{equation}
\Theta_{i j, k}=\frac{2 \pi i}{N} (M_{i k j} - M_{kij}+M_{j k i} - M_{kji}),
\end{equation}
as described by Equation 13 of Ref. \cite{Wang2014}.

Now we want to compare this to our expression for the braiding, Equation \ref{Equation_three_loop_braiding_result_1D} in the case where the 2-cocycle is symmetric (and so the projective irreps are 1d). We found the phase $e^{i \theta}=\alpha_2^{k,c}(a) \alpha_1^{k,a}( c)$, so we want to raise this to the power of $N$ and use the form of the cocycle given in Equation \ref{Equation_dim_red_cocycle_form}. To do so we use an iterative procedure. Using the composition rule Equation \ref{Equation_projective_irrep_basis} for projective representations, we know that
\begin{equation}
	\alpha_1^{k,a}(c^{n+1})=[c^n,c]_{k,a}^{-1} \alpha_1^{k,a}( c^n) \alpha_1^{k,a}( c),
\end{equation}
where $c^n$ is $c$ applied $n$ times via the group multiplication. Therefore,
\begin{equation}
	\alpha_1^{k,a}(c^N)= \big(\prod_{n=1}^{N-1} [c^n,c]_{k,a}^{-1} \big) \alpha_1^{k,a}(c)^N.
\end{equation}
Then, because $G=\mathbb{Z}_N^W$, $c^N=1_G$ and so $\alpha_1^{k,a}(c^N)= \alpha_1^{k,a}(1_G)=1$. This means that
$$\alpha_1^{k,a}(c)^N= \big(\prod_{n=1}^{N-1} [c^n,c]_{k,a} \big),$$
which does not depend on the choice of projective representation $\alpha_1^{k,a}$ at all. Similarly, 
$$\alpha_2^{k,c}(a)^N = \big(\prod_{n=1}^{N-1} [a^n,a]_{k,c}\big).$$

Now, using the form of the 4-cocycle from Equation \ref{Equation_cocycle_form} and the corresponding 2-cocycle described in Equation \ref{Equation_dim_red_2_cocycle}, we have
\begin{equation}
	[c^n,c]_{k,a} =\exp(\frac{2 \pi i}{N^2} \sum_{i,j} P_{ij}^{k}a_i ((c^n)_j + c_j - [(c^n)_j + c_j])).
\end{equation}
Now, note that $(c^n)_j + c_j - [(c^n)_j + c_j]$ is the difference between regular addition and addition modulo $N$. Because both $(c^n)_j$ and $c_j$ are in the range $[0, N-1]$, this difference is either 0 or $N$. Specifically, it is $N$ whenever $(c^n)_j +c_j$ ``overflows" $N$. With that in mind,
\begin{align*}
	\big(&\prod_{n=1}^{N-1} [c^n,c]_{k,a} \big)\\ &= \prod_{n=1}^{N-1} \exp(\frac{2 \pi i}{N^2} \sum_{i,j} P_{ij}^{k}a_i ((c^n)_j + c_j - [(c^n)_j + c_j]))\\
	&=\exp( \sum_{n=1}^{N-1}\frac{2 \pi i}{N^2} \sum_{i,j} P_{ij}^{k}a_i ((c^n)_j + c_j - [(c^n)_j + c_j]))\\
	&=\exp( \frac{2 \pi i}{N^2} \sum_{i,j} P_{ij}^{k}a_i \sum_{n=1}^{N-1} ((c^n)_j + c_j - [(c^n)_j + c_j])).
\end{align*}
Then $\sum_{n=1}^{N-1} ((c^n)_j + c_j - [(c^n)_j + c_j])$ contains a contribution of $N$ for each time $(c^n)_j + c_j$ exceeds $N$. This happens $c_j$ times. For example, if $c_j=1$ then we get one factor of $N$ from the $N-1$th term:
\begin{align*}
(c^{N-1})_j + c_j - [(c^{N-1})_j + c_j]&= N-1 +1 - [N-1 +1]\\
&= N - 0 =N.
\end{align*}

As another example, if $c_j=2$ then we get two contributions of $N$, one from the $(N-1)/2$ (if $N$ is odd) or $N/2 -1$ (if $N$ is even) term and one from the $N-1$ term. For the latter, note that $(c^n)_j$ refers to group multiplication, so $(c^{N-1})_j =N-2$ in that case, not $2N-2$. This means that we generally obtain
\begin{align*}
	\big(\prod_{n=1}^{N-1} [c^n,c]_{k,a} \big) &= \exp( \frac{2 \pi i}{N^2} \sum_{i,j} P_{ij}^{k}a_i Nc_j)\\
	&=\exp( \frac{2 \pi i}{N} \sum_{i,j} P_{ij}^{k}a_i c_j).
\end{align*}
A similar expression holds for the other term:
\begin{align*}
	\big(\prod_{n=1}^{N-1} [a^n,a]_{k,c} \big) &= \exp( \frac{2 \pi i}{N^2} \sum_{i,j} P_{ij}^{k}c_i Na_j)\\
	&=\exp( \frac{2 \pi i}{N} \sum_{i,j} P_{ij}^{k}c_i a_j)\\
	&=\exp( \frac{2 \pi i}{N} \sum_{i,j} P_{ji}^{k}a_i c_j),
\end{align*}
where in the last term we use the fact that $i$ and $j$ are dummy indices in the same set to swap them. This means that the braiding phase is
\begin{align}
	e^{iN \theta} = \exp( \frac{2 \pi i}{N} \sum_{i,j} (P_{ij}^{k}+P_{ji}^{k})a_i c_j).
\end{align}
This agrees with the result from Ref. \cite{Wang2014}, Equation S25 in the Supplemental Material \cite{Wang2014Sup} of that paper, once we multiply that result by $N$ and drop the term that is a multiple of $2 \pi$ (see Equation \ref{Equation_Wang2014_braiding_1}). When we take the special case where the three loops each carry unit flux in one of the copies of $\mathbb{Z}_N$ and trivial flux in the other copies, so that $a_i = \delta_{i, i_a}$, $c_j= \delta_{j, j_c}$ and $k_l= \delta_{l, l_k}$ for some indices $i_a, j_c, l_k$, this becomes
\begin{align*}
	e^{iN \theta} &= \exp( \frac{2 \pi i}{N} (P_{i_aj_c}^{k}+P_{j_ci_a}^{k})).
\end{align*}
Substituting the expression given in Equation \ref{Equation_P_definition} for $P_{ij}^k$, with $k_l= \delta_{l, l_k}$, we see that
\begin{align*}
	e^{iN \theta} &= \exp( \frac{2 \pi i}{N} (M_{i_a l_k j_c} - M_{l_ki_aj_c}+M_{j_c l_k i_a} - M_{l_kj_ci_a})),
\end{align*}
from which we see that $N\theta$ matches the tensor $\Theta_{i_a j_c, l_k}$ given in Equation 13 of Ref. \cite{Wang2014}.

\subsection{Four-loop braiding and the non-Abelian nature of three-loop braiding}

\begin{figure}
	\centering
	\includegraphics[width=\linewidth]{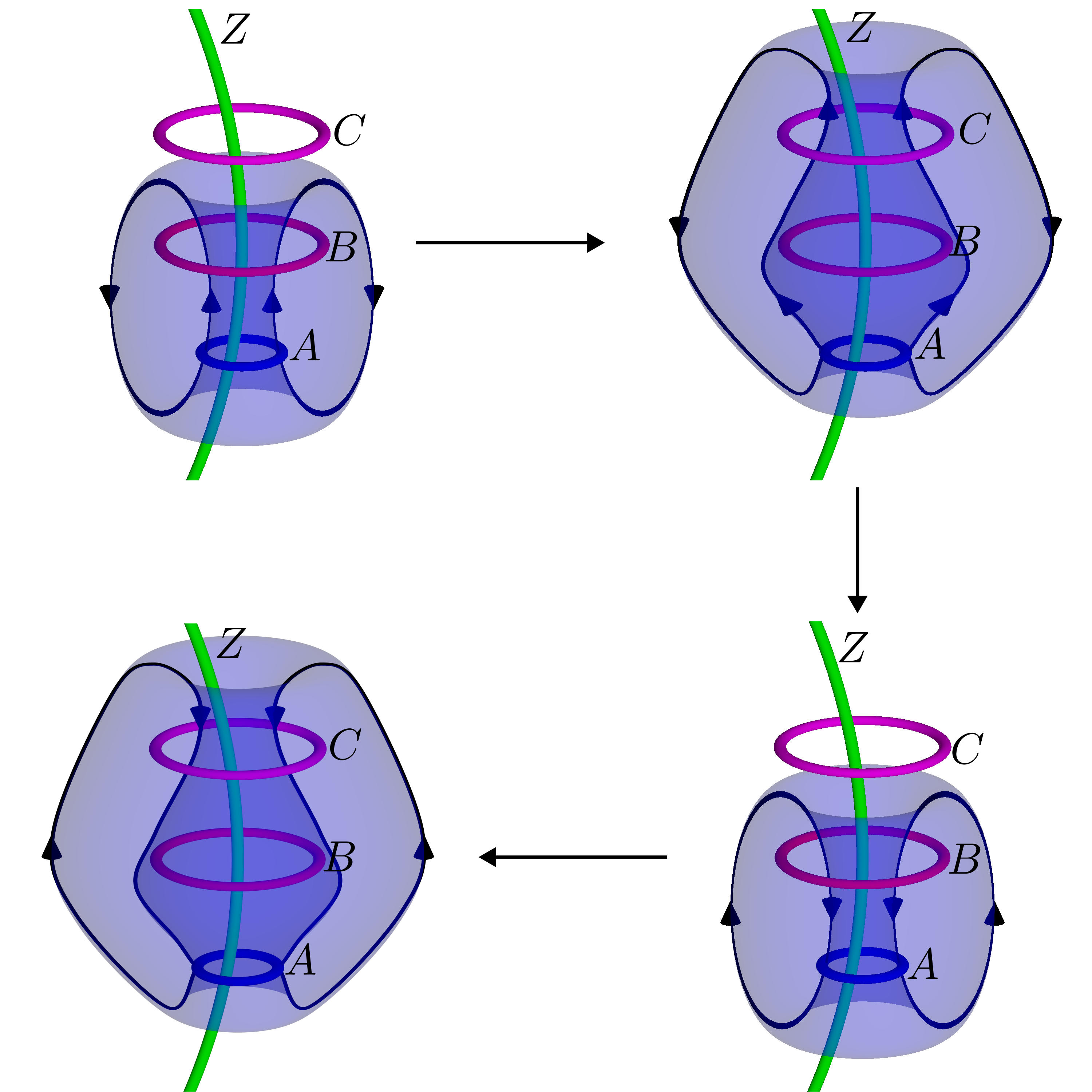}
	\caption{Four-loop braiding can be expressed as a series of three-loop braids, shown in each picture. Here the membranes represent the motion of the lower (blue) loop $A$ in each step, with the black arrows representing the paths traced by two points on opposite ends of the loop to highlight the direction of the motion. The third motion is the first in reverse (and similar for second and fourth), so this motion measures the non-Abelian nature of three-loop braiding. We may worry that in the second and fourth step the loop is pulled over the middle (red) loop $B$, but this does not apply a transformation in this case (as long as we do not braid through the loop at the other end of the membrane operator that produces the $B$).}
	\label{Figure_four_loop_braiding}
\end{figure}

One interesting aspect of this model is that the three-loop braiding is non-Abelian, even though the underlying group $G$ is Abelian. One way of examining this further is through a more complicated braiding process, called four-loop braiding, which is analogous to a commutator of three-loop braiding processes \cite{Tiwari2017,Wang2020, Putrov2017, Wang2015a, Zhang2021}. In this process, there are three participating loops $A$, $B$ and $C$, all of which are linked to the same base loop $Z$. We first braid $A$ through $B$, then braid $A$ through $C$. Next, we braid $A$ through $B$ in reverse, then through $C$ in reverse. This is illustrated in Figure \ref{Figure_four_loop_braiding}. From the perspective of $A$, we can write this process schematically as $BCB^{-1}C^{-1}$, where each letter indicates $A$ braiding with the corresponding loop and an inverse indicates the reverse process. If the braiding were Abelian, the resulting transformation would be trivial because each individual braid is performed and then reversed. However, because the forwards and reverse processes are interrupted by braiding with another loop, the transformation can be non-trivial if the three-loop braiding is non-Abelian. 

We can calculate the transformation under this process by applying the transformations under the individual three-loop braiding moves simultaneously. Given that the flux of the base loop $Z$ is $z$, we can denote the initial state of the loop $A$ by $(a, \alpha^{z,a}_A, i_A)$, where $a$ and $\alpha^{z,a}$ are the flux and charge, respectively, which will remain unchanged by the braiding process. On the other hand, $i_A$ is an internal label, corresponding to one of the matrix indices of the membrane operator (one, because only one is affected by the braiding relation, as described by Equation \ref{Equation_three_loop_braiding_result_1}). Similarly, we denote the initial states of $B$ and $C$ by $(b, \alpha^{z,b}_B, j_B)$ and $(c, \alpha^{z,c}_C, j_C)$, respectively. For $B$ and $C$ we use the symbol $j$ rather than $i$ to reflect the fact that Equation \ref{Equation_three_loop_braiding_result_1} describes braiding between a loop $A$ at one end of a membrane operator and another loop ($B$ or $C$) at the other end of another membrane operator. Then applying the braiding relation between $A$ and $B$ we would obtain
$$(a, \alpha^{z,a}_A, i_A) \rightarrow \sum_{n_A} [\alpha^{z,a}_A(b)]_{i_A n_A} (a, \alpha^{z,a}_A, n_A)$$
$$(b, \alpha^{z,b}_B, j_B)  \rightarrow \sum_{n_B} (b, \alpha^{z,b}_B, n_B) [\alpha^{z,b}_B(a)]_{n_B j_B}.$$
Then applying the relation between $A$ (in its new state) and $C$ we find that
$$(a, \alpha^{z,a}_A, n_A) \rightarrow \sum_{p_A} [\alpha^{z,a}_A(c)]_{n_A p_A} (a, \alpha^{z,a}_A, p_A)$$
$$(c, \alpha^{z,c}_C, j_C)  \rightarrow \sum_{p_C} (c, \alpha^{z,c}_C, p_C) [\alpha^{z,c}_C(a)]_{p_C j_C}.$$
This means that under these two transformations, the state of $A$ has become
\begin{align*}
\sum_{p_A}\sum_{n_A} &[\alpha^{z,a}_A(b)]_{i_A n_A}  [\alpha^{z,a}_A(c)]_{n_A p_A} (a, \alpha^{z,a}_A, p_A) \\ &= \sum_{p_A} [\alpha^{z,a}_A(b)\alpha^{z,a}_A(c)]_{i_A p_A }(a, \alpha^{z,a}_A, p_A).
\end{align*}

Next we apply the reverse braiding relation between $A$ and $B$, obtaining the following result:
$$(a, \alpha^{z,a}_A, p_A) \rightarrow \sum_{q_A} [\alpha^{z,a}_A(b)^{-1}]_{p_A q_A} (a, \alpha^{z,a}_A, q_A)$$
$$(b, \alpha^{z,b}_B, n_B) \rightarrow \sum_{q_B}  (b, \alpha^{z,b}_B, q_B) [\alpha^{z,b}_B(a)^{-1}]_{q_B n_B}.$$

In total, the new state of $B$ is 
\begin{align*}
\sum_{q_B} \sum_{n_B}& (b, \alpha^{z,b}_B, q_B) [\alpha^{z,b}_B(a)^{-1}]_{q_B n_B}	[\alpha^{z,b}_B(a)]_{n_B j_B}\\
&=  \sum_{q_B} (b, \alpha^{z,b}_B, q_B) [\alpha^{z,b}_B(a)^{-1} \alpha^{z,b}_B(a) ]_{q_B j_B}\\
&= \sum_{q_B} (b, \alpha^{z,b}_B, q_B) \delta_{q_B j_B}\\
&=  (b, \alpha^{z,b}_B, j_B),
\end{align*}
from which we see that $B$ is left unaffected.

Finally, we apply the reverse braiding relation between $A$ and $C$ to obtain
$$(a, \alpha^{z,a}_A, q_A) \rightarrow \sum_{r_A} [\alpha^{z,a}_A(c)^{-1}]_{q_A r_A} (a, \alpha^{z,a}_A, r_A)$$
$$(c, \alpha^{z,c}_C, p_C) \rightarrow \sum_{r_C} (c, \alpha^{z,c}_C, r_C) [\alpha^{z,c}_C(a)^{-1}]_{r_C p_C}.$$
Similar to $B$, $C$ is left invariant by the total motion:
\begin{align*}
(c, \alpha^{z,c}_C&, j_C)  \\
\rightarrow& \sum_{p_C} \sum_{r_C} (c, \alpha^{z,c}_C, r_C)  [\alpha^{z,c}_C(a)^{-1}]_{r_C p_C} [\alpha^{z,c}_C(a)]_{p_C j_C} \\
&= (c, \alpha^{z,c}_C, j_C).
\end{align*}

On the other hand, $A$ can undergo a non-trivial transformation:
\begin{align*}
	&(a, \alpha^{z,a}_A, i_A) \rightarrow\\
	& \sum_{n_A,p_a, q_A, r_A} [\alpha^{z,a}_A(b)]_{i_A n_A} [\alpha^{z,a}_A(c)]_{n_A p_A} \\
	& \hspace{2cm} [\alpha^{z,a}_A(b)^{-1}]_{p_A q_A} [\alpha^{z,a}_A(c)^{-1}]_{q_A r_A} (a, \alpha^{z,a}_A, r_A)\\
	&= \sum_{r_A} [\alpha^{z,a}_A(b)\alpha^{z,a}_A(c)\alpha^{z,a}_A(b)^{-1}\alpha^{z,a}_A(c)^{-1}]_{i_A r_A} (a, \alpha^{z,a}_A, r_A).
\end{align*}

We can simplify the product of matrices by using the properties of projective representations and the fact that the group $G$ is Abelian. We have
$$\alpha^{z,a}_A(b)\alpha^{z,a}_A(c) = [b,c]_{z,a} \alpha^{z,a}_A(bc)$$
from the composition rule for projective representations. Then, because $G$ is Abelian, we can reverse the order of multiplication of the group elements:
$$[b,c]_{z,a} \alpha^{z,a}_A(bc)= [b,c]_{z,a} \alpha^{z,a}_A(cb)$$
and separate the contributions from $b$ and $c$ again to obtain
\begin{align*}
\alpha^{z,a}_A(b)\alpha^{z,a}_A(c) &= \frac{[b,c]_{z,a}}{[c,b]_{z,a}} \alpha^{z,a}_A(c) \alpha^{z,a}_A(b) \\
&= \eta^{z,a}(b,c)  \alpha^{z,a}_A(c) \alpha^{z,a}_A(b). 
\end{align*}

Inserting this into the matrix product from the four-loop braiding relation, we obtain
\begin{align*}
	\alpha^{z,a}_A&(b)\alpha^{z,a}_A(c)\alpha^{z,a}_A(b)^{-1}\alpha^{z,a}_A(c)^{-1}\\
	&=  \eta^{z,a}(b,c) 	\alpha^{z,a}_A(c) \alpha^{z,a}_A(b)\alpha^{z,a}_A(b)^{-1}\alpha^{z,a}_A(c)^{-1}\\
	&= \eta^{z,a}(b,c)  \mathbb{I}.
\end{align*}
From this, we see that the transformation of $A$ under the four-loop braiding process is just a phase:
\begin{align*}
	&(a, \alpha^{z,a}_A, i_A) \\
	&\rightarrow \sum_{r_A} \eta^{z,a}(b,c) \delta_{i_A, r_A} (a, \alpha^{z,a}_A, r_A) = \eta^{z,a}(b,c) (a, \alpha^{z,a}_A, i_A).
\end{align*}
Because the other loops transform trivially, the total effect of the four-loop braiding relation is just this phase:
\begin{equation}
\theta_{4}(a,b,c,z) = \eta^{z,a}(b,c),
\end{equation}
which only depends on the fluxes of the four loops (and the 4-cocycle), not on their charges. Notably, this relation has a high degree of symmetry between its indices $a$, $b$, $c$ and $z$: $\eta$ is the 1-cocycle obtained from the underlying 4-cocycle by applying the slant product three times, and swapping any two of its indices results in inverting the phase.

\section{Fusion Rules for Cylindrical Membranes}
\label{Section_fusion_membranes}

Having considered the braiding relations of the loop-like excitations, we now consider the fusion rules. These describe how two excitations can be combined into a single one. One way to demonstrate the fusion rules is to apply two membrane operators on the same membrane and show how these can be expressed as a single membrane operator (or sum of membrane operators in the non-Abelian case). Ideally, we would use an open membrane operator, so that the excitations themselves fuse together. However, in this model, the membrane operators are not well-defined near their boundaries or other plaquette excitations, as we discussed in Section \ref{Section_membrane_operators}, which means that bringing excitations close together could result in additional boundary operators. Instead, we will use closed membrane operators, which do not produce plaquette excitations but still carry the fusion information. We are particularly interested in the fusion rules of the loop-like excitations with a given base loop (we cannot fuse excitations that have different base loops). To examine the fusion rule of such excitations, we use the same geometry shown in Figures \ref{Figure_braiding_direct_membranes}, \ref{Figure_braiding_upper_wedge} and \ref{Figure_braiding_lower_wedge}, except that we also apply periodic boundary conditions in the $x$ direction and we only apply membrane operators on the horizontal (red) membrane $m_1$. If we apply two membrane operators $ F^{a, \alpha_1^{k,a}, i_1, j_1}(m_1) $ and $F^{b, \alpha_2^{k,b}, i_2, j_2}(m_1)$, we have
\begin{align}
& F^{b, \alpha_2^{k,b}, i_2, j_2}(m_1)  F^{a, \alpha_1^{k,a}, i_1, j_1}(m_1) \ket{\psi}\notag \\
& =C_0^b(m_1) \theta_D^b(m_1) \theta_S^{b, \alpha_2^{k,b}, i_2, j_2}(m_1)C_0^a(m_1) \theta_D^a(m_1) \notag \\ & \hspace{1cm} \theta_S^{a, \alpha_1^{k,a}, i_1, j_1}(m_1)\ket{\psi} \notag\\
 &= C_0^b(m_1) C_0^a(m_1) (C_0^a(m_1) :\theta_D^b(m_1)) \theta_D^a(m_1) \notag \\ & \hspace{1cm}(C_0^a(m_1) :\theta_S^{b, \alpha_2^{k,b}, i_2, j_2}(m_1)) \theta_S^{a, \alpha_1^{k,a}, i_1, j_1}(m_1)\ket{\psi}.\label{Equation_fusion_intermediate_1}
 \end{align}

Using Equations \ref{Equation_braiding_dual_phase_upper_1} and \ref{Equation_braiding_dual_phase_lower_1}, which describe the dual phase for the upper and lower wedges of the affected region respectively, we have 
$$\theta_D^a(m_1) = [a, y', x']_k [a,x,y]_k^{-1} [x,a,y]_k.$$
In this case, we take $m_1$ to be a closed membrane, which we achieve by applying additional periodic boundary conditions in the $x$ direction. This results in $y=y'$ directly from the boundary condition and $x=x'$ from flatness in the region. Therefore,
\begin{equation}
\theta_D^a(m_1) = [a, y, x]_k [a,x,y]_k^{-1} [x,a,y]_k = [a,y]_{k,x}.
\end{equation}
Because the other membrane operator is applied on the same membrane $m_1$, its dual phase has the same form (but with $a$ replaced by $b$). However, because $C_0^a(m_1)$ acts before the phase $\theta_D^b(m_1)$, we should also replace $y$ with $ay$ to obtain
\begin{equation}
 (C_0^a(m_1) :\theta_D^b(m_1))= [b,ay]_{k,x}.
\end{equation}
Putting these two phases together, we have
\begin{align*}
\theta_D^a(m_1) (C_0^a(m_1) :\theta_D^b(m_1))&= [a,y]_{k,x} [b,ay]_{k,x}\\
&= [b,a]_{k,x} [ba,y]_{k,x},
\end{align*}
where we used the 2-cocycle condition, Equation \ref{Equation_2_cocycle_condition}, for the latter equality. For the surface weights, we have
$$\theta_S^{a, \alpha_1^{k,a}, i_1, j_1}(m_1) = [\alpha_1^{k,a}(x)]_{i_1 j_1}$$
and
$$C_0^a(m_1) :\theta_S^{b, \alpha_2^{k,b}, i_2, j_2}(m_1)= [\alpha_2^{k,b}(x)]_{i_2 j_2}.$$

Inserting these results into Equation \ref{Equation_fusion_intermediate_1}, we see that
\begin{align*}
	F^{b, \alpha_2^{k,b}, i_2, j_2}&(m_1) F^{a, \alpha_1^{k,a}, i_1, j_1}(m_1) \ket{\psi}\\
	&  = C_0^b(m_1) C_0^a(m_1) [b,a]_{k,x} [ba,y]_{k,x} \\ 
 & \hspace{1.5cm}[\alpha_1^{k,a}(x)]_{i_1 j_1} [\alpha_2^{k,b}(x)]_{i_2 j_2} \ket{\psi}.
\end{align*}
We note that the fusion rule for the untwisted membrane operators $C_0^b(m_1)$ and $C_0^a(m_1)$ is just
$$ C_0^b(m_1) C_0^a(m_1)= C_0^{ba}(m_1),$$
because each untwisted membrane operator just multiplies the edges cut by its dual membrane by $a^{\pm 1}$ or $b^{\pm 1}$. Similarly, we note that the 2-cocycle $[ba,y]_{k,x}$ is just the dual phase we would get for the membrane operator with label $ba$. Therefore,
\begin{align}
	&F^{b, \alpha_2^{k,b}, i_2, j_2}(m_1) F^{a, \alpha_1^{k,a}, i_1, j_1}(m_1) \ket{\psi} \notag \\
	&= C_0^{ba}(m_1) \theta_D^{ba}(m_1) [b,a]_{k,x} [\alpha_1^{k,a}(x)]_{i_1 j_1} [\alpha_2^{k,b}(x)]_{i_2 j_2} \ket{\psi}. \label{Equation_fusion_intermediate_2}
\end{align}

Now we claim that $[b,a]_{k,x} \alpha_1^{k,a}(x) \otimes \alpha_2^{k,b}(x)$ defines a (generally reducible) $\beta_{k,ba}$-projective representation of $G$, which we call $\alpha_T^{k,ba}$:

\begin{equation}
\alpha_T^{k,ba}(x)=[b,a]_{k,x} \alpha_1^{k,a}(x) \otimes \alpha_2^{k,b}(x) \label{Equation_fusion_rep}
\end{equation}

 To see this, note that
\begin{align*}
&\alpha_T^{k,ba}(x) \alpha_T^{k,ba}(y)\\
&= [b,a]_{k,x} [b,a]_{k,y} \big(\alpha_1^{k,a}(x)\alpha_1^{k,a}(y)\big) \otimes \big( \alpha_2^{k,b}(x)\alpha_2^{k,b}(y) \big)\\
&= [b,a]_{k,x} [b,a]_{k,y} \big( [x,y]_{k,a}\alpha_1^{k,a}(xy)\big) \otimes \big( [x,y]_{k,b} \alpha_2^{k,b}(y) \big),
\end{align*}
where we used the defining relations for the two projective irreps $\alpha_1^{k,a}$ and $\alpha_2^{k,b}$.
Then we can insert the identity in the form
$$\frac{[x,y]_{k,ba}}{[x,y]_{k,ba}} \frac{[b,a]_{k,xy}}{[b,a]_{k,xy}},$$
to obtain
\begin{align*}
&\alpha_T^{k,ba}(x) \alpha_T^{k,ba}(y) \\
&= [b,a]_{k,x} [b,a]_{k,y} [x,y]_{k,a} [x,y]_{k,b} \frac{[x,y]_{k,ba}}{[x,y]_{k,ba}} \frac{[b,a]_{k,xy}}{[b,a]_{k,xy}} \\
& \hspace{1cm} \big( \alpha_1^{k,a}(xy)\big) \otimes \big( \alpha_2^{k,b}(y) \big)\\
&=[b,a]_{k,x} [b,a]_{k,y} [x,y]_{k,a} [x,y]_{k,b} \frac{[x,y]_{k,ba}}{[x,y]_{k,ba}} \\
& \hspace{1cm}\frac{1}{[b,a]_{k,xy}} \alpha_T^{k,ba}(xy),
\end{align*}
where we used the definition of $\alpha_T^{k,ba}$ for the last equality. Then we note that $[x,y]_{k,ba}$ is the phase that we expect if $\alpha_T^{k,ba}$ is indeed a $\beta_{k,ba}$-projective representation. We can therefore write the above relation as
\begin{align*}
	\alpha_T^{k,ba}(x) \alpha_T^{k,ba}(y) &= \theta_R [x,y]_{k,ba} \alpha_T^{k,ba}(xy),
\end{align*}
where
\begin{equation}
\theta_R = \frac{[b,a]_{k,x} [b,a]_{k,y} [x,y]_{k,a} [x,y]_{k,b} }{[x,y]_{k,ba} [b,a]_{k,xy}} \label{Equation_fusion_remnant_cocycle}
\end{equation}
is the additional phase compared to the required phase for the projective representation relation. However, by writing these 2-cocycles in terms of the underlying 3-cocycles (using Equation \ref{Equation_2_cocycle_definition}), we can show that this additional phase is just equal to 1. We have
\begin{align*}
	&\theta_R= \frac{[b,a]_{k,x} [b,a]_{k,y} [x,y]_{k,a} [x,y]_{k,b} }{[x,y]_{k,ba} [b,a]_{k,xy}}\\&= \frac{[x,b,a]_{k}[b,a,x]_{k}[ y, b, a]_{k} [b,a, y]_{k} [a,x, y]_{k} [x, y, a]_{k}} {[b, x, a]_{k} [ b, y, a]_{k} [x, a, y]_{k} } \\
 & \hspace{1cm} \frac{[b,x,y]_{k}[x,y,b]_{k} [b, xy, a]_{k} [x, ba, y]_{k}}{[x, b, y]_{k} [xy,b, a]_{k}[ b, a, xy]_{k} [ba, x, y]_{k} [x, y, ba]_{k}}.
\end{align*}
Applying the 3-cocycle condition Equation \ref{Equation_3_cocycle_condition}, we have $$\frac{[x, y, b]_k [y,b, a]_k}{[xy,b,a]_k [x, y, ba]_k}= [x, by, a]_k^{-1}$$ and $$\frac{[b,a,x]_k [a,x,y]_k}{[ba,x,y]_k[b,a,xy]_k}= [b,ax,y]_k^{-1},$$ so
\begin{align*}
	\theta_R	&= \frac{[x,b,a]_{k} [b,a, y]_{k} [b,x, y]_{k} [x,y,a]_{k} [b, xy, a]_{k} [x, ba, y]_{k}} {[b, x, a]_{k} [ b, y, a]_{k} [x, b, y]_{k} [x, a, y]_{k} [x,by,a]_{k} [b,ax,y]_{k}}. 
\end{align*}
Then applying the 3-cocycle condition $[x,b,a]_{k} [b,a, y]_{k} [x,ba,y]_k = [xb,a,y]_k [x, b, ay]_k$ and $[b,x,y]_k[x,y,a]_k[b,xy,a]_k= [bx, y, a]_k [b, x, ya]_k$, so we obtain
\begin{align*}
	\theta_R = \frac{[xb, a, y]_{k} [x,b, ay]_{k} [bx, y, a]_{k} [b, x, ya]_{k}} {[b, x, a]_{k} [ b, y, a]_{k} [x, b, y]_{k} [x, a, y]_{k} [x,by,a]_{k} [b,ax,y]_{k}}.
\end{align*}
Finally, using the 3-cocycle condition we find $$\frac{[x,b,ya]_k [xb,y,a]_k}{[b, y, a]_k [x, b, y]_k[ x, by, a]_k}=1$$ and $$\frac{[xb,a,y]_k [b, x, ay]_k}{[b,x,a]_k[x,a,y]_k [b, xa, y]_k}=1,$$ allowing us to remove the last remaining cocycles. That is,
$$\theta_R=\frac{[b,a]_{k,x} [b,a]_{k,y} [x,y]_{k,a} [x,y]_{k,b} }{[x,y]_{k,ba} [b,a]_{k,xy}} =1.$$
Therefore, we have
\begin{align*}
	\alpha_T^{k,ba}(x) \alpha_T^{k,ba}(y) &= [x,y]_{k,ba} \alpha_T^{k,ba}(xy)
\end{align*}
and $\alpha_T^{k,ba}$ is indeed a projective representation. The fusion relation Equation \ref{Equation_fusion_intermediate_2} therefore becomes
\begin{align}
	F^{b, \alpha_2^{k,b}, i_2, j_2}&(m_1) F^{a, \alpha_1^{k,a}, i_1, j_1}(m_1) \ket{\psi}\notag \\
	&= C_0^{ba}(m_1) \theta_D^{ba}(m_1) [\alpha_T^{k,ba}(x)]_{i_1j_1 i_2 j_2} \ket{\psi} \label{Equation_fusion_result}.
\end{align}

The expression on the right-hand side of Equation \ref{Equation_fusion_result} has the same form as the membrane operators, consisting of an untwisted part (with flux label $ba$), a dual phase $\theta_D^{ba}(m_1)$ and a surface weight. However, this surface weight is given by a matrix element $[\alpha_T^{k,ba}(x)]_{i_1j_1 i_2 j_2}$ of the projective representation $\alpha_T^{k,ba}$, where
\begin{equation}
[\alpha_T^{k,ba}(x)]_{i_1j_1 i_2 j_2}= [b,a]_{k,x} [\alpha_1^{k,a}(x)]_{i_1j_1} [\alpha_2^{k,b}(x)]_{i_2 j_2}.
\end{equation}
This matrix element is determined by four indices, but this is simply because it is constructed from a tensor product of the two projective irreps $\alpha_1^{k,a}$ and $\alpha_2^{k,b}$. A two-index object can be obtained by grouping the indices $i_1$ and $i_2$ as well as $j_1$ and $j_2$ if desired. Regardless, we can define the membrane operator with this weight as $F^{ba, \alpha_T^{k,ba}, i_1, i_2, j_1, j_2}(m_1)$, so that the fusion relation becomes
\begin{align}
	F^{b, \alpha_2^{k,b}, i_2, j_2}(m_1)& F^{a, \alpha_1^{k,a}, i_1, j_1}(m_1) \ket{\psi} \notag \\
	&= F^{ba, \alpha_T^{k,ba}, i_1, i_2, j_1, j_2}(m_1)\ket{\psi}. \label{Equation_fusion_result2}
\end{align}
We see that the flux labels $b$ and $a$ fuse to a total flux of $ba$, while the total charge label is $\alpha_T^{k,ba}$. If $\alpha_1^{k,a}$ and $\alpha_2^{k,b}$ are 1-dimensional, the matrix indices drop out and $\alpha_T^{k,ba}(xy)$ is a 1-dimensional projective irrep, meaning that it is the final fusion product. Otherwise, $\alpha_T^{k,ba}(xy)$ is generally reducible, with the constituent irreps being the possible fusion products. This result has the same structure found in Ref. \cite{Bullivant2019}, where the tube algebra was used to determine the fusion rules for the excitations.

It is instructive to consider how this result simplifies if we take the base loop to be trivial (by taking $k=1_G$), thereby considering the fusion of two unlinked loops. In this case, $\alpha_1$ and $\alpha_2$ are linear irreps. Then

$$\alpha_T^{k,ba}(x)=[b,a]_{k,x} \alpha_1^{k,a}(x) \otimes \alpha_2^{k,b}(x)$$
becomes
$$\alpha_T(x)=[b,a]_{1_G,x} \alpha_1(x) \otimes \alpha_2(x) = \alpha_1(x) \otimes \alpha_2(x),$$
which is the usual fusion of irreps. Indeed, for Abelian $G$ these irreps are 1d, so this fusion is simple multiplication.

	\section{Topological Charge}
	\label{Section_topological_charge}
	
	Topological charge is a quantity that is conserved, without the need for a symmetry, on the level of the Hilbert space (although the Hamiltonian picks out a set of charges that are relevant for the model and in particular picks the vacuum charge as the ground state). The charge within a region can only be changed by moving charge out of that region, which can be detected by an operator on the surface of that region. The topological charge measurement operator is therefore a surface operator, which should also satisfy some additional properties. In particular, it should be topological, as deforming the measurement operator without crossing any excitations should leave the enclosed charge unaffected. The measurement operator should also not produce any excitations. The operators that satisfy these conditions are closed ribbon and membrane operators \cite{Huxford2023,Bombin2008}. In order to construct a general charge measurement operator, we first choose a measurement surface and then apply all the independent closed ribbon and membrane operators on that surface, following the method used in 2+1d in Ref. \cite{Bombin2008} and 3+1d in Ref. \cite{Huxford2023} for different models. For example, for a torus surface, we would apply a closed magnetic membrane operator over the surface itself and closed electric ribbon operators on the two cycles of that torus (as any other closed ribbon operator can be deformed and split into closed ribbon operators around those two cycles, or else deformed to nothing). On the other hand, a spherical surface has no non-contractible loops, so any closed ribbon operator applied on that surface is trivial. From this, we see that the charges that can be measured by a sphere and by a torus are different (corresponding to point-like charge and loop-like or link-like charge, respectively).

	We are mostly interested in the charge measured by a toroidal measurement surface (corresponding to link-like or loop-like charge), although we will also briefly discuss the charge measured by a sphere (corresponding to point-like charge). More complicated measurement surfaces, with more handles, are also possible, though we will not consider them here. First, we examine a spherical measurement surface $S$. The only operator we can apply on the surface which does not leave excitations is a magnetic membrane operator, so the general measurement operator has the form
	$$\sum_{h \in G} a_h C^h(S),$$
	where $a_h$ are coefficients. Here $C^h(S)$ is the magnetic membrane operator, including the untwisted membrane operator, the dual phase and the surface weight. Unlike for a torus, where there are multiple reference diagrams that can be assigned different values, the surface weight is entirely determined by the graphical rules, meaning that the membrane operator is specified solely by its flux. The individual operators $C^h(S)$ form a basis for the space of measurement operators, but we want to construct basis operators that are orthogonal projectors (because a measurement of charge should correspond to a projection operator). To do so, we need to know the algebra satisfied by the spherical membrane operators. Because the spherical membrane operators are equivalent to a product of vertex transforms (as we show in Section \ref{Section_spherical_membranes} of the Supplemental Material), they obey the same algebra as the vertex transforms: $C^h(S) C^k(S) = C^{hk}(S)$. This means that we can easily construct projectors using the ordinary linear irreps of $G$:
	\begin{equation}
	P^R(S) = \frac{1}{|G|} \sum_{g \in G} R(g) C^g(S). \label{Equation_spherical_projector}
	\end{equation}
	This indicates that the charge measured by a spherical membrane, which corresponds to point-like charge, is labeled by regular irreps of $G$. This agrees with our intuition that the point-like charges are the same as in the untwisted case because the electric ribbon operators are the same (as we discussed in Section \ref{Section_ribbon_operators}).

	Notably, the number of these projectors (and so the number of point-like particle types) matches the ground state degeneracy of the model on the manifold $S^2 \times S^1$. This ground state degeneracy can be calculated from Equation 42 in Ref. \cite{Wan2015}, which describes the ground state degeneracy on the 3-torus, by taking two of the cycles of the 3-torus to have the trivial label $1_G$. To see this, note that the 3-torus can be written as $T^2 \times S^1$, where $T^2$ is the 2-torus and can be represented by a square with opposite sides identified. Taking the sides of that square (the cycles of the 2-torus) to have trivial label allows us to collapse these edges to points without altering the allowed states, giving us a 2-sphere. Then the total manifold is $S^2 \times S^1$ and its degeneracy is equal to that of the 3-torus with two of the cycle labels to be taken to be trivial algebraically. For the 3-torus, Wan et al. \cite{Wan2015} find the following basis for the ground states:
	\begin{align*}
	\big\{\frac{1}{|G|} \sum_{x \in G} \eta^{k,g}(h,x) &\ket{xkx^{-1}, xgx^{-1}, xhx^{-1}} \\ & \hspace{2cm}| \ k,g, h \text{ commute} \big\},
	\end{align*}
	where $k$, $g$ and $h \in G$ are the labels for the three independent cycles of the 3-torus. Taking two of these, say $k$ and $g$, to be trivial, we obtain
	$$\big\{\frac{1}{|G|} \sum_{x \in G} \eta^{1_G,1_G}(h,x) \ket{1_G, 1_G, xhx^{-1}} \ | \ h \in G \big\},$$
	where $h$ of course commutes with the identity element. Significantly, $\eta^{1_G,1_G}(h,x)$ is equal to unity, which is important because a non-trivial $\eta$ could lead to some terms vanishing when we sum over $x$ \cite{Wan2015}. Then we see that the basis for $S^2 \times S^1$ is just given by
	$$\big\{\frac{1}{|G|} \sum_{x \in G} \ket{xhx^{-1}}  |h \in G \big\},$$
	meaning that there is one basis state per conjugacy class of $G$ (or just the elements, when $G$ is Abelian). This can also be verified by a direct calculation. These conjugacy classes are in one-to-one correspondence with the irreps of $G$, which are the labels for the projectors in Equation \ref{Equation_spherical_projector}. This equivalence between the ground state degeneracy and the number of charges measured by a sphere (i.e., point-like charges) is analogous to how the the number of types of topological charge measured by a circular surface in 2+1d (i.e., the number of particle types) matches the ground state degeneracy for a 2-torus for (modular) 2+1d topological theories. This equality matches a general rule for topological phases \cite{Wang2015, Wen2017}, which we expect from the relationship between the Hamiltonian models and topological quantum field theories. We note that the number of point-like charges is also equal to the number of types of pure unlinked loops, which are the unlinked loops that carry no point-like charge (these are the pure fluxes, of which there are $|G|$ for Abelian $G$).

	Next, we consider the toroidal measurement surface, which carries additional complexity. In this case, we can apply ribbon operators around the two cycles $c_1$ and $c_2$ of the torus in addition to the membrane operator over the torus surface. This suggests that the general measurement operator has the form
	$$\sum_{h, g_1, g_2 \in G} a_{g_1, g_2,h} C^h(m) \delta(\hat{g}(c_1), g_1) \delta(\hat{g}(c_2), g_2),$$
	but this is not quite right. To understand this, recall that the magnetic membrane operator includes a surface weight which must transform appropriately under various diagrammatic moves in order for the overall membrane operator to commute with the vertex transforms (and in order to be topological). These diagrammatic moves allow us to calculate the surface weight in terms of certain reference diagrams, which we must assign a value manually. However, we will now show that not all combinations of $h$, $g_1$ and $g_2$ are consistent with these diagrammatic moves. We can think of this as an obstruction to closing a cylindrical membrane into a torus: when we try there may be some unavoidable vertex excitations where we join the two ends (or equivalently, the place where we join them is not topological, because we cannot ordinarily move the boundary of a membrane operator without affecting its action, so additional conditions may be necessary to ensure that the join between the ends becomes ``seamless").

	\begin{figure}
		\centering
		\includegraphics[width=\linewidth]{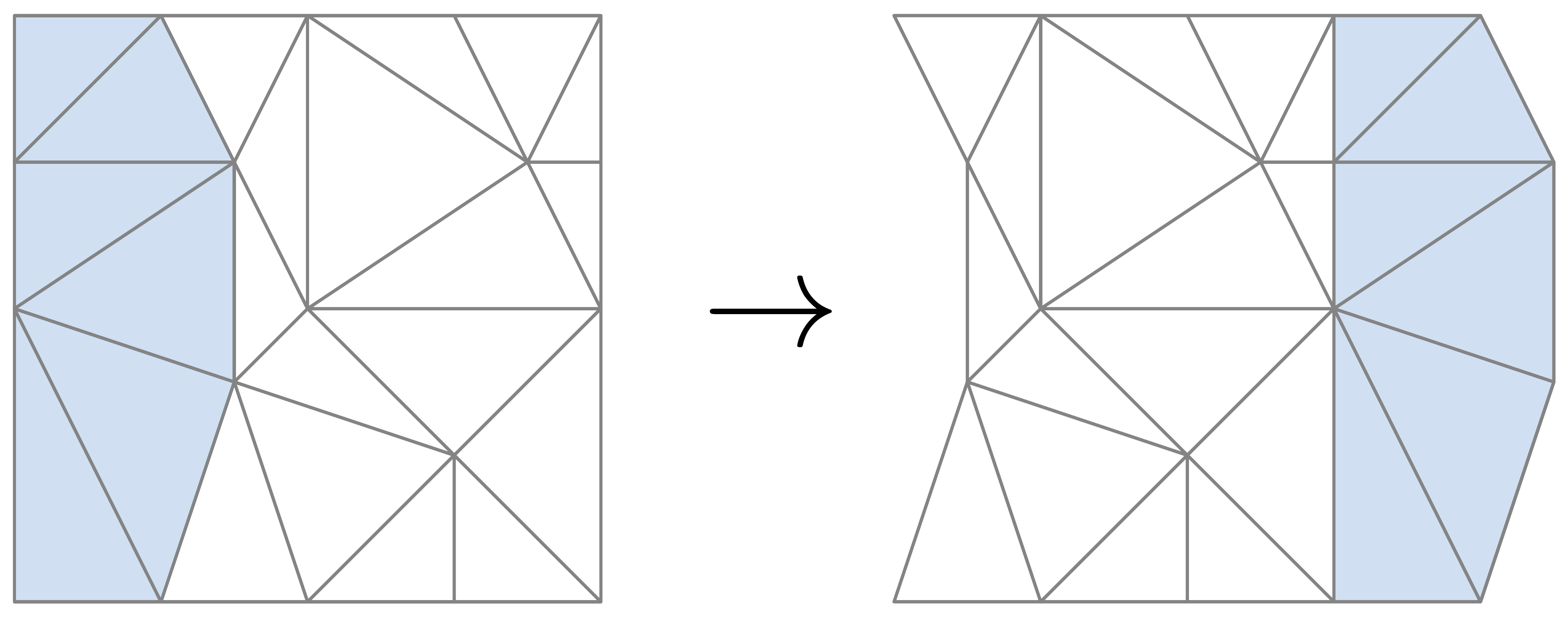}
		\caption{This is an example of a rectangle representing a torus surface (opposite edges are identified). However, the same torus can be represented in multiple ways by choosing different positions for the edges of the square. Here, we show one example of how shifting the ``view" of the torus appears to change the diagram (we shaded the region that is shifted from one diagram to the other).}
		\label{Figure_torus_shift_view}
	\end{figure}

\begin{figure}
		\centering
		\includegraphics[width=\linewidth]{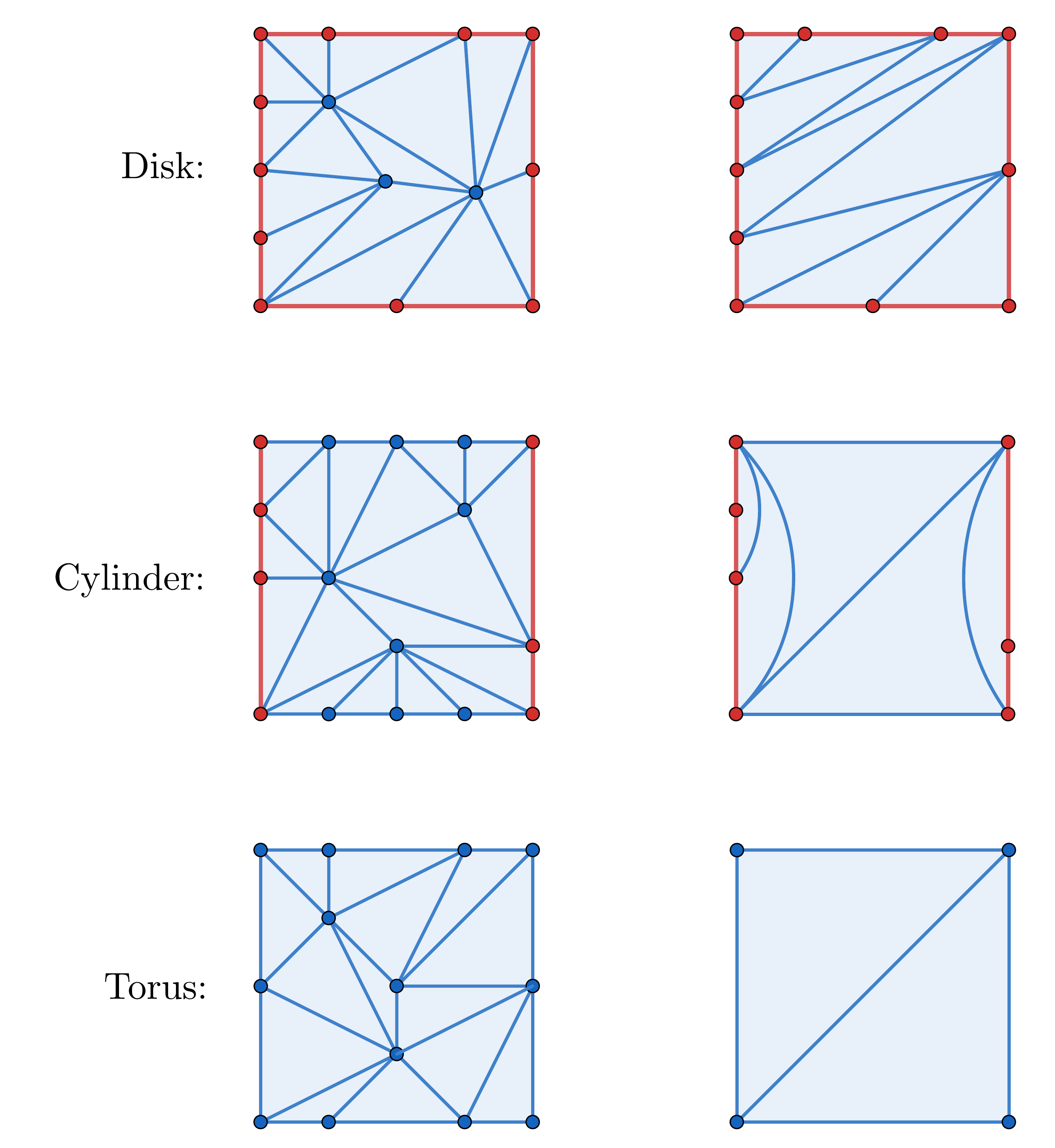}
		\caption{The extent to which we can reduce a diagram using the bistellar flips depends on the topology of the underlying surface. We can represent a disk by a rectangle with no edges identified (top line), an open cylinder by a rectangle with the top and bottom edges identified (middle line) and a torus by a rectangle with both sets of opposite edges identified (bottom line). We can remove internal vertices and vertices on the identified edges (the blue vertices) but not vertices on the boundary (the red vertices). In addition, even for the torus we cannot remove the last vertex (although any vertex can be the last one remaining).}
		\label{Figure_example_reduction}
	\end{figure}

	To see the additional conditions, note that we can represent a torus by a square with opposite ends identified, such as shown in Figure \ref{Figure_torus_shift_view}. However, the torus is the fundamental object and the square is just a representation of it, so we can always shift perspective, which appears to change the square (such as the shift shown in Figure \ref{Figure_torus_shift_view}). By applying the diagrammatic rules given in Section \ref{Section_membrane_operators}, we can simplify this surface diagram. In particular, we can remove vertices, with the ability to remove vertices from the 2d diagram corresponding to the commutation of the overall membrane operator with vertex transforms at that vertex. Note that, in the case of open membrane operators, the vertices on the boundary cannot be removed by diagrammatic moves and so the membrane operator does not commute with vertex transforms on the boundary. When we close an open membrane (e.g., closing the square into a torus) requiring the surface weight to satisfy the diagrammatic moves at the boundary (where the membrane closes up) allows us to remove the vertices at the boundary and ensures that the membrane commutes with transforms at the boundary. This also reduces the number of independent reference diagrams. For example, the reference diagrams for the cylinder can have any number of vertices on the two ends, whereas for the torus, we can simplify these diagrams further so that there is only one vertex (see Figure \ref{Figure_example_reduction} for examples). 
	
\begin{figure}
		\centering
		\includegraphics[width=\linewidth]{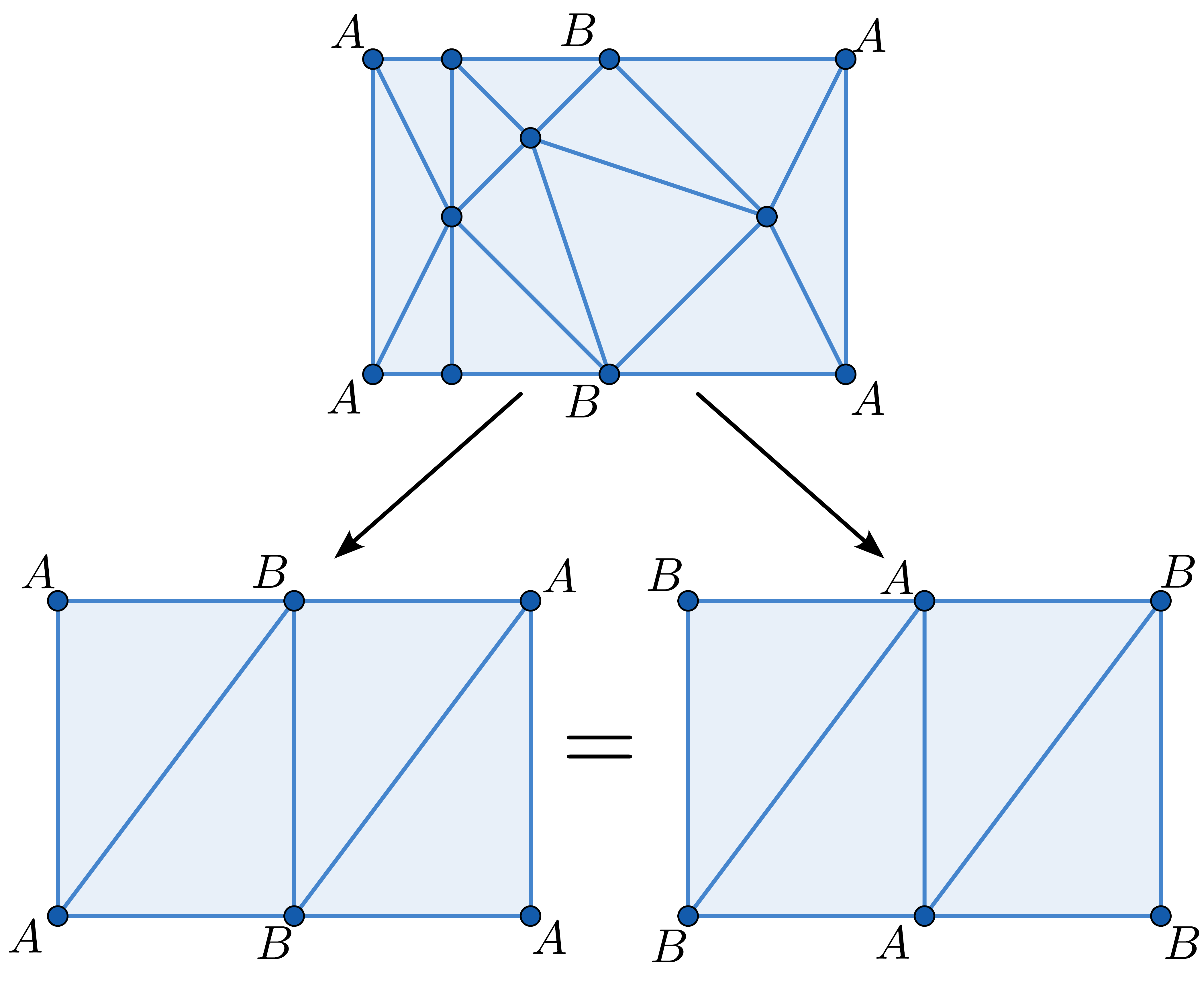}
		\caption{We can reduce a general diagram representing the torus, such as the one at the top of the image, to a diagram only involving two vertices $A$ and $B$. Because the diagram represents a torus, we can look at the torus from a different angle without affecting the label assigned to it assuming that there are no privileged vertices. This corresponds to shifting the diagram with periodic boundary conditions, so the two diagrams at the bottom of the figure represent the same torus. These diagrams should therefore be assigned the same value by the membrane operator, assuming that there are no privileged vertices on the torus (which correspond to vertices whose energy terms do not commute with the membrane operator).}
		\label{Figure_torus_partial_reduction}
	\end{figure}

	This suggests that for the torus there is only one type of reference diagram to consider and so precisely one flux membrane operator $C^h(m)$ for each flux label $h$ and cycle labels $g_1$ and $g_2$. However, there is still one vertex in the diagram which we cannot remove. How do we know that the membrane operator commutes with the vertex transform at that vertex? We can demonstrate this by showing that the surface weight does not depend on which vertex is left over. Consider two vertices $A$ and $B$. By reducing the surface diagram to a reference diagram only including $A$ we show that the membrane operator commutes with all other vertex transforms including the one at $B$. By reducing the surface diagram to one only including $B$ we show that it commutes with all other vertex transforms including the one at $A$. If the phase is the same in both cases, this means that the membrane transforms commute with all vertex transforms (including the ones at both $A$ and $B$), or equivalently that we can move the location of the left-over vertex freely without changing the action of the membrane operator. To this end, we first consider reducing the diagram to one only including $A$ and $B$, as shown in Figure \ref{Figure_torus_partial_reduction}. This reduction will accumulate some phase, but we want to compare the relative phase from reducing to only $A$ or only $B$, so the phase accumulated up to this point does not matter.

\begin{figure*}
		\centering
		\includegraphics[width=0.8\linewidth]{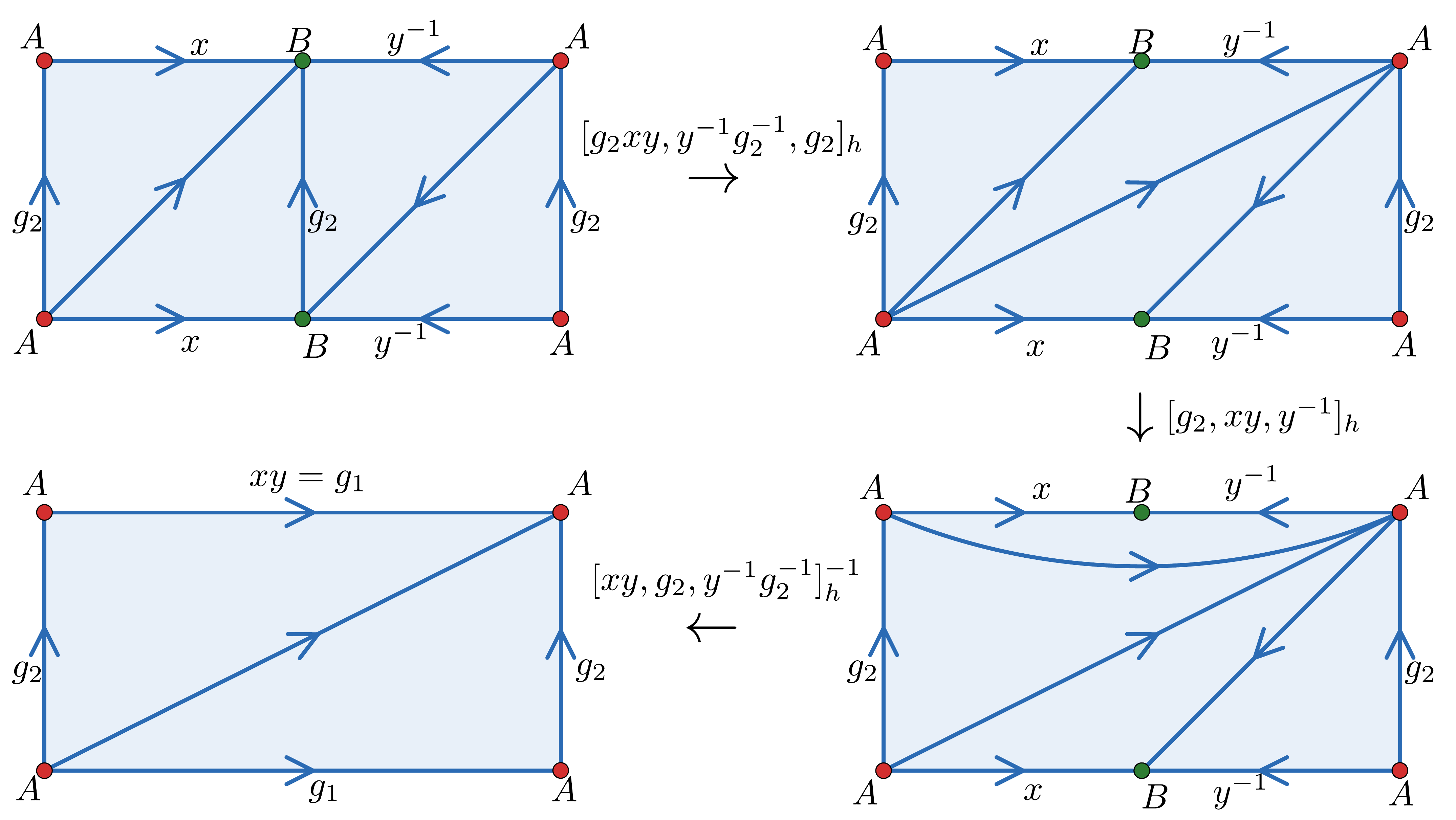}
		\caption{We can reduce the diagram involving only $A$ and $B$ from Figure \ref{Figure_torus_partial_reduction} to one involving only $A$ through the steps shown here. The cocycles next to each arrow represent the phase gained during each step.}
		\label{Figure_torus_reduction_A}
	\end{figure*}

\begin{figure*}
		\centering
		\includegraphics[width=0.8\linewidth]{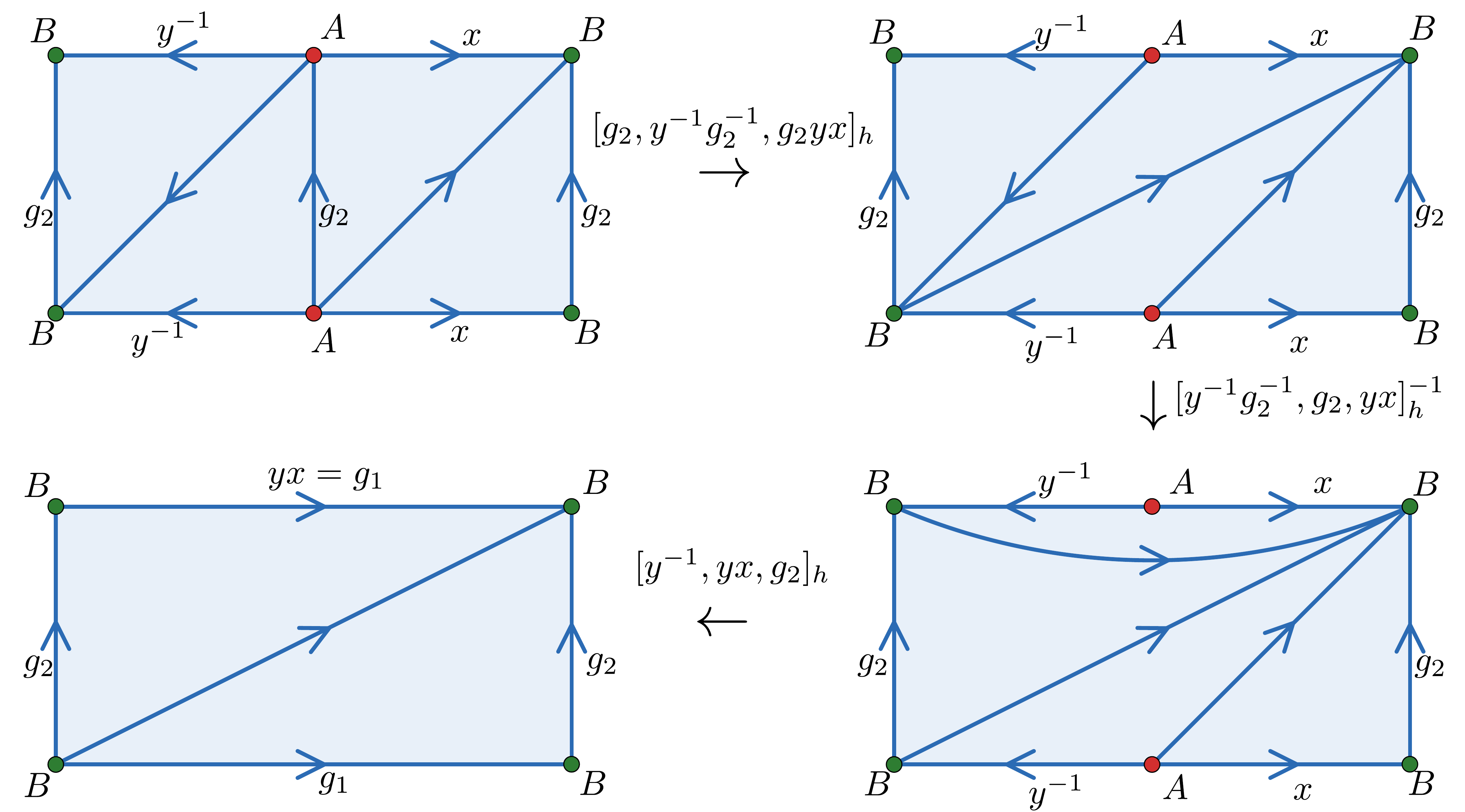}
		\caption{We can also reduce the diagram involving only $A$ and $B$ from Figure \ref{Figure_torus_partial_reduction} to one involving only $B$. The resulting reference diagram looks the same as the one from Figure \ref{Figure_torus_reduction_A}, so the phase gained during the reduction should be the same if the final vertex does not matter.}
		\label{Figure_torus_reduction_B}
	\end{figure*}

	From the diagram with only $A$ and $B$, if we reduce to only $A$ we accumulate the phase and reference diagram as shown in Figure \ref{Figure_torus_reduction_A}. If we reduce to only $B$ we accumulate the phase shown in Figure \ref{Figure_torus_reduction_B} (and obtain the same final reference diagram). The phase from the reduction to $A$ can be written as
	\begin{align*}
		[xyg_2,y^{-1}g_2^{-1},g_2]_h& [g_2, xy, y^{-1}]_h [xy, g_2, y^{-1}g_2^{-1}]_h^{-1} \\
  &= [xy, y^{-1}, g_2]_h [g_2, y^{-1}g_2^{-1}, g_2]_h  \\ & \hspace{1cm} [xy,g_2,y^{-1}]_h^{-1} [g_2, xy, y^{-1}]_h \\
		&= [xy,y^{-1}]_{h, g_2} [g_2, y^{-1}g_2^{-1}, g_2]_h,
	\end{align*}
	where the first equality comes from an application of the 3-cocycle condition Equation \ref{Equation_3_cocycle_condition} and the second comes from the definition of the 2-cocycle given in Equation \ref{Equation_2_cocycle_definition}. Similarly, the phase from the reduction to $B$ can be written as
	\begin{align*}
		[g_2, y^{-1}g_2^{-1}, g_2 yx]_h &[y^{-1}g_2^{-1}, g_2, yx]_h^{-1} [y^{-1}, yx, g_2]_h \\
  &= [g_2, y^{-1}, yx]_h [g_2, y^{-1}g_2^{-1},g_2]_h \\
  & \hspace{1cm} [y^{-1}, g_2, yx]_h^{-1} [y^{-1},yx,g_2]_h\\
		&= [y^{-1},yx]_{h, g_2} [g_2, y^{-1}g_2^{-1},g_2]_h.
	\end{align*}
	The ratio of these phases is then
	$$\theta_{AB}= \frac{ [xy,y^{-1}]_{h, g_2}}{[y^{-1},xy]_{h, g_2}}.$$
	Using the fact that the total group element for the horizontal cycle is given by $xy=g_1$, this becomes
	$$\theta_{AB}= \frac{ [g_1,y^{-1}]_{h, g_2}}{[y^{-1},g_1]_{h, g_2}}.$$
	which is not generally unity. Requiring it to be unity for all $y$ gives us an additional condition on the labels $g_1$, $g_2$ and $h$:
	\begin{equation}
		[g_1,y^{-1}]_{h, g_2} = [y^{-1},g_1]_{h, g_2} \forall y \in G.
	\end{equation}

	This same condition appears in the ground state degeneracy calculation in Ref. \cite{Wan2015}, where an element $h\in G$ is said to be $\beta_{g_1,g_2}$-regular if $[h,x]_{g_1,g_2}=[x,h]_{g_1,g_2}$ for all $x \in G$ (for Abelian $G$). This seems like an asymmetric condition on the three elements, but we can show that if $h$ is $\beta_{g_1, g_2}$-regular then $g_2$ is $\beta_{g_1, h}$-regular. This is because
	\begin{align*}
		\frac{[h,x]_{g_1, g_2}}{[x,h]_{g_1,g_2}}&= \frac{ [g_2,h,x]_{g_1} [h, x, g_2]_{g_1} [x, g_2, h]_{g_1}}{[h,g_2, x]_{g_1} [g_2, x, h]_{g_1} [x, h, g_2]_{g_1}} \\
		&= \frac{[x,g_2]_{g_1, h} }{[g_2, x]_{g_1, h}},
	\end{align*}
	so 
 \begin{align*}
	[h,x]_{g_1, g_2} = [x&,h]_{g_1,g_2} \ \forall x \in G \\
 &\iff [x,g_2]_{g_1, h} =[g_2, x]_{g_1, h} \  \forall x \in G.
 \end{align*}
	That is, $h$ is $\beta_{g_1, g_2}$-regular if and only if $g_2$ is $\beta_{g_1, h}$-regular. Additionally, the twisted 2-cocycle satisfies the permutation condition $[h,x]_{g_1, g_2}= [h,x]_{g_2, g_1}^{-1}$ \cite{Wan2015}. Using this, if an element is $\beta_{g_1, g_2}$-regular it is also $\beta_{g_2, g_1}$-regular. Following the same reasoning as before, we can then show that $h$ is $\beta_{g_1, g_2}$-regular if and only if $g_1$ is $\beta_{g_2, h}$-regular.

	In our case, it is convenient to put the condition on $h$ and restrict $h$ to be $\beta_{g_1, g_2}$-regular while allowing $g_1$ and $g_2$ to be general elements of $G$ (although we could restrict $g_1$ or $g_2$ instead and obtain the same triples $(g_1, g_2, h)$ when we take all possible combinations). With this in mind, the resulting measurement operators take the form
	$$\sum_{g_1,g_2 \in G} \sum_{h \in G_{g_1, g_2}} a_{g_1,g_2, h} C^h(m)\delta(\hat{g}(c_1), g_1) \delta(\hat{g}(c_2), g_2).$$
	
	The surface weight for $C^h(m)$ can be calculated from the relevant reference diagram, which is defined by the labels $g_1$ and $g_2$, using the diagrammatic rules. This means that we could have a different surface weight for each $g_1$ and $g_2$ (and each $h$). However, the coefficients $a_{g_1,g_2, h}$ already allow us to have a different weight for each value of these labels. Therefore, we can define the phase associated to the reference diagram in the bottom-left of Figure \ref{Figure_torus_reduction_A} to be 1 for any value of $g_1$, $g_2$ and $h$, without missing any allowed operators. A simple basis for this space of measurement operators is given by the operators
	\begin{equation}
		T^{g_1,g_2,h}(m)=C^h(m)\delta(\hat{g}(c_1), g_1) \delta(\hat{g}(c_2), g_2),
	\end{equation}
	where there is one such operator for each $g_1, g_2 \in G$ and $h \in G_{g_1, g_2}$.

	Now, just as we did for the spherical measurement operators, we want to take a basis that consists of orthogonal projection operators. Each projector is labeled by group elements $g_1$ and $g_2$ of $G$ as well as a $\beta_{g_1, g_2}$-projective irrep $\alpha^{g_1,g_2}$ of $G$. These irreps satisfy
	$$\alpha^{g_1,g_2}(x) \alpha^{g_1,g_2}(y)= [x,y]_{g_1, g_2}\alpha^{g_1,g_2}(xy).$$
	The projectors are given by
	\begin{align}
		&P^{g_1, g_2, \alpha^{g_1,g_2}}(m) \notag  \\
  & \hspace{0.5cm} = \frac{1}{\sqrt{|G||G_{g_1, g_2}|}} \sum_{h \in G_{g_1, g_2}} \chi_{\alpha^{g_1,g_2}}(h) T^{g_1, g_2, h}(m) \notag \\
		& \hspace{0.5cm} = \frac{1}{\sqrt{|G||G_{g_1, g_2}|}} \sum_{h \in G_{g_1, g_2}} \chi_{\alpha^{g_1,g_2}}(h)C^h(m) \notag \\
  & \hspace{3cm}\delta(\hat{g}(c_1), g_1) \delta(\hat{g}(c_2),g_2), \label{Equation_charge_projectors}
	\end{align}
	where $\chi_{\alpha^{g_1,g_2}}$ is the character for irrep $\alpha^{g_1,g_2}$ (i.e., the trace of the representative matrices). We prove that these are indeed orthogonal projectors and can be obtained from the previous basis of operators by an invertible transformation, in the Supplemental material in Section \ref{Section_supplement_topological_charge}.

	We see that the projector, and therefore the associated topological charge, is labeled by the fluxes $g_1$ and $g_2$ around the two handles of the torus in addition to a projective representation that depends on these two fluxes. The simplest type of excitation that would have both fluxes non-trivial would be a link of two magnetic flux tubes. Note that when one of these fluxes is trivial, corresponding to unlinked excitations, the projective irrep $\alpha^{g_1,g_2}$ becomes a linear irrep (because $[x,y]_{g_1, 1_G}=[x,y]_{1_G,g_2}=1$). In that case, we find that the topological charge is labeled by the flux and a linear irrep charge, which matches our expectation for the untwisted case. We therefore see that the cocycle only becomes relevant when considering the charge of linked objects, so studying unlinked objects is insufficient to determine the topological order of a system.
	
	We also note that the labels for the projectors in Equation \ref{Equation_charge_projectors}, namely two group elements $g_1$ and $g_2$ and a $\beta_{g_1, g_2}$-projective irrep $\alpha^{g_1,g_2}$, match the labels of the ground state basis for the 3-torus given in Ref. \cite{Wan2015} in the case of Abelian groups. This indicates that the number of topological charges measured by a 2-torus surface matches the ground state degeneracy on the 3-torus, similar to how the number of charges measured by a sphere (corresponding to point-like charge) matches the ground state degeneracy on $S^2 \times S^1$, as we discussed earlier. It is notable that the toroidal charge measurement operators in this case measure the flux of both handles of the torus, indicating that the correspondence between particle types and the ground state degeneracy of the 3-torus requires counting linked excitations in addition to unlinked loops.

	\section{Conclusion}

	We have constructed the ribbon and membrane operators which produce the topological excitations in the twisted lattice gauge theory model in 3+1d for Abelian groups. In particular we examined the membrane operators which produce flux tubes linked to an existing base loop, finding that these are well described by projective irreps in addition to the flux labels. We used these operators to study the three-loop braiding relations previously studied using indirect methods and found that the same projective irreps which label the membrane operators also describe these braiding relations, matching previous results. In addition, we found the fusion rules which describe how two excitations combine into a third, in terms of the projective irreps. Finally, we constructed the projectors to definite topological charges for toroidal and spherical surfaces, finding that the number of charges measured by a torus match the ground state degeneracy of the model on a 3-torus. Because the torus measures the flux about two handles, this implies that the ground state degeneracy does not match the number of simple loop-like excitations, but instead the number of loop-like excitations that may be linked to different kinds of existing base loops. The allowed loop-like excitations are different depending on the base loop, so this enhances the number of types of excitation compared to what we may naively expect.
	
	There are several interesting directions along which to expand this work. The most obvious avenue is to consider non-Abelian groups. While we can construct the membrane operators for the flux tubes in such cases, it is more difficult to find a good basis for these membrane operators that respects their topological charge. In addition, when calculating the braiding relations of two flux tubes while linked to a third loop, we must account for both the conjugation of the flux labels of the two loops as well as the transformation of the charges attached to them. For non-Abelian groups, we know from the untwisted case that the membranes will be different before and after the two membranes intersect. This means that we need to be able to define the membrane operators near their boundaries, which we were able to avoid in the Abelian case.
	
	Another possibility would be to consider models with emergent fermions. While Dijkgraaf-Witten is conjectured to include all topological phases in 3+1d without emergent fermions, the model must be generalized somewhat to allow for such fermions \cite{Lan2019}. There has already been significant study of such models via field theory \cite{Wang2019, Zhang2023a}, by adding terms that transmute the self-statistics of the emergent excitations. It is not immediately clear how such excitations could be constructed using ribbon and membrane operators in a 3+1d lattice model. 
	
	Another feature realized in some field theory models \cite{Chan2018, Zhang2021, Zhang2023, Zhang2023a, Lan2019}, but not the Abelian twisted lattice gauge theory model, is non-trivial Borromean rings braiding. This braiding motion describes a point-like particle braiding with two loop-like excitations, such that the path of motion of the point-like particle (with this path being a closed loop) and the two loops together form Borromean rings. In this formation, the linking number between any two of the three rings is zero and yet the rings cannot be disentangled. This motion measures the non-Abelian nature of point-loop braiding and is trivial in the twisted lattice gauge theory model when the group $G$ ia Abelian. However, it can be realized for the field theory models mentioned previously, even with Abelian $G$. In this case non-trivial Borromean rings braiding has limited compatibility with three-loop and four-loop braiding \cite{Zhang2021}. It would be interesting to construct a lattice model which has this feature, to see how the excitations differ in such a case. In a lattice realization with non-trivial Borromean ring braiding, it seems likely that linking two loop-like excitations may cause them to be connected by a linking string, restricting their motion and causing the limited compatibility with three-loop and four-loop braiding.
		
	Finally, it would be interesting to further study topological charge and the topological charge projectors in this setting or more generally. The toroidal charge measurement operators depend on the sets of doubly-twisted 2-cocycles derived from the underlying 4-cocycle of the model. However, these sets of 2-cocycles seem to be unique for each equivalence class of 4-cocycles \cite{Wang2015a}. Therefore, the twisted lattice gauge theory models are in one-to-one correspondence with the torus charges. It is not known to the authors whether there is a reason that this should be the case and that no more general charges (for example, from higher genus surfaces) are needed to distinguish the models, although it makes intuitive sense that more general objects can be built from the simpler links and point-like excitations.
	
\section*{Acknowledgements}

We thank S. Pace for helpful discussions about gauge theory. This work was supported by the Natural Science and Engineering Council of Canada (NSERC) Discovery Grant No. RGPIN-2023-03296 and the Center for Quantum Materials at the University of Toronto (J.H. and Y.B.K.). Our collaboration is a part of the effort in the Advanced Study Group on ``Entanglement and Dynamics in Quantum Matter" in the Center for Theoretical Physics of Complex Systems at the Institute for Basic Science (D.X.N. and Y.B.K.). D.X.N. is supported by Grant No. IBS-R024-D1.

\bibliography{referencesTwistedLatticeGaugeTheory}{}

\pagebreak

\onecolumngrid

\begin{center}
	\textbf{Supplemental Material for ``Twisted Lattice Gauge Theory: Membrane Operators, Three-loop Braiding and Topological Charge"}
\end{center}

\setcounter{equation}{0}
\setcounter{figure}{0}
\setcounter{table}{0}
\makeatletter
\renewcommand{\theequation}{S\arabic{equation}}
\renewcommand{\thefigure}{S\arabic{figure}}
\setcounter{section}{0}
\renewcommand{\thesection}{S-\Roman{section}}

\section{Spherical Membrane Operators}
\label{Section_spherical_membranes}

\subsection{Motivation from Vertex Transforms}
\label{Section_spherical_membrane_approach}
In Section \ref{Section_spherical_membranes} and \ref{Section_more_general_membranes} we will give further information about the motivation and construction of the membrane operators, including proof that they satisfy the requirement of being topological. In Section \ref{Section_spherical_membranes} we will provide motivation for the form of the membrane operators, by considering the case of a spherical membrane.

In commuting projector models, the closed membrane operators are often generated by terms in the Hamiltonian. This ensures the topological nature of the membrane operators, because they can be deformed over a small region by applying such a term, which must act trivially on the ground state due to the structure of the Hamiltonian (see, e.g., Ref. \cite{HuxfordPaper3}). Here we show that the same approach works for the twisted lattice gauge theory model. Specifically, we will show that the membrane operators applied on a (topologically) spherical surface can be generated by applying vertex transforms in the region enclosed by the membrane (at least in the subspace of the Hilbert space where there are no plaquette excitations enclosed by the surface). We can then generalize this approach to open membranes and membranes of more general topology (such as tori), by constructing membrane operators that look locally like the spherical membrane operators and so look locally like a series of vertex transforms. These are the membrane operators that we described in Section \ref{Section_membrane_operators} of the main text.
\begin{figure}[H]
	\centering
	\includegraphics[width=0.4\linewidth]{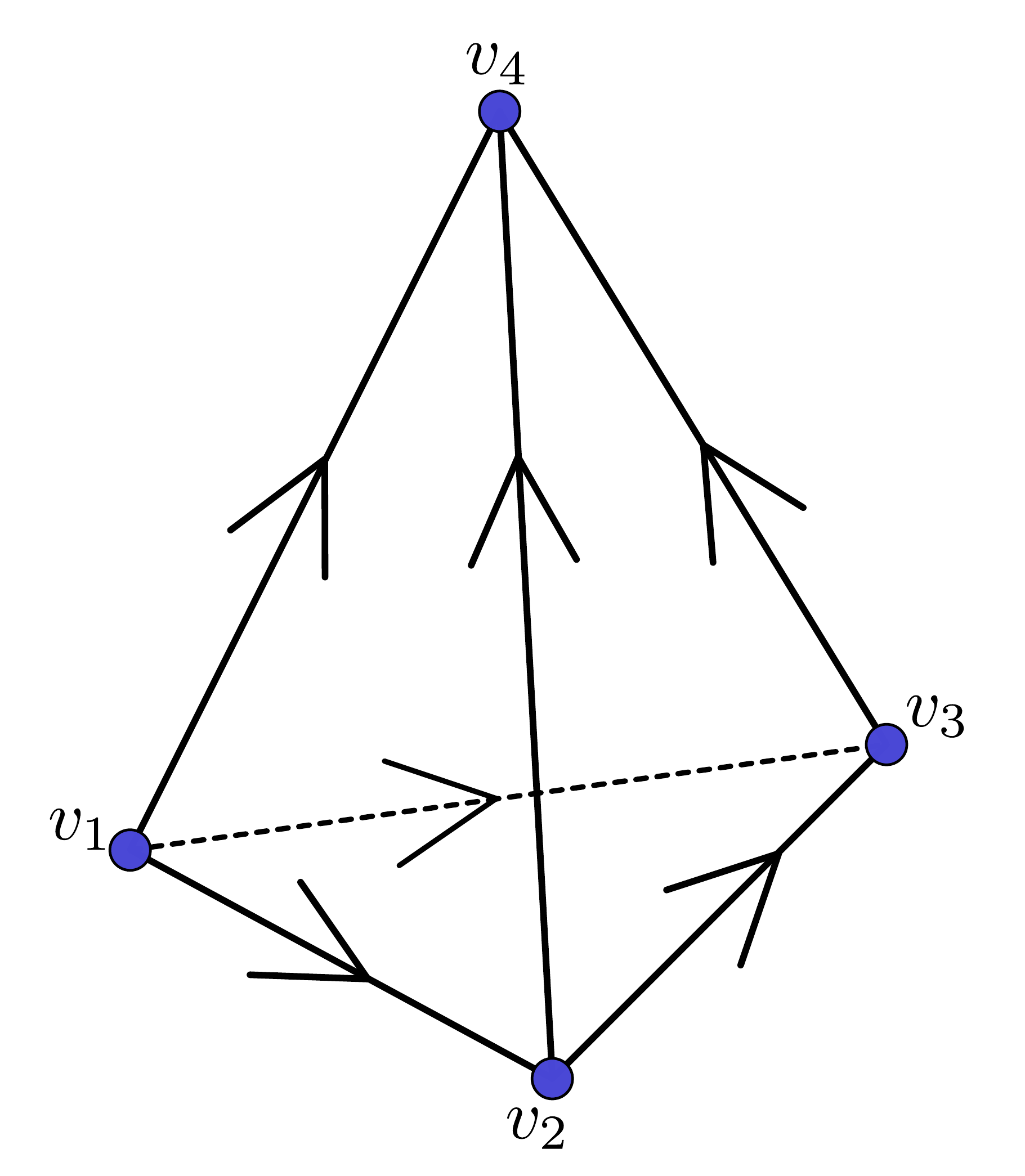}
	\caption{A simple example of a tetrahedron. Applying an appropriate vertex transform on each vertex leaves the edge labels invariant but gains an overall phase.}
	\label{Figure_tetrahedron_example}
\end{figure}

Before we construct the spherical membranes, we can consider a simple example of a single tetrahedron. Consider applying a vertex transform on every vertex of the tetrahedron shown in Figure \ref{Figure_tetrahedron_example}. We denote the original edge label for the edge between $v_i$ and $v_j$ ($j >i$) by $g_{ij}$. Because of flatness, which is preserved by the vertex transforms, we only need to keep track of the edge labels $g_{12}$, $g_{23}$ and $g_{34}$ (the other labels are fixed by the flatness condition). We write the initial state of the tetrahedron as $\ket{g_{12}, g_{23}, g_{34}}$. Then, we apply the vertex transforms
$$A_4^{g_{34}^{-1}g_{23}^{-1}g_{12}^{-1}hg_{12}g_{23}g_{34}}A_3^{g_{23}^{-1}g_{12}^{-1}hg_{12}g_{23}}A_2^{g_{12}^{-1}hg_{12}}A_1^h, $$

where we allow $G$ to be non-Abelian to demonstrate that a similar approach works for non-Abelian groups as well. We then obtain
\begin{align}
	&A_4^{g_{34}^{-1}g_{23}^{-1}g_{12}^{-1}hg_{12}g_{23}g_{34}}A_3^{g_{23}^{-1}g_{12}^{-1}hg_{12}g_{23}}A_2^{g_{12}^{-1}hg_{12}}A_1^h \ket{g_{12}, g_{23}, g_{34}} \notag \\
	&= [h, g_{12}, g_{23}, g_{34}] \ A_4^{g_{34}^{-1}g_{23}^{-1}g_{12}^{-1}hg_{12}g_{23}g_{34}}A_3^{g_{23}^{-1}g_{12}^{-1}hg_{12}g_{23}}A_2^{g_{12}^{-1}hg_{12}} \ket{hg_{12},g_{23},g_{34}} \notag \\
	&= [h, g_{12}, g_{23}, g_{34}] [hg_{12} (g_{12}^{-1}hg_{12})^{-1}, g_{12}^{-1}hg_{12}, g_{23}, g_{34}]^{-1} \notag \\
	& \hspace{0.7cm} A_4^{g_{34}^{-1}g_{23}^{-1}g_{12}^{-1}hg_{12}g_{23}g_{34}}A_3^{g_{23}^{-1}g_{12}^{-1}hg_{12}g_{23}} \ket{hg_{12}(g_{12}^{-1}hg_{12})^{-1},g_{12}^{-1}hg_{12}g_{23},g_{34}} \notag \\
	&=[h, g_{12}, g_{23}, g_{34}] [g_{12}, g_{12}^{-1}hg_{12}, g_{23}, g_{34}]^{-1} [g_{12},g_{12}^{-1}hg_{12}g_{23} (g_{23}^{-1}g_{12}^{-1}hg_{12}g_{23})^{-1}, g_{23}^{-1}g_{12}^{-1}hg_{12}g_{23}, g_{34}] \notag \\
	& \hspace{0.7cm} A_4^{g_{34}^{-1}g_{23}^{-1}g_{12}^{-1}hg_{12}g_{23}g_{34}} \ket{g_{12},g_{12}^{-1}hg_{12}g_{23}(g_{23}^{-1}g_{12}^{-1}hg_{12}g_{23})^{-1},(g_{23}^{-1}g_{12}^{-1}hg_{12}g_{23})g_{34}} \notag \\
	&=[h, g_{12}, g_{23}, g_{34}] [g_{12}, g_{12}^{-1}hg_{12}, g_{23}, g_{34}]^{-1} [g_{12},g_{23} , g_{23}^{-1}g_{12}^{-1}hg_{12}g_{23}, g_{34}] \notag \\ & \hspace{0.7cm} [g_{12}, g_{23}, (g_{23}^{-1}g_{12}^{-1}hg_{12}g_{23})g_{34}(g_{34}^{-1}g_{23}^{-1}g_{12}^{-1}hg_{12}g_{23}g_{34})^{-1}, g_{34}^{-1}g_{23}^{-1}g_{12}^{-1}hg_{12}g_{23}g_{34} ]^{-1} \notag \\ & \hspace{1cm} \ket{g_{12},g_{23},(g_{23}^{-1}g_{12}^{-1}hg_{12}g_{23})g_{34}(g_{34}^{-1}g_{23}^{-1}g_{12}^{-1}hg_{12}g_{23}g_{34})^{-1}} \notag \\
	&=[h, g_{12}, g_{23}, g_{34}] [g_{12}, g_{12}^{-1}hg_{12}, g_{23}, g_{34}]^{-1} [g_{12},g_{23} , g_{23}^{-1}g_{12}^{-1}hg_{12}g_{23}, g_{34}] [g_{12}, g_{23}, g_{34}, g_{34}^{-1}g_{23}^{-1}g_{12}^{-1}hg_{12}g_{23}g_{34} ]^{-1} \notag \\
	& \hspace{0.7cm} \ket{g_{12},g_{23},g_{34}}.
\end{align}

We see that all of the edge labels on the tetrahedron are preserved (as we would expect from similar results on untwisted gauge theory). On the other hand, other edges attached to the four vertices will generally not be left invariant: those edges with only one vertex on the tetrahedron will have their labels changed. In addition, we obtain a phase
$$ [h, g_{12}, g_{23}, g_{34}] [g_{12}, g_{12}^{-1}hg_{12}, g_{23}, g_{34}]^{-1} [g_{12},g_{23} , g_{23}^{-1}g_{12}^{-1}hg_{12}g_{23}, g_{34}] [g_{12}, g_{23}, g_{34}, g_{34}^{-1}g_{23}^{-1}g_{12}^{-1}hg_{12}g_{23}g_{34} ]^{-1}.$$

Comparing to Ref. \cite{Wan2015}, we see that this expression is just the inverse of the slant product
\begin{equation}
	[g_{12}, g_{23}, g_{34}]_h:= \frac{[g_{12}, g_{12}^{-1}hg_{12}, g_{23},g_{34}] [g_{12}, g_{23},g_{34},(g_{12}g_{23}g_{34})^{-1}h(g_{12}g_{23}g_{34}) ]}{[h, g_{12}, g_{23}, g_{34}] [g_{12}, g_{23}, (g_{12}g_{23})^{-1}hg_{12}g_{23}, g_{34}]}. \label{Equation_twisted_cocycle_definition}
\end{equation}
This slant product obeys the twisted 3-cocycle condition:
\begin{equation}
	\frac{[x,y,z]_{w^{-1}uw} [w, xy, z]_{u} [w, x,y]_u}{[wx,y,z]_u [w,x,yz]_u}=1, \label{Equation_twisted_cocycle_condition}
\end{equation}
which becomes the regular 3-cocycle condition for Abelian $G$ (because $w^{-1}uw=u$ in that case), along with the normalization condition
\begin{equation}
	[1_G,y,z]_u = [x,1_G,z]_u = [x,y,1_G]_u =[x,y,z]_{1_G}=1. \label{Equation_twisted_normalization}
\end{equation}

We claim that if we apply a similar series of vertex transforms in a solid region, we will obtain a similar phase for each tetrahedron wholly contained within that region, in addition to an action on the boundary of the region. Note that unlike the untwisted case, where the action on the bulk is completely trivial, there is a phase associated to the bulk (i.e., the tetrahedra in the solid region). We will show that this apparent bulk phase can be written in terms of boundary variables, becoming the surface weight that we discussed in Section \ref{Section_membrane_operators} of the main text.

\subsection{Evaluation of the Vertex Transform Product}
\label{Section_membranes_from_vertex_transforms}
We will first prove that the product of vertex transforms in a region results in a phase associated to the slant product for every tetrahedron in the bulk, and we also describe the action on the boundary. At this point, we will restrict to cases where $G$ is Abelian, to match our discussion in the main text (although the construction of membrane operators can be extended to the non-Abelian case). In this case, the conjugation in previous expressions is trivial and we just apply a vertex transform with label $h$ on each vertex in the region. That is, we apply the operator
\begin{equation}
	\hat{O}^h(S)= \prod_{v \in S} A_v^h
\end{equation}
onto the (3d) spatial region $S$ of a state satisfying flatness. We claim that this product of vertex transforms has the same action as a membrane operator on the boundary of $S$.

To proceed, we split each vertex transform into two parts, the multiplication action on the edge labels, which we call $A_v^h(0)$ (because it is equivalent to an untwisted gauge transform), and the phase from the attached tetrahedra:
$$A_v^h= A_v^h(0) \prod_{\text{tetrahedra }t \ni v} \theta^h_{v,t},$$
where
\begin{equation}
	\theta^h_{v,t}= \begin{cases} [h, \hat{g}_{v_1(t) v_2(t)}, \hat{g}_{v_2(t)v_3(t)}, \hat{g}_{v_3(t)v_4(t)}]^{\epsilon(t)} & \text{if $v=v_1(t)$} \\
		[ \hat{g}_{v_1(t) v_2(t)}h^{-1}, h , \hat{g}_{v_2(t)v_3(t)}, \hat{g}_{v_3(t)v_4(t)}]^{-\epsilon(t)} & \text{if $v=v_2(t)$} \\
		[ \hat{g}_{v_1(t) v_2(t)}, \hat{g}_{v_2(t)v_3(t)}h^{-1}, h, \hat{g}_{v_3(t)v_4(t)}]^{\epsilon(t)} & \text{if $v=v_3(t)$} \\
		[ \hat{g}_{v_1(t) v_2(t)}, \hat{g}_{v_2(t)v_3(t)}, \hat{g}_{v_3(t)v_4(t)}h^{-1}, h]^{-\epsilon(t)} & \text{if $v=v_4(t)$.} \label{Equation_vertex_tetrahedron_factor}
	\end{cases}
\end{equation}
Then
$$\prod_{v \in S} A_v^h= \big( \prod_{v \in S} A_v^h(0)\prod_{\text{tetrahedra }t \ni v} \theta^h_{v,t}\big).$$

We aim to move all of the phase factors $\theta^h_{v,t}$ to the right of all of the $A_v^h(0)$, so that they act directly on the state, and then group all phases corresponding to a given tetrahedron (rather than grouping them by vertex). The phase factors commute with each-other because they are diagonal in the configuration basis. However, these phase factors do not generally commute with the untwisted vertex transforms. Specifically, the phase factor for a particular tetrahedron does not commute with any vertex transform on the tetrahedron. This is because the tetrahedron phase factor in Equation \ref{Equation_vertex_tetrahedron_factor} depends on the edges on that tetrahedron. Each edge label is only changed by the vertex transforms at either end of the edge, and if the edge is on a particular tetrahedron then so are the vertices at the end of that edge. Therefore, a tetrahedron phase factor $\theta^h_{v,t}$ fails to commute with vertex transforms $A_{v'}^h(0)$ for vertices $v'$ on the same tetrahedron $t$, but commutes with transforms for vertices outside the tetrahedron. It would be quite demanding to consider every possible string of vertex transforms to find the general transformation of the phase factor as we commute it past the untwisted vertex transforms. Thankfully, we can simplify this process. Because the overall vertex transforms $A_v^h$ all commute with each-other \cite{Wan2015}, we are free to order the product over vertices in any way we desire. In particular, we choose to apply them in order of ascending index. This means that the index decreases as we go from the left of the expression to the right.

At this point, we divide the tetrahedra into two types. Firstly, we consider tetrahedra wholly within the region $S$, by which we mean that every vertex on the tetrahedron is in $S$. This means that all four vertex transforms appear in the product, with the transform $v_1(t)$ on the right, then $v_2(t)$ and so on (possibly with transforms from vertices not on the tetrahedron in between). Because the phase factors only fail to commute with vertex transforms from other vertices on the tetrahedron, we just need to consider how to commute them past those transforms. For $\theta^h_{v_1(t),t}$, its transform is to the right of the other vertex transforms because of our ordering, so that we can move it to the right with no effect. For $\theta^h_{v_2(t),t}$, we need to move it past $A_{v_1(t)}^h$. More generally, we need to consider an expression of the form $\theta^h_{v_n(t), t} \prod_{j <n} A_{v_j(t)}^h$. The commutation relation between the phase factor and the transforms derives from the action of the vertex transform on the edge labels:
\begin{align*}
	\hat{g}_{v_j v_{j+1}}A_{v_j}^h &= A_{v_j}^h h\hat{g}_{v_j v_{j+1}}\\
	\hat{g}_{v_j v_{j+1}}A_{v_{j+1}}^h &= A_{v_{j+1}}^h \hat{g}_{v_j v_{j+1}}h^{-1}.
\end{align*}
Defining $\theta^h_{v,t}(\hat{g}_{v_1(t)v_2(t)},\hat{g}_{v_2(t)v_3(t)},\hat{g}_{v_3(t)v_4(t)}) := \theta^h_{v,t}$ to make the dependence of $\theta^h_{v,t}$ on the edge labels apparent, the phase factors transform as
\begin{equation}
	\theta^h_{v,t}(\hat{g}_{v_1(t)v_2(t)},\hat{g}_{v_2(t)v_3(t)},\hat{g}_{v_3(t)v_4(t)})A_{v'}^h=A_{v'}^h \cdot \begin{cases} \theta^h_{v,t}(h\hat{g}_{v_1(t)v_2(t)},\hat{g}_{v_2(t)v_3(t)},\hat{g}_{v_3(t)v_4(t)}) & \text{if $v'=v_1(t)$} \\
		\theta^h_{v,t}(\hat{g}_{v_1(t)v_2(t)}h^{-1},h\hat{g}_{v_2(t)v_3(t)},\hat{g}_{v_3(t)v_4(t)}) & \text{if $v'=v_2(t)$}\\
		\theta^h_{v,t}(\hat{g}_{v_1(t)v_2(t)},\hat{g}_{v_2(t)v_3(t)}h^{-1},h\hat{g}_{v_3(t)v_4(t)}) & \text{if $v'=v_3(t)$}\\
		\theta^h_{v,t}(\hat{g}_{v_1(t)v_2(t)},\hat{g}_{v_2(t)v_3(t)},\hat{g}_{v_3(t)v_4(t)}h^{-1}) & \text{if $v'=v_4(t)$}\\
		\theta^h_{v,t}(\hat{g}_{v_1(t)v_2(t)},\hat{g}_{v_2(t)v_3(t)},\hat{g}_{v_3(t)v_4(t)}) & \text{otherwise.}
	\end{cases}
\end{equation}
Then, we can use this to evaluate $\theta^h_{v_n(t),t} \prod_{j <n} A_{v_j(t)}^h(0)$ (recall that there may be other vertex transforms and phase factors between the operators, but they will commute with $\theta^h_{v_n(t),t}$).

If $n=1$ there is no need for commutation and we obtain the phase
\begin{equation}
	\theta^h_{v_1(t),t}(\hat{g}_{v_1(t)v_2(t)},\hat{g}_{v_2(t)v_3(t)},\hat{g}_{v_3(t)v_4(t)})= [h, \hat{g}_{v_1(t) v_2(t)}, \hat{g}_{v_2(t)v_3(t)}, \hat{g}_{v_3(t)v_4(t)}]^{\epsilon(t)} .
\end{equation}

If $n=2$ we get
\begin{align}
	\theta^h_{v_2(t),t}(\hat{g}_{v_1(t)v_2(t)},\hat{g}_{v_2(t)v_3(t)},\hat{g}_{v_3(t)v_4(t)})A_{v_1(t)}^h(0) &= A_{v_1(t)}^h(0)
	\theta^h_{v_2(t),t}(h\hat{g}_{v_1(t)v_2(t)},\hat{g}_{v_2(t)v_3(t)},\hat{g}_{v_3(t)v_4(t)}) \notag\\
	&= A_{v_1(t)}^h(0) [ h\hat{g}_{v_1(t) v_2(t)}h^{-1}, h , \hat{g}_{v_2(t)v_3(t)}, \hat{g}_{v_3(t)v_4(t)}]^{-\epsilon(t)} \notag\\
	&= A_{v_1(t)}^h(0) [ \hat{g}_{v_1(t) v_2(t)}, h , \hat{g}_{v_2(t)v_3(t)}, \hat{g}_{v_3(t)v_4(t)}]^{-\epsilon(t)}.
\end{align}

If $n=3$ we get
\begin{align}
	\theta^h_{v_3(t),t}(\hat{g}_{v_1(t)v_2(t)},\hat{g}_{v_2(t)v_3(t)},\hat{g}_{v_3(t)v_4(t)})&A_{v_2(t)}^h(0) A^h_{v_1(t)}(0) \notag \\
	&= A_{v_2(t)}^h(0)A_{v_1(t)}^h(0)
	\theta^h_{v_3(t),t}(h\hat{g}_{v_1(t)v_2(t)}h^{-1},h\hat{g}_{v_2(t)v_3(t)},\hat{g}_{v_3(t)v_4(t)}) \notag\\
	&=A_{v_2(t)}^h(0)A_{v_1(t)}^h(0)
	\theta^h_{v_3(t),t}(\hat{g}_{v_1(t)v_2(t)},h\hat{g}_{v_2(t)v_3(t)},\hat{g}_{v_3(t)v_4(t)}) \notag\\
	&=A_{v_2(t)}^h(0) A_{v_1(t)}^h(0)[ \hat{g}_{v_1(t) v_2(t)}, h\hat{g}_{v_2(t)v_3(t)}h^{-1}, h, \hat{g}_{v_3(t)v_4(t)}]^{\epsilon(t)} \notag\\
	&=A_{v_2(t)}^h(0) A_{v_1(t)}^h(0) [ \hat{g}_{v_1(t) v_2(t)}, \hat{g}_{v_2(t)v_3(t)}, h, \hat{g}_{v_3(t)v_4(t)}]^{\epsilon(t)}.
\end{align}

Finally, if $n=4$ we get
\begin{align}
	\theta^h_{v_4(t),t}(\hat{g}_{v_1(t)v_2(t)},\hat{g}_{v_2(t)v_3(t)},&\hat{g}_{v_3(t)v_4(t)})A_{v_3(t)}^h(0)A_{v_2(t)}^h(0)A_{v_1(t)}^h(0) \notag \\
	&= A_{v_3(t)}^h(0)A_{v_2(t)}^h(0)A_{v_1(t)}^h(0)
	\theta^h_{v_4(t),t}(h\hat{g}_{v_1(t)v_2(t)}h^{-1},h\hat{g}_{v_2(t)v_3(t)}h^{-1},h\hat{g}_{v_3(t)v_4(t)}) \notag\\
	&=A_{v_3(t)}^h(0)A_{v_2(t)}^h(0)A_{v_1(t)}^h(0) [ h\hat{g}_{v_1(t) v_2(t)}h^{-1}, h\hat{g}_{v_2(t)v_3(t)}h^{-1}, h\hat{g}_{v_3(t)v_4(t)}h^{-1}, h]^{-\epsilon(t)} \notag \\
	&=A_{v_3(t)}^h(0)A_{v_2(t)}^h(0)A_{v_1(t)}^h(0) [ \hat{g}_{v_1(t) v_2(t)}, \hat{g}_{v_2(t)v_3(t)}, \hat{g}_{v_3(t)v_4(t)}, h]^{-\epsilon(t)}. 
\end{align}

Putting this together, we see that moving all four of the phase factors associated to a tetrahedron $t$ past the vertex transforms gives us the total phase
\begin{align}
	&[h, \hat{g}_{v_1(t) v_2(t)}, \hat{g}_{v_2(t)v_3(t)}, \hat{g}_{v_3(t)v_4(t)}]^{\epsilon(t)} [ \hat{g}_{v_1(t) v_2(t)}, h , \hat{g}_{v_2(t)v_3(t)}, \hat{g}_{v_3(t)v_4(t)}]^{-\epsilon(t)} \notag \\
	& \hspace{2cm}[ \hat{g}_{v_1(t) v_2(t)}, \hat{g}_{v_2(t)v_3(t)}, h, \hat{g}_{v_3(t)v_4(t)}]^{\epsilon(t)}[ \hat{g}_{v_1(t) v_2(t)}, \hat{g}_{v_2(t)v_3(t)}, \hat{g}_{v_3(t)v_4(t)}, h]^{-\epsilon(t)}\notag \\
	&\hspace{9cm} = [\hat{g}_{v_1(t) v_2(t)}, \hat{g}_{v_2(t)v_3(t)}, \hat{g}_{v_3(t)v_4(t)}]_h^{- \epsilon(t)}.
\end{align}

This is the phase factor we obtain for each tetrahedron in the bulk (i.e., for tetrahedra where every vertex is in $S$). However we still have to consider the phase factors from tetrahedra for which not all vertices are in $S$, which are the boundary tetrahedra. We can characterize them by the subset of their vertices which are in $S$, meaning that we have to find the phase for 14 cases:
\begin{align*}
	&\set{v_1(t)},\set{v_2(t)},\set{v_3(t)},\set{v_4(t)}, \set{v_1(t),v_2(t)}, \set{v_1(t),v_3(t)}, \set{v_1(t),v_4(t)}, \set{v_2(t),v_3(t)}, \set{v_2(t),v_4(t)},\\
	& \set{v_3(t),v_4(t)}, \set{v_1(t),v_2(t), v_3(t)}, \set{v_1(t),v_2(t), v_4(t)}, \set{v_1(t),v_3(t), v_4(t)}, \set{v_2(t),v_3(t), v_4(t)}.
\end{align*}

We can work out the phase associated to each boundary tetrahedron in the same way as for the bulk ones, by commuting them past the lower indexed vertices which are in $S$. For example, for $\set{v_1(t),v_2(t), v_3(t)}$ we have to commute the phase factor from $v_2$ past the transform on $v_1$ and the factor from $v_3$ past $v_2$ and $v_1$ (just as we did for these factors for the bulk tetrahedra), but for $\set{v_1(t),v_2(t), v_4(t)}$ the factor from $v_4$ is only commuted past $v_1$ and $v_2$ because $v_3$ is not in $S$. These phase factors are exactly the dual phase contributions that we describe in Section \ref{Section_membrane_operators} of the main text, and the results for each possible set of vertices are given in Table \ref{Table_Dual_Phase} in the main text.

Now, we have moved all of the phase factors to the right. Finally, we have the product of untwisted gauge transforms $\prod_{v \in S} A_v^h(0)$. For any edge in the region $S$, there is a vertex transform on both ends of the edge. This leads to the edge label gaining a factor of $h$ (from the vertex at the start of the edge, which we call the source) and a factor of $h^{-1}$ (from the vertex at the end of the edge, which we call the target). These factors cancel, leaving the edge label unaffected. On the other hand, edges which have only one vertex in $S$ only gain one of these factors. This means that edges which points away from $S$ gain a factor of $h$, while edges that point towards $S$ gain a factor of $h^{-1}$. This is exactly the action of a magnetic membrane operator $C_0^h( \delta S)$ from untwisted lattice gauge theory (i.e., lattice gauge theory with a trivial 4-cocycle), with the direct membrane being the boundary of $S$ and the dual membrane cutting through the edges with one vertex in $S$ (see, e.g., Ref. \cite{HuxfordPaper3}):
\begin{equation}
	\prod_{v \in S} A_v^h(0)= C_0^h( \delta S).
\end{equation}

Combining these results, we see that the total action of the vertex transforms in $S$ when acting on the flat subspace can be written as
\begin{equation}
	\prod_{v \in S} A_v^h = C^h_0(\delta S) \prod_{t \in \delta S} \hat{\theta}_D^h(t) \prod_{t \in \text{bulk}(S)} [\hat{g}_{v_1(t) v_2(t)}, \hat{g}_{v_2(t)v_3(t)}, \hat{g}_{v_3(t)v_4(t)}]_h^{- \epsilon(t)} \label{Equation_prod_vertices_result},
\end{equation}
where the $\hat{\theta}_D^h(t)$ are the dual phase factors from the boundary tetrahedra, given in Table \ref{Table_Dual_Phase} in the main text. We see that the right-hand side of Equation \ref{Equation_prod_vertices_result} takes the form of a boundary operator multiplied by a bulk factor $\prod_{t \in \text{bulk}(S)} [\hat{g}_{v_1(t) v_2(t)}, \hat{g}_{v_2(t)v_3(t)}, \hat{g}_{v_3(t)v_4(t)}]_h^{- \epsilon(t)}$. If the product of vertex transforms in $S$ is indeed equivalent to a magnetic membrane operator on the boundary of $S$, we need to show that this bulk factor can be written in terms of surface variables (so that it becomes the surface weight discussed in Section \ref{Section_membrane_operators} of the main text).

\subsection{Evaluation of the Bulk Factor}
\label{Section_bulk_factor}

We now want to evaluate the factor $\prod_{t \in \text{bulk}(S)} [\hat{g}_{v_1(t) v_2(t)}, \hat{g}_{v_2(t)v_3(t)}, \hat{g}_{v_3(t)v_4(t)}]_h^{- \epsilon(t)}$. In order to do that, we wish to first prove that it is independent of the details in the bulk and only depends on the surface variables. We will do this in several steps, to demonstrate that the bulk factor is invariant under changes to the indexing of interior vertices as well as altering the triangulation in the bulk.

\subsubsection{Proof that the bulk factor is invariant under re-indexing of interior vertices}
\label{Section_bulk_factor_reindexing}
\begin{figure}[h]
	\centering
	\includegraphics[width=0.9\linewidth]{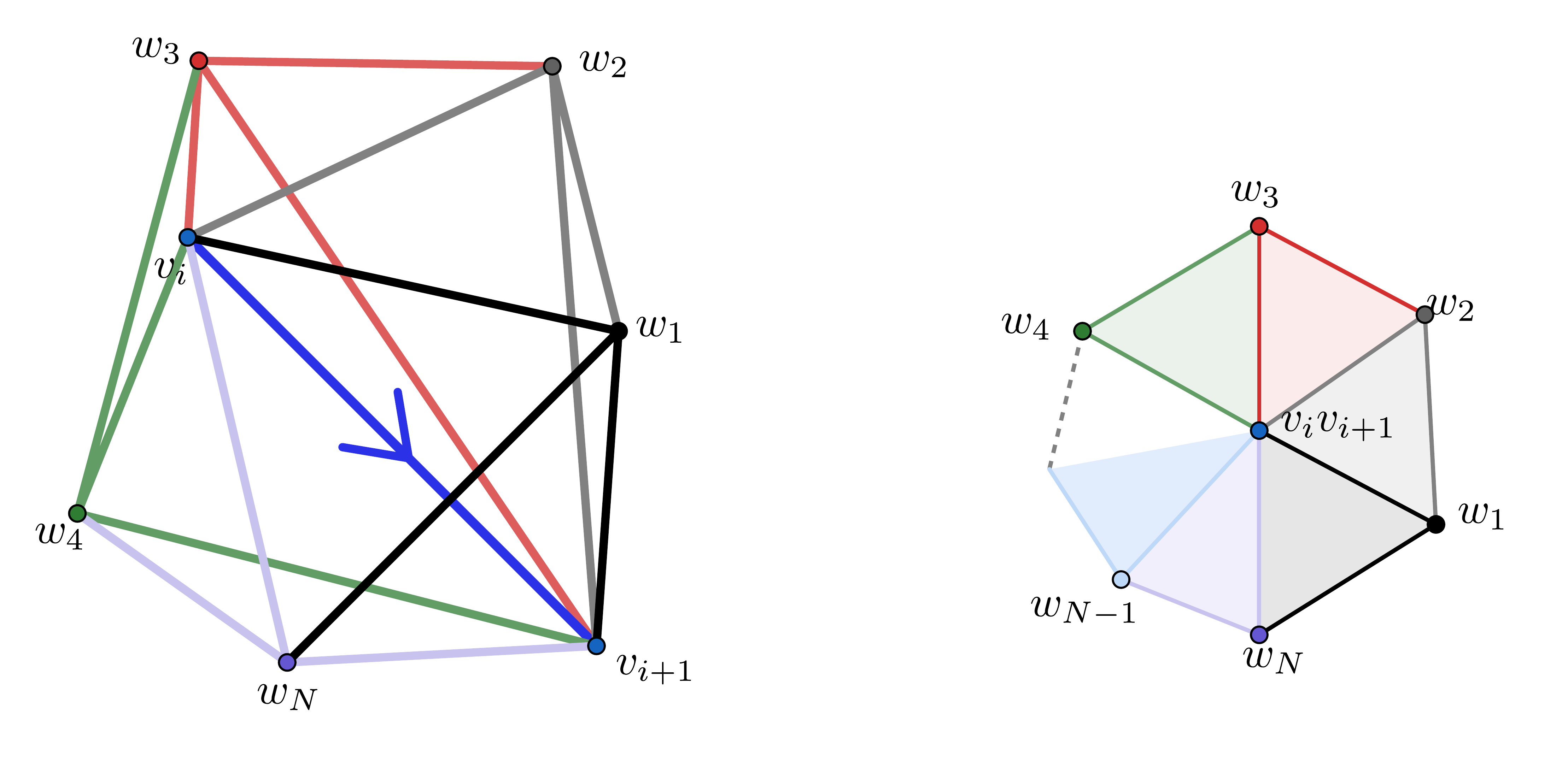}
	\caption{In the left image, we show an example of the tetrahedra that can surround the edge $v_i v_{i+1}$. In the right image we show a schematic view, with the edge $v_iv_{i+1}$ pointing out of the page.}
	\label{Figure_tetrahedra_around_edge}
\end{figure}

If the bulk factor in fact only depends on the surface, then the indexing of the vertices in the bulk should be irrelevant. Showing that this is the case is also useful because it will allow us to choose the indexing of vertices for other proofs, rather than having to consider every possible case. To prove this invariance under re-indexing, consider taking a vertex $v_i$ in the interior of the region $S$ (i.e., not one on the boundary, or equivalently one for which every attached tetrahedron is wholly in the region $S$) and slowly increasing its index. This has no effect on the phase factor until it increases past the index of one of its neighbours, which we call $v_{i+1}$. This is because each tetrahedron phase factor only depends on the ordering of vertices on that tetrahedron, and all vertices on a tetrahedron are adjacent. Because we slowly increased the index, only the ordering of those two vertices (among the neighbours of $v_i$) has changed. This affects the phase factors from the tetrahedra which contain both $v_i$ and $v_{i+1}$, or equivalently, the tetrahedra that contain the edge $[v_i v_{i+1}]$. A schematic picture of those tetrahedra is shown in Figure \ref{Figure_tetrahedra_around_edge}. We denote each such tetrahedron by $\set{v_i, v_{i+1}, w_j, w_{j+1}}$, where $j=1,...,N$ increases as we circle around the edge and has no relation to the index of the vertices involved. We define $w_{N+1}=w_1$ due to the circular transversal around the edge.

Now consider how the phase factor for one of those tetrahedra changes as we increase the index of $v_i$ past $v_{i+1}$. There are many different cases depending on the relative indices of $v_i$, $v_{i+1}$, $w_j$ and $w_{j+1}$. By construction, $v_{i+1}$ has a higher index than $v_i$ and no (relevant) vertex has an index between theirs. Therefore, calling the tetrahedron $t$, we can have $v_i, v_{i+1}=v_1(t), v_2(t)$ or $v_i, v_{i+1}=v_2(t), v_3(t)$ or $v_i, v_{i+1}=v_3(t), v_4(t)$. Accounting also for the relative indices of $w_j$ and $w_{j+1}$, we then have six cases:

\begin{align*}
	\text{Case 1: } & (v_1(t), v_2(t),v_3(t),v_4(t))=(v_i, v_{i+1}, w_j, w_{j+1})\\
	\text{Case 2: } &(v_1(t), v_2(t),v_3(t),v_4(t))=(v_i, v_{i+1}, w_{j+1}, w_{j})\\
	\text{Case 3: } &(v_1(t), v_2(t),v_3(t),v_4(t))=(w_j, v_i, v_{i+1}, w_{j+1})\\
	\text{Case 4: } &(v_1(t), v_2(t),v_3(t),v_4(t))=(w_{j+1}, v_i, v_{i+1}, w_j)\\
	\text{Case 5: } &(v_1(t), v_2(t),v_3(t),v_4(t))=( w_j, w_{j+1}, v_i, v_{i+1})\\
	\text{Case 6: } &(v_1(t), v_2(t),v_3(t),v_4(t))=( w_{j+1}, w_j, v_i, v_{i+1}).
\end{align*}

We now consider how the phase factor for each case changes under re-indexing. In each case, we will denote the tetrahedron after re-indexing by $t'$ and will write the phase factor for $t'$ in terms of that for $t$.

\hfill

\noindent \textbf{Case 1:}

\noindent Re-indexing swaps the relative indices of $v_i$ and $v_{i+1}$. This means that the ordered vertices of the new tetrahedron are:

$$(v_1(t'),v_2(t'),v_3(t'),v_4(t'))=(v_{i+1},v_i,w_j,w_{j+1})=(v_2(t),v_1(t),v_3(t),v_4(t)).$$

The orientation of the new tetrahedron is $\epsilon(t')=-\epsilon(t)$, because a single pair of vertices is swapped to obtain $t'$ from $t$. In addition, $\epsilon(t)=+1$ for Case 1 (see Figure \ref{Figure_tetrahedra_orientation} in the main text for the definition of the orientation of a tetrahedron). Therefore the phase factor for the new tetrahedron is
\begin{align*}
	\theta(t')&= [\hat{g}_{v_1(t')v_2(t')}, \hat{g}_{v_2(t')v_3(t')}, \hat{g}_{v_3(t')v_4(t')}]_h^{- \epsilon(t')}\\
	&= [\hat{g}_{v_{i+1}v_i}, \hat{g}_{v_i w_j}, \hat{g}_{w_j w_{j+1}}]_h.
\end{align*}
Here we define $\hat{g}_{v_{i+1}v_i}$, which is not the label of an edge in the original lattice (because the original edge points from $v_i$ to $v_{i+1}$), to be equal to $\hat{g}_{v_{i}v_{i+1}}^{-1}$. In other words, if we reverse the orientation of an edge, we must invert its label (as we expect for lattice gauge theory). The phase factor for the old tetrahedron is
\begin{align}
	\theta(t)&= [\hat{g}_{v_1(t)v_2(t)}, \hat{g}_{v_2(t)v_3(t)}, \hat{g}_{v_3(t)v_4(t)}]_h^{- \epsilon(t)} \notag \\
	&= [\hat{g}_{v_{i}v_{i+1}}, \hat{g}_{v_{i+1} w_j}, \hat{g}_{w_j w_{j+1}}]_h^{-1}.
\end{align}
Therefore,
\begin{equation}
	\frac{\theta(t')}{\theta(t)}= [\hat{g}_{v_{i}v_{i+1}}^{-1}, \hat{g}_{v_i w_j}, \hat{g}_{w_j w_{j+1}}]_h [\hat{g}_{v_{i}v_{i+1}}, \hat{g}_{v_{i+1} w_j}, \hat{g}_{w_j w_{j+1}}]_h .
\end{equation}

The slant product $[x,y,z]_u$ satisfies the twisted cocycle condition \cite{Wan2015} in Equation \ref{Equation_twisted_cocycle_condition}, which for $G$ Abelian becomes the ordinary 3-cocycle condition:
\begin{equation}
	\frac{[x,y,z]_u [w, xy,z]_u [w,x,y]_u}{[wx,y,z]_u [w,x, yz]_u}=1. \label{Equation_3_cocycle_condition_supplement}
\end{equation}
Identifying $w=\hat{g}_{v_{i}v_{i+1}}^{-1}$, $x=\hat{g}_{v_{i}v_{i+1}}$, $y=\hat{g}_{v_{i+1} w_j}$, $z= \hat{g}_{w_j w_{j+1}}$ and $u=h$, we see that
\begin{align*}
	\frac{\theta(t')}{\theta(t)}&= [\hat{g}_{v_{i}v_{i+1}}^{-1},\hat{g}_{v_{i}v_{i+1}} \hat{g}_{v_{i+1} w_j} , \hat{g}_{w_j w_{j+1}}]_h [\hat{g}_{v_{i}v_{i+1}}, \hat{g}_{v_{i+1} w_j}, \hat{g}_{w_j w_{j+1}}]_h \\
	&= [w, xy, z]_u [x,y,z]_u,
\end{align*}
where flatness ensures that $\hat{g}_{v_{i}v_{i+1}} \hat{g}_{v_{i+1} w_j} = \hat{g}_{v_{i} w_j}$. Therefore,
\begin{align}
	\frac{\theta(t')}{\theta(t)}&= \frac{[wx,y,z]_u [w,x,yz]_u}{[w,x,y]_u} \notag\\
	&= \frac{[1_G, \hat{g}_{v_{i+1} w_j}, \hat{g}_{w_j w_{j+1}}]_h [\hat{g}_{v_{i}v_{i+1}}^{-1},\hat{g}_{v_{i}v_{i+1}} , \hat{g}_{v_{i+1} w_{j+1}}]_h}{[\hat{g}_{v_{i}v_{i+1}}^{-1},\hat{g}_{v_{i}v_{i+1}} , \hat{g}_{v_{i+1} w_{j}}]_h} \notag \\
	&=\frac{ [\hat{g}_{v_{i}v_{i+1}}^{-1},\hat{g}_{v_{i}v_{i+1}} , \hat{g}_{v_{i+1} w_{j+1}}]_h}{[\hat{g}_{v_{i}v_{i+1}}^{-1},\hat{g}_{v_{i}v_{i+1}} , \hat{g}_{v_{i+1} w_{j}}]_h},
\end{align}
where we used the normalisation condition Equation \ref{Equation_twisted_normalization} to remove the slant product involving $1_G$.

\hfill

\noindent \textbf{Case 2:}

\noindent In this case, the phase factor for the original tetrahedron is

\begin{align*}
	\theta(t)&= [\hat{g}_{v_1(t)v_2(t)}, \hat{g}_{v_2(t)v_3(t)}, \hat{g}_{v_3(t)v_4(t)}]_h^{- \epsilon(t)}\\
	&= [\hat{g}_{v_{i}v_{i+1}}, \hat{g}_{v_{i+1} w_{j+1}}, \hat{g}_{w_{j+1} w_{j}}]_h^{+1}
\end{align*}
and the phase factor for the new tetrahedron is
\begin{align*}
	\theta(t')&= [\hat{g}_{v_1(t')v_2(t')}, \hat{g}_{v_2(t')v_3(t')}, \hat{g}_{v_3(t')v_4(t')}]_h^{- \epsilon(t')}\\
	&= [\hat{g}_{v_{i+1}v_{i}}, \hat{g}_{v_{i} w_{j+1}}, \hat{g}_{w_{j+1} w_{j}}]_h^{-1}.
\end{align*}
Therefore,
\begin{align*}
	\frac{\theta(t')}{\theta(t)}&=\big([\hat{g}_{v_{i}v_{i+1}}^{-1}, \hat{g}_{v_{i} w_{j+1}}, \hat{g}_{w_{j+1} w_{j}}]_h [\hat{g}_{v_{i}v_{i+1}}, \hat{g}_{v_{i+1} w_{j+1}}, \hat{g}_{w_{j+1} w_{j}}]_h\big)^{-1}.
\end{align*}
Using the cocycle condition as before, we obtain
\begin{equation}
	\frac{\theta(t')}{\theta(t)}=\frac{ [\hat{g}_{v_{i}v_{i+1}}^{-1},\hat{g}_{v_{i}v_{i+1}} , \hat{g}_{v_{i+1} w_{j+1}}]_h}{[\hat{g}_{v_{i}v_{i+1}}^{-1},\hat{g}_{v_{i}v_{i+1}} , \hat{g}_{v_{i+1} w_{j}}]_h},
\end{equation}
which is the same expression as for Case 1.

\hfill

\noindent \textbf{Case 3:}

\noindent In this case, the phase factor for the original tetrahedron is
\begin{align*}
	\theta(t)&= [\hat{g}_{v_1(t)v_2(t)}, \hat{g}_{v_2(t)v_3(t)}, \hat{g}_{v_3(t)v_4(t)}]_h^{- \epsilon(t)}\\
	&= [\hat{g}_{w_j v_i}, \hat{g}_{v_{i}v_{i+1}}, \hat{g}_{v_{i+1} w_{j+1}}]_h^{-1}
\end{align*}
and the phase factor for the new one is
\begin{align*}
	\theta(t')&= [\hat{g}_{v_1(t')v_2(t')}, \hat{g}_{v_2(t')v_3(t')}, \hat{g}_{v_3(t')v_4(t')}]_h^{- \epsilon(t')}\\
	&= [\hat{g}_{w_j v_{i+1}}, \hat{g}_{v_{i+1}v_{i}}, \hat{g}_{v_{i} w_{j+1}}]_h^{+1}.
\end{align*}
Therefore
\begin{align*}
	\frac{\theta(t')}{\theta(t)}&= [\hat{g}_{w_j v_i}, \hat{g}_{v_{i}v_{i+1}}, \hat{g}_{v_{i+1} w_{j+1}}]_h [\hat{g}_{w_j v_{i+1}}, \hat{g}_{v_{i}v_{i+1}}^{-1}, \hat{g}_{v_{i} w_{j+1}}]_h.
\end{align*}

We can then use the cocycle condition again, with $w= \hat{g}_{w_jv_{i+1}}$, $x= \hat{g}_{v_{i}v_{i+1}}^{-1}$, $y= \hat{g}_{v_{i}v_{i+1}}$ and $z=\hat{g}_{v_{i+1} w_{j+1}}$, to obtain 
\begin{align}
	\frac{\theta(t')}{\theta(t)}&= [\hat{g}_{w_j v_i}, \hat{g}_{v_{i}v_{i+1}}, \hat{g}_{v_{i+1} w_{j+1}}]_h [\hat{g}_{w_j v_{i+1}}, \hat{g}_{v_{i}v_{i+1}}^{-1}, \hat{g}_{v_{i} w_{j+1}}]_h \notag \\
	&= [wx, y, z]_h [w, x, yz]_h \notag \\
	&= [x,y,z]_h [w, xy,z]_h [w,x,y]_h \notag \\
	&=[\hat{g}_{v_{i}v_{i+1}}^{-1}, \hat{g}_{v_{i}v_{i+1}}, \hat{g}_{v_{i+1} w_{j+1}} ]_h [\hat{g}_{w_jv_{i+1}}, 1_G, \hat{g}_{v_{i+1} w_{j+1}}]_h [\hat{g}_{w_jv_{i+1}}, \hat{g}_{v_{i}v_{i+1}}^{-1}, \hat{g}_{v_{i}v_{i+1}}]_h \notag \\
	&= [\hat{g}_{v_{i}v_{i+1}}^{-1}, \hat{g}_{v_{i}v_{i+1}}, \hat{g}_{v_{i+1} w_{j+1}} ]_h [\hat{g}_{w_jv_{i+1}}, \hat{g}_{v_{i}v_{i+1}}^{-1}, \hat{g}_{v_{i}v_{i+1}}]_h,
\end{align}
where we used the normalization condition to remove the cocycle involving the identity element.

\hfill

\noindent \textbf{Case 4:}

\noindent Having described a few cases in detail so far, we will be more brief for the remaining three. For Case 4, we have
\begin{equation*}
	\frac{\theta(t')}{\theta(t)}= \big( [g_{w_{j+1}v_{i+1}, g_{v_{i+1}v_i}}g_{v_i w_j}]_h[g_{w_{j+1}v_{i}, g_{v_{i}v_{i+1}}}g_{v_{i+1} w_j}]_h \big)^{-1}.
\end{equation*}
Using the cocycle condition with $w= \hat{g}_{w_{j+1}v_{i+1}}$, $x= \hat{g}_{v_{i}v_{i+1}}^{-1}$, $y= \hat{g}_{v_{i}v_{i+1}}$ and $z=\hat{g}_{v_{i+1} w_{j}}$ we obtain

\begin{equation}
	\frac{\theta(t')}{\theta(t)}= \big([\hat{g}_{v_{i}v_{i+1}}^{-1}, \hat{g}_{v_{i}v_{i+1}}, \hat{g}_{v_{i+1} w_{j}} ]_h [\hat{g}_{w_{j+1}v_{i+1}}, \hat{g}_{v_{i}v_{i+1}}^{-1}, \hat{g}_{v_{i}v_{i+1}}]_h \big)^{-1}.
\end{equation}

\hfill

\noindent \textbf{Case 5:}

\noindent We have 
\begin{equation*}
	\frac{\theta(t')}{\theta(t)}= \big([\hat{g}_{w_j w_{j+1}} , \hat{g}_{w_{j+1} v_{i+1}}, \hat{g}_{v_{i+1}v_i}]_h [\hat{g}_{w_j w_{j+1}} , \hat{g}_{w_{j+1} v_{i}}, \hat{g}_{v_{i}v_{i+1}}]_h \big)^{+1}.
\end{equation*}
Using the cocycle condition with $w= \hat{g}_{w_j w_{j+1}}$, $x= \hat{g}_{w_{j+1}v_{i+1}}$, $y=\hat{g}_{v_{i} v_{i+1}}^{-1}$ and $z= \hat{g}_{v_i v_{i+1}}$ we obtain
\begin{equation}	
	\frac{\theta(t')}{\theta(t)}= \frac{[\hat{g}_{w_j v_{i+1}}, \hat{g}_{v_{i} v_{i+1}}^{-1}, \hat{g}_{v_i v_{i+1}}]_h} {[\hat{g}_{w_{j+1}v_{i+1}}, \hat{g}_{v_{i} v_{i+1}}^{-1}, \hat{g}_{v_i v_{i+1}}]_h}.
\end{equation}

\hfill

\noindent \textbf{Case 6:}

\noindent Finally, for Case 6, we have

\begin{equation*}
	\frac{\theta(t')}{\theta(t)}= \big([\hat{g}_{w_{j+1} w_{j}} , \hat{g}_{w_{j} v_{i+1}}, \hat{g}_{v_{i+1}v_i}]_h [\hat{g}_{w_{j+1} w_{j}} , \hat{g}_{w_{j} v_{i}}, \hat{g}_{v_{i}v_{i+1}}]_h \big)^{-1}.
\end{equation*}
Using the cocycle condition with $w= \hat{g}_{w_{j+1} w_{j}}$, $x= \hat{g}_{w_{j}v_{i+1}}$, $y=\hat{g}_{v_i v_{i+1}}^{-1}$ and $z= \hat{g}_{v_i v_i+1}$ we obtain
\begin{equation}
	\frac{\theta(t')}{\theta(t)}= \frac{[\hat{g}_{w_j v_{i+1}}, \hat{g}_{v_{i} v_{i+1}}^{-1}, \hat{g}_{v_i v_{i+1}}]_h} {[\hat{g}_{w_{j+1}v_{i+1}}, \hat{g}_{v_{i} v_{i+1}}^{-1}, \hat{g}_{v_i v_{i+1}}]_h},
\end{equation}
which is the same expression as for Case 5.

Next, notice that the expression for every case can be written in the following form:
\begin{equation}
	\frac{\theta(t')}{\theta(t)}(\set{w_j, w_{j+1}}) = \frac{ \eta(w_{j+1})}{\eta(w_j)},
\end{equation}	
where 
\begin{equation}
	\eta(w_j)= \begin{cases}
		[\hat{g}_{w_j v_i+1}, \hat{g}_{v_i v_{i+1}}^{-1}, \hat{g}_{v_i v_i+1}^{-1}]_h & \text{if $w_j < v_i$} \\
		[\hat{g}_{v_{i}v_{i+1}}^{-1},\hat{g}_{v_{i}v_{i+1}} , \hat{g}_{v_{i+1} w_{j}}]_h & \text{if $w_j > v_{i+1}$}.
	\end{cases}
\end{equation}

Here $w_j > v_{i+1}$ means that $w_j$ has a greater index. Then the total phase for all tetrahedra around the edge, and therefore the total phase change for the re-indexing procedure, is
\begin{align}
	\prod_{j=1}^N \frac{\theta(t')}{\theta(t)}(\set{w_j, w_{j+1}}) &= \prod_{j=1}^N \frac{ \eta(w_{j+1})}{\eta(w_j)}=1,
\end{align}
where $N$ is the total number of tetrahedra around the edge $[v_i v_{i+1}]$ and $w_{N+1} := w_1$. We see that there is no overall change to the phase under re-indexing as we increase its index beyond that of the vertex with the next lowest index. By repeating this process (and the inverse process of decreasing the index) we are free to change the index on any interior vertex without affecting the bulk phase factor.

\subsubsection{Proof that the bulk factor is invariant under bistellar flips}

\begin{figure}[h]
	\centering
	\includegraphics[width=0.8\linewidth]{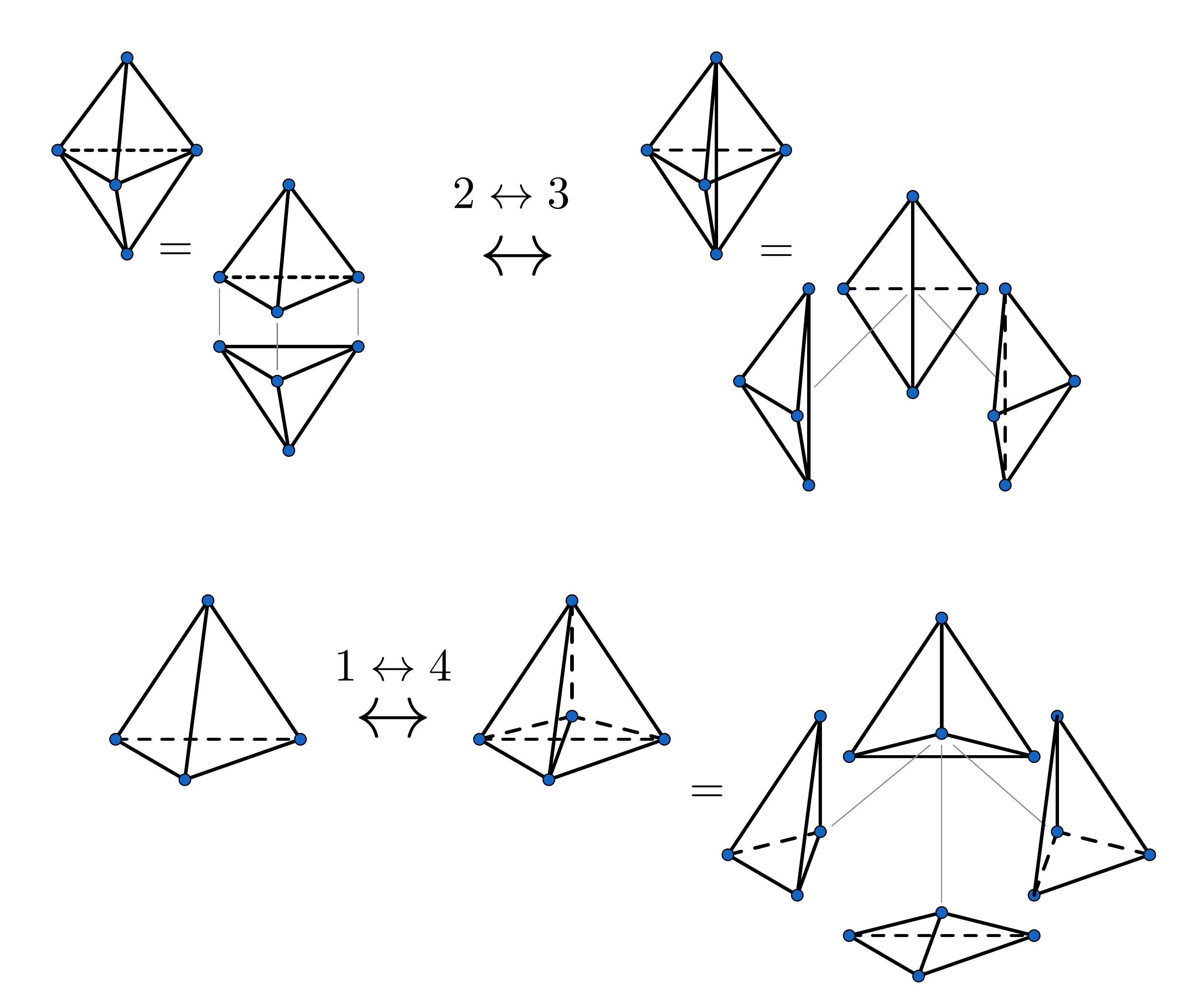}
	\caption{There are two bistellar flips (the $2 \leftrightarrow 3$ and $1 \leftrightarrow 4$ moves) which can be used to relate equivalent triangulations.}
	\label{Figure_bistellar_flips}
\end{figure}

In 3d, two triangulations that represent the same manifold can be related by bistellar flips (also called Pachner moves) \cite{Pachner1990, Pachner1991, Casali1995}. We therefore wish to show that the bulk phase factor is invariant under the application of these moves in the interior of $S$ (i.e., for moves which preserve the boundary of $S$). In 3d, there are two bistellar flips: the $2 \leftrightarrow 3$ move and the $1 \leftrightarrow 4$ move. These are illustrated in Figure \ref{Figure_bistellar_flips}. 

First, consider the $2 \leftrightarrow 3$ move. Unfortunately, because one or more of the vertices in this diagram may be on the surface of $S$, we cannot use the re-indexing procedure (which is only valid for vertices in the interior of $S$) and must consider all possible indices for the five vertices involved. We denote the lowest indexed vertex in the diagram by $v_1(T)$, the next lowest by $v_2(T)$ and so on up to $v_5(T)$. Each tetrahedron involved in either side of the $2 \leftrightarrow 3$ move can be obtained by deleting one vertex from the picture and attaching edges between all remaining vertices. Denoting such a tetrahedron by $T - v_i(T)$, the phase associated to each tetrahedron is

\begin{align*}
	\theta(T-v_1(T))&= [\hat{g}_{v_2(T)v_3(T)}, \hat{g}_{v_3(T)v_4(T)}, \hat{g}_{v_4(T)v_5(T)}]_h^{ - \epsilon(T- v_1(T))}\\
	\theta(T-v_2(T))&= [\hat{g}_{v_1(T)v_3(T)}, \hat{g}_{v_3(T)v_4(T)}, \hat{g}_{v_4(T)v_5(T)}]_h^{ - \epsilon(T- v_2(T))}\\
	\theta(T-v_3(T))&= [\hat{g}_{v_1(T)v_2(T)}, \hat{g}_{v_2(T)v_4(T)}, \hat{g}_{v_4(T)v_5(T)}]_h^{ - \epsilon(T- v_3(T))}\\
	\theta(T-v_4(T))&= [\hat{g}_{v_1(T)v_2(T)}, \hat{g}_{v_2(T)v_3(T)}, \hat{g}_{v_3(T)v_5(T)}]_h^{ - \epsilon(T- v_4(T))}\\
	\theta(T-v_5(T))&= [\hat{g}_{v_1(T)v_2(T)}, \hat{g}_{v_2(T)v_3(T)}, \hat{g}_{v_3(T)v_4(T)}]_h^{ - \epsilon(T- v_5(T))}.
\end{align*}

We want to show that the product of the phases for the two tetrahedra before the move is the same as the product of the phases for the three tetrahedra after the move. Equivalently, we want to show that the product of all of the phases for the double multiplied by the inverse of the phases for the triple is equal to unity. There are generally many cases to consider, but rather than consider them all we want to show that all of the cocycles corresponding to tetrahedra with odd vertices removed ($v_1(T), v_3(T)$ and $v_5(T)$) will all have the same overall exponent ($\pm 1$) in the total product, and will have the opposite exponent compared to the cocycles from the tetrahedra with even vertices ($v_2(T)$ and $v_4(T)$) removed.

\begin{figure}[h]
	\centering
	\includegraphics[width=0.9\linewidth]{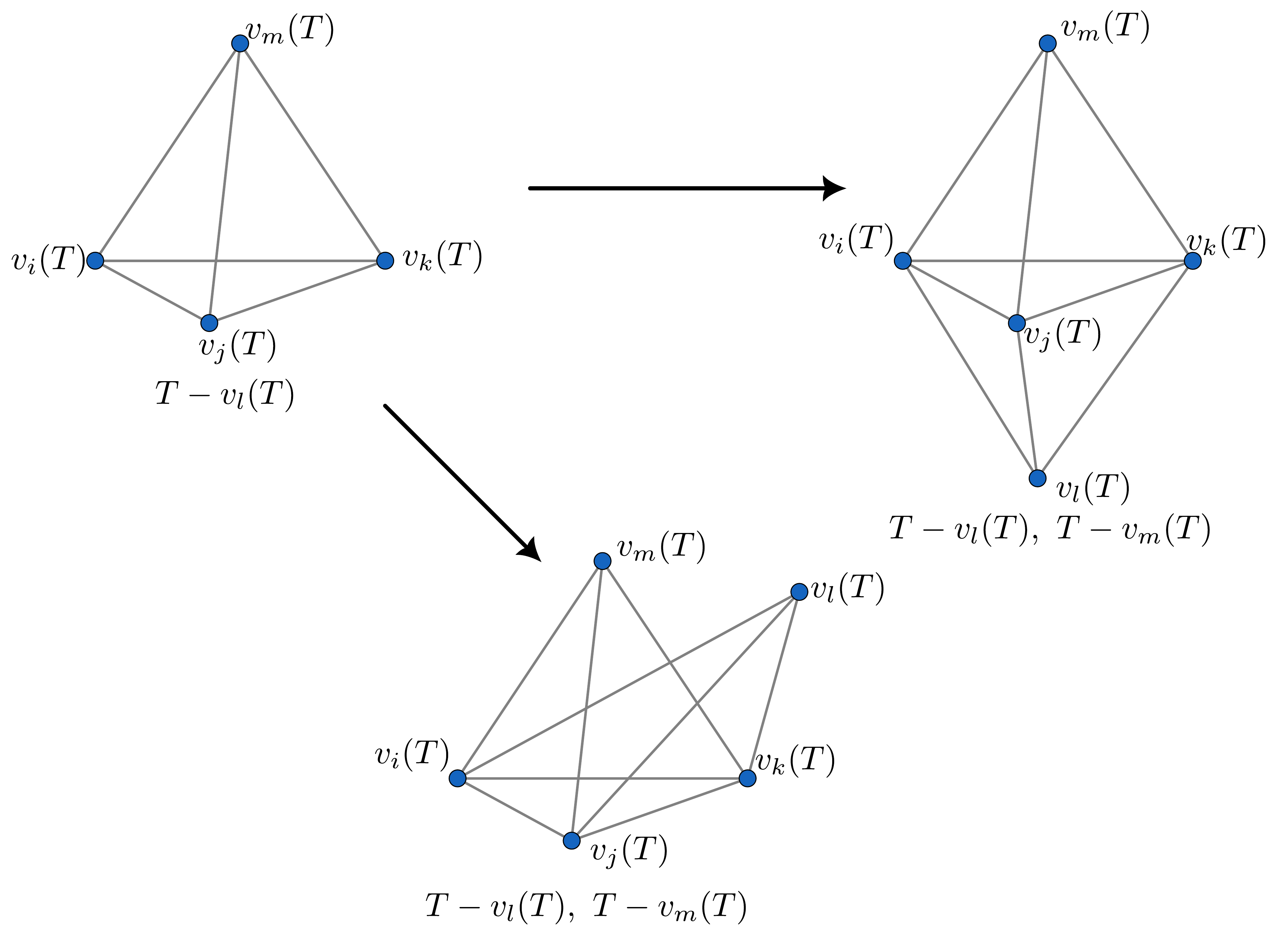}
	\caption{Given a tetrahedron $T-v_l(T) = \set{v_i(T), v_j(T), v_k(T), v_m(T)}$, there are two ways to attach the tetrahedron $T-v_m(T)$ that shares the face $\set{v_i(T),v_j(T),v_K(T)}$. In the first (upper path) the two tetrahedra do not intersect and both belong to the same side of the bistellar flip. In the second (lower path) they do intersect and belong to opposite sides of the bistellar flip.}
	\label{Figure_2_3_flip_orientation}
\end{figure}

To see this, first consider a single tetrahedron $T-v_l(T)$, as shown in Figure \ref{Figure_2_3_flip_orientation}. Now we add the vertex $v_l(T)$ and consider another tetrahedron, $T-v_m(T)$. We label the other three vertices as $v_i(T)$, $v_j(T)$ and $v_k(T)$, so that $i,j,k,l,m$ is some permutation of $1,2,3,4,5$. There are two ways to attach the new tetrahedron, $T-v_m(T)$, to the first one, $T-v_l(T)$. Either it can be attached along the shared face $\set{v_i(T), v_j(T), v_k(T)}$ without intersection, in which case both tetrahedrons belong to either the double or the triple (the same side of the bistellar flip), or it can be attached so that it intersects with the first tetrahedron, in which case they belong to opposite sides of the bistellar flip. Consider the former case first. Recall that the orientation of a tetrahedron can be determined using the right-hand rule. With the thumb pointing towards the highest-indexed vertex on the tetrahedron, the orientation is positive if the remaining vertices increase in index with the circulation of the fingers and negative otherwise. Swapping the index of a pair of vertices reverses the orientation. For the tetrahedron $T-v_l(T)$, if $v_m(T)$ is the highest indexed vertex and the indices of the remaining vertices increase in the order $v_i(T)$, $v_j(T)$, $v_k(T)$, then the tetrahedron satisfies the right-hand rule (see Figure \ref{Figure_2_3_flip_orientation}) and so has positive orientation. Therefore, the orientation $\epsilon(T-v_l(T))$ of the tetrahedron for a generic ordering of vertices is given by
$$\epsilon(T-v_l(T)) = \text{sgn}( v_i(T), v_j(T), v_k(T), v_m(T)),$$
where $\text{sgn}( v_i(T), v_j(T), v_k(T), v_m(T))$ is the signature of the permutation required to put the vertices in ascending order of index. Now consider the other tetrahedron $T-v_m(T)$. Because the new vertex $v_l(T)$ is on the opposite side of the face $\set{v_i(T,)v_j(T),v_k(T)}$ compared to $v_m(T)$, if $v_l(T)$ is the highest indexed vertex then its orientation is positive if the indices of the remaining vertices increase as $v_j(T), v_i(T), v_k(T)$ (instead of $v_i(T)$, $v_j(T)$, $v_k(T)$). This means that the orientation of $T-v_m(T)$ is given by
\begin{align*}
	\epsilon(T-v_m(T)) &= \text{sgn}( v_j(T), v_i(T), v_k(T), v_l(T))\\
	&= - \text{sgn} ( v_i(T), v_j(T), v_k(T), v_l(T))
\end{align*}
for a general ordering of vertices.

Now suppose that $m$ and $l$ are both odd (or both even). Then there are an odd number of vertices with index between them (e.g., if $m=3$ and $l=5$, then $v_4$ is between them). This means that an odd number of additional swaps is required to order $(v_i(T), v_j(T), v_k(T), v_m(T))$ compared to $(v_i(T), v_j(T), v_k(T), v_l(T))$. For example, if $m=3$ and $l=5$, then after ordering $(v_i(T), v_j(T), v_k(T))$ into $(v_1(T), v_2(T), v_4(T))$, then for $v_m(T)=v_5(T)$, we have $(v_1(T), v_2(T), v_4(T),v_m(T))=(v_1(T), v_2(T), v_4(T),v_5(T))$ which is already in ascending order. On the other hand, for $v_l(T)=v_3(T)$ we have $(v_1(T), v_2(T), v_4(T),v_l(T))=(v_1(T), v_2(T), v_4(T),v_3(T))$ which needs one additional swap. On the other hand, if one of $m$ and $l$ is odd and the other is even then they are separated by an even number of vertices and so have the same signature. This means that
\begin{align}
	\epsilon(T-v_m(T))& = - \text{sgn} ( v_i(T), v_j(T), v_k(T), v_l(T)) \notag \\
	&= \text{sgn} ( v_i(T), v_j(T), v_k(T), v_m(T)) \times (-1)^{m-l}.
\end{align}
That is, if $T-v_l(T)$ and $T-v_m(T)$ belong to the same side of the bistellar flip, they appear with the same power if $l$ and $m$ are both odd or both even and the opposite power otherwise.

Next, consider the case where the two tetrahedra belong to opposite sides of the bistellar flip. In that case, they appear on the same side of the face $\set{v_i(T),v_j(T),v_k(T)}$. Therefore, $\epsilon(T-v_m(T)) = \text{sgn}( v_i(T), v_j(T), v_k(T), v_l(T))$ rather than $-\text{sgn}( v_i(T), v_j(T), v_k(T), v_l(T))$. This just introduces another relative minus sign compared to the case where the tetrahedra are on the same side of the bistellar flip. This cancels with the minus sign from the tetrahedra belonging to the opposite side of the bistellar flip that comes from dividing the phases corresponding to the state before the flip by the phase after the flip. Therefore, the same conclusion holds: if $l$ and $m$ are both odd or both even, they appear with the same power, otherwise they appear with opposite power. This means that the phase gained from the bistellar flip takes the form
\begin{align*}
	&\nu(2 \rightarrow 3) \\
	&= \big(\frac{[\hat{g}_{v_2(T)v_3(T)}, \hat{g}_{v_3(T)v_4(T)}, \hat{g}_{v_4(T)v_5(T)}]_h [\hat{g}_{v_1(T)v_2(T)}, \hat{g}_{v_2(T)v_4(T)}, \hat{g}_{v_4(T)v_5(T)}]_h [\hat{g}_{v_1(T)v_2(T)}, \hat{g}_{v_2(T)v_3(T)}, \hat{g}_{v_3(T)v_4(T)}]_h }{[\hat{g}_{v_1(T)v_3(T)}, \hat{g}_{v_3(T)v_4(T)}, \hat{g}_{v_4(T)v_5(T)}]_h [\hat{g}_{v_1(T)v_2(T)}, \hat{g}_{v_2(T)v_3(T)}, \hat{g}_{v_3(T)v_5(T)}]_h } \big)^{\pm 1}.
\end{align*}
Using the cocycle condition Equation \ref{Equation_twisted_cocycle_condition}, with $w=\hat{g}_{v_1(T)v_2(T)}$, $x=\hat{g}_{v_2(T)v_3(T)}$, $y=\hat{g}_{v_3(T)v_4(T)}$ and $z=\hat{g}_{v_4(T)v_5(T)}$, we see that $\nu(2 \rightarrow 3) =1$. In other words, there is no net phase gain from this bistellar flip.

\begin{figure}[h]
	\centering
	\includegraphics[width=0.7\linewidth]{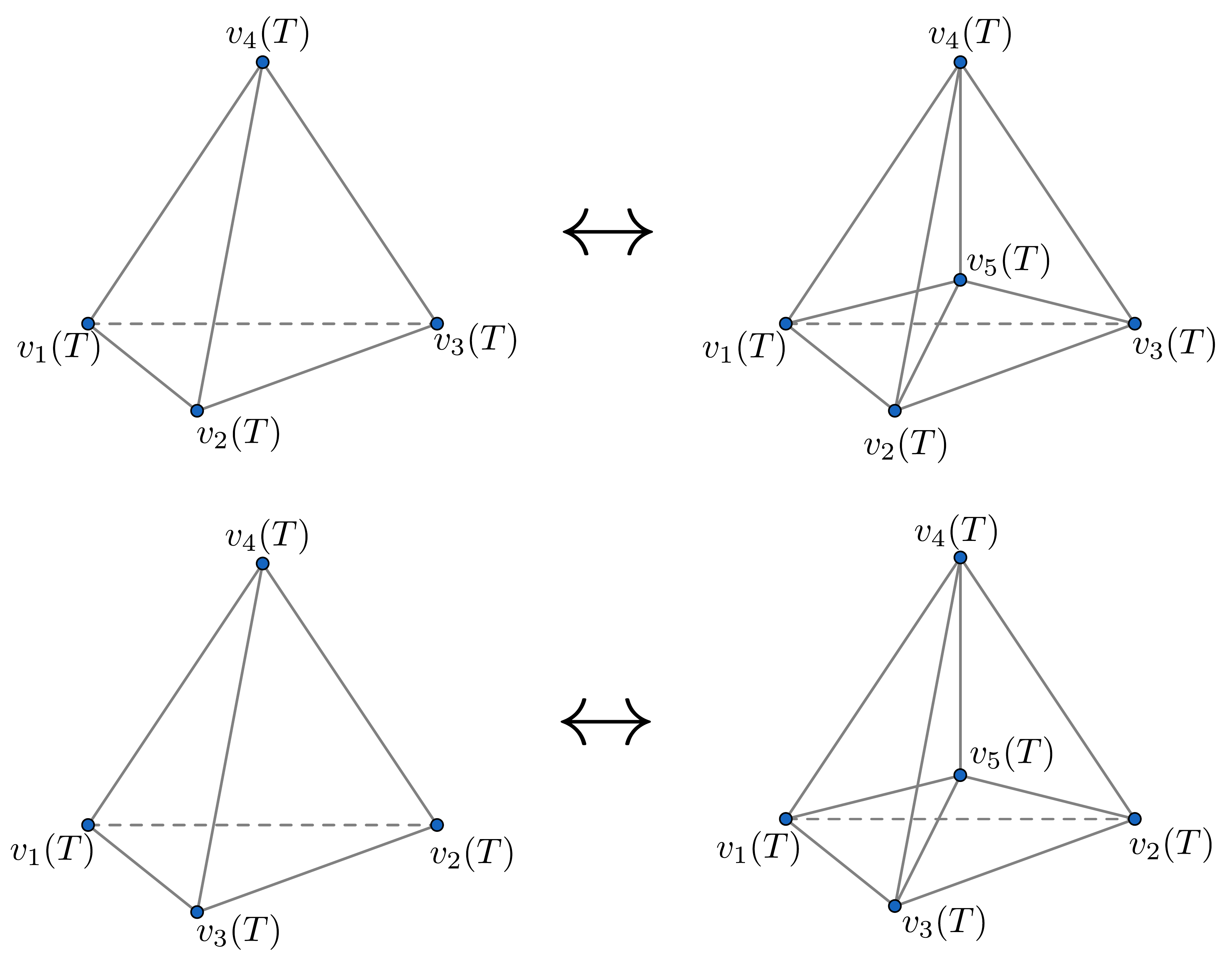}
	\caption{For the $1 \leftrightarrow 4$ bistellar flip, we can use the re-indexing procedure to choose the index for the internal vertex. This leaves two possible orientations for the system (up to rotation), with the positive orientation shown on the top line and the negative orientation shown on the bottom.}
	\label{1_4_flip_orientation}
\end{figure}

Next, consider the $1 \leftrightarrow 4$ move. This time, because the additional vertex after the move from $1 \rightarrow 4$ is in the middle of the original tetrahedron, it is always an interior vertex and we can freely re-index it. We choose it to be the largest indexed vertex of the system, $v_5(T)$. This means that we only need to consider the two orientations of the tetrahedron before the move. This gives us the two cases shown in Figure \ref{1_4_flip_orientation}. For the first case, the quadruple (the right-hand side of the bistellar flip) has the phase
\begin{align*}
	\theta(4)&= [\hat{g}_{v_2(T) v_3(T)}, \hat{g}_{v_3(T)v_4(T)}, \hat{g}_{v_4(T)v_5(T)}]_h^{+1} [\hat{g}_{v_1(T) v_3(T)}, \hat{g}_{v_3(T)v_4(T)}, \hat{g}_{v_4(T)v_5(T)}]_h^{-1} \\
	& \hspace{2cm} [\hat{g}_{v_1(T) v_2(T)}, \hat{g}_{v_2(T)v_4(T)}, \hat{g}_{v_4(T)v_5(T)}]_h^{+1} [\hat{g}_{v_1(T) v_2(T)}, \hat{g}_{v_2(T)v_3(T)}, \hat{g}_{v_3(T)v_5(T)}]_h^{-1} .
\end{align*}

Using the cocycle move $w=\hat{g}_{v_1(T)v_2(T)}$, $x=\hat{g}_{v_2(T)v_3(T)}$, $y=\hat{g}_{v_3(T)v_4(T)}$ and $z=\hat{g}_{v_4(T)v_5(T)}$, we see that this is equal to
\begin{equation}
	\theta(4)=[\hat{g}_{v_1(T)v_2(T)}, \hat{g}_{v_2(T)v_3(T)} , \hat{g}_{v_3(T)v_4(T)}]_h^{-1} = \theta(1),
\end{equation}
where $\theta(1)$ is the phase of the single tetrahedron on the other side of the bistellar flip.

For the second case, shown in the lower part of Figure \ref{1_4_flip_orientation}, we get the same result. For the quadruple, we have the phase
\begin{align*}
	\theta(4)&= [\hat{g}_{v_2(T) v_3(T)}, \hat{g}_{v_3(T)v_4(T)}, \hat{g}_{v_4(T)v_5(T)}]_h^{-1} [\hat{g}_{v_1(T) v_3(T)}, \hat{g}_{v_3(T)v_4(T)}, \hat{g}_{v_4(T)v_5(T)}]_h^{+1} \\
	& \hspace{2cm} [\hat{g}_{v_1(T) v_2(T)}, \hat{g}_{v_2(T)v_4(T)}, \hat{g}_{v_4(T)v_5(T)}]_h^{-1} [\hat{g}_{v_1(T) v_2(T)}, \hat{g}_{v_2(T)v_3(T)}, \hat{g}_{v_3(T)v_5(T)}]_h^{+1} \\
	&=[\hat{g}_{v_1(T)v_2(T)}, \hat{g}_{v_2(T)v_3(T)} , \hat{g}_{v_3(T)v_4(T)}]_h^{+1} = \theta(1), 
\end{align*}
which is the phase of the single tetrahedron. We therefore see that the phase is unchanged under the bistellar flip. This tells us that we can re-triangulate the interior of $S$ without affecting the phase factor.

\subsubsection{Proof that the bulk factor is invariant under interior vertex transforms}

The final ingredient we will need to find a simple expression for the bulk factor is to show that the bulk phase is invariant under vertex transforms in the interior of the region $S$ (i.e., in $S$ and not on the boundary of $S$). This is just a result of the total operator (the dual phase, bulk phase and untwisted membrane operator) being a product of vertex transforms. This means that the bulk phase only differs from the product of vertex transforms by the untwisted membrane operator and the dual phase, both of which only act on the boundary of the region $S$. That is, using Equation \ref{Equation_prod_vertices_result} the bulk phase can be written as

$$\theta(\text{bulk})= \theta(\text{dual})^{-1} (C_0^h(\partial S))^{-1} \prod_{v' \in S} A_{v'}^h.$$

The vertex transform $A_v^x$ will commute with $\theta(\text{surface})^{-1} (C_0^h(\partial S))^{-1}$ because it does not affect the surface variables (due to the vertex being in the interior). It will also commute with $\prod_{v' \in S} A_{v'}^h$ because vertex transforms commute when $G$ is Abelian \cite{Wan2015}. Therefore, the vertex transform $A_v^x$ commutes with the bulk phase factor. This means that we are free to apply vertex transforms in the interior before evaluating the phase. Because the phase part of $A_v^x$ automatically commutes with the bulk phase (because both are diagonal in the configuration basis, where each edge is labeled by a group element), we can just apply the untwisted vertex transform $A_v^x(0)$ for convenience.

\subsubsection{Statement of the bulk factor}

\begin{figure}[h]
	\centering
	\includegraphics[width=0.5\linewidth]{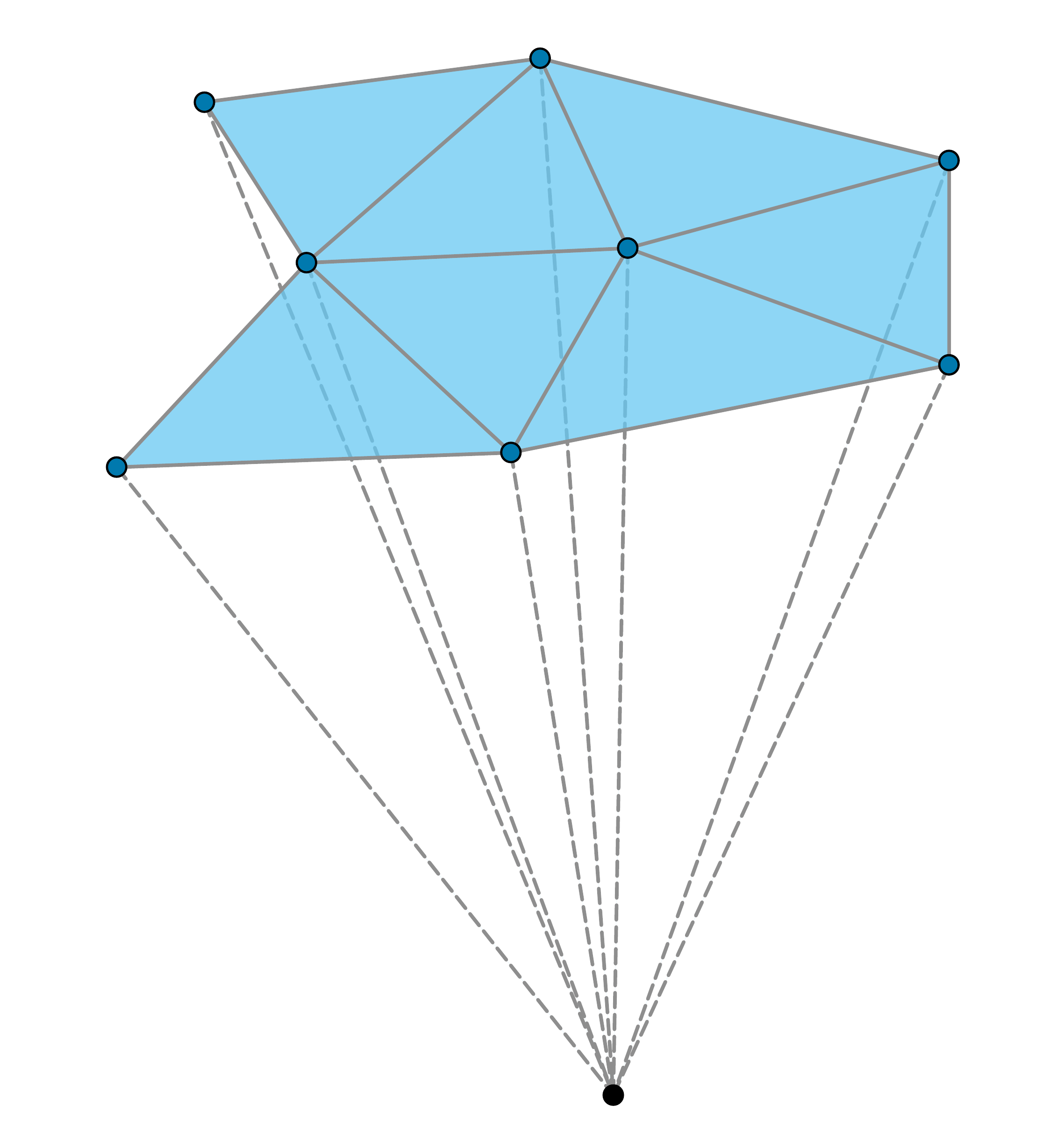}
	\caption{Perhaps the simplest way to complete a spherical surface to a solid region is to add a single internal vertex attached to all vertices on the surface. Here we show a fragment of that surface rather than the whole sphere for clarity. The triangles on the surface become tetrahedrons when the internal vertex is added.}
	\label{Figure_simple_completion}
\end{figure}

We can now use the results from the previous sections to evaluate the bulk factor. We have shown that the bulk factor is invariant under re-indexing, re-triangulation and vertex transforms in the interior of the region $S$. This means that any valid interior of the correct topology that is compatible with the surface will give the same bulk factor. This means that the bulk factor only depends on the surface, allowing the series of vertex transforms to be written as a membrane operator on the surface. For the spherical membrane, we can evaluate the bulk phase, which becomes the surface weight of the membrane operator, by choosing a simple interior. In particular, for a contractible spherical surface we choose the interior to be a single vertex which is attached to every vertex on the surface, as shown in Figure \ref{Figure_simple_completion}. We take the index of this vertex to be larger than any vertex on the surface, so that all of the edges connecting the surface to the interior vertex point inwards. The labels of the edges connecting the interior vertex to the surface are fixed up to an overall vertex transform. Any such choice will give the same factor, as we showed in the previous section. To have a definitive choice, we choose a particular vertex on the surface, which we call the start-point, and give the edge from that vertex to the interior vertex the identity label $1_G$. Then the label from an arbitrary vertex $v_j$ to the interior vertex $v_{\text{interior}}$ must satisfy $g(s.p-v_j)g(v_j-v_{\text{interior}})= g(s.p-v_{\text{interior}})$ by flatness, where $(s.p-v_j)$ is any path from the $s.p$ to $v_j$ on the surface (all choices give the same result due to flatness). Then, using $ g(s.p-v_{\text{interior}})=1_G$, we have $g(v_j-v_{\text{interior}})= g(s.p-v_j)^{-1}$. This is equivalent to gluing the interior vertex to the start-point.

The bulk factor is then
\begin{equation}
	\theta(\text{bulk}) = \prod_{\text{triangles on surface}} [g_{v_1(t) v_2(t)}, g_{v_2(t)v_3(t)}, g(s.p-v_3(t))^{-1}]_h^{\sigma(t)},
\end{equation}
where $\sigma(t)=+1$ if the indices of the vertices on the triangle ascend in the anticlockwise direction with respect to the outwards normal of the surface (and $-1$ for clockwise). Equivalently, $\sigma(t)= - \epsilon(t \cup v_{\text{internal}})$, where $t \cup v_{\text{internal}}$ is the tetrahedron formed by attaching the triangle to the internal vertex. This statement of the bulk factor depends only on the variables on the surface. Indeed, the bulk factor is exactly the surface weight which we described in Section \ref{Section_membrane_operators}. For a spherical surface, there is no choice of reference diagram and the surface weight is entirely defined by the flux label $h$. Combining the results from Section \ref{Section_membranes_from_vertex_transforms} and \ref{Section_bulk_factor}, we see that the product of vertex transforms that we started with can be written as

\begin{align}
	\prod_{v \in S} A_v^h &= C^h_0(\delta S) \prod_{t \in \delta S} \hat{\theta}_D^h(t) \prod_{\text{triangles on surface}} [g_{v_1(t) v_2(t)}, g_{v_2(t)v_3(t)}, g(s.p-v_3(t))^{-1}]_h^{\sigma(t)} \notag \\
	&= C^h_0(\delta S) \hat{\theta}_D^h( \delta S) \theta_S^h(\delta S) \label{Equation_prod_vertices_result_2},
\end{align}
which is the magnetic membrane operator on the boundary of $S$, as we claimed. We see that a spherical magnetic membrane operator is equivalent to a product of vertex transforms. This means that it does not excite any energy terms, because the energy terms commute with vertex transforms. It also implies that the membrane operator can be deformed without affecting its action on the ground state. We can deform the membrane operator on $\partial S$ by adding or removing vertices to the region $S$, which means applying additional vertex transforms. These vertex transforms act trivially on the ground state and so do not change the action of the membrane operator on the ground state. Indeed, the entire membrane operator acts trivially on the ground state, reflecting the fact that a sphere can be contracted to nothing. On the other hand, if the membrane operator acts on a more general state, its action might not be trivial. However, it still has the topological property, meaning can be deformed through an unexcited region (i.e., as long as the deformation does not cause the membrane to pass through any excitations). We will prove this when we discuss more general membranes in Section \ref{Section_more_general_membranes}.

\subsection{Graphical rules}

In Section \ref{Section_membrane_operators} of the main text, we described how the surface weight for a general membrane operator can be calculated using graphical rules. Then, in Section \ref{Section_spherical_membranes} we explained how the surface weight for a spherical membrane can be derived from the phases associated to tetrahedra enclosed by the membrane. Here we will discuss how these two pictures are related. Given a spherical surface $C$, its surface weight can be written as a product over tetrahedra in the region $R$ enclosed by the surface as
$$\theta_S(C)= \prod_{t \in R} \theta_t.$$
We can then separate out the contribution from one tetrahedron $t_r$ near the boundary to obtain
$$\theta_S(C)= \theta_{t_r} \prod_{t \in R'} \theta_t,$$
where $R'$ is the region $R$ but with $t_r$ removed. Then the boundary of $R'$ is some new surface $C'$, which will typically be spherical if we remove a tetrahedron from the boundary. Then
\begin{equation}
	\theta_S(C)= \theta_{t_r} \theta_S(C'). \label{Equation_surface_phase_deform_1}
\end{equation}
That is, we can relate the surface weight of the membrane $C$ to the surface weight of the membrane obtained by removing a tetrahedron from the enclosed region. Equivalently, $C'$ is the membrane obtained by deforming $C$ over $t_r$. As we described in Section \ref{Section_membrane_operators} of the main text, these deformations over a tetrahedron induce the 2d bistellar flips moves shown in Figure \ref{Figure_3d_deform_to_2d_move}, giving us the graphical rules described by Figure \ref{Figure_2d_graphical_rules}.

While Equation \ref{Equation_surface_phase_deform_1} refers to the specific tetrahedron $t_r$, $\theta_S(C)$ does not actually depend on the region $R$ enclosed by $C$, only on the surface itself (as we proved in Section \ref{Section_bulk_factor}). This means that we can use the graphical rules to evaluate the surface weight by simplifying the surface, even if the physical lattice does not contain the requisite tetrahedra (we can construct a ``virtual" bulk). Indeed, as we described in Section \ref{Section_membrane_operators} of the main text and prove in Section \ref{Section_more_general_membranes}, we can take the graphical rules as a starting point to define the surface weight, even for more general surfaces than spheres.

\section{Membranes of General Topology}
\label{Section_more_general_membranes}

In Section \ref{Section_membrane_operators} of the main text, we defined the magnetic membrane operators in terms of an untwisted operator, a dual phase and a surface weight. In this section, we will prove that these are indeed appropriate membrane operators, not just for spherical surfaces. To do so, we will show that they have the properties required by membrane operators. Namely, they are topological, meaning that they can be deformed (while keeping their boundary fixed) without changing their action, as long as they are not deformed over any existing excitations. Additionally, they do not excite any energy terms, except at their boundaries.

As stated in Equation \ref{Equation_membrane_operator_definition} of the main text, a magnetic membrane operator with flux label $h$ is given by
\begin{equation}
	F^{h, \vec{v}}(m)=C_0^h(m) \theta_D^h(m) \theta_S^{h, \vec{v}}(m),
\end{equation}
where $\vec{v}$ assigns a weight to the possible reference diagrams (which are a minimal set of diagrams that we can reduce the surface to by using the graphical rules). We start by showing that such an operator is topological. To do this, we will show that deforming the membrane operator is equivalent to applying vertex transforms, which act trivially on states where the vertices are not excited. That is, we will show that for two membranes $m$ and $m'$ related by deformation, we have
\begin{equation}
	F^{h, \vec{v}}(m')=F^{h, \vec{v}}(m) \prod_{v \in R} A_v^{h^{\pm 1}},
\end{equation}
where $R$ is the region over which we deform the membrane $m$ to obtain $m'$. Note that one example of this is that spherical membrane operators are equivalent to a series of vertex transforms, as we showed in Section \ref{Section_spherical_membranes}, and so can be deformed to nothing by applying the inverse of this series of transforms.

Firstly, we must consider how the action of the membrane operator changes as we deform it. Consider pulling the direct membrane over a single tetrahedron (larger deformations can then be built out of these smaller moves). The graphical rules tell us how the surface (diagrammatic) phase changes during this process. However we also need to consider the dual phase, which depends on the position of the dual membrane. Recall that for each tetrahedron cut by the dual membrane, the membrane operator applies a phase which depends on which vertices on the tetrahedron are on the direct membrane (see Section \ref{Section_membrane_operators} in the main text).

\begin{figure}
	\centering
	\includegraphics[width=0.7\linewidth]{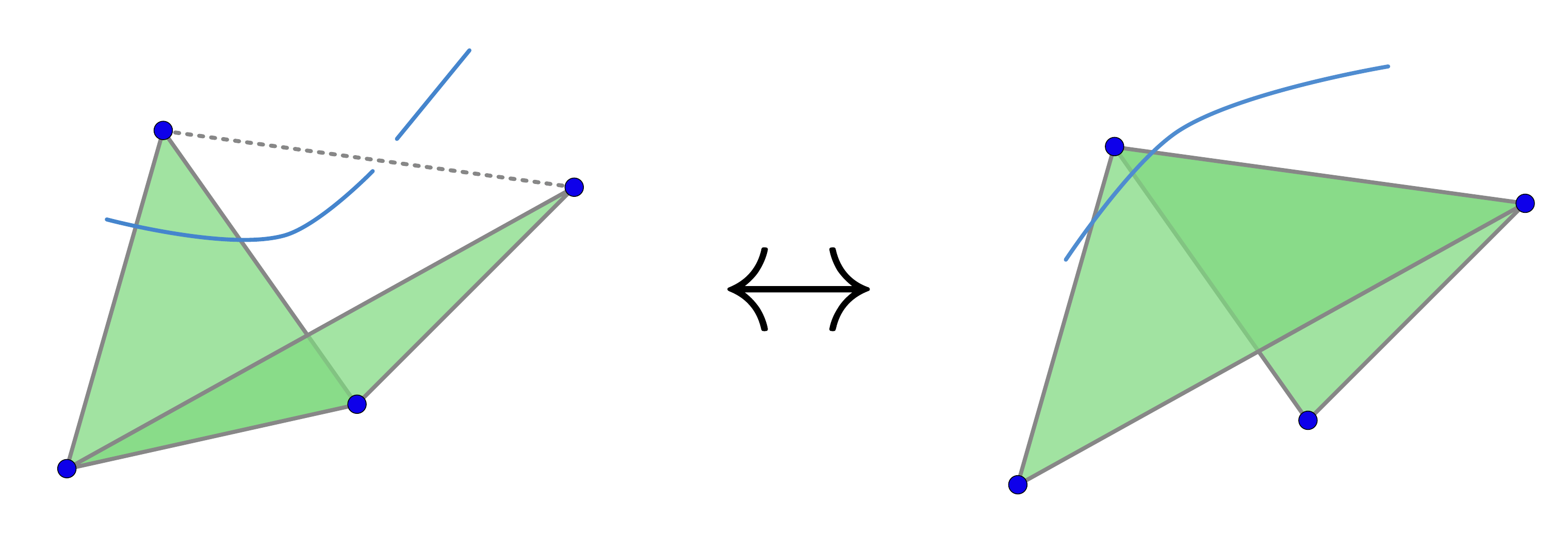}
	\caption{The $2 \leftrightarrow 2$ bistellar flip corresponds to pulling the membrane over a tetrahedron which has two faces on the direct membrane (the green ones). The blue arc represents part of the dual membrane (on the left side, it cuts the tetrahedron, whereas on the right side, it does not).}
	\label{Figure_membrane_deformation_1}
\end{figure}

The first category of local deformations are the ones that induce a $2 \leftrightarrow 2$ bistellar flip on the surface. These involve boundary tetrahedra where two of the faces are on the direct membrane already. Such a move is shown in Figure \ref{Figure_membrane_deformation_1}.
For the case on the left side of Figure \ref{Figure_membrane_deformation_1}, the tetrahedron contributes to the dual phase because it is cut by the dual membrane. Because every vertex of the tetrahedron is on the direct membrane, this phase is given by $[g_{v_1(t)v_2(t)}, g_{v_2(t)v_3(t)}, g_{v_3(t)v_4(t)}]_h^{- \epsilon(t)}$, where $h$ is the label of the membrane operator. On the right-hand side, the dual membrane does not cut the tetrahedron so there is no such dual phase. However, the graphical rules tell us that the surface weight should be multiplied by the phase of the tetrahedron relative to the left-hand side (because we have to deform the membrane over that tetrahedron to get back to the original surface). This means that the phase associated to that tetrahedron is simply moved between the surface weight and the dual phase, but the overall phase is preserved by the deformation. That is, the product of the surface weight and dual phase is preserved by the deformation. This just leaves the action of the untwisted membrane operator, which multiplies the edges cut by the dual membrane by $h$ or $h^{-1}$ depending on which vertex on the edge lies on the direct membrane. Specifically, if the source of the edge lies on the direct membrane the edge label is multiplied by $h$, while if the target is on the direct membrane the label is multiplied by $h^{-1}$. The only edge that may be affected by this is the dotted edge shown in the left side of Figure \ref{Figure_membrane_deformation_1}, because this edge is cut by the dual membrane on the left side of the deformation but not on the right. However, this edge has both source and target on the direct membrane, meaning that the edge is multiplied by both $h$ and $h^{-1}$ and so is unaffected by the membrane operator even though it is cut by the dual membrane. Therefore the action of the untwisted membrane operator is also unaffected by the deformation. Therefore the action of the membrane operator is preserved by this deformation of the direct and dual membranes.

\begin{figure}
	\centering
	\includegraphics[width=0.7\linewidth]{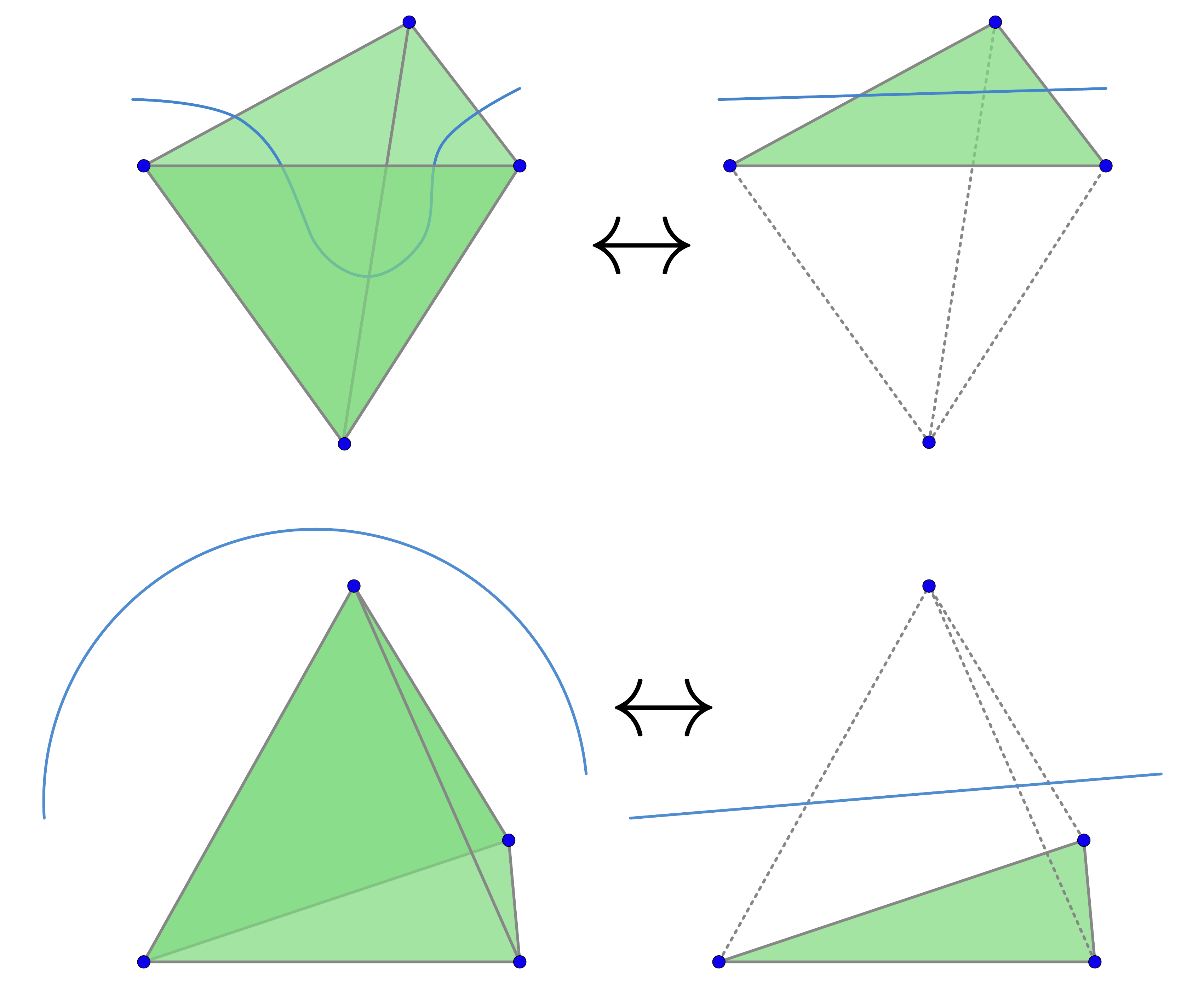}
	\caption{For the $1 \leftrightarrow 3$ move, we must consider two cases. In the first (top) we deform the direct membrane over a boundary tetrahedron (one cut by the dual membrane, represented by the blue line). In the second (bottom) we deform the direct membrane over a ``bulk" tetrahedron (one ``below" the direct membrane, meaning on the other side compared to the dual membrane).}
	\label{Figure_membrane_deformation_2}
\end{figure}

Now consider the type of deformation which results in the $1 \leftrightarrow 3 $ bistellar flip. In this case, we deform the membrane operator over a tetrahedron where either one or three of the faces were originally on the direct membrane. For simplicity, we consider the $3 \rightarrow 1$ move. That is, we assume that three faces are on the direct membrane prior to deformation and one face is on it afterwards. To treat the reverse direction we simply need to reverse this process. Given that we consider the $3 \rightarrow 1$ move, we must differentiate between two cases, one where the tetrahedron in question is initially ``above" the direct membrane, meaning that it is cut by the dual membrane prior to deformation (as shown in the top line of Figure \ref{Figure_membrane_deformation_2}) and one where the tetrahedron is initially ``below" the direct membrane, meaning that is is not cut by the dual membrane before the deformation, only afterward (as shown in the bottom line of Figure \ref{Figure_membrane_deformation_2}). 

In the first case, prior to deformation all four vertices of the tetrahedron are on the direct membrane. This means that the dual phase associated to that tetrahedron is $[g_{v_1(t)v_2(t)}, g_{v_2(t)v_3(t)}, g_{v_3(t)v_4(t)}]_h^{- \epsilon(t)}$. Just as in the $2 \leftrightarrow 2$ case we considered previously, when we deform the membrane over that tetrahedron this dual phase is removed because the dual membrane no longer cuts the tetrahedron, but the surface (diagrammatic) phase gains a compensating phase due to the graphical rules. Therefore, the action of the membrane operator is preserved by the deformation.

The second case is more complicated. Prior to the deformation, the tetrahedron is not cut by the dual membrane, meaning that it does not contribute to the dual phase. On the other hand, after the deformation it is cut by the dual membrane and so it contributes to the dual phase. Furthermore, the deformation removes a vertex from the direct membrane and causes it to move to the other side of the dual membrane. This may affect the dual phase from other tetrahedra that are attached to that vertex, because the dual phase from a tetrahedron depends on which of its vertices lie on the direct membrane. Finally, the surface phase is changed by the deformation and this change can be calculated from the graphical rules. To determine how these factors change the total membrane operator, we can consider this entire process in two steps.

Step 1: We add the ``bulk" tetrahedron (the one ``underneath" the direct membrane) to the list of boundary tetrahedra (those cut by the dual membrane), with all four of its vertices treated as being on the direct membrane. We also change the surface weight to only include the single face from the tetrahedron (i.e., perform the $3 \rightarrow 1$ move). This does not change the overall phase or the multiplication action on the edges, because the additional contribution from the tetrahedron to the dual phase is balanced by the change to the surface weight (just as we discussed for the $2 \leftrightarrow 2$ moves).

Step 2: Remove the contribution from the additional vertex (the one which is on the direct membrane before the deformation, but not afterwards) to the dual phase, by recalculating the phase for all boundary tetrahedra that include that vertex.

The first step does not affect the action of the membrane operator, so we need to examine this second step. We claim that the change to the action of the membrane operator from that step is equivalent to the action of a vertex transform $A_v^{h^{-1}}$ on the vertex that we delete from the membrane (applied before the membrane operator). That is, we claim that if $m$ is the membrane before the deformation, $m'$ is the membrane after deformation and $w$ is the additional vertex, the membrane operator satisfies $F^{h, \vec{v}}(m')=F^{h, \vec{v}}(m) A_w^{h^{- 1}}$. To establish this, first note that $A_w^{h^{-1}}=(A_w^h)^{-1} = \big(\prod_{t \ni w} (\theta_{w,t}^h)^{-1} \big)A_w^{h^{-1}}(0)$, where $A_w^{h^{-1}}(0)$ is an untwisted vertex transform. Then, expanding the membrane operator into the untwisted part, the dual phase and the surface weight using Equation \ref{Equation_membrane_operator_definition} of the main text, we have
\begin{align*}
	F^{h, \vec{v}}(m) A_w^{h^{- 1}} &= C_0^h(m) \theta_D^h(m) \theta_S^{h, \vec{v}}(m) A_w^{h^{- 1}} \\
	&= C_0^h(m) \theta_D^h(m) \theta_S^{h, \vec{v}}(m) \big(\prod_{t \ni w} {\theta_{w,t}^h}^{-1} \big) A_w^{h^{- 1}}(0).
\end{align*}
Then we can insert the identity in the form $A_w^{h^{-1}}(0) A_w^h(0)$ to obtain
\begin{align*}
	F^{h, \vec{v}}(m) A_w^{h^{- 1}} &= C_0^h(m)A_w^{h^{-1}}(0) A_w^h(0)\theta_D^h(m) \theta_S^{h, \vec{v}}(m) \big(\prod_{t \ni w} {\theta_{w,t}^h}^{-1} \big) A_w^{h^{- 1}}(0).
\end{align*}
We do this because $A_w^h(0)\theta_D^h(m) \theta_S^{h, \vec{v}}(m) \big(\prod_{t \ni w} {\theta_{w,t}^h}^{-1} \big) A_w^{h^{- 1}}(0)$ is then a pure phase or weight (in the configuration basis), which we want to show corresponds to the dual phase and surface weight of the deformed membrane operator. On the other hand, $C_0^h(m)A_w^{h^{-1}}(0)$ contains no phase, only a multiplicative action on the edges, and we claim this is equal to the untwisted part of the deformed membrane operator.

First, note that the vertex transform correctly alters the untwisted part of the membrane operator: the direct multiplicative action on the edges cut by the dual membrane. The vertex transform multiplies any adjacent edge which points away from it by $h^{-1}$ and any adjacent edge which points towards it by $h$. This cancels the action of the membrane operator (which multiplies edges that are cut by the dual membrane by $h$ if they point away from the direct membrane and by $h^{-1}$ if they point towards it) for all edges that are both attached to $w$ and were cut by the dual membrane (note that $w$ is initially on the direct membrane, so any edges which point towards $w$ also point towards the direct membrane initially). On the other hand, any edges that are adjacent to $w$ but which were not originally cut by the dual membrane (because they lie on the direct membrane) only experience the action of the vertex transform. This reproduces the action of the membrane after deformation, because any such edge which points towards $w$ (and so is multiplied by $h$ by the vertex transform) points away from the direct membrane after the deformation (as $w$ is now above the direct membrane) and so would be multiplied by $h$ by the deformed membrane operator. Similarly, any such edge which points away from $w$ would be multiplied by $h^{-1}$ by the vertex transform, and similarly would be multiplied by $h^{-1}$ by the membrane operator because the edge points towards the deformed direct membrane.

We therefore see that applying the vertex transform in sequence with the undeformed membrane operator correctly reproduces the direct multiplication action of the deformed membrane operator, but we also need to show that it reproduces the phase of the deformed membrane operator. That is, we want to show that $$A_w^h(0)\theta_D^h(m) \theta_S^{h, \vec{v}}(m) \big(\prod_{t \ni w} {\theta_{w,t}^h}^{-1} \big) A_w^{h^{- 1}}(0) = \theta_D^h(m') \theta_S^{h, \vec{v}}(m').$$ As we stated earlier, by using the graphical rules we can extract the phase corresponding to the tetrahedron over which we deform the membrane from the surface phase, to obtain $\theta_S^{h, \vec{v}}(m) = \theta_D^h(t_{\text{extra}})\theta_S^{h, \vec{v}}(m')$. Therefore,
$$A_w^h(0)\theta_D^h(m) \theta_S^{h, \vec{v}}(m) \big(\prod_{t \ni w} {\theta_{w,t}^h}^{-1} \big) A_w^{h^{- 1}}(0) = A_w^h(0)\theta_D^h(m)\theta_D^h(t_{\text{extra}}) \theta_S^{h, \vec{v}}(m') \big(\prod_{t \ni w} {\theta_{w,t}^h}^{-1} \big) A_w^{h^{- 1}}(0) .$$
Because the deformed direct membrane does not include $w$, $\theta_S^{h, \vec{v}}(m')$ now commutes with the vertex transform, giving us
$$A_w^h(0)\theta_D^h(m) \theta_S^{h, \vec{v}}(m) \big(\prod_{t \ni w} {\theta_{w,t}^h}^{-1} \big) A_w^{h^{- 1}}(0) = \theta_S^{h, \vec{v}}(m') A_w^h(0)\theta_D^h(m)\theta_D^h(t_{\text{extra}}) \big(\prod_{t \ni w} {\theta_{w,t}^h}^{-1}\big) A_w^{h^{- 1}}(0) .$$
Having extracted the surface weight, we now want to show that
$$A_w^h(0)\theta_D^h(m)\theta_D^h(t_{\text{extra}}) \big(\prod_{t \ni w} {\theta_{w,t}^h}^{-1} \big) A_w^{h^{- 1}}(0) $$
is equivalent to the dual phase of the deformed membrane operator. Recall from Section \ref{Section_membrane_operators} of the main text and Section \ref{Section_spherical_membranes} of the Supplemental Material that the dual phase for each boundary tetrahedron is the same phase that would be obtained from that tetrahedron by applying a vertex transform on each vertex on the direct membrane (in ascending index order): $$\theta_D^h(t) = \big(\prod_{v \in t \cap m} A_v^h(0)^{-1}\big) \big( \prod_{v \in t \cap m} A_v^h(0) \theta_{v,t}^h \big).$$
Because each tetrahedron dual phase commutes with vertex transforms outside of the tetrahedron, we can extend the product over vertices to all vertices in the direct membrane $m$:
$$\theta_D^h(t) = \big(\prod_{v \in m} A_v^h(0)^{-1}\big) \big( \prod_{v \in m} A_v^h(0) \theta_{v,t}^h\big),$$
where we define $\theta_{v,t}^h$ to be 1 if $v$ is not on the tetrahedron $t$. The total dual phase is then a product over all of these terms. Because the vertex terms $A_v^h(0)$ which act to the right of a given term $\theta_{v',t}^h$ in the product $ \prod_{v \in m} A_v^h(0) \theta_{v,t}^h$ depend on $v'$ and not $t$, we can collect the terms corresponding to each vertex. This is the inverse of the process where we separated out the dual phase of the spherical membrane into the factors corresponding to each tetrahedron in Section \ref{Section_membranes_from_vertex_transforms}. By doing this, we can write the total dual phase in the same way as for an individual tetrahedron:
$$\theta_D^h(m) = \big(\prod_{v \in m} A_v^h(0)^{-1}\big) \big( \prod_{v \in m} A_v^h(0) \prod_{t \in D(v,m) }\theta_{v,t}^h\big),$$
where $D(v,m)$ is the set of tetrahedra adjacent to $v$ and cut by the dual membrane of $m$. Therefore, we have
\begin{align*}
	A_w^h(0)&\theta_D^h(m)\theta_D^h(t_{\text{extra}}) \big(\prod_{t \ni w} {\theta_{w,t}^h}^{-1} \big) A_w^{h^{- 1}}(0)\\
	&= A_w^h(0)\big(\prod_{v \in m} A_v^h(0)^{-1}\big) \big( \prod_{v \in m} A_v^h(0) \prod_{t \in D(v,m) } \theta_{v,t}^h \big)\theta_D^h(t_{\text{extra}}) \big(\prod_{t \ni w} {\theta_{w,t}^h}^{-1} \big) A_w^{h^{- 1}}(0). 
\end{align*}
We can then write $\theta_D^h(t_{\text{extra}})$ as
$$\theta_D^h(t_{\text{extra}}) = \big(\prod_{v \in m} A_v^h(0)^{-1}\big) \big( \prod_{v \in m} A_v^h(0) \theta_{v, t_{\text{extra}}}^h \big),$$
using the fact that all vertices on $t_{\text{extra}}$ are already in the direct membrane of $m$. This allows us to incorporate $\theta_D^h(t_{\text{extra}})$ into the rest of the dual phase, just like we combined the dual phase of the other tetrahedra, by including the extra tetrahedron in $D(v,m)$, to obtain the new set of tetrahedra $D(v,m)'$. After doing this, $D(v,m)'$ contains tetrahedra cut by the dual membrane of the deformed membrane $m'$, except it may include extra tetrahedra that are attached to the vertex $w$ which will be removed from the direct membrane, but not any other vertices on $m$ (these erroneous tetrahedra will be removed by the next step, where $w$ is removed from the product over vertices).
\begin{equation}
	A_w^h(0)\theta_D^h(m)\theta_D^h(t_{\text{extra}}) \big(\prod_{t \ni w} {\theta_{w,t}^h}^{-1}\big) A_w^{h^{- 1}}(0) = A_w^h(0)\big(\prod_{v \in m} A_v^h(0)^{-1}\big) \big( \prod_{v \in m} A_v^h(0) \prod_{t \in D(v,m) '} \theta_{v,t}^h \big)\big(\prod_{t \ni w} {\theta_{w,t}^h}^{-1}\big) A_w^{h^{- 1}}(0). 
\end{equation}
We can then cancel the vertex transform $\big(\prod_{t \ni w} {\theta_{w,t}^h}^{-1} \big) A_w^{h^{- 1}}(0)$ with the corresponding term $A_v^h(0) \prod_{t \in D(v,m) '} \theta_{v,t}^h$ for $v=w$ on the right of this expression, along with cancelling $A_w^h(0)$ with $A_w^h(0)^{-1}$. This is the same as deleting $w$ from the list of vertices on $m$, matching the procedure described in Step 2 and so matching the deformed membrane operator:
\begin{equation}
	A_w^h(0)\theta_D^h(m)\theta_D^h(t_{\text{extra}}) \big(\prod_{t \ni w} {\theta_{w,t}^h}^{-1}\big) A_w^{h^{- 1}}(0) = \big(\prod_{v \in m'} A_v^h(0)^{-1}\big) \big( \prod_{v \in m'} A_v^h(0) \prod_{t \in D(v,m') } \theta_{v,t}^h \big)= \theta_D^h(m').
\end{equation}
In particular, note that the factors corresponding to any tetrahedra that only had vertex $w$ on the direct membrane are removed (because $w$ is removed from the product over vertices), which matches the action of the deformed membrane operator because these tetrahedra are no longer cut by the dual membrane. There is a small subtlety to the cancellation of factors corresponding to $w$, however because the inverse transform is applied to the right of the expression (before the membrane operator). However, the product over vertices for $\big( \prod_{v \in m} A_v^h(0) \prod_{t \in D(v,m) '} \theta_{v,t}^h \big)$ is applied in a fixed order, namely in order of ascending vertex, and so the relevant term cannot be freely moved to the right to cancel with the factors from $A_w^{h^{-1}}$. If $w$ is the lowest indexed vertex on $t$, we can directly cancel the right-most term of the product, which will be $A_w^h(0)\theta_{w,t}$ with ${\theta_{w,t}^h}^{-1} A_w^h(0)^{-1}$. However, if $w$ is not the lowest indexed vertex, we cannot just cancel the terms because there will be other terms from the product in between them and they cannot be directly cancelled. The individual terms $A_v^h(0) \theta_{v,t}$ do not commute with each-other. The full vertex transforms $A_v^h(0)\prod_{t \ni v} \theta_{v,t}$ do commute \cite{Wan2015}, but the dual phase does not include all of these factors because the dual membrane only cuts some of the tetrahedra, those on one side of the direct membrane. However, the support of the vertex transform $(A_w^h)^{-1}$ only lies on that same side of the direct membrane. This means that the missing part of the other vertex transforms, which is outside the support of $(A_w^h)^{-1}$, cannot affect the commutation relation. Therefore we can commute $(A_w^h)^{-1}$ to be adjacent to the appropriate factor of $A_w^h$ and cancel its contributions as described by Step 2. Applying the vertex transform therefore deforms the membrane operator as claimed.

Recall that we considered the deformation corresponding to the $3 \rightarrow 1$ move, but have not yet considered the reverse move $1 \rightarrow 3$. To go in the reverse direction (i.e. to go leftwards in the bottom line of Figure \ref{Figure_membrane_deformation_2}) we would reverse Steps 1 and 2: we would first add the additional vertex to the list of vertices on the direct membrane and then move the tetrahedron from the boundary to the ``bulk". This would be equivalent to applying the vertex transform $A_v^h$ rather than the inverse transform.

The fact that deforming the membrane is at most equivalent to applying vertex transforms, which act trivially on the ground state, means that the membrane operator is topological on the ground state. This also means that the membrane operator does not excite energy terms (except on the boundary of the membrane) because the position of the membrane (when applied on the ground state) cannot be detected by any operator, including the energy terms. However, the boundary must be kept fixed when deforming it, so terms on the boundary can and will be excited. In particular, if we close a membrane by bringing its edges together (for example, by closing a cylinder into a torus), we may need to apply additional conditions to ensure that the graphical rules are satisfied at the ``seam" where the membrane joins itself (because the open membrane would not have to satisfy the graphical rules at the boundary). If we do not do this, the seam may host excitations and cannot be deformed using the topological property of the membrane operator.

\section{Topological Charge}
\label{Section_supplement_topological_charge}

In Section \ref{Section_topological_charge} in the main text, we described the topological charge measurement projectors, but did not prove that they are orthogonal projectors or span the basis of measurement operators. In this section we will prove these results. Recall that each projector is labeled by group elements $g_1$ and $g_2$ of $G$ as well as a $\beta_{g_1, g_2}$-projective irrep $\alpha^{g_1,g_2}$ of $G$. These irreps satisfy
$$\alpha^{g_1,g_2}(x) \alpha^{g_1,g_2}(y)= [x,y]_{g_1, g_2}\alpha^{g_1,g_2}(xy).$$
The projectors are then given by
\begin{align}
	P^{g_1, g_2, \alpha^{g_1,g_2}}(m) &= \frac{1}{\sqrt{|G||G_{g_1, g_2}|}} \sum_{h \in G_{g_1, g_2}} \chi_{\alpha^{g_1,g_2}}(h) T^{g_1, g_2, h}(m) \notag \\
	&= \frac{1}{\sqrt{|G||G_{g_1, g_2}|}} \sum_{h \in G_{g_1, g_2}} \chi_{\alpha^{g_1,g_2}}(h)C^h(m) \delta(\hat{g}(c_1), g_1) \delta(\hat{g}(c_2),g_2). \label{Equation_charge_projectors_2}
\end{align}
Here $G_{g_1,g_2}$ is the subgroup of $G$ consisting of $\beta_{g_1,g_2}$-regular elements (where an element $h\in G$ is $\beta_{g_1,g_2}$-regular if $[h,x]_{g_1,g_2}=[x,h]_{g_1,g_2}$ for all $x \in G$) and $\chi_{\alpha^{g_1,g_2}}$ is the character for irrep $\alpha^{g_1,g_2}$ (i.e., the trace of the representative matrices). In Section \ref{Section_topological_charge} of the main text, we showed that the $T^{g_1,g_2,h}(m)$ operators, for the set of triples $g_1$, $ g_2$ in $G$ and $h $ in $G_{g_1,g_2}$, form a basis for the topological measurement operators (although they are not projectors). The map to the projectors is then a change of basis for the space of measurement operators. This change of basis is invertible as required, with the inverse transformation
\begin{align}
	T^{g_1, g_2, k}(m) &= \sqrt{\frac{|G_{g_1,g_2}|}{|G|}} \sum_{\alpha^{g_1,g_2} \in A^{g_1,g_2}(G) } \chi_{\alpha^{g_1,g_2}}(k)^* P^{g_1, g_2, \alpha^{g_1,g_2}}(m),
\end{align}
where $A^{g_1,g_2}(G)$ is the set of $\beta_{g_1,g_2}$-projective irreps of $G$. We can verify that this is the inverse mapping by direct substitution:
\begin{align*}
	\sqrt{\frac{|G_{g_1,g_2}|}{|G|}}& \sum_{\alpha^{g_1,g_2} \in A^{g_1,g_2}(G) } \chi_{\alpha^{g_1,g_2}}(k)^* P^{g_1, g_2, \alpha^{g_1,g_2}}(m) \\
	&= \sqrt{\frac{|G_{g_1,g_2}|}{|G|}} \sum_{\alpha^{g_1,g_2} \in A^{g_1,g_2}(G) } \chi_{\alpha^{g_1,g_2}}(k)^* \frac{1}{\sqrt{|G||G_{g_1, g_2}|}} \sum_{h \in G_{g_1, g_2}} \chi_{\alpha^{g_1,g_2}}(h)C^h(m) \delta(\hat{g}(c_1), g_1) \delta(\hat{g}(c_2),g_2)\\
	&=\sqrt{\frac{|G_{g_1,g_2}|}{|G|}} \frac{1}{\sqrt{|G||G_{g_1, g_2}|}} \sum_{h \in G_{g_1, g_2}} \big(\sum_{\alpha^{g_1,g_2} \in A^{g_1,g_2}(G) } \chi_{\alpha^{g_1,g_2}}(k)^* \chi_{\alpha^{g_1,g_2}}(h) \big)C^h(m) \delta(\hat{g}(c_1), g_1) \delta(\hat{g}(c_2),g_2).
\end{align*}
Then, using the second part of Equation 54 in Ref. \cite{Wan2015}, which is analogous to the column orthogonality for characters of linear irreps, we have
$$\big(\sum_{\alpha^{g_1,g_2} \in A^{g_1,g_2}(G) } \chi_{\alpha^{g_1,g_2}}(k)^* \chi_{\alpha^{g_1,g_2}}(h) \big)= |G| \delta(h,k)$$
and so
\begin{align*}
	\sqrt{\frac{|G_{g_1,g_2}|}{|G|}} \sum_{\alpha^{g_1,g_2} \in A^{g_1,g_2}(G) }& \chi_{\alpha^{g_1,g_2}}(k)^* P^{g_1, g_2, \alpha^{g_1,g_2}}(m)\\
	&= \sqrt{\frac{|G_{g_1,g_2}|}{|G|}} \frac{1}{\sqrt{|G||G_{g_1, g_2}|}} \sum_{h \in G_{g_1, g_2}} \big(|G| \delta(h,k) \big) C^h(m) \delta(\hat{g}(c_1), g_1) \delta(\hat{g}(c_2),g_2)\\
	&= \sqrt{\frac{|G_{g_1,g_2}|}{|G|}} \frac{\sqrt{|G|}}{\sqrt{|G_{g_1, g_2}|}} C^k(m) \delta(\hat{g}(c_1), g_1) \delta(\hat{g}(c_2),g_2)\\
	&= T^{g_1, g_2, k}(m),
\end{align*}
as we claimed. This establishes an invertible mapping between the original basis and the new set of operators, so the new operators form a valid basis.

Before we show that the operators in Equation \ref{Equation_charge_projectors_2} are indeed orthogonal projectors, we must discuss some of the properties of the $\beta_{g_1, g_2}$-projective irreps. Note that in the projector we only sum over elements $h \in G_{g_1, g_2}$ (recall from Section \ref{Section_topological_charge} of the main text that having $h$ outside the subgroup $G_{g_1, g_2}$ of $\beta_{g_1, g_2}$-regular elements of $G$ would lead to vertex excitations). However, we can extend the sum to all elements of $G$ because the character is zero for elements outside of this subgroup (as shown in Ref. \cite{Wan2015}). In addition, the representative matrices are diagonal for elements inside the subgroup. To see these facts, consider the expression $\alpha^{g_1,g_2}(x) \alpha^{g_1,g_2}(y) \alpha^{g_1,g_2}(x)^{-1}$, which is a similarity transformation on the matrix $\alpha^{g_1,g_2}(y)$. From Equation \ref{Equation_proj_inverse} in the main text, we have
\begin{equation}
	\alpha^{g_1,g_2}(x)^{-1} = [x,x^{-1}]_{g_1, g_2}^{-1} \alpha^{g_1,g_2}(x^{-1}). \label{Equation_projective_irrep_inverse}
\end{equation}
Then, 
\begin{align*}
	\alpha^{g_1,g_2}(x) \alpha^{g_1,g_2}(y) \alpha^{g_1,g_2}(x)^{-1}&= [x,x^{-1}]_{g_1, g_2}^{-1} 	\alpha^{g_1,g_2}(x) \alpha^{g_1,g_2}(y) \alpha^{g_1,g_2}(x^{-1})\\
	&= [x,x^{-1}]_{g_1, g_2}^{-1} [x,y]_{g_1, g_2} \alpha^{g_1,g_2}(xy) \alpha^{g_1,g_2}(x^{-1})\\
	&= [x,x^{-1}]_{g_1, g_2}^{-1} [x,y]_{g_1, g_2} [xy,x^{-1}]_{g_1, g_2} \alpha^{g_1,g_2}(xyx^{-1})\\
	&=[x,x^{-1}]_{g_1, g_2}^{-1} [x,y]_{g_1, g_2} [yx,x^{-1}]_{g_1, g_2} \alpha^{g_1,g_2}(y),
\end{align*}
where for the last equality we used the Abelian nature of $G$. Then, applying the 2-cocycle condition (Equation \ref{Equation_2_cocycle_condition} in the main text), we have
$$ [yx,x^{-1}]_{g_1, g_2} [x,x^{-1}]_{g_1, g_2}^{-1} = [y,x]_{g_1, g_2}^{-1} [y, xx^{-1}]_{g_1, g_2}= [y,x]_{g_1, g_2}^{-1}. $$
Therefore,
\begin{align*}
	\alpha^{g_1,g_2}(x) \alpha^{g_1,g_2}(y) \alpha^{g_1,g_2}(x)^{-1}&= \frac{ [x,y]_{g_1, g_2}}{[y,x]_{g_1, g_2}} \alpha^{g_1,g_2}(y).
\end{align*}
If $y$ is an element of $G_{g_1, g_2}$, then by definition $\frac{ [x,y]_{g_1, g_2}}{[y,x]_{g_1, g_2}} =1$ for all $x \in G$ and so 
$$	\alpha^{g_1,g_2}(x) \alpha^{g_1,g_2}(y) \alpha^{g_1,g_2}(x)^{-1} = \alpha^{g_1,g_2}(y).$$
That is, the matrix $\alpha^{g_1,g_2}(y)$ commutes with all $\alpha^{g_1,g_2}(x)$ and so Schur's lemma, which still holds for projective irreps \cite{Melnikov2022}, implies that $\alpha^{g_1,g_2}(y)$ is a scalar multiple of the identity matrix. On the other hand, if $y$ is not in $G_{g_1, g_2}$ then by definition there is some $x \in G$ for which $\frac{ [x,y]_{g_1, g_2}}{[y,x]_{g_1, g_2}} \neq 1$ and is instead some non-trivial phase. Denoting this phase by $\theta$, we have
$$\alpha^{g_1,g_2}(x) \alpha^{g_1,g_2}(y) \alpha^{g_1,g_2}(x)^{-1} = \theta \alpha^{g_1,g_2}(y).$$
Taking the trace of both sides, we have
$$\text{Tr}(\alpha^{g_1,g_2}(x) \alpha^{g_1,g_2}(y) \alpha^{g_1,g_2}(x)^{-1} ) = \theta \text{Tr}(\alpha^{g_1,g_2}(y)).$$
Using the cyclic property of the trace and the definition of the character as the trace of the matrix representation, this implies that
$$\chi_{\alpha^{g_1,g_2}}(y) = \theta \chi_{\alpha^{g_1,g_2}}(y),$$
for some phase $\theta \neq 1$. This means that $\chi_{\alpha^{g_1,g_2}}(y)$ must be equal to zero. The character is therefore zero for group elements outside of $G_{g_1, g_2}$ \cite{Wan2015}. This means that we can extend the sum over elements $h \in G_{g_1, g_2}$ in the expression for the projector to a sum over all elements $h \in G$ without affecting the operator at all.

These characters satisfy the orthogonality condition \cite{Wan2015, Melnikov2022} 
$$\frac{1}{|G|}\sum_{g \in G} \chi_{\alpha^{g_1,g_2}}(g)^* \chi_{\gamma^{g_1,g_2}} (g) = \delta(\alpha^{g_1,g_2}, \gamma^{g_1,g_2}).$$
Because the characters are zero for elements outside $G_{g_1, g_2}$, we can restrict the sum to this subgroup to obtain
$$\frac{1}{|G|}\sum_{g \in G_{g_1, g_2}} \chi_{\alpha^{g_1,g_2}}(g)^* \chi_{\gamma{g_1,g_2}}(g) = \delta(\alpha^{g_1,g_2}, \gamma{g_1,g_2}).$$
This is interesting because the matrix representation for an element $g$ inside $G_{g_1, g_2}$ is just a multiple of the identity. Denoting the dimension of the irrep $\alpha^{g_1,g_2}$ by $|\alpha^{g_1,g_2}|$, we then have $\chi_{\alpha^{g_1,g_2}}(g) = |\alpha^{g_1,g_2}| [\alpha^{g_1,g_2}(g)]_{11}.$ Denoting $[\alpha^{g_1,g_2}(g)]_{11}$ by $\alpha_R(g)$ (and similarly denoting $[\gamma^{g_1,g_2}(g)]_{11}$ by $\gamma_R(g)$), we obtain
\begin{equation}
	\frac{1}{|G|}\sum_{g \in G_{g_1, g_2}} |\alpha^{g_1,g_2}| |\gamma^{g_1,g_2}| \alpha_R(g)^* \gamma_R(g) = \delta(\alpha^{g_1,g_2}, \gamma^{g_1,g_2}). \label{Equation_projective_character_orthogonality_reduced}
\end{equation}

This looks just like the orthogonality relation for a $\beta_{g_1 g_2}$-projective irrep of $G_{g_1, g_2}$ (rather than of $G$). Indeed $\alpha_R$ and $\gamma_R$ must be such projective irreps because, for $x,y \in G_{g_1, g_2}$
\begin{align*}
	\alpha^{g_1,g_2}(x)\alpha^{g_1,g_2}(y) &= [x,y]_{g_1, g_2} \alpha^{g_1,g_2}(xy)\\
	& \implies\alpha_R(x) I \alpha_R(y) I= [x,y]_{g_1, g_2} \alpha_R(xy) I \\
	& \implies \alpha_R(x) \alpha_R(y) = [x,y]_{g_1, g_2} \alpha_R(xy),
\end{align*}
meaning that they are projective representations, and they are irreducible because they are 1d. These projective irreps of the subgroup must also satisfy their own orthogonality condition:
$$\frac{1}{|G_{g_1, g_2}|}\sum_{g \in G_{g_1, g_2}} \alpha_R(g)^* \gamma_R(g) = \delta(\alpha_R, \gamma_R).$$

Furthermore, the projective irreps $\alpha^{g_1,g_2}$ of the full group are in one-to-one correspondence with projective irreps $\alpha_R$ of the subgroup. To see this, suppose there were two distinct irreps $\alpha^{g_1,g_2}$ and $\gamma^{g_1,g_2}$ such that $\alpha_R= \gamma_R$. Then, from the orthogonality condition Equation \ref{Equation_projective_character_orthogonality_reduced} on $\alpha^{g_1,g_2}$ and $\gamma^{g_1,g_2}$, we obtain
$$\frac{1}{|G|}\sum_{g \in G_{g_1, g_2}} |\alpha^{g_1,g_2}| |\gamma^{g_1,g_2}| |\alpha_R(g)|^2= 0,$$
which cannot be true. Therefore, each $\alpha^{g_1,g_2}$ is associated to a unique $\alpha_R$. Furthermore, the number of projective irreps $\alpha^{g_1,g_2}$ is equal to the number of elements of $G_{g_1, g_2}$, as we showed when changing the basis for our measurement operators. But the number of projective irreps of the subgroup is also equal to the number of elements of $G_{g_1,g_2}$ by the same reasoning. Each $\alpha^{g_1,g_2}$ is associated to a unique $\alpha_R$ and the numbers of projective irreps of the subgroup and group are the same, so each irrep of the subgroup $G_{g_1, g_2}$ must be related to a projective irrep $\alpha^{g_1,g_2}$ of $G$ in this way.

Because of this one-to-one relation, we also have $\delta(\alpha^{g_1,g_2}, \gamma^{g_1,g_2})= \delta(\alpha_R, \gamma_R)$. Inserting this into Equation \ref{Equation_projective_character_orthogonality_reduced} gives us
$$\frac{1}{|G|}\sum_{g \in G_{g_1, g_2}} |\alpha^{g_1,g_2}| |\gamma^{g_1,g_2}| \alpha_R(g)^* \gamma_R(g) = \delta(\alpha_R, \gamma_R),$$
which for $\alpha^{g_1,g_2} = \gamma^{g_1,g_2}$ becomes
$$\frac{1}{|G|}\sum_{g \in G_{g_1, g_2}} |\alpha^{g_1,g_2}|^2 |\alpha_R(g)|^2=1$$
and because $\alpha_R$ is a phase this means that
$$\frac{1}{|G|}\sum_{g \in G_{g_1, g_2}} |\alpha^{g_1,g_2}|^2=\frac{|G_{g_1, g_2}|}{|G|}|\alpha^{g_1,g_2}|^2=1.$$
Therefore, 
\begin{equation}
	|\alpha^{g_1,g_2}| = \sqrt{\frac{|G|}{|G_{g_1, g_2}|}}. \label{Equation_size_proj_irrep}
\end{equation}
This tells us that the dimension of each $\beta_{g_1,g_2}$-projective irrep is the same for a given pair $g_1,g_2$ (for Abelian groups).

With this in mind, we can now show that the projectors given in Equation \ref{Equation_charge_projectors_2} are indeed orthogonal projectors. Because the measurement operators are topological, their algebra will not depend on local details and we can use a simple geometry to calculate it. To this end, we use the same geometry that we utilized for fusion in Section \ref{Section_fusion_membranes} in the main text. We apply two measurement operators on membrane $m_1$, which we will just call $m$ here. That is, we consider the product
\begin{align*}
	P^{g_1, g_2, \alpha^{g_1,g_2}}(m) &P^{g_1', g_2', \gamma^{g_1',g_2'}}(m)\\
	&= \frac{1}{\sqrt{|G||G_{g_1, g_2}|}} \sum_{h \in G_{g_1, g_2}} \chi_{\alpha^{g_1,g_2}}(h)T^{g_1, g_2,h}(m)\frac{1}{\sqrt{|G||G_{g_1', g_2'}|}} \sum_{k \in G_{g_1', g_2'}} \chi_{\gamma^{g_1',g_2'}}(k)T^{g_1', g_2', k}(m).
\end{align*}
In order to combine these into one operator, we consider the product of $T$ operators:
\begin{align*}
	T^{g_1, g_2, h}(m) T^{g_1', g_2', k}(m)&=C^h(m)\delta(\hat{g}(c_1), g_1) \delta(\hat{g}(c_2), g_2) C^k(m) \delta(\hat{g}(c_1),g_1') \delta(\hat{g}(c_2), g_2') \\
	&= \delta(g_1, g_1') \delta(g_2, g_2') C^h(m) C^k(m) \delta(\hat{g}(c_1), g_1) \delta(\hat{g}(c_2), g_2) .
\end{align*}
This expression is only non-zero when $g_1=g_1'$ and $g_2=g_2'$, so we can freely replace $g_1'$ with $g_1$ and $g_2'=g_2$ outside of these Kronecker delta expressions.

To combine the magnetic components of the measurement operators, we can use the algebra for the magnetic membrane operators we described in Section \ref{Section_fusion_membranes} in the main text. Using Equation \ref{Equation_fusion_intermediate_2} from that section, but substituting in the appropriate labels (and taking the surface weight to be 1), we have
\begin{align*}
	C^h(m) C^k(m) 	&= C_0^{hk}(m) [h,k]_{g_2, g_1} 	\theta^{hk}_B(m)\\
	&= [h,k]_{g_2, g_1} C^{hk}(m).
\end{align*}

This algebra for the magnetic membrane operators gives us
\begin{align}
	T^{g_1, g_2, h}(m) T^{g_1', g_2', k}(m)&=C^h(m)\delta(\hat{g}(c_1), g_1) \delta(\hat{g}(c_2), g_2) C^k(m) \delta(\hat{g}(c_1),g_1') \delta(\hat{g}(c_2), g_2') \notag\\
	&= \delta(g_1, g_1') \delta(g_2, g_2') [h,k]_{g_2, g_1} C^{hk}(m) \delta(\hat{g}(c_1), g_1) \delta(\hat{g}(c_2), g_2) \notag\\
	&= \delta(g_1, g_1') \delta(g_2, g_2') [h,k]_{g_2, g_1} T^{g_1, g_2, hk}(m) \label{Equation_charge_group_basis_algebra}.
\end{align}

Then the product of projectors can be written as
\begin{align*}
	P^{g_1, g_2, \alpha^{g_1,g_2}}(m) &P^{g_1', g_2', \gamma^{g_1',g_2'}}(m)\\
	&= \frac{1}{\sqrt{|G||G_{g_1, g_2}|}} \sum_{h \in G_{g_1, g_2}} \chi_{\alpha^{g_1,g_2}}(h)T^{g_1, g_2,h}(m)\frac{1}{\sqrt{|G||G_{g_1', g_2'}|}} \sum_{k \in G_{g_1', g_2'}} \chi_{\gamma^{g_1,g_2}}(k)T^{g_1', g_2', k}(m)\\
	&=\frac{1}{|G||G_{g_1, g_2}|} \sum_{h \in G_{g_1, g_2}} \sum_{k \in G_{g_1, g_2}} \chi_{\alpha^{g_1,g_2}}(h) \chi_{\gamma^{g_1,g_2}}(k)C^k(m) \delta(g_1, g_1') \delta(g_2, g_2') [h,k]_{g_1, g_2}^{-1} T^{g_1, g_2, hk}(m)\\
	&= \frac{|\alpha^{g_1,g_2}| |\gamma^{g_1,g_2}|} {|G||G_{g_1, g_2}|} \sum_{h \in G_{g_1, g_2}} \delta(g_1, g_1') \delta(g_2, g_2') \sum_{k \in G_{g_1, g_2}} [h,k]_{g_1, g_2}^{-1} \alpha_R(h) \gamma_R(k) T^{g_1, g_2, hk}(m) .
\end{align*}
Then, changing dummy variables from $h$ to $h'=hk$ (using the fact that $G_{g_1, g_2}$ is a subgroup), we find
\begin{align*}
	&P^{g_1, g_2, \alpha^{g_1,g_2}}(m) P^{g_1', g_2', \gamma^{g_1',g_2'}}(m)\\
	&= \frac{|\alpha^{g_1,g_2}| |\gamma^{g_1,g_2}|} {|G||G_{g_1, g_2}|} \delta(g_1, g_1') \delta(g_2, g_2') \sum_{h'=hk \in G_{g_1, g_2}} \sum_{k \in G_{g_1, g_2}} [h'k^{-1},k]_{g_1, g_2}^{-1} \alpha_R(h'k^{-1}) \gamma_R(k) T^{g_1, g_2, h'}(m) \\
	&= \frac{|\alpha^{g_1,g_2}| |\gamma^{g_1,g_2}|}{|G||G_{g_1, g_2}|} \delta(g_1, g_1') \delta(g_2, g_2') \sum_{h'=hk \in G_{g_1, g_2}} \sum_{k \in G_{g_1, g_2}} [h'k^{-1},k]_{g_1, g_2}^{-1} [h', k^{-1}]_{g_1, g_2}^{-1}\alpha_R(h') \alpha_R(k^{-1}) \gamma_R(k) T^{g_1, g_2, h'}(m).
\end{align*}

Next we use Equation \ref{Equation_projective_irrep_inverse} to write $\alpha_R(k^{-1}) = [k^{-1},k]_{g_1, g_2}\alpha_R(k)^{-1} = [k^{-1},k]_{g_1, g_2}\alpha_R(k)^*$ (note that $[k^{-1},k]_{g_1, g_2} = [k,k^{-1}]_{g_1, g_2}$ for any $k \in G$ from the 2-cocycle condition). Therefore,
\begin{align*}
	P^{g_1, g_2, \alpha^{g_1,g_2}}(m) P^{g_1', g_2', \gamma^{g_1',g_2'}}(m)	&= \frac{|\alpha^{g_1,g_2}| |\gamma^{g_1,g_2}|}{|G||G_{g_1, g_2}|} \delta(g_1, g_1') \delta(g_2, g_2') \sum_{h'=hk \in G_{g_1, g_2}} \sum_{k \in G_{g_1, g_2}} [h'k^{-1},k]_{g_1, g_2}^{-1} [h', k^{-1}]_{g_1, g_2}^{-1} \\
	& \hspace{0.5cm} \alpha_R(h') [k^{-1},k]_{g_1, g_2}\alpha_R(k)^{-1} \gamma_R(k) T^{g_1, g_2, h'}(m).
\end{align*}

Using the 2-cocycle condition (Equation \ref{Equation_2_cocycle_condition} in the main text), $$ [h'k^{-1},k]_{g_1, g_2}^{-1} [h', k^{-1}]_{g_1, g_2}^{-1}[k^{-1},k]_{g_1, g_2} = [h', k^{-1}k]_{g_1, g_2}^{-1}=1.$$ Then we obtain
\begin{align*}
	P^{g_1, g_2, \alpha^{g_1,g_2}}(m) P^{g_1', g_2', \gamma^{g_1',g_2'}}(m)	&= \frac{|\alpha^{g_1,g_2}| |\gamma^{g_1,g_2}|}{|G||G_{g_1, g_2}|} \delta(g_1, g_1') \delta(g_2, g_2') \sum_{h' \in G_{g_1, g_2}} \alpha_R(h') \sum_{k \in G_{g_1, g_2}} \alpha_R(k)^{-1} \gamma_R(k)T^{g_1, g_2, h'}(m)\\
	&= \frac{|\alpha^{g_1,g_2}| |\gamma^{g_1,g_2}|}{|G||G_{g_1, g_2}|} \delta(g_1, g_1') \delta(g_2, g_2') \sum_{h' \in G_{g_1, g_2}} \alpha_R(h') |G_{g_1, g_2}| \delta(\alpha^{g_1,g_2}, \gamma^{g_1,g_2}) T^{g_1, g_2, h'}(m),
\end{align*}
using the orthogonality condition for projective irreps of $G_{g_1, g_2}$. Therefore,
\begin{align*}
	P^{g_1, g_2, \alpha^{g_1,g_2}}(m) P^{g_1', g_2', \gamma^{g_1',g_2'}}(m)	&=\frac{|\alpha^{g_1,g_2}|^2 |G_{g_1, g_2}|}{|G||G_{g_1, g_2}|} \delta(g_1, g_1') \delta(g_2, g_2') \delta( \alpha^{g_1,g_2}, \gamma^{g_1,g_2}) \sum_{h' \in G_{g_1, g_2}} \alpha_R(h') T^{g_1, g_2, h'}(m)\\
	&=\frac{|\alpha^{g_1,g_2}| }{|G|} \delta(g_1, g_1') \delta(g_2, g_2') \delta( \alpha^{g_1,g_2}, \gamma^{g_1,g_2}) \sum_{h' \in G_{g_1, g_2}} \chi_{\alpha^{g_1,g_2}}(h') T^{g_1, g_2, h'}(m).
\end{align*}

Substituting $|\alpha^{g_1,g_2}| = \sqrt{\frac{|G|}{|G_{g_1, g_2}|}}$ this becomes
\begin{align*}
	P^{g_1, g_2, \alpha^{g_1,g_2}}(m) P^{g_1', g_2', \gamma^{g_1',g_2'}}(m)	&=\frac{1}{\sqrt{|G||G_{g_1, g_2}|}} \delta(g_1, g_1') \delta(g_2, g_2') \delta( \alpha^{g_1,g_2}, \gamma^{g_1,g_2}) \sum_{h' \in G_{g_1, g_2}} \chi_{\alpha^{g_1,g_2}}(h') T^{g_1, g_2, h'}(m)\\
	&= \delta(g_1, g_1') \delta(g_2, g_2') \delta( \alpha^{g_1,g_2}, \gamma^{g_1,g_2})	P^{g_1, g_2, \alpha^{g_1,g_2}}(m) .
\end{align*}
Therefore, the $P^{g_1, g_2, \alpha^{g_1,g_2}}(m) $ are indeed orthogonal projectors.

We also wish to show that the projectors are complete in the space of measurement operators. To see this, note that

\begin{align*}
	\sum_{g_1, g_2 \in G} &\sum_{\alpha^{g_1,g_2} \in A^{g_1,g_2}(G) }	P^{g_1, g_2, \alpha^{g_1,g_2}}(m) 	=\sum_{g_1, g_2 \in G} \sum_{\alpha^{g_1,g_2} \in A^{g_1,g_2}(G) } \frac{1}{\sqrt{|G||G_{g_1, g_2}|}} \sum_{h \in G_{g_1, g_2}} \chi_{\alpha^{g_1,g_2}}(h)T^{g_1, g_2, h}(m)\\
	&= \sum_{g_1, g_2 \in G} \frac{1}{\sqrt{|G||G_{g_1, g_2}|}} \sum_{h \in G_{g_1, g_2}} |\alpha^{g_1,g_2}| \big(\sum_{\alpha_R\in A^{g_1,g_2}(G_{g_1,g_2})} \alpha_R(h) \big)C^h(m) \delta(\hat{g}(c_1), g_1) \delta(\hat{g}(c_2),g_2)\\
	&= \sum_{g_1, g_2 \in G} \frac{1}{\sqrt{|G||G_{g_1, g_2}|}} \sum_{h \in G_{g_1, g_2}} \sqrt{\frac{|G|}{|G_{g_1, g_2}|}} \big(\sum_{\alpha_R\in A^{g_1,g_2}(G_{g_1,g_2})} \alpha_R(h) \alpha_R(1_G) \big)C^h(m) \delta(\hat{g}(c_1), g_1) \delta(\hat{g}(c_2),g_2),
\end{align*}
where we inserted $\alpha_R(1_G)=1$. The projective irreps (here of $G_{g_1, g_2}$) satisfy the completeness relation \cite{Wan2015}
$$\frac{1}{|G_{g_1, g_2}|}\sum_{\alpha_R\in A^{g_1,g_2}(G_{g_1,g_2})} \alpha_R(h) \alpha_R(x) = \delta(h, x).$$
Therefore, we have
\begin{align*}
	\sum_{g_1, g_2 \in G}& \sum_{\alpha^{g_1,g_2} \in A^{g_1,g_2}(G) }	P^{g_1, g_2, \alpha^{g_1,g_2}}(m) \\
	&= \sum_{g_1, g_2 \in G} \frac{1}{\sqrt{|G||G_{g_1, g_2}|}} \sum_{h \in G_{g_1, g_2}} \sqrt{\frac{|G|}{|G_{g_1, g_2}|}} \big( |G_{g_1, g_2}|\delta(h, 1_G) \big)C^h(m) \delta(\hat{g}(c_1), g_1) \delta(\hat{g}(c_2),g_2)\\
	&= \sum_{g_1, g_2 \in G} \frac{1}{\sqrt{|G||G_{g_1, g_2}|}}\sqrt{\frac{|G|}{|G_{g_1, g_2}|}} |G_{g_1, g_2}|C^{1_G}(m) \delta(\hat{g}(c_1), g_1) \delta(\hat{g}(c_2),g_2)\\
	&=C^{1_G}(m) = I,
\end{align*}
so the projectors are complete.

Another useful result is that the one-to-one correspondence between projective irreps of $G$ and projective irreps of $G_{g_1, g_2}$ means that we can express the projectors in a different way as 
\begin{equation}
	P^{g_1, g_1, \alpha^{g_1,g_2}}(m) = P_R^{g_1, g_2, \alpha_R}(m) := \frac{1}{|G_{g_1, g_2}|} \sum_{h \in G_{g_1, g_2}} \alpha_R(h)C^h(m) \delta(\hat{g}(c_1), g_1) \delta(\hat{g}(c_2),g_2).
\end{equation}

\subsection{Example: Topological Charge of a linked loop}

\begin{figure}
	\centering
	\includegraphics[width=0.7\linewidth]{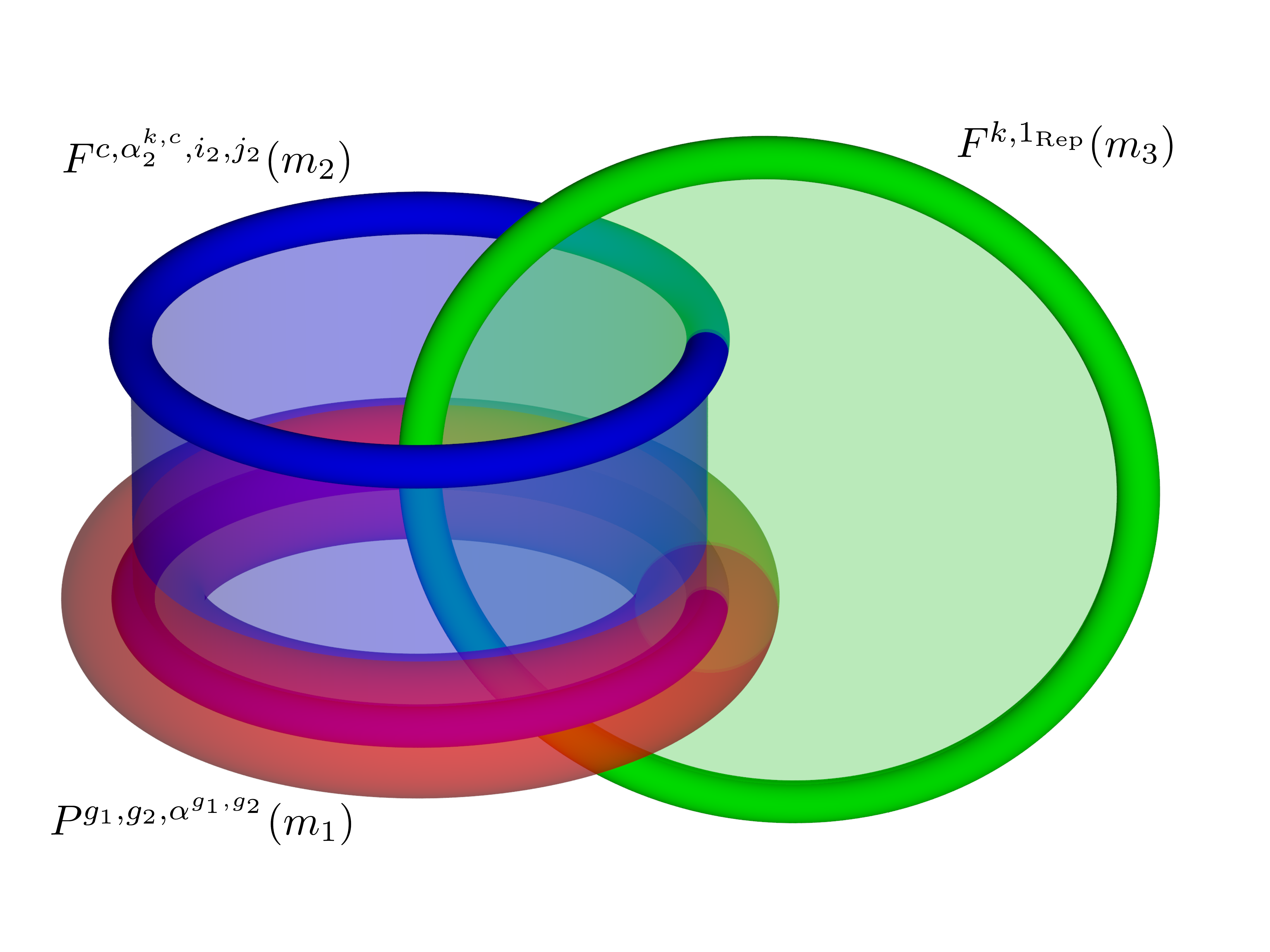}
	
	\caption{We consider applying a toroidal measurement operator (red) enclosing the loop excitation at one end of a membrane operator (blue) linked to a base loop (green).}
	\label{Figure_torus_charge_measurement}
\end{figure}

As an example of topological charge measurement, we can consider the charge of a single loop-like excitation linked to a base loop. To do so, we consider the geometry shown in Figure \ref{Figure_torus_charge_measurement}. In this case, we consider the state where we have a base loop of label $k$ on the boundary of some membrane operator $F^{k,1_{\text{Rep}}}(m_3)$ and then a cylindrical membrane operator $F^{c, \alpha_2^{k,c},i_2,j_2}(m_2)$ linked with that excitation. This gives us a state 
\begin{equation}
	\ket{\psi_{\text{initial}}}= F^{c, \alpha_2^{k,c},i_2,j_2}(m_2)F^{k,1_{\text{Rep}}}(m_3) \ket{GS} . 
\end{equation}
Then we apply a toroidal measurement operator, $P^{g_1, g_2, \alpha^{g_1,g_2}}(m_1)$, which encloses the loop-like excitation at one end of the cylinder. If this excitation has well-defined charge, the topological measurement operator will either give us zero (in which case the excitation does not have the charge labels described by the measurement operator) or return the original state (in which case the excitation does possess the charge described by the measurement operator).

\begin{figure}
	\centering
	\includegraphics[width=0.6\linewidth]{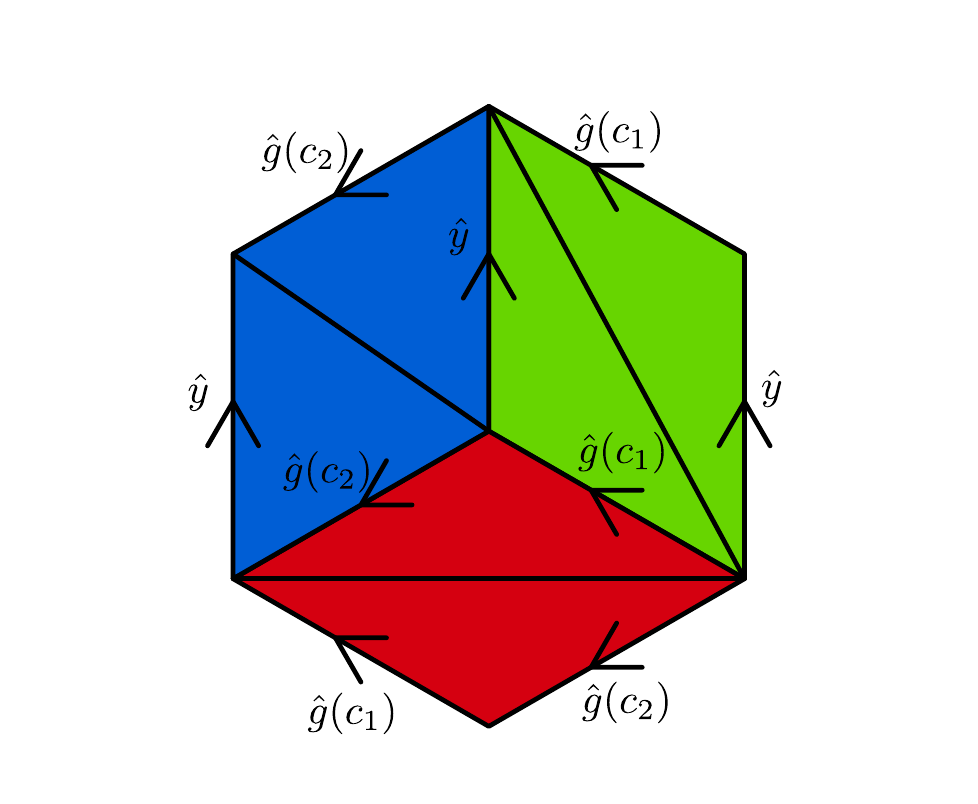}
	\caption{Here we give the simplest possible membranes that realize the geometry shown in Figure \ref{Figure_torus_charge_measurement}. The red, blue and green surfaces are the direct membranes for the corresponding membrane operators shown in Figure \ref{Figure_torus_charge_measurement} (or in the case of the base loop, a membrane operator that would produce it). There are periodic boundary conditions in the $c_1$ and $c_2$ directions, so these are closed loops. Note that these boundary conditions, together with flatness on the membranes themselves, imply that the parallel edges in the figure all have the same label. The dual membranes of the three operators all intersect in the cube-shaped (or thickened torus after boundary conditions) region bounded on three faces by these direct membranes.}
	\label{Figure_measurement_direct_membranes}
\end{figure}

After applying the measurement operator, the state is

\begin{align*}
	\ket{\psi_{\text{result}}} &= P^{g_1, g_2, \alpha^{g_1,g_2}}(m_1) F^{c, \alpha_2^{k,c},i_2,j_2}(m_2)F^{k,1_{\text{Rep}}}(m_3) \ket{GS} \\
	&= \frac{1}{\sqrt{|G| |G_{g_1,g_2}|}} \sum_{h \in G_{g_1,g_2}} \chi_{\alpha^{g_1,g_2}}(h) C_0^h(m_1) \theta_D^h(m_1) \delta(\hat{g}(c_1),g_1) \delta(\hat{g}(c_2),g_2) C_0^c(m_2) \theta_D^c(m_2) [\alpha_2^{k,c}(\hat{y})]_{i_2,j_2} \\
	& \hspace{0.5cm}C_0^k(m_3) \theta_D^k(m_3) \ket{GS}.
\end{align*}

Firstly, we can commute $\delta(\hat{g}(c_1), g_1)$ and $\delta(\hat{g}(c_2), g_2)$ from the measurement operator to act directly on the ground state. To do this, note that 

$$\hat{g}(c_1) C_0^c(m_2) = C_0^c(m_2) \hat{g}(c_1)c^{-1}$$

and

$$\hat{g}(c_2)C_0^k(m_3)=C_0^k(m_3) k \hat{g}(c_2).$$

This means that

\begin{align*}
	\ket{\psi_{\text{result}}} &= \frac{1}{\sqrt{|G| |G_{g_1,g_2}|}} \sum_{h \in G_{g_1,g_2}} \chi_{\alpha^{g_1,g_2}}(h) C_0^h(m_1) \theta_D^h(m_1) C_0^c(m_2) \theta_D^c(m_2) [\alpha_2^{k,c}(\hat{y})]_{i_2,j_2} C_0^k(m_3) \theta_D^k(m_3)\\
	& \hspace{1cm} \delta(\hat{g}(c_1)c^{-1},g_1) \delta(k\hat{g}(c_2),g_2)\ket{GS}.
\end{align*}

In the ground state $\hat{g}(c_1)=\hat{g}(c_2)=1_G$ because $c_1$ and $c_2$ are contractible loops that satisfy flatness in the ground state. Therefore

$$\delta(\hat{g}(c_1)c^{-1},g_1) \delta(k\hat{g}(c_2),g_2)\ket{GS}= \delta(c^{-1},g_1) \delta(k,g_2)\ket{GS}.$$

From this we see that the two flux labels of the topological charge carried by the excitation are $k$ (corresponding to the base loop) and $c^{-1}$ corresponding to the excitation itself as we may expect. Note that while the flux label of this excitation is $c^{-1}$, the label of the excitation at the other end of the cylinder would be $c$. 

Now that $\delta(c^{-1},g_1) \delta(k,g_2)$ are numbers rather than operators we can bring them to the front of the state, so

\begin{align}
	\ket{\psi_{\text{result}}} &= \frac{\delta(c^{-1},g_1) \delta(k,g_2)}{\sqrt{|G| |G_{g_1,g_2}|}} \sum_{h \in G_{g_1,g_2}} \chi_{\alpha^{g_1,g_2}}(h) C_0^h(m_1) \theta_D^h(m_1) C_0^c(m_2) \theta_D^c(m_2) [\alpha_2^{k,c}(\hat{y})]_{i_2,j_2} C_0^k(m_3) \theta_D^k(m_3) \ket{GS}.
\end{align}

Next, we want to move the magnetic membrane operator from the measurement operator to the right to act directly on the ground state. Firstly we move the dual phase $\theta_D^h(m_1)$ to obtain

\begin{align}
	\ket{\psi_{\text{result}}} &= \frac{\delta(c^{-1},g_1) \delta(k,g_2)}{\sqrt{|G| |G_{g_1,g_2}|}} \sum_{h \in G_{g_1,g_2}} \chi_{\alpha^{g_1,g_2}}(h) C_0^h(m_1) C_0^c(m_2) \theta_D^c(m_2) ( C_0^c(m_2):\theta_D^h(m_1) ) \notag \\
	& \hspace{1cm} [\alpha_2^{k,c}(\hat{y})]_{i_2,j_2} C_0^k(m_3) \theta_D^k(m_3) \ket{GS}. \label{Equation_measurement_result_intermediate_1}
\end{align}

For this example, we use the same geometry that we did for the braiding relation (see Figures \ref{Figure_braiding_direct_membranes}, \ref{Figure_braiding_upper_wedge} and \ref{Figure_braiding_lower_wedge} in the main text) with $m_1$ and $m_2$ being the same as in that case. However we apply an additional periodic boundary condition along $x$ compared to the braiding case. Because of this, we can use the results from our previous calculation (see Equation \ref{Equation_braiding_dual_phase_ratio} in the main text) to write

\begin{align*}
	\theta_D^c(m_2) ( C_0^c(m_2):&\theta_D^h(m_1) ) = (\theta_D^{\text{ratio}})^{-1} \theta_D^h(m_1) (C_0^h(m_1): \theta_D^C(m_2))\\
	&= \frac{[\hat{g}(c_1), \hat{y}]_{k,h}}{[\hat{g}(c_1)c^{-1}, \hat{y}]_{k,h} [\hat{g}(c_1)\hat{y}c^{-1},c]_{k,h}} \frac{[\hat{g}(c_1)c^{-1}, h\hat{y}]_{k,c}}{[\hat{g}(c_1)c^{-1}, \hat{y}]_{k,c}[\hat{g}(c_1)\hat{y}c^{-1},h]_{k,c}} \theta_D^h(m_1) (C_0^h(m_1): \theta_D^C(m_2)).
\end{align*}

Here we can replace $\hat{g}(c_1)$ with $1_G$, because it is $1_G$ in the ground state and $C_0^k(m_3) $ does not affect its value. Therefore,
\begin{align*}
	\theta_D^c(m_2) ( C_0^c(m_2):\theta_D^h(m_1) ) &=\frac{1}{[c^{-1}, \hat{y}]_{k,h} [\hat{y}c^{-1},c]_{k,h}} \frac{[c^{-1}, h\hat{y}]_{k,c}}{[c^{-1}, \hat{y}]_{k,c}[\hat{y}c^{-1},h]_{k,c}} \theta_D^h(m_1) (C_0^h(m_1): \theta_D^C(m_2)).
\end{align*}

Substituting this into Equation \ref{Equation_measurement_result_intermediate_1}, we obtain

\begin{align*}
	\ket{\psi_{\text{result}}} &= \frac{\delta(c^{-1},g_1) \delta(k,g_2)}{\sqrt{|G| |G_{g_1,g_2}|}} \sum_{h \in G_{g_1,g_2}} \chi_{\alpha^{g_1,g_2}}(h) C_0^h(m_1) C_0^c(m_2) \frac{1}{[c^{-1}, \hat{y}]_{k,h} [\hat{y}c^{-1},c]_{k,h}} \frac{[c^{-1}, h\hat{y}]_{k,c}}{[c^{-1}, \hat{y}]_{k,c}[\hat{y}c^{-1},h]_{k,c}} \\
	& \hspace{1cm} \theta_D^h(m_1) (C_0^h(m_1): \theta_D^C(m_2)) [\alpha_2^{k,c}(\hat{y})]_{i_2,j_2} C_0^k(m_3) \theta_D^k(m_3) \ket{GS}\\
	&= \frac{\delta(c^{-1},g_1) \delta(k,g_2)}{\sqrt{|G| |G_{g_1,g_2}|}} \sum_{h \in G_{g_1,g_2}} \chi_{\alpha^{g_1,g_2}}(h)C_0^c(m_2) \theta_D^C(m_2) C_0^h(m_1) \frac{1}{[c^{-1}, \hat{y}]_{k,h} [\hat{y}c^{-1},c]_{k,h}} \frac{[c^{-1}, h\hat{y}]_{k,c}}{[c^{-1}, \hat{y}]_{k,c}[\hat{y}c^{-1},h]_{k,c}} \\
	& \hspace{1cm} \theta_D^h(m_1) [\alpha_2^{k,c}(\hat{y})]_{i_2,j_2} C_0^k(m_3) \theta_D^k(m_3) \ket{GS},
\end{align*}

where we used the fact that the untwisted membrane operators $C_0^h(m_1)$ and $C_0^{c}(m_2)$ commute for Abelian $G$. Next we aim to move the surface weight for $m_2$ left past the measurement operator. We have 
$$C_0^h(m_1)[\alpha_2^{k,c}(\hat{y})]_{i_2,j_2} = [\alpha_2^{k,c}(h^{-1}\hat{y})]_{i_2,j_2} C_0^h(m_1).$$
However, as we want to collect the additional quantities from this commutation with the other cocycle factors, it is more convenient to separate the factor of $h^{-1}$ from the irrep before we perform the commutation. To do this, we write
$$[\alpha_2^{k,c}(\hat{y})]_{i_2,j_2}=[\alpha_2^{k,c}(h^{-1} h\hat{y})]_{i_2,j_2}.$$
Then, using Equation \ref{Equation_projective_irrep_basis_matrix} from the main text, which defines the projective irrep composition rule, we obtain
$$[\alpha_2^{k,c}(h^{-1} h\hat{y})]_{i_2,j_2} = [h^{-1},h\hat{y}]^{-1}_{k,c} \sum_{l_2=1}^{|\alpha_2^{k,c}|} [\alpha_2^{k,c}(h^{-1})]_{i_2, l_2} [\alpha_2^{k,c}(h\hat{y})]_{l_2,j_2}. $$
We can then commute $[\alpha_2^{k,c}(h\hat{y})]_{l_2,j_2}$ past $C_0^h(m_1)$ using 
$$C_0^h(m_1)[\alpha_2^{k,c}(h\hat{y})]_{l_2,j_2} = [\alpha_2^{k,c}(\hat{y})]_{l_2,j_2}C_0^h(m_1),$$
while leaving the other factors in place. This gives us
\begin{align*}
	\ket{\psi_{\text{result}}} &= \frac{\delta(c^{-1},g_1) \delta(k,g_2)}{\sqrt{|G| |G_{g_1,g_2}|}} \sum_{h \in G_{g_1,g_2}} \chi_{\alpha^{g_1,g_2}}(h)\sum_{l_2=1}^{|\alpha_2^{k,c}|} C_0^c(m_2) \theta_D^C(m_2) [\alpha_2^{k,c}(\hat{y})]_{l_2,j_2} C_0^h(m_1) \frac{1}{[c^{-1}, \hat{y}]_{k,h} [\hat{y}c^{-1},c]_{k,h}} \\
	& \hspace{1cm} \frac{[c^{-1}, h\hat{y}]_{k,c}}{[c^{-1}, \hat{y}]_{k,c}[\hat{y}c^{-1},h]_{k,c}} \frac{[\alpha_2^{k,c}(h^{-1})]_{i_2,l_2}}{[h^{-1},h\hat{y}]_{k,c}} \theta_D^h(m_1) C_0^k(m_3) \theta_D^k(m_3) \ket{GS}.
\end{align*}

At this point, we note that $ C_0^c(m_2) \theta_D^C(m_2) [\alpha_2^{k,c}(\hat{y})]_{l_2,j_2} $ is the complete membrane operator $F^{c, \alpha_2^{k,c},l_2, j_2}(m_2)$. We can further simplify the equation by applying the 2-cocycle condition (Equation \ref{Equation_2_cocycle_condition} in the main text) to obtain

$$\frac{[c^{-1}, \hat{y}h]_{k,c}}{[c^{-1},\hat{y}]_{k,c} [c^{-1}\hat{y},h]_{k,c}}= \frac{1}{[\hat{y},h]_{k,c}}.$$
Therefore the measurement result becomes
\begin{align*}
	\ket{\psi_{\text{result}}} &= \frac{\delta(c^{-1},g_1) \delta(k,g_2)}{\sqrt{|G| |G_{g_1,g_2}|}} \sum_{h \in G_{g_1,g_2}} \chi_{\alpha^{g_1,g_2}}(h)\sum_{l_2=1}^{|\alpha_2^{k,c}|} F^{c, \alpha_2^{k,c},l_2, j_2}(m_2)C_0^h(m_1) \frac{1}{[c^{-1}, \hat{y}]_{k,h} [\hat{y}c^{-1},c]_{k,h}} \frac{1}{[\hat{y},h]_{k,c}} \\
	& \hspace{1cm} \frac{[\alpha_2^{k,c}(h^{-1})]_{i_2,l_2}}{[h^{-1},h\hat{y}]_{k,c}} \theta_D^h(m_1) C_0^k(m_3) \theta_D^k(m_3) \ket{GS}.
\end{align*}

We can use the 2-cocycle condition to further simplify the remaining cocycles, which are
$$\frac{1}{[c^{-1}, \hat{y}]_{k,h} [\hat{y}c^{-1},c]_{k,h}} \frac{1}{[\hat{y},h]_{k,c} [h^{-1},h\hat{y}]_{k,c}}.$$
Firstly, 
$$[h^{-1},h\hat{y}]_{k,c}= \frac{[h, \hat{y}]_{k,c}}{[h^{-1},h]_{k,c}}$$
using the cocycle condition and normalization condition. Then, using Equation \ref{Equation_1_cocycle_permutation_2} from the braiding calculation in the main text,
$$\frac{[h,\hat{y}]_{k,c}}{[\hat{y},h]_{k,c}}= \frac{[\hat{y},c]_{k,h}}{[c, \hat{y}]_{k,h}} = \frac{[c^{-1},\hat{y}]_{k,h}}{[\hat{y}, c^{-1}]_{k,h}}. $$
This means that the remaining cocycles are
$$\frac{[c^{-1},\hat{y}]_{k,h}}{[c^{-1}, \hat{y}]_{k,h} [\hat{y}c^{-1},c]_{k,h} [\hat{y}, c^{-1}]_{k,h}} \frac{1}{[h^{-1},h]_{k,c}}= \frac{1}{ [\hat{y}c^{-1},c]_{k,h} [\hat{y}, c^{-1}]_{k,h}} \frac{1}{[h^{-1},h]_{k,c}}.$$

Applying the 2-cocycle and normalization conditions again
$$\frac{1}{[\hat{y}c^{-1},c]_{k,h} [\hat{y}, c^{-1}]_{k,h}} =\frac{1}{[c^{-1},c]_{k,h}},$$
which just leaves us with the cocycles 
$$\frac{1}{[h^{-1},h]_{k,c}}\frac{1}{[c^{-1},c]_{k,h}}.$$
Note that these cocycles only depend on the labels $k$, $h$ and $c$ of the membrane operators and not on any operators like $\hat{y}$. Furthermore, we can remove the cocycle $\frac{1}{[h^{-1},h]_{k,c}}$ using the projective matrix representation $[\alpha_2^{k,c}(h^{-1})]_{i_2,l_2}$. From Equation \ref{Equation_proj_inverse} in the main text, we know that
$$\frac{\alpha_2^{k,c}(h^{-1}) }{[h^{-1},h]_{k,c}}= \alpha_2^{k,c}(h)^{-1},$$
where the inverse on the latter expression is on the matrix not the group element. Using these simplifications, the state resulting from the charge measurement is given by
\begin{align*}
	&\ket{\psi_{\text{result}}} \\
	&= \frac{\delta(c^{-1},g_1) \delta(k,g_2)}{\sqrt{|G| |G_{g_1,g_2}|}} \sum_{h \in G_{g_1,g_2}} \chi_{\alpha^{g_1,g_2}}(h)\sum_{l_2=1}^{|\alpha_2^{k,c}|} \frac{ [\alpha_2^{k,c}(h)^{-1}]_{i_2,l_2}}{[c^{-1},c]_{k,h}} F^{c, \alpha_2^{k,c},l_2, j_2}(m_2)C_0^h(m_1) \theta_D^h(m_1) C_0^k(m_3) \theta_D^k(m_3) \ket{GS}.
\end{align*}

\begin{figure}
	\centering
	\includegraphics[width=0.8\linewidth]{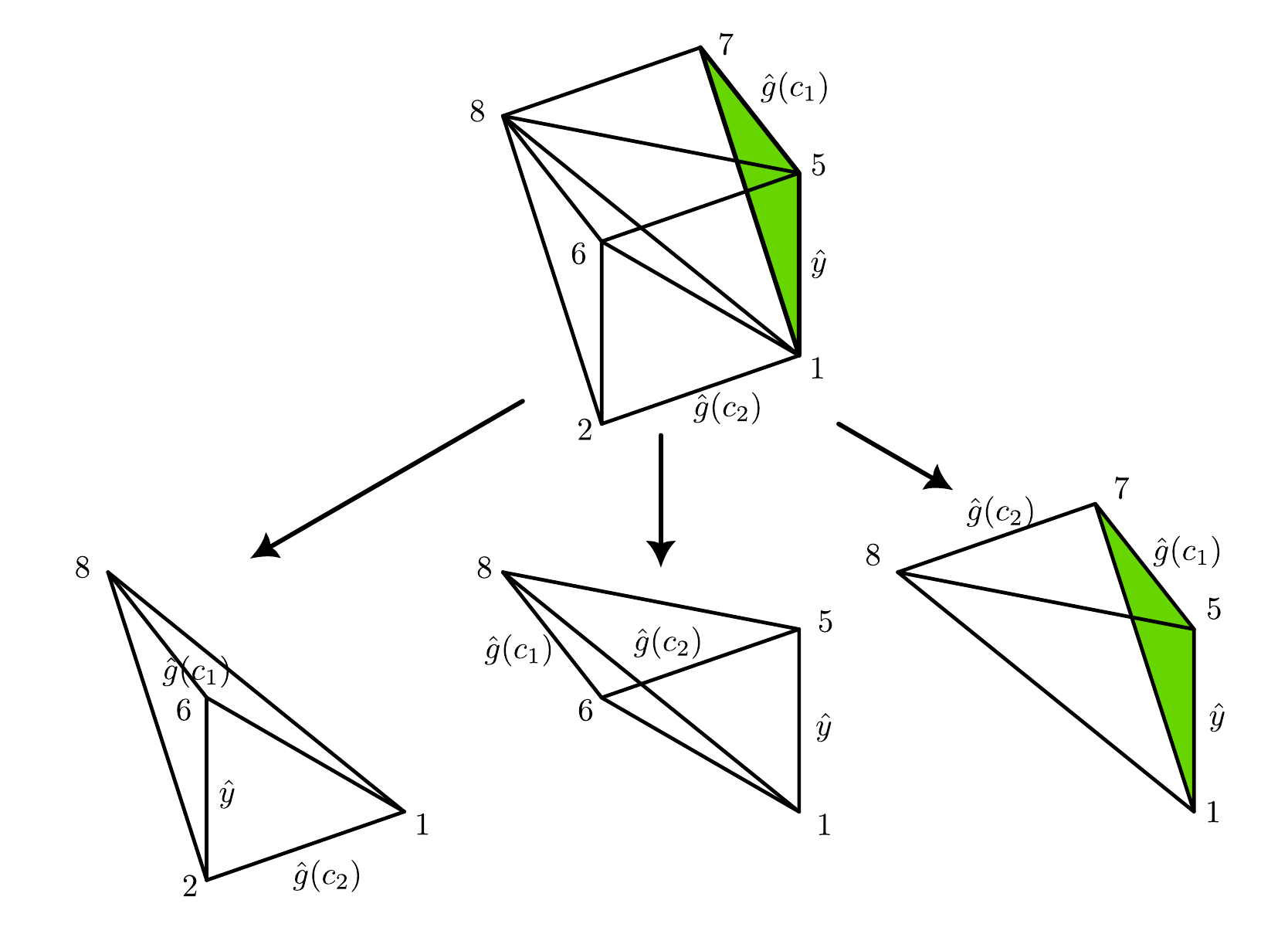}
	\caption{The intersection region for the three membranes can be divided into two wedges, of which this is the upper one.}
	\label{Figure_measurement_upper_wedge}
\end{figure}

\begin{figure}
	\centering
	\includegraphics[width=0.8\linewidth]{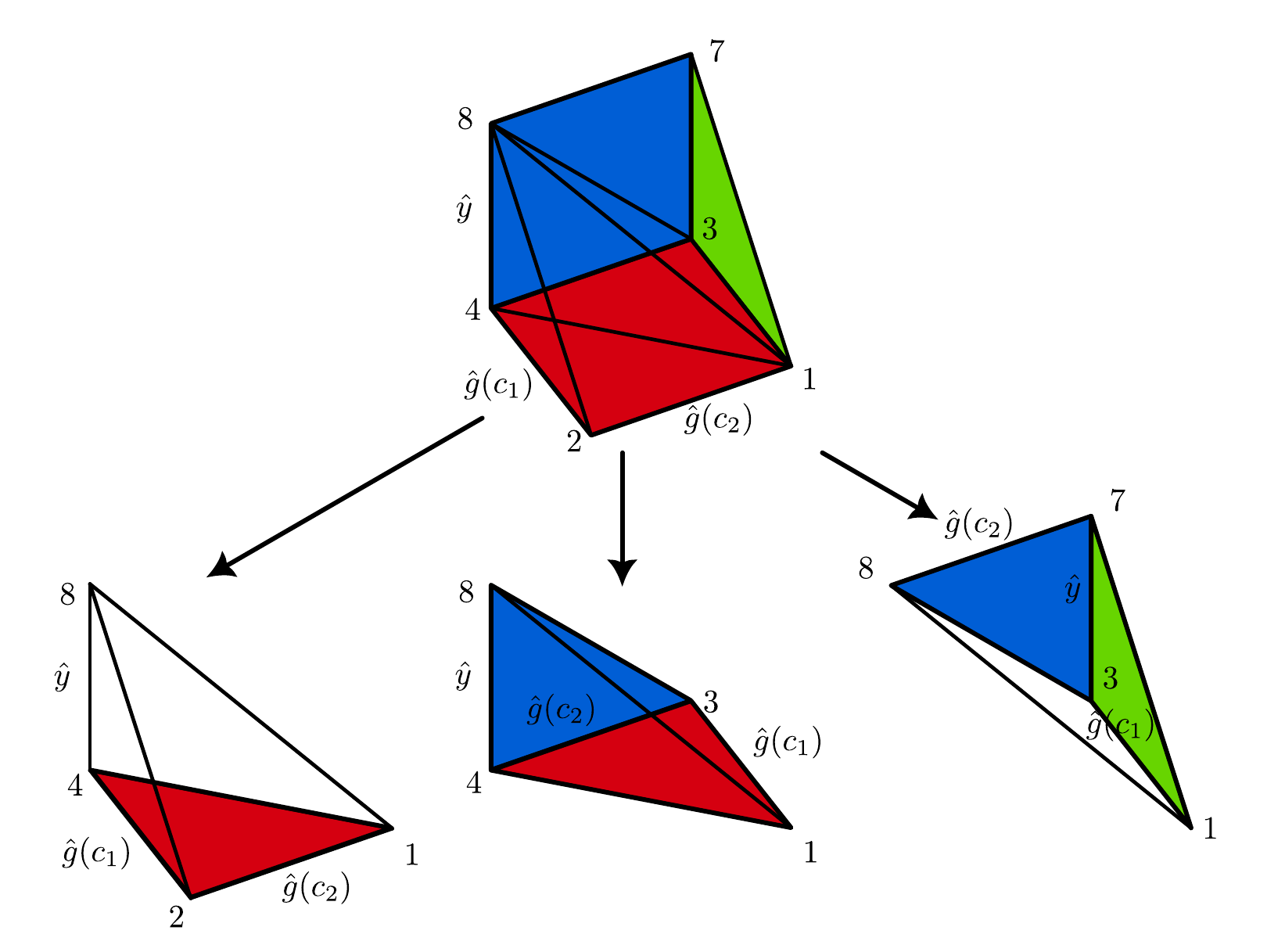}
	\caption{This is the lower wedge of the intersection region.}
	\label{Figure_measurement_lower_wedge}
\end{figure}

Next, we wish to commute the remaining operators on $m_1$ past the membrane operators on $m_3$ which produce the base loop. To do so, we must first obtain expressions for the dual phases on the geometry given in Figure \ref{Figure_measurement_direct_membranes}. Note that the periodic boundary conditions in the two horizontal directions, along with flatness in the cube region itself, means that the parallel edges have equal labels. As usual, we split the intersection region into two wedges, shown in Figures \ref{Figure_measurement_upper_wedge} and \ref{Figure_measurement_lower_wedge}. Then, using the formulae given in Table \ref{Table_Dual_Phase} for the contributions to the dual phase for each tetrahedron, the contribution to $\theta^h_D(m_1)$ from the upper wedge is
$$\theta_D^h(m_1; \text{upper})=[h, \hat{g}(c_2), \hat{y}, \hat{g}(c_1)] [\hat{g}(c_2),h, \hat{y}, \hat{g}(c_1)]^{-1} [h,\hat{y}, \hat{g}(c_2), \hat{g}(c_1)]^{-1} [h, \hat{y}, \hat{g}(c_1), \hat{g}(c_2)]= [h, \hat{y}, \hat{g}(c_1)]_{\hat{g}(c_2)}.$$
However, regardless of whether this dual phase acts before or after $C_0^h(m_1) $ and $C_0^k(m_3)$, $\hat{g}(c_1)=1_G$ in this state (as neither operator affects the label and it is $1_G$ in the ground state from flatness). Therefore the contribution to $\theta_D^h(m_1) $ from the upper wedge is just equal to the identity. Similarly, the contribution to the dual phase $\theta_D^k(m_3)$ (where the direct membrane of $m_3$ has vertices 1, 3, 5 and 7 on it) from the upper wedge is just
\begin{align*}
	\theta_D^k(m_3; \text{upper})&= [k, \hat{g}(c_2), \hat{y}, \hat{g}(c_1)] [k, \hat{y}, \hat{g}(c_2), \hat{g}(c_1)]^{-1} [\hat{y},k,\hat{g}(c_2), \hat{g}(c_1)][k, \hat{y}, \hat{g}(c_1), \hat{g}(c_2)] \\
	& \hspace{1cm}[\hat{y}, k, \hat{g}(c_1), \hat{g}(c_2)]^{-1} [\hat{y}, \hat{g}(c_1),k, \hat{g}(c_2)]\\
	&=1,
\end{align*}
where the latter equality holds because $\hat{g}(c_1)=1_G$ in the relevant states.

Then the contributions from the lower wedge are also equal to the identity, because $\hat{g}(c_1)=1_G$. For $m_1$ we have
\begin{align*}
	\theta^h_D(m_1; \text{lower}) &= [h, \hat{g}(c_2), \hat{g}(c_1),\hat{y}]^{-1} [\hat{g}(c_2), h, \hat{g}(c_1), \hat{y}] [\hat{g}(c_2), \hat{g}(c_1),h, \hat{y}]^{-1} [h, \hat{g}(c_1), \hat{g}(c_2), \hat{y}]\\
	& \hspace{1cm} [\hat{g}(c_1),h,\hat{g}(c_2), \hat{y}]^{-1} [\hat{g}(c_1), \hat{g}(c_2),h, \hat{y}] [h, \hat{g}(c_1), \hat{y}, \hat{g}(c_2)]^{-1} [\hat{g}(c_1),h, \hat{y}, \hat{g}(c_2) ]\\
	&=1
\end{align*}
and for $m_3$ we have
\begin{align*}
	\theta^k_D(m_3; \text{lower}) &= [k, \hat{g}(c_2), \hat{g}(c_1), \hat{y}]^{-1} [k, \hat{g}(c_1), \hat{g}(c_2),y] [\hat{g}(c_1), k, \hat{g}(c_2), \hat{y}]^{-1} [k, \hat{g}(c_1), \hat{y}, \hat{g}(c_2)]^{-1} \\
	& \hspace{1cm} [\hat{g}(c_1), k , \hat{y}, \hat{g}(c_2)] [\hat{g}(c_1), \hat{y}, k, \hat{g}(c_2)]^{-1}\\
	&=1.
\end{align*}

We see that the dual phase for both operators is just equal to the identity and so commutes with $C_0^h(m_1)$ and $C_0^k(m_3)$. $C_0^h(m_1)$ and $C_0^k(m_3)$ also commute with each other so we can just commute $C_0^h(m_1)\theta^h_D(m_1)$ past $C_0^k(m_3)\theta^k_D(m_3)$ to act directly on the ground state. Then the state after measurement can be written as
\begin{align*}
	&\ket{\psi_{\text{result}}} \\
	&= \frac{\delta(c^{-1},g_1) \delta(k,g_2)}{\sqrt{|G| |G_{g_1,g_2}|}} \sum_{h \in G_{g_1,g_2}} \chi_{\alpha^{g_1,g_2}}(h)\sum_{l_2=1}^{|\alpha_2^{k,c}|} \frac{ [\alpha_2^{k,c}(h)^{-1}]_{i_2,l_2}}{[c^{-1},c]_{k,h}} F^{c, \alpha_2^{k,c},l_2, j_2}(m_2) C_0^k(m_3) \theta_D^k(m_3) C_0^h(m_1) \theta_D^h(m_1) \ket{GS}.
\end{align*}

Now $C_0^h(m_1) \theta_D^h(m_1)$ is a closed membrane operator acting directly on the ground state so it can be contracted to nothing: $C_0^h(m_1) \theta_D^h(m_1)\ket{GS}= \ket{GS}$. Note that $C_0^h(m_1) \theta_D^h(m_1)$ is missing the surface weight, but this would be 1 anyway due to the surface diagram being trivial (all edges are $1_G$ when acting on the ground state for the simple geometry in this case). This leaves us with

\begin{align*}
	\ket{\psi_{\text{result}}} &= \frac{\delta(c^{-1},g_1) \delta(k,g_2)}{\sqrt{|G| |G_{g_1,g_2}|}} \sum_{h \in G_{g_1,g_2}} \chi_{\alpha^{g_1,g_2}}(h)\sum_{l_2=1}^{|\alpha_2^{k,c}|} \frac{ [\alpha_2^{k,c}(h)^{-1}]_{i_2,l_2}}{[c^{-1},c]_{k,h}} F^{c, \alpha_2^{k,c},l_2, j_2}(m_2) C_0^k(m_3) \theta_D^k(m_3)\ket{GS}.
\end{align*}

At this point, the label $h$ only appears in numbers like $[\alpha_2^{k,c}(h)^{-1}]_{i_2,l_2}$, not in operators. This means that we can evaluate the sum over $h$, which we do using the orthogonality of projective irreps. However there are some subtleties in doing this. $\alpha^{g_1,g_2}$ is a $\beta_{g_1,g_2}$-projective irrep. With the Kronecker delta $\delta(c^{-1},g_1) \delta(k,g_2)$, this means that $\alpha^{g_1,g_2}=\alpha^{c^{-1},k}$ is a $\beta_{c^{-1},k}$-projective irrep for any contributing (non-zero) terms. On the other hand, $\alpha_2^{k,c}(h)$ is a $\beta_{k,c}$-projective irrep. Because these are different types of irrep, we cannot immediately apply orthogonality. However, we claim that
\begin{equation}
	\gamma^{k,c}(x)= \frac{1}{[c^{-1},c]_{k,x}} \alpha^{c^{-1},k}(x) \label{Equation_irrep_charge_transformation}
\end{equation}
defines a $\beta_{k,c}(h)$-projective irrep. To see this note that 
\begin{align*}
	\gamma^{k,c}(x) \gamma^{k,c}(y) &= \frac{1}{[c^{-1},c]_{k,x}} \alpha^{c^{-1},k}(x) \frac{1}{[c^{-1},c]_{k,y}} \alpha^{c^{-1},k}(y)\\
	&= \frac{1}{[c^{-1},c]_{k,x} [c^{-1},c]_{k,y} } \alpha^{c^{-1},k}(x) \alpha^{c^{-1},k}(y)\\
	&=\frac{1}{[c^{-1},c]_{k,x} [c^{-1},c]_{k,y} } [x,y]_{c^{-1},k} \alpha^{c^{-1},k}(xy).
\end{align*}
We then insert the identity in the form $$\frac{[c^{-1},c]_{k,xy}}{[c^{-1},c]_{k, xy}} \frac{[x,y]_{k,c}}{[x,y]_{k,c}}$$
to obtain
\begin{align*}
	\gamma^{k,c}(x) \gamma^{k,c}(y) &= \frac{1}{[c^{-1},c]_{k,x} [c^{-1},c]_{k,y} } [x,y]_{c^{-1},k}\frac{[c^{-1},c]_{k,xy}}{[c^{-1},c]_{k, xy}} \frac{[x,y]_{k,c}}{[x,y]_{k,c}} \alpha^{c^{-1},k}(xy)\\
	&=\frac{[c^{-1},c]_{k,xy} [x,y]_{c^{-1},k} }{[c^{-1},c]_{k,x} [c^{-1},c]_{k,y} [x,y]_{k,c}} [x,y]_{k,c} \frac{1}{[c^{-1},c]_{k, xy}} \alpha^{c^{-1},k}(xy).
\end{align*}

We can then recognize $\frac{1}{[c^{-1},c]_{k, xy}} \alpha^{c^{-1},k}(xy)$ as $	\gamma^{k,c}(xy)$ and $[x,y]_{k,c} $ as the 2-cocycle that should appear if $\gamma^{k,c}$ is indeed a $\beta_{k,c}$-projective representation. The remaining part is
$$\frac{[c^{-1},c]_{k,xy} [x,y]_{c^{-1},k} }{[c^{-1},c]_{k,x} [c^{-1},c]_{k,y} [x,y]_{k,c}},$$
which we claim is equal to one. First note that $[x,y]_{c^{-1},k}=[x,y]_{k, c^{-1}}^{-1}$ from the permutation property (Equation \ref{Equation_2_cocycle_permutation} in the main text). Then 
$$\frac{[c^{-1},c]_{k,xy} }{[c^{-1},c]_{k,x} [c^{-1},c]_{k,y} [x,y]_{k,c} [x,y]_{k,c^{-1}} }$$
is the same expression as $\theta_R^{-1}$ from Equation \ref{Equation_fusion_remnant_cocycle} in the main text, for $a=c$ and $b=c^{-1}$, except that Equation \ref{Equation_fusion_remnant_cocycle} has the additional factor $[x,y]_{k, ba}$. However in this case $ba=1_G$ so this additional factor is just the identity and the two expressions are the same. We showed that the expression in Equation \ref{Equation_fusion_remnant_cocycle} is equal to the identity, so this expression is as well. Therefore we just have
\begin{align*}
	\gamma^{k,c}(x) \gamma^{k,c}(y) &= [x,y]_{k,c} \gamma^{k,c}(xy)
\end{align*}
and so $ \gamma^{k,c}(x)=\frac{1}{[c^{-1},c]_{k,x}} \alpha^{c^{-1},k}(x)$ is indeed a $\beta_{k,c}$-projective representation. Furthermore, if $\alpha^{c^{-1},k}$ is irreducible then so is $\gamma^{k,c}$ because the matrices in the two representations only differ by phases (although these phases are different for each matrix) which do not affect the ability to bring the matrices to block diagonal form. By introducing $\gamma^{k,c}$ we can write the state as
\begin{align*}
	\ket{\psi_{\text{result}}} &= \frac{\delta(c^{-1},g_1) \delta(k,g_2)}{\sqrt{|G| |G_{c^{-1},k}|}} \sum_{l_2=1}^{|\alpha_2^{k,c}|} \sum_{h \in G_{c^{-1},k}} \chi_{\gamma^{k,c}}(h) [\alpha_2^{k,c}(h)^{-1}]_{i_2,l_2} F^{c, \alpha_2^{k,c},l_2, j_2}(m_2) C_0^k(m_3) \theta_D^k(m_3)\ket{GS}.
\end{align*}

Note that the sum over $h$ is still over the subgroup $G_{c^{-1},k}$ rather than $G_{k,c}$. However, we claim that these two subgroups are the same. Recall that $G_{k,c}$ is defined as the subgroup containing the elements $h$ which satisfy $[h,x]_{k,c}=[x,h]_{k,c}$ for all $x \in G$. We already saw (see Section \ref{Section_topological_charge} of the main text) that $G_{c^{-1},k}=G_{k,c^{-1}}$. Therefore, we just need to show that $G_{k,c}=G_{k,c^{-1}}$. We have
\begin{align*}
	\frac{[h,x]_{k,c}}{[x,h]_{k,c}} &=\frac{[c,h]_{k,x}}{[h,c]_{k,x}}\\
	&=\frac{[h,c^{-1}]_{k,x}}{[c^{-1},h]_{k,x}},
\end{align*}
as we showed in Equation \ref{Equation_1_cocycle_permutation_2} in the main text. Then $\frac{[h,c^{-1}]_{k,x}}{[c^{-1},h]_{k,x}}=\frac{[x,h]_{k,c^{-1}}}{[h,x]_{k,c^{-1}}}$. Therefore if $[h,x]_{k,c}=[x,h]_{k,c}$ for all $x \in G$ then $[h,x]_{k,c^{-1}}=[x,h]_{k,c^{-1}}$ for all $x \in G$ (and vice-versa). This means that $G_{k,c}=G_{k,c^{-1}}=G_{c^{-1},k}$. This allows us to write 
\begin{align*}
	\ket{\psi_{\text{result}}} &= \frac{\delta(c^{-1},g_1) \delta(k,g_2)}{\sqrt{|G| |G_{k,c}|}} \sum_{l_2=1}^{|\alpha_2^{k,c}|} \sum_{h \in G_{k,c}} \chi_{\gamma^{k,c}}(h) [\alpha_2^{k,c}(h)^{-1}]_{i_2,l_2} F^{c, \alpha_2^{k,c},l_2, j_2}(m_2) C_0^k(m_3) \theta_D^k(m_3) \ket{GS}.
\end{align*}
Then the character $\chi_{\gamma^{k,c}}(h) $ is zero for $h$ outside of the subgroup $G_{k,c}$, so we can extend the sum to the entire group. Finally, we can use the unitarity of the projective irreps along with the orthogonality condition \cite{Melnikov2022}
\begin{equation}
	\sum_{g \in G} [\gamma^{k,c}(g)]_{ij} [\alpha_2^{k,c}(g)]_{kl}^* = \frac{|G|}{|\gamma|} \delta(\gamma^{k,c}, \alpha_2^{k,c}) \delta_{ik} \delta_{jl},
\end{equation}
where $|\gamma|$ is the dimension of the irrep $\gamma$, to obtain 
\begin{align*}
	\ket{\psi_{\text{result}}} &= \frac{\delta(c^{-1},g_1) \delta(k,g_2)}{\sqrt{|G| |G_{k,c}|}} \sum_{l_2=1}^{|\alpha_2^{k,c}|} \big( \sum_{h \in G} \sum_{q} [\gamma^{k,c}(h)]_{q,q} [\alpha_2^{k,c}(h)]_{l_2, i_2}^* \big)F^{c, \alpha_2^{k,c},l_2, j_2}(m_2) C_0^k(m_3) \theta_D^k(m_3) \ket{GS}\\
	&= \frac{\delta(c^{-1},g_1) \delta(k,g_2)}{\sqrt{|G| |G_{k,c}|}} \sum_{l_2=1}^{|\alpha_2^{k,c}|} \big(\sum_q \frac{|G|}{|\alpha_2^{k,c}|} \delta(\gamma^{k,c}, \alpha_2^{k,c}) \delta_{q, i_2} \delta_{q, l_2}\big)F^{c, \alpha_2^{k,c},l_2, j_2}(m_2) C_0^k(m_3) \theta_D^k(m_3) \ket{GS}\\
	&= \frac{\delta(c^{-1},g_1) \delta(k,g_2)}{\sqrt{|G| |G_{k,c}|}} \sum_{l_2=1}^{|\alpha_2^{k,c}|} \big( \frac{|G|}{|\alpha_2^{k,c}|} \delta(\gamma^{k,c}, \alpha_2^{k,c}) \delta_{i_2, l_2}\big)F^{c, \alpha_2^{k,c},l_2, j_2}(m_2) C_0^k(m_3) \theta_D^k(m_3)\ket{GS}\\
	&= \delta(c^{-1},g_1) \delta(k,g_2)\delta(\gamma^{k,c}, \alpha_2^{k,c}) \sqrt{\frac{|G|}{ |G_{k,c}|}} \frac{1}{|\alpha_2^{k,c}|} F^{c, \alpha_2^{k,c},i_2, j_2}(m_2) C_0^k(m_3) \theta_D^k(m_3) \ket{GS}.
\end{align*}

Now we recall that the dimension of the $\beta_{k,c}$-projective irreps is just given by $|\alpha_2^{k,c}|= \sqrt{\frac{|G|}{|G_{k,c}}}$ for an Abelian group (see Equation \ref{Equation_size_proj_irrep}) and so we are just left with
\begin{equation}
	\ket{\psi_{\text{result}}} = \delta(c^{-1},g_1) \delta(k,g_2)\delta(\gamma^{k,c}, \alpha_2^{k,c}) F^{c, \alpha_2^{k,c},i_2, j_2}(m_2) C_0^k(m_3) \theta_D^k(m_3) \ket{GS}.
\end{equation}
We can recognize $ F^{c, \alpha_2^{k,c},i_2, j_2}(m_2) C_0^k(m_3) \theta_D^k(m_3) \ket{GS}$ as the pre-measurement state $	\ket{\psi_{\text{initial}}}$. This means that the measurement operator $P^{g_1, g_2, \alpha^{g_1,g_2}}(m_1)$ gives back the original state for $g_1=c^{-1}$, $g_2=k$ and $\gamma^{g_2, g_1^{-1}}=\alpha_2^{k,c}$, where $\gamma^{g_2, g_1^{-1}}$ is related to $\alpha^{g_1,g_2}$ by Equation \ref{Equation_irrep_charge_transformation} (equivalently, $\alpha^{c^{-1},k}(x)=[c^{-1},c]_{k,x}\alpha_2^{k,c}(x)$). Otherwise, the measurement operator annihilates the state. This implies that the excitation carries the topological charge labeled by the triple $g_1$, $g_2$ and $\alpha^{g_1,g_2}$.

\end{document}